%% file: Planck_inflation_driver2014.tex
\documentclass[a4paper, traditabstract, longauth]{aa} 

\usepackage{fixltx2e}
\usepackage{amsmath}
\usepackage{graphicx}
\usepackage{color}
\usepackage[breaklinks, colorlinks, citecolor=blue]{hyperref}
\usepackage{natbib}
\usepackage{longtable,lscape}
\usepackage{txfonts}
\usepackage{amsfonts}
\usepackage{amssymb}
\usepackage{newlfont}
\usepackage{textcomp}
\usepackage{times}
\usepackage{units}
\usepackage{mathbbol}
\usepackage{amsbsy}
\usepackage{multirow}
\usepackage{xcolor}
\usepackage{tikz}

\defcitealias{planck2013-p11}{PCP13}
\defcitealias{planck2013-p08}{PPL13}
\defcitealias{planck2013-p17}{PCI13}
\defcitealias{pb2015}{BKP}

\input Planck.tex

\begin{document}

\title{\textit{Planck} 2015 results. XX. Constraints on inflation}

\input A24_Inflation_authors_and_institutes.tex

{\input macros.tex

\abstract
{
\input abstract.tex
}

}


\authorrunning{Planck Collaboration}
\titlerunning{Constraints on inflation}

\maketitle


\section{Introduction \label{sec:introduction}}

{
\input section_one.tex
}

\section{What new information does polarization provide? \label{sec:paradigm}}

{
\input section_two.tex
}

\section{Methodology \label{sec:methodology}}

{
\input section_three.tex
}

\section{Constraints on the primordial spectrum of curvature perturbations \label{sec:updates}}

{

\input section_four.tex
}

\section{Constraints on tensor modes \label{sec:modelcomp}}

{

\input section_five.tex

}

\section{Implications for single-field slow-roll inflation \label{sec:numerical}}

{

\input section_six.tex

}

\section{Reconstruction of the potential and analysis beyond slow-roll approximation \label{sec:pot}}

{

\input section_seven.tex
}

\section{$\mathcal{P}(k)$ reconstruction \label{sec:pk}}

{
\input section_eight_1.tex 
}

{

\input section_eight_2.tex
}

{
\input section_eight_3.tex
}

{
\input section_eight_4.tex
}

\section{Search for parameterized features \label{sec:oscillations}}

{

\input section_nine.tex

}

\section{Implications of {\it Planck} bispectral constraints on inflationary models \label{sec:putfnl}}

{
\input section_ten.tex
}

\section{Constraints on isocurvature modes \label{sec:iso}}

{
\input section_eleven.tex 
}

\section{Statistical anisotropy and inflation \label{sec:anisotropy}}

{
\input section_twelve.tex
}

\section{Combination with BICEP2/Keck Array-\Planck\ cross-correlation \label{sec:bicep2}}

{
\input section_thirteen.tex
}

\section{Conclusions \label{sec:conclusions}}

{
\input conclusions.tex

}

\begin{acknowledgements}
The Planck Collaboration acknowledges the support of: ESA; CNES and CNRS/INSU-IN2P3-INP (France); 
ASI, CNR, and INAF (Italy); NASA and DoE (USA); STFC and UKSA (UK); CSIC, MINECO, JA, and RES (Spain); 
Tekes, AoF, and CSC (Finland); DLR and MPG (Germany); CSA (Canada); DTU Space (Denmark); SER/SSO 
(Switzerland); RCN (Norway); SFI (Ireland); FCT/MCTES (Portugal); ERC and PRACE (EU). A description of 
the Planck Collaboration and a list of its members, indicating which technical or scientific activities they 
have been involved in, can be found at \href{http://www.cosmos.esa.int/web/planck/planck-collaboration}{http://www.cosmos.esa.int/web/planck/planck-collaboration}.
This research used resources of the National Energy Research Scientific Computing Center,
a DOE Office of Science User Facility supported by the Office of Science of the U.S.\ Department
of Energy under Contract No. DE-AC02-05CH11231.
Part of this work was undertaken at the STFC DiRAC HPC Facilities at the University of Cambridge, funded by UK BIS National E-infrastructure capital grants.
We gratefully acknowledge 
the IN2P3 Computer Center 
(\url{http://cc.in2p3.fr}) for providing a significant 
amount of the computing resources and services needed for this work.
\end{acknowledgements}

\begin{raggedright}
\bibliographystyle{aat}

\bibliography{Planck_bib,Inflation_references}

\end{raggedright}

\end{document}

%% file: Planck.tex
\def\setsymbol#1#2{\expandafter\def\csname #1\endcsname{#2}}
\def\getsymbol#1{\csname #1\endcsname}

\def\Planck{\textit{Planck}}

\newbox\tablebox    \newdimen\tablewidth
\def\leaderfil{\leaders\hbox to 5pt{\hss.\hss}\hfil}
\def\endPlancktable{\tablewidth=\columnwidth 
    $$\hss\copy\tablebox\hss$$
    \vskip-\lastskip\vskip -2pt}

\def\tablenote#1 #2\par{\begingroup \parindent=0.8em
    \abovedisplayshortskip=0pt\belowdisplayshortskip=0pt
    \noindent
    $$\hss\vbox{\hsize\tablewidth \hangindent=\parindent \hangafter=1 \noindent
    \hbox to \parindent{$^#1$\hss}\strut#2\strut\par}\hss$$
    \endgroup}
\def\doubleline{\vskip 3pt\hrule \vskip 1.5pt \hrule \vskip 5pt}

\def\L2{\ifmmode L_2\else $L_2$\fi}

\def\DeltaT{\ifmmode \Delta T\else $\Delta T$\fi}
\def\deltat{\ifmmode \Delta t\else $\Delta t$\fi}
\def\fknee{\ifmmode f_{\rm knee}\else $f_{\rm knee}$\fi}
\def\Fmax{\ifmmode F_{\rm max}\else $F_{\rm max}$\fi}
\def\solar{\ifmmode{\rm M}_{\mathord\odot}\else${\rm M}_{\mathord\odot}$\fi}
\def\Msolar{\ifmmode{\rm M}_{\mathord\odot}\else${\rm M}_{\mathord\odot}$\fi}
\def\Lsolar{\ifmmode{\rm L}_{\mathord\odot}\else${\rm L}_{\mathord\odot}$\fi}
\def\inv{\ifmmode^{-1}\else$^{-1}$\fi}
\def\mo{\ifmmode^{-1}\else$^{-1}$\fi}
\def\sup#1{\ifmmode ^{\rm #1}\else $^{\rm #1}$\fi}
\def\expo#1{\ifmmode \times 10^{#1}\else $\times 10^{#1}$\fi}
\def\,{\thinspace}
\def\lsim{\mathrel{\raise .4ex\hbox{\rlap{$<$}\lower 1.2ex\hbox{$\sim$}}}}
\def\gsim{\mathrel{\raise .4ex\hbox{\rlap{$>$}\lower 1.2ex\hbox{$\sim$}}}}

\def\simprop{\mathrel{\raise .4ex\hbox{\rlap{$\propto$}\lower 1.2ex\hbox{$\sim$}}}}
\def\deg{\ifmmode^\circ\else$^\circ$\fi}
\def\pdeg{\ifmmode $\setbox0=\hbox{$^{\circ}$}\rlap{\hskip.11\wd0 .}$^{\circ}
          \else \setbox0=\hbox{$^{\circ}$}\rlap{\hskip.11\wd0 .}$^{\circ}$\fi}
\def\arcs{\ifmmode {^{\scriptstyle\prime\prime}}
          \else $^{\scriptstyle\prime\prime}$\fi}
\def\arcm{\ifmmode {^{\scriptstyle\prime}}
          \else $^{\scriptstyle\prime}$\fi}
\newdimen\sa  \newdimen\sb
\def\parcs{\sa=.07em \sb=.03em
     \ifmmode \hbox{\rlap{.}}^{\scriptstyle\prime\kern -\sb\prime}\hbox{\kern -\sa}
     \else \rlap{.}$^{\scriptstyle\prime\kern -\sb\prime}$\kern -\sa\fi}
\def\parcm{\sa=.08em \sb=.03em
     \ifmmode \hbox{\rlap{.}\kern\sa}^{\scriptstyle\prime}\hbox{\kern-\sb}
     \else \rlap{.}\kern\sa$^{\scriptstyle\prime}$\kern-\sb\fi}
\def\ra[#1 #2 #3.#4]{#1\sup{h}#2\sup{m}#3\sup{s}\llap.#4}
\def\dec[#1 #2 #3.#4]{#1\deg#2\arcm#3\arcs\llap.#4}
\def\deco[#1 #2 #3]{#1\deg#2\arcm#3\arcs}
\def\rra[#1 #2]{#1\sup{h}#2\sup{m}}

\def\dots{\relax\ifmmode \ldots\else $\ldots$\fi}
\def\WHzsr{\ifmmode $W\,Hz\mo\,sr\mo$\else W\,Hz\mo\,sr\mo\fi}
\def\mHz{\ifmmode $\,mHz$\else \,mHz\fi}
\def\GHz{\ifmmode $\,GHz$\else \,GHz\fi}
\def\mKs{\ifmmode $\,mK\,s$^{1/2}\else \,mK\,s$^{1/2}$\fi}
\def\muKs{\ifmmode \,\mu$K\,s$^{1/2}\else \,$\mu$K\,s$^{1/2}$\fi}
\def\muKRJs{\ifmmode \,\mu$K$_{\rm RJ}$\,s$^{1/2}\else \,$\mu$K$_{\rm RJ}$\,s$^{1/2}$\fi}
\def\muKHz{\ifmmode \,\mu$K\,Hz$^{-1/2}\else \,$\mu$K\,Hz$^{-1/2}$\fi}
\def\MJysr{\ifmmode \,$MJy\,sr\mo$\else \,MJy\,sr\mo\fi}
\def\MJysrmK{\ifmmode \,$MJy\,sr\mo$\,mK$_{\rm CMB}\mo\else \,MJy\,sr\mo\,mK$_{\rm CMB}\mo$\fi}
\def\microns{\ifmmode \,\mu$m$\else \,$\mu$m\fi}

\def\muK{\ifmmode \,\mu$K$\else \,$\mu$\hbox{K}\fi}
\def\microK{\ifmmode \,\mu$K$\else \,$\mu$\hbox{K}\fi}
\def\muW{\ifmmode \,\mu$W$\else \,$\mu$\hbox{W}\fi}
\def\kms{\ifmmode $\,km\,s$^{-1}\else \,km\,s$^{-1}$\fi}
\def\kmsMpc{\ifmmode $\,\kms\,Mpc\mo$\else \,\kms\,Mpc\mo\fi}
\providecommand{\sorthelp}[1]{}

%% file: A24_Inflation_authors_and_institutes.tex
\author{\small
Planck Collaboration: P.~A.~R.~Ade\inst{99}
\and
N.~Aghanim\inst{66}
\and
M.~Arnaud\inst{82}
\and
F.~Arroja\inst{74, 88}
\and
M.~Ashdown\inst{78, 6}
\and
J.~Aumont\inst{66}
\and
C.~Baccigalupi\inst{97}
\and
M.~Ballardini\inst{54, 56, 37}
\and
A.~J.~Banday\inst{112, 11}
\and
R.~B.~Barreiro\inst{73}
% ...
}
\institute{\small
APC, AstroParticule et Cosmologie, Universit\'{e} Paris Diderot, CNRS/IN2P3, CEA/lrfu, Observatoire de Paris, Sorbonne Paris Cit\'{e}, 10, rue Alice Domon et L\'{e}onie Duquet, 75205 Paris Cedex 13, France\goodbreak
\and
Aalto University Mets\"{a}hovi Radio Observatory and Dept of Radio Science and Engineering, P.O. Box 13000, FI-00076 AALTO, Finland\goodbreak
\and
African Institute for Mathematical Sciences, 6-8 Melrose Road, Muizenberg, Cape Town, South Africa\goodbreak
\and
Agenzia Spaziale Italiana Science Data Center, Via del Politecnico snc, 00133, Roma, Italy\goodbreak
\and
Aix Marseille Universit\'{e}, CNRS, LAM (Laboratoire d'Astrophysique de Marseille) UMR 7326, 13388, Marseille, France\goodbreak
\and
Astrophysics Group, Cavendish Laboratory, University of Cambridge, J J Thomson Avenue, Cambridge CB3 0HE, U.K.\goodbreak
\and
Astrophysics \& Cosmology Research Unit, School of Mathematics, Statistics \& Computer Science, University of KwaZulu-Natal, Westville Campus, Private Bag X54001, Durban 4000, South Africa\goodbreak
\and
Atacama Large Millimeter/submillimeter Array, ALMA Santiago Central Offices, Alonso de Cordova 3107, Vitacura, Casilla 763 0355, Santiago, Chile\goodbreak
\and
CGEE, SCS Qd 9, Lote C, Torre C, 4$^{\circ}$ andar, Ed. Parque Cidade Corporate, CEP 70308-200, Bras\'{i}lia, DF, Brazil\goodbreak
\and
CITA, University of Toronto, 60 St. George St., Toronto, ON M5S 3H8, Canada\goodbreak
\and
CNRS, IRAP, 9 Av. colonel Roche, BP 44346, F-31028 Toulouse cedex 4, France\goodbreak
\and
CRANN, Trinity College, Dublin, Ireland\goodbreak
\and
California Institute of Technology, Pasadena, California, U.S.A.\goodbreak
\and
Centre for Theoretical Cosmology, DAMTP, University of Cambridge, Wilberforce Road, Cambridge CB3 0WA, U.K.\goodbreak
\and
Centro de Estudios de F\'{i}sica del Cosmos de Arag\'{o}n (CEFCA), Plaza San Juan, 1, planta 2, E-44001, Teruel, Spain\goodbreak
\and
Computational Cosmology Center, Lawrence Berkeley National Laboratory, Berkeley, California, U.S.A.\goodbreak
\and
Consejo Superior de Investigaciones Cient\'{\i}ficas (CSIC), Madrid, Spain\goodbreak
\and
DSM/Irfu/SPP, CEA-Saclay, F-91191 Gif-sur-Yvette Cedex, France\goodbreak
\and
DTU Space, National Space Institute, Technical University of Denmark, Elektrovej 327, DK-2800 Kgs. Lyngby, Denmark\goodbreak
\and
D\'{e}partement de Physique Th\'{e}orique, Universit\'{e} de Gen\`{e}ve, 24, Quai E. Ansermet,1211 Gen\`{e}ve 4, Switzerland\goodbreak
\and
Dark Cosmology Centre, Niels Bohr Institute, University of Copenhagen, Juliane Maries Vej 30, 2100 Copenhagen, Denmark\goodbreak
\and
Departamento de Astrof\'{i}sica, Universidad de La Laguna (ULL), E-38206 La Laguna, Tenerife, Spain\goodbreak
\and
Departamento de F\'{\i}sica, Universidad de Oviedo, Avda. Calvo Sotelo s/n, Oviedo, Spain\goodbreak
\and
Department of Astronomy and Astrophysics, University of Toronto, 50 Saint George Street, Toronto, Ontario, Canada\goodbreak
\and
Department of Astrophysics/IMAPP, Radboud University Nijmegen, P.O. Box 9010, 6500 GL Nijmegen, The Netherlands\goodbreak
\and
Department of Physics \& Astronomy, University of British Columbia, 6224 Agricultural Road, Vancouver, British Columbia, Canada\goodbreak
\and
Department of Physics and Astronomy, Dana and David Dornsife College of Letter, Arts and Sciences, University of Southern California, Los Angeles, CA 90089, U.S.A.\goodbreak
\and
Department of Physics and Astronomy, University College London, London WC1E 6BT, U.K.\goodbreak
\and
Department of Physics and Astronomy, University of Sussex, Brighton BN1 9QH, U.K.\goodbreak
\and
Department of Physics, Florida State University, Keen Physics Building, 77 Chieftan Way, Tallahassee, Florida, U.S.A.\goodbreak
\and
Department of Physics, Gustaf H\"{a}llstr\"{o}min katu 2a, University of Helsinki, Helsinki, Finland\goodbreak
\and
Department of Physics, Princeton University, Princeton, New Jersey, U.S.A.\goodbreak
\and
Department of Physics, University of California, Berkeley, California, U.S.A.\goodbreak
\and
Department of Physics, University of California, Santa Barbara, California, U.S.A.\goodbreak
\and
Department of Physics, University of Illinois at Urbana-Champaign, 1110 West Green Street, Urbana, Illinois, U.S.A.\goodbreak
\and
Dipartimento di Fisica e Astronomia G. Galilei, Universit\`{a} degli Studi di Padova, via Marzolo 8, 35131 Padova, Italy\goodbreak
\and
Dipartimento di Fisica e Astronomia, ALMA MATER STUDIORUM, Universit\`{a} degli Studi di Bologna, Viale Berti Pichat 6/2, I-40127, Bologna, Italy\goodbreak
\and
Dipartimento di Fisica e Scienze della Terra, Universit\`{a} di Ferrara, Via Saragat 1, 44122 Ferrara, Italy\goodbreak
\and
Dipartimento di Fisica, Universit\`{a} La Sapienza, P. le A. Moro 2, Roma, Italy\goodbreak
\and
Dipartimento di Fisica, Universit\`{a} degli Studi di Milano, Via Celoria, 16, Milano, Italy\goodbreak
\and
Dipartimento di Fisica, Universit\`{a} degli Studi di Trieste, via A. Valerio 2, Trieste, Italy\goodbreak
\and
Dipartimento di Matematica, Universit\`{a} di Roma Tor Vergata, Via della Ricerca Scientifica, 1, Roma, Italy\goodbreak
\and
Discovery Center, Niels Bohr Institute, Blegdamsvej 17, Copenhagen, Denmark\goodbreak
\and
Discovery Center, Niels Bohr Institute, Copenhagen University, Blegdamsvej 17, Copenhagen, Denmark\goodbreak
\and
European Southern Observatory, ESO Vitacura, Alonso de Cordova 3107, Vitacura, Casilla 19001, Santiago, Chile\goodbreak
\and
European Space Agency, ESAC, Planck Science Office, Camino bajo del Castillo, s/n, Urbanizaci\'{o}n Villafranca del Castillo, Villanueva de la Ca\~{n}ada, Madrid, Spain\goodbreak
\and
European Space Agency, ESTEC, Keplerlaan 1, 2201 AZ Noordwijk, The Netherlands\goodbreak
\and
Gran Sasso Science Institute, INFN, viale F. Crispi 7, 67100 L'Aquila, Italy\goodbreak
\and
HGSFP and University of Heidelberg, Theoretical Physics Department, Philosophenweg 16, 69120, Heidelberg, Germany\goodbreak
\and
Helsinki Institute of Physics, Gustaf H\"{a}llstr\"{o}min katu 2, University of Helsinki, Helsinki, Finland\goodbreak
\and
INAF - Osservatorio Astronomico di Padova, Vicolo dell'Osservatorio 5, Padova, Italy\goodbreak
\and
INAF - Osservatorio Astronomico di Roma, via di Frascati 33, Monte Porzio Catone, Italy\goodbreak
\and
INAF - Osservatorio Astronomico di Trieste, Via G.B. Tiepolo 11, Trieste, Italy\goodbreak
\and
INAF/IASF Bologna, Via Gobetti 101, Bologna, Italy\goodbreak
\and
INAF/IASF Milano, Via E. Bassini 15, Milano, Italy\goodbreak
\and
INFN, Sezione di Bologna, Via Irnerio 46, I-40126, Bologna, Italy\goodbreak
\and
INFN, Sezione di Roma 1, Universit\`{a} di Roma Sapienza, Piazzale Aldo Moro 2, 00185, Roma, Italy\goodbreak
\and
INFN, Sezione di Roma 2, Universit\`{a} di Roma Tor Vergata, Via della Ricerca Scientifica, 1, Roma, Italy\goodbreak
\and
INFN/National Institute for Nuclear Physics, Via Valerio 2, I-34127 Trieste, Italy\goodbreak
\and
IPAG: Institut de Plan\'{e}tologie et d'Astrophysique de Grenoble, Universit\'{e} Grenoble Alpes, IPAG, F-38000 Grenoble, France, CNRS, IPAG, F-38000 Grenoble, France\goodbreak
\and
IUCAA, Post Bag 4, Ganeshkhind, Pune University Campus, Pune 411 007, India\goodbreak
\and
Imperial College London, Astrophysics group, Blackett Laboratory, Prince Consort Road, London, SW7 2AZ, U.K.\goodbreak
\and
Infrared Processing and Analysis Center, California Institute of Technology, Pasadena, CA 91125, U.S.A.\goodbreak
\and
Institut N\'{e}el, CNRS, Universit\'{e} Joseph Fourier Grenoble I, 25 rue des Martyrs, Grenoble, France\goodbreak
\and
Institut Universitaire de France, 103, bd Saint-Michel, 75005, Paris, France\goodbreak
\and
Institut d'Astrophysique Spatiale, CNRS, Univ. Paris-Sud, Universit\'{e} Paris-Saclay, B\^{a}t. 121, 91405 Orsay cedex, France\goodbreak
\and
Institut d'Astrophysique de Paris, CNRS (UMR7095), 98 bis Boulevard Arago, F-75014, Paris, France\goodbreak
\and
Institut f\"ur Theoretische Teilchenphysik und Kosmologie, RWTH Aachen University, D-52056 Aachen, Germany\goodbreak
\and
Institute for Space Sciences, Bucharest-Magurale, Romania\goodbreak
\and
Institute of Astronomy, University of Cambridge, Madingley Road, Cambridge CB3 0HA, U.K.\goodbreak
\and
Institute of Theoretical Astrophysics, University of Oslo, Blindern, Oslo, Norway\goodbreak
\and
Instituto de Astrof\'{\i}sica de Canarias, C/V\'{\i}a L\'{a}ctea s/n, La Laguna, Tenerife, Spain\goodbreak
\and
Instituto de F\'{\i}sica de Cantabria (CSIC-Universidad de Cantabria), Avda. de los Castros s/n, Santander, Spain\goodbreak
\and
Istituto Nazionale di Fisica Nucleare, Sezione di Padova, via Marzolo 8, I-35131 Padova, Italy\goodbreak
\and
Jet Propulsion Laboratory, California Institute of Technology, 4800 Oak Grove Drive, Pasadena, California, U.S.A.\goodbreak
\and
Jodrell Bank Centre for Astrophysics, Alan Turing Building, School of Physics and Astronomy, The University of Manchester, Oxford Road, Manchester, M13 9PL, U.K.\goodbreak
\and
Kavli Institute for Cosmological Physics, University of Chicago, Chicago, IL 60637, USA\goodbreak
\and
Kavli Institute for Cosmology Cambridge, Madingley Road, Cambridge, CB3 0HA, U.K.\goodbreak
\and
Kazan Federal University, 18 Kremlyovskaya St., Kazan, 420008, Russia\goodbreak
\and
LAL, Universit\'{e} Paris-Sud, CNRS/IN2P3, Orsay, France\goodbreak
\and
LERMA, CNRS, Observatoire de Paris, 61 Avenue de l'Observatoire, Paris, France\goodbreak
\and
Laboratoire AIM, IRFU/Service d'Astrophysique - CEA/DSM - CNRS - Universit\'{e} Paris Diderot, B\^{a}t. 709, CEA-Saclay, F-91191 Gif-sur-Yvette Cedex, France\goodbreak
\and
Laboratoire Traitement et Communication de l'Information, CNRS (UMR 5141) and T\'{e}l\'{e}com ParisTech, 46 rue Barrault F-75634 Paris Cedex 13, France\goodbreak
\and
Laboratoire de Physique Subatomique et Cosmologie, Universit\'{e} Grenoble-Alpes, CNRS/IN2P3, 53, rue des Martyrs, 38026 Grenoble Cedex, France\goodbreak
\and
Laboratoire de Physique Th\'{e}orique, Universit\'{e} Paris-Sud 11 \& CNRS, B\^{a}timent 210, 91405 Orsay, France\goodbreak
\and
Lawrence Berkeley National Laboratory, Berkeley, California, U.S.A.\goodbreak
\and
Lebedev Physical Institute of the Russian Academy of Sciences, Astro Space Centre, 84/32 Profsoyuznaya st., Moscow, GSP-7, 117997, Russia\goodbreak
\and
Leung Center for Cosmology and Particle Astrophysics, National Taiwan University, Taipei 10617, Taiwan\goodbreak
\and
Max-Planck-Institut f\"{u}r Astrophysik, Karl-Schwarzschild-Str. 1, 85741 Garching, Germany\goodbreak
\and
McGill Physics, Ernest Rutherford Physics Building, McGill University, 3600 rue University, Montr\'{e}al, QC, H3A 2T8, Canada\goodbreak
\and
National University of Ireland, Department of Experimental Physics, Maynooth, Co. Kildare, Ireland\goodbreak
\and
Nicolaus Copernicus Astronomical Center, Bartycka 18, 00-716 Warsaw, Poland\goodbreak
\and
Niels Bohr Institute, Blegdamsvej 17, Copenhagen, Denmark\goodbreak
\and
Niels Bohr Institute, Copenhagen University, Blegdamsvej 17, Copenhagen, Denmark\goodbreak
\and
Nordita (Nordic Institute for Theoretical Physics), Roslagstullsbacken 23, SE-106 91 Stockholm, Sweden\goodbreak
\and
Optical Science Laboratory, University College London, Gower Street, London, U.K.\goodbreak
\and
SISSA, Astrophysics Sector, via Bonomea 265, 34136, Trieste, Italy\goodbreak
\and
SMARTEST Research Centre, Universit\`{a} degli Studi e-Campus, Via Isimbardi 10, Novedrate (CO), 22060, Italy\goodbreak
\and
School of Physics and Astronomy, Cardiff University, Queens Buildings, The Parade, Cardiff, CF24 3AA, U.K.\goodbreak
\and
School of Physics and Astronomy, University of Nottingham, Nottingham NG7 2RD, U.K.\goodbreak
\and
Simon Fraser University, Department of Physics, 8888 University Drive, Burnaby BC, Canada\goodbreak
\and
Sorbonne Universit\'{e}-UPMC, UMR7095, Institut d'Astrophysique de Paris, 98 bis Boulevard Arago, F-75014, Paris, France\goodbreak
\and
Space Research Institute (IKI), Russian Academy of Sciences, Profsoyuznaya Str, 84/32, Moscow, 117997, Russia\goodbreak
\and
Space Sciences Laboratory, University of California, Berkeley, California, U.S.A.\goodbreak
\and
Special Astrophysical Observatory, Russian Academy of Sciences, Nizhnij Arkhyz, Zelenchukskiy region, Karachai-Cherkessian Republic, 369167, Russia\goodbreak
\and
Stanford University, Dept of Physics, Varian Physics Bldg, 382 Via Pueblo Mall, Stanford, California, U.S.A.\goodbreak
\and
Sub-Department of Astrophysics, University of Oxford, Keble Road, Oxford OX1 3RH, U.K.\goodbreak
\and
Sydney Institute for Astronomy, School of Physics A28, University of Sydney, NSW 2006, Australia\goodbreak
\and
The Oskar Klein Centre for Cosmoparticle Physics, Department of Physics,Stockholm University, AlbaNova, SE-106 91 Stockholm, Sweden\goodbreak
\and
Theory Division, PH-TH, CERN, CH-1211, Geneva 23, Switzerland\goodbreak
\and
UPMC Univ Paris 06, UMR7095, 98 bis Boulevard Arago, F-75014, Paris, France\goodbreak
\and
Universit\'{e} de Toulouse, UPS-OMP, IRAP, F-31028 Toulouse cedex 4, France\goodbreak
% ...
}

%% file: macros.tex
\def\reff@jnl#1{{\rm#1\/}}
\def\apj{\reff@jnl{ApJ}}       
\def\apjs{\reff@jnl{ApJS}}     
\def\aaps{\reff@jnl{A\&AS}}    
\def\mnras{\reff@jnl{MNRAS}}   
\def\prd{\reff@jnl{Phys.\ Rev.\ D}}    

\newcommand{\Nside}{\ensuremath{N_{\mathrm{side}}}} 
\newcommand{\Npix}{\ensuremath{N_{\mathrm{pix}}}}   
\newcommand{\Ntau}{\ensuremath{N_{\tau}}}   
\newcommand{\vA}{\mathbf{A}}
\newcommand{\va}{\mathbf{a}}
\newcommand{\vB}{\mathbf{B}}
\newcommand{\vM}{\mathbf{M}}
\newcommand{\vN}{\mathbf{N}}
\newcommand{\vP}{\mathbf{P}}
\newcommand{\vS}{\mathbf{S}}
\newcommand{\vX}{\mathbf{X}}
\newcommand{\vY}{\mathbf{Y}}
\newcommand{\vd}{\mathbf{d}}
\newcommand{\vn}{\mathbf{n}}
\newcommand{\vs}{\mathbf{s}}
\newcommand{\vC}{\mathbf{C}}
\newcommand{\vI}{\mathbf{I}}
\newcommand{\vt}{\mathbf{t}}
\newcommand{\vE}{\mathbf{E}}
\newcommand{\vx}{\mathbf{x}}
\newcommand{\vphi}{\mathbf{\phi}}
\newcommand{\veta}{\mathbf{\eta}}
\newcommand{\vshat}{\vec{\hat{s}}}
\newcommand{\vxhat}{\vec{\hat{x}}}
\newcommand{\vnu}{\vn_{u}}
\newcommand{\vNu}{\vN_{u}}
\newcommand{\tr}{^{\mathrm{T}}} 
\newcommand{\beq}{\begin{equation}}
\newcommand{\eeq}{\end{equation}}
\newcommand{\rsht}{{MASTER}}
\newcommand{\tC}{\widetilde{C}}
\newcommand{\tN}{\widetilde{N}}
\renewcommand{\r}{{\bf{r}}}
\newcommand{\n}{{\bf{n}}}
\renewcommand{\k}{{\bf{k}}}
\newcommand{\fsky}{{f_{\rm sky}}}
\newcommand{\fzero}{{F^{(0)}}}
\newcommand{\fzerol}{{F^{(0)}_{\ell}}}
\newcommand{\npix}{{N_{\rm pix}}}
\newcommand{\nbins}{{n_{\rm bins}}}
\newcommand{\lmax}{{\ell_{\rm max}}}
\newcommand{\nmc}{{N_{\rm MC}}}
\newcommand{\nmcs}{{N_{\rm MC}^{\rm (s)}}}
\newcommand{\nmcn}{{N_{\rm MC}^{\rm (n)}}}
\newcommand{\nmcsn}{{N_{\rm MC}^{\rm (s+n)}}}
\newcommand{\ntau}{{N_{\tau}}}
\newcommand{\nfft}{{N_{\rm FFT}}}
\newcommand{\Cltheory}{{C_\ell^{\rm th}}}
\newcommand{\Ctheory}{{C^{\rm th}}}
\newcommand{\VEV}[1]{\langle#1\rangle}
\newcommand{\boom}{{\sc{BOOMERanG}}}
\newcommand{\bldb}{{Boom-LDB}}
\newcommand{\wjjj}[6]
{{
\left(
\begin{array}{lcr} #1 & #2 & #3 \\#4 & #5 & #6 \end{array}
\right)
}}
\newcommand{\niter}{{N_{\rm iter}}}
\newcommand{\col}[2]{\left[\begin{array}{c}{#1}\\{#2}\end{array}\right]}
\newcommand{\mat}[4]{\left[\begin{array}{cc}{#1}&{#2}\\{#3}&{#4}\end{array}
\right]}
\newcommand{\vsh}{\hat{\vs}}
\newcommand{\vxh}{\hat{\vx}}
\newcommand{\Nmap}{\vN_{\mathrm{map}}}
\newcommand{\Noff}{\vN_{\mathrm{off}}}
\newcommand{\Nmapz}{\Nmap^{(0)}}
\newcommand{\thalf}{\tfrac{1}{2}}
\newcommand{\be}{\begin{equation}}
\newcommand{\ee}{\end{equation}}
\newcommand{\bea}{\begin{eqnarray}}
\newcommand{\eea}{\end{eqnarray}}
\def\nn{\nonumber}
\def\L{\mathcal{L}}

\renewcommand{\dbltopfraction}{1.0}
\renewcommand{\textfraction}{0}

\newenvironment{myitem}%
{\begin{enumerate}\setlength{\itemsep}{0mm}}%
{\end{enumerate}}
\newenvironment{myenum}%
{\begin{enumerate}\setlength{\itemsep}{0mm}}%
{\end{enumerate}}

\def\nd#1#2{{d #1 \over d #2}}
\def\pd#1#2{{\upartial #1 \over \upartial #2}}
\def\spd#1#2#3{{\upartial ^2 #1 \over \upartial #2 \upartial #3}}
\def\sspd#1#2{{\upartial ^2 #1 \over \upartial #2^2}}
\def\tfrac#1#2{{\textstyle\frac{#1}{#2}}}
\def\vect#1{{\mathbf{#1}}}

\newcommand{\bc}{\begin{center}}
\newcommand{\ec}{\end{center}}
\newcommand{\bi}{\begin{itemize}}
\newcommand{\ei}{\end{itemize}}
\newcommand{\ben}{\begin{enumerate}}
\newcommand{\een}{\end{enumerate}}
\newcommand{\R}{\Re\textrm{e}}
\newcommand{\I}{\Im\textrm{m}}
\newcommand{\Ab}{\boldsymbol{A}}
\newcommand{\Mb}{\boldsymbol{M}}
\newcommand{\Tb}{\boldsymbol{T}}

 \newcommand{\mtc}[1]{\mathcal{#1}}
 \newcommand{\Pow}{{\mathcal P}}
 \newcommand{\nad}{n_\mathrm{ad}}
 \newcommand{\nadI}{n_\mathrm{ar}}
 \newcommand{\nadII}{n_\mathrm{as}}
 \newcommand{\niso}{n_\mathrm{iso}}
 \newcommand{\ncor}{n_\mathrm{cor}}
 \newcommand{\etzz}{\eta_{\sigma\sigma}}
 \newcommand{\etzs}{\eta_{\sigma s}}
 \newcommand{\etss}{\eta_{ss}}
 \newcommand{\veps}{\varepsilon}
 \newcommand{\jvc}[1]{{{\textcolor{red}{JV: #1 endJV.}}}}
\newcommand{\ff}[1]{{{\textcolor{blu}{#1}}}}

\newfont{\gwpfont}{cmssq8 scaled 1000}
\newcommand{\rexcess}{{\gwpfont REXCESS}}

\def\xmm{{\it XMM-Newton}}
\def\Mv {M_\mathrm{500}}
\def\msol {\mathrm{M}_{\odot}}
\def\YX {Y_\mathrm{X}} 
\def \Rv {R_{500}} 
\def\keV {\mathrm{keV}} 

\newcommand{\chisq}{\Delta \chi^2_\mathrm{eff}}
\newcommand{\kpiv}{k_{\mathrm{pivot}}}
\newcommand{\Mpl}{M_{\mathrm{pl}}}
\newcommand{\aend}{a_{\mathrm{end}}}
\newcommand{\aeq}{a_{\mathrm{eq}}}
\newcommand{\arh}{a_{\mathrm{th}}}
\newcommand{\wprim}{w_{\mathrm{prim}}}
\newcommand{\wint}{w_{\mathrm{int}}}
\newcommand{\rhorh}{\rho_{\mathrm{th}}}
\newcommand{\rhoeq}{\rho_{\mathrm{eq}}}
\newcommand{\rhoend}{\rho_{\mathrm{end}}}
\newcommand{\vend}{V_{\mathrm{end}}}
\newcommand{\ModeCode}{{\tt ModeCode}}
\newcommand{\MultiNest}{{\tt MultiNest}}
\newcommand{\CAMB}{{\tt CAMB}}
\newcommand{\CosmoMC}{{\tt CosmoMC}}
\newcommand{\codename}{{\ModeCode}}

\def\gtorder{\mathrel{\raise.3ex\hbox{$>$}\mkern-14mu
             \lower0.6ex\hbox{$\sim$}}}
\def\ltorder{\mathrel{\raise.3ex\hbox{$<$}\mkern-14mu
             \lower0.6ex\hbox{$\sim$}}}

%% file: abstract.tex
We present the implications for cosmic inflation of the \Planck\ measurements of the 
cosmic microwave background (CMB) anisotropies in both temperature and polarization based
on the full \Planck\ survey, which includes more than twice the integration time 
of the nominal survey used for the 2013 Release papers.
The \Planck\ full mission temperature data and a first release of polarization data on
large angular scales measure the spectral index of curvature perturbations to be 
$n_\mathrm{s} =  0.968 \pm 0.006$ and tightly constrain its scale
dependence to $d n_\mathrm{s}/d \ln k =-0.003 \pm 0.007$ when combined with the \Planck\ lensing likelihood. 
When the \Planck\ high-$\ell$ polarization data is included, the results 
are consistent and uncertainties are further reduced.  
The upper bound on the tensor-to-scalar ratio 
is $r_{0.002} <  0.11$ (95\,\%\ CL).
This upper limit is consistent with the {\it B}-mode polarization constraint 
$r <  0.12$ (95\,\%\ CL)
obtained from a joint
analysis of the BICEP2/Keck Array and \Planck\ data.
These results imply that 
$V(\phi) \propto \phi^2$ and natural inflation are now
disfavoured compared to models predicting a smaller tensor-to-scalar ratio, such as $R^2$ inflation. 
We search for several physically motivated deviations from a simple power-law spectrum 
of curvature perturbations, including those motivated by a reconstruction of the 
inflaton potential not relying on the slow-roll approximation. We find 
that such models are not preferred, either according to a Bayesian model comparison
or according to a frequentist 
simulation-based analysis.
Three independent methods reconstructing the primordial power spectrum 
consistently recover a featureless and smooth ${\cal P}_{\cal{R}}(k)$ over the range of scales $0.008\,\mathrm{Mpc}^{-1}
\lesssim k \lesssim 0.1\,\mathrm{Mpc}^{-1}$. 
At large scales, each method finds deviations from a power law, connected to a deficit at 
multipoles $\ell \approx 20$--$40$ in the temperature power spectrum, but at 
an uncompelling statistical significance owing to the large cosmic variance present at these multipoles. 
By combining power spectrum and 
non-Gaussianity bounds, we constrain models with generalized Lagrangians,
including Galileon models and axion monodromy models. The \Planck\ data are 
consistent with adiabatic primordial perturbations, and the estimated values for the 
parameters of the base $\Lambda$CDM model are not significantly altered when
more general initial conditions are admitted. In correlated mixed adiabatic and isocurvature 
models, the 95\,\% CL upper bound for the non-adiabatic contribution to the observed CMB 
temperature variance is $|\alpha_{\text{non-adi}}|<$ 1.9\,\%,
4.0\,\%, and 2.9\,\% for cold dark matter (CDM), neutrino density, and neutrino velocity isocurvature 
modes, respectively. We have tested inflationary models producing an anisotropic 
modulation of the primordial curvature power spectrum finding that the dipolar 
modulation in the CMB temperature field induced by a CDM isocurvature perturbation is 
not preferred at a statistically significant 
level. We also establish tight constraints on a possible
quadrupolar modulation of the curvature perturbation. These results are consistent 
with the \Planck\ 2013 analysis based on the nominal mission data and further
constrain slow-roll single-field inflationary models, as expected from the increased precision 
of \Planck\ data using the full set of observations.

%% file: section_one.tex
\def\gtorder{\mathrel{\raise.3ex\hbox{$>$}\mkern-14mu
             \lower0.6ex\hbox{$\sim$}}}
\def\ltorder{\mathrel{\raise.3ex\hbox{$<$}\mkern-14mu
             \lower0.6ex\hbox{$\sim$}}}

The precise measurements by \Planck
\footnote{\Planck\ (\url{http://www.esa.int/Planck}) is a project of the European Space Agency 
(ESA) with instruments provided by two scientific consortia funded by ESA member states and led by 
Principal Investigators from France and Italy, telescope reflectors provided through a collaboration between 
ESA and a scientific consortium led and funded by Denmark, and additional contributions from NASA (USA).}
of the cosmic microwave background (CMB) anisotropies covering the entire sky and 
over a broad range of scales, from the largest visible down to a resolution of approximately $5'$, provide 
a powerful probe of cosmic inflation, as detailed in the \Planck\ 2013 inflation paper 
\citep[][hereafter \citetalias{planck2013-p17}]{planck2013-p17}.
In the 2013 results, the robust detection of the departure of the scalar spectral 
index from exact scale invariance, i.e., $n_\mathrm{s} < 1$, at more than $5\,\sigma$
confidence, as well as the lack of the observation of any statistically significant 
running of the spectral index, were found to be consistent with simple slow-roll models of inflation. 
Single-field inflationary models with a standard kinetic term were also found to be compatible with
the new tight upper bounds on the 
primordial non-Gaussianity parameters $f_\mathrm{NL}$ reported in \citet{planck2013-p19}.
No evidence of isocurvature perturbations as generated in multi-field inflationary models 
\citepalias{planck2013-p17} or by cosmic strings or topological defects was found \citep{planck2013-p20}.
The \Planck\ 2013 results overall favoured the simplest inflationary models. However, we noted 
an amplitude deficit for multipoles $\ell \lsim 40$ whose statistical significance relative
to the six-parameter
base $\Lambda$ cold dark matter ($\Lambda$CDM) model is only about $2\,\sigma$, 
as well as other anomalies on large angular scales but also without compelling statistical significance 
\citep{planck2013-p09}.  The constraint on the tensor-to-scalar ratio, $r < 0.12$ at 95\,\% CL,
inferred from the temperature power spectrum alone, combined with the determination
of $n_\mathrm{s}$, 
suggested models with concave potentials.

This paper updates the implications 
for inflation in the light of the \Planck\ full mission temperature and polarization data. 
The \Planck\  2013 cosmology results included only the nominal mission,
comprising the first 14 months of the data taken, and used only the temperature data.  
However, the full mission includes the full 29 months of scientific data taken by the
cryogenically cooled high frequency instrument (HFI) (which ended when
the ${}^3$He/${}^4$He supply for the final stage of the cooling chain ran out) and the approximately
four years of data taken by the low frequency instrument (LFI), which covered a longer
period than the HFI because the LFI did not rely
on cooling down to 100\,mK for its operation. 
For a detailed discussion of the new likelihood and a comparison with the 2013 likelihood,
we refer the reader to \citet{planck2014-a13} and \citet{planck2014-a15}, but we mention here some 
highlights of the differences between the 2013 and 2015 data processing and likelihoods: 
(1) Improvements in the data processing such as beam characterization and absolute calibration at each frequency
result in a better removal of systematic effects and (2) the 2015 temperature high-$\ell$ likelihood uses half-mission cross-power spectra 
over more of the sky, owing to less aggressive Galactic cuts. 
The use of polarization information in the 2015 likelihood release contributes to the constraining
power of \Planck\ in two principal ways: (1) The measurement of the $E$-mode polarization at large angular 
scales (presently based on the 70\,GHz channel) constrains the reionization 
optical depth, $\tau ,$ independently 
of other estimates using ancillary data; and (2) the measurement 
of the $TE$ and $EE$ spectra at $\ell \ge 30$ at the same frequencies used for the 
$TT$ spectra (100, 143, and 217\,GHz) helps break parameter degeneracies, 
particularly for extended cosmological models (beyond the baseline six-parameter model).    
A full analysis of the \Planck\ low-$\ell $ polarization is still in progress and will be the 
subject of another forthcoming set of \Planck\ publications. 

The \Planck\ 2013 results have sparked a revival of interest in several aspects of 
inflationary models. We mention here a few examples without the ambition to be exhaustive.
A lively debate arose on the conceptual problems of some of the inflationary models favoured 
by the \Planck\ 2013 data
\citep{Ijjas:2013vea,Guth:2013sya,Linde:2014nna,Ijjas:2014nta}.
The interest in the $R^2$ inflationary model originally proposed by
\cite{Starobinsky:1980te} increased, since its predictions for cosmological 
fluctuations \citep{Mukhanov:1981xt,Starobinsky:1983zz}
are compatible with the \Planck\ 2013 results \citepalias{planck2013-p17}.
It has been shown that supergravity motivates a potential similar to
the Einstein gravity conformal representation of the $R^2$ inflationary model in different contexts
\citep{Ellis:2013xoa,Ellis:2013nxa,Buchmuller:2013zfa,Farakos:2013cqa,Ferrara:2013wka}.
A similar potential can also be generated by spontaneous breaking of 
conformal symmetry \citep{Kallosh:2013lkr}.

The constraining power of \Planck\ also motivated a comparison between large numbers 
of inflationary models \citep{Martin:2014vha} and stimulated different perspectives on how best to compare 
theoretical inflationary predictions with observations based on the parameterized 
dependence of the Hubble parameter on the scale factor during inflation \citep{Mukhanov:2013tua,Binetruy:2014zya,Garcia-Bellido:2014gna}.
The interpretation of the asymmetries on large angular scales \citep{planck2013-p09} 
also prompted a reanalysis of the primordial dipole modulation 
\citep{Lyth:2013vha,liddle2013cosmic,kanno2013viable} of curvature perturbations during inflation.

Another recent development has been the renewed interest in possible tensor modes generated during inflation,
sparked by the BICEP2 results \citep{bicep2014B,bicep2014A}. 
The BICEP2 team suggested that the $B$-mode polarization signal detected 
at $50 < \ell < 150$ at a single frequency
(150\,GHz) might be of primordial origin. However, a crucial step in this possible 
interpretation was excluding
an explanation based on polarized thermal dust emission from our Galaxy. 
The BICEP2 team put forward a number of
models to estimate the likely contribution from dust, but at the time relevant observational data were
lacking, and this modelling involved a high degree of extrapolation.
If dust polarization were negligible in the observed patch of 380\,deg$^2$, 
this interpretation would lead to a tensor-to-scalar ratio of $r = 0.2^{+0.07}_{-0.05}$ for a 
scale-invariant spectrum.
A value of $r \approx 0.2$, as suggested by \cite{bicep2014A}, 
would have obviously changed the \Planck\ 2013 perspective according to which slow-roll inflationary models
are favoured, and such a high value of $r$ would also have required a strong running of the 
scalar spectral index, or some other modification from a simple power-law spectrum,
to reconcile the contribution of gravitational waves to temperature anisotropies at low multipoles
with the observed $TT$ spectrum. 

The interpretation of the $B$-mode signal in terms of gravitational waves {\em alone} presented in
\cite{bicep2014A} was later cast in
doubt by \Planck\ measurements of dust polarization at 353\,GHz \citep{planck2014-XIX,planck2014-XX,
planck2014-XXI,planck2014-XXII}.
The \Planck\ measurements characterized the frequency dependence
of intensity and polarization of the Galactic dust emission, and moreover showed that
the polarization fraction is higher than expected in regions of low dust emission.
With the help of the \Planck\ measurements of Galactic dust properties \citep{planck2014-XIX},
it was shown that the interpretation of the $B$-mode polarization signal in terms of a
primordial tensor signal plus a lensing contribution
was not statistically preferred to an explanation 
based on the expected dust signal at 150\,GHz plus a lensing
contribution \citep[see also][]{Flauger:2014qra,Mortonson:2014bja}.
Subsequently, \cite{planck2014-XXX} extrapolated the \Planck\ $B$-mode power
spectrum of dust polarization at 353\,GHz over the multipole range $40 < \ell < 120$ to 150\,GHz,
showing that the $B$-mode polarization signal detected by BICEP2 could be entirely due to dust.

More recently, a BICEP2/Keck Array-\Planck\ (BKP) joint analysis 
\citep[][herafter \citetalias{pb2015}]{pb2015} combined
the high-sensitivity $B$-mode maps from BICEP2 and Keck Array
with the \Planck\ maps at higher frequencies where dust emission dominates.
A study of the cross-correlations of all these maps
in the BICEP2 field found the absence
of any statistically significant evidence for primordial
gravitational waves, setting an upper limit of $r < 0.12$ at 95\,\% CL \citepalias{pb2015}. 
Although this upper limit is numerically almost identical to the \Planck\ 2013 result obtained combining 
the nominal mission temperature data with WMAP polarization to remove parameter degeneracies
\citep{planck2013-p11,planck2013-p17},
the BKP upper bound is much more robust against modifications of the inflationary model, since $B$ modes are
insensitive to the shape of the predicted scalar anisotropy pattern.
In Sect.~\ref{sec:bicep2} we explore how the recent BKP analysis 
constrains inflationary models.

This paper is organized as follows.
Section~\ref{sec:paradigm} briefly reviews the additional information on the primordial cosmological fluctuations
encoded in the polarization angular power spectrum.
Section~\ref{sec:methodology} describes the statistical methodology as well as the \Planck\ and other likelihoods
used throughout the paper. Sections~\ref{sec:updates} and \ref{sec:modelcomp} 
discuss the \Planck\ 2015 constraints on scalar and tensor fluctuations,
respectively. Section~\ref{sec:numerical} is dedicated to constraints on the slow-roll parameters and 
provides a Bayesian comparison of selected
slow-roll inflationary models. In Sect.~\ref{sec:pot} we reconstruct the inflaton potential and the Hubble parameter
as a Taylor expansion of the inflaton in the observable range without relying on the slow-roll approximation.
The reconstruction of the curvature perturbation power spectrum is presented in Sect.~\ref{sec:pk}.
The search for parameterized features is presented in Sect.~\ref{sec:oscillations}, and 
combined constraints from the \Planck\ 2015 power spectrum and primordial non-Gaussianity derived in
\citet{planck2014-a19} are presented in Sect.~\ref{sec:putfnl}.
The analysis of isocurvature perturbations combined and correlated with curvature perturbations
is presented in Sect.~\ref{sec:iso}.
In Sect.~\ref{sec:anisotropy} we study the implications of relaxing
the assumption of statistical isotropy of the primordial fluctuations.
We discuss two examples
of {\em anisotropic} inflation in light of the tests of isotropy performed in 
\citet{planck2014-a18}.  
Section~\ref{sec:conclusions} presents some concluding remarks.

%% file: section_two.tex
\begin{figure}[!h]
\begin{center}
\includegraphics[width=0.45\textwidth]{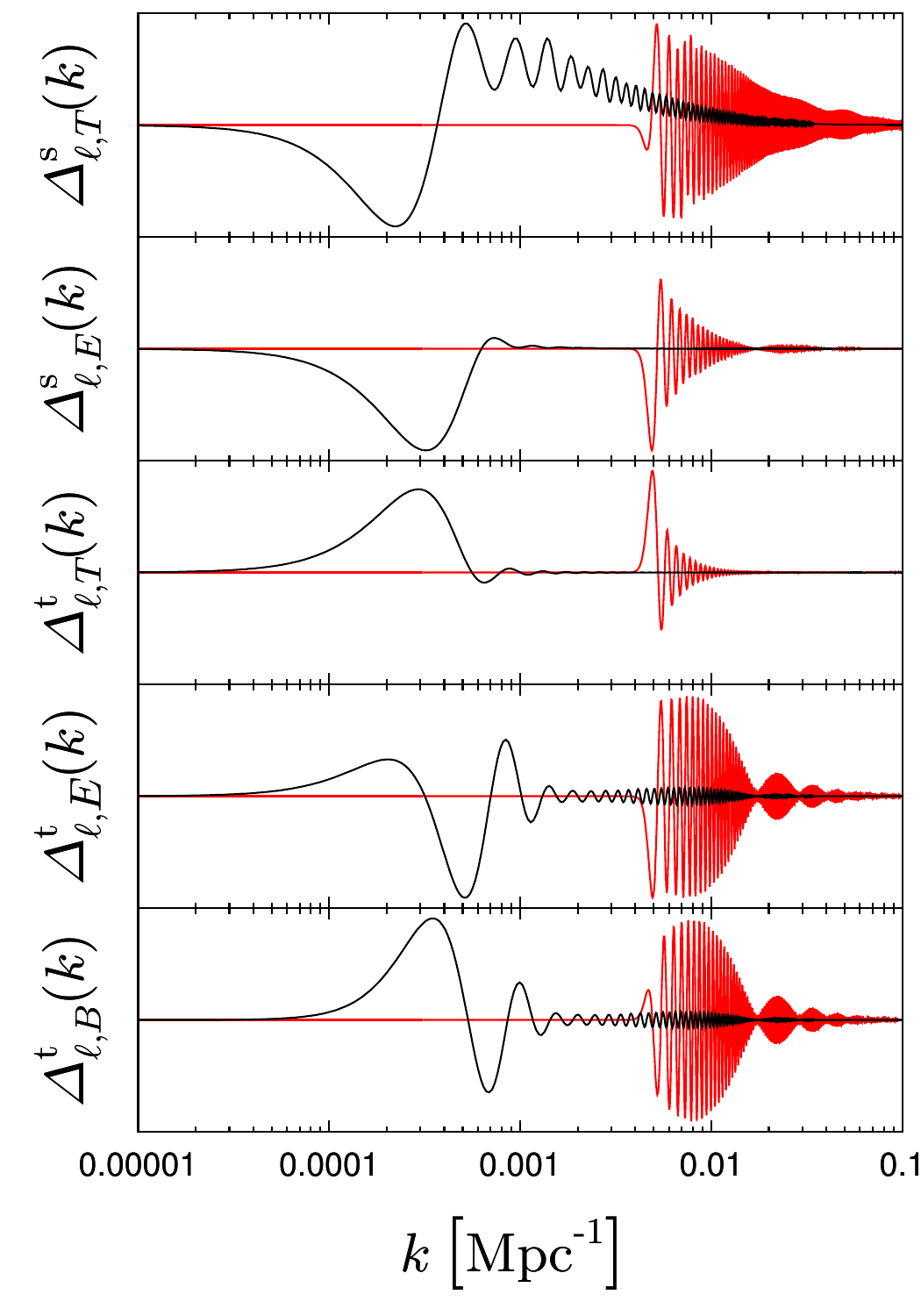}
\end{center}
\caption{Comparison of transfer functions for the scalar and tensor modes.
The CMB transfer functions
$\Delta _{\ell ,\mathcal{A}}^\mathrm{s}(k)$ 
and  
$\Delta _{\ell ,\mathcal{A}}^\mathrm{t}(k)$, 
where $\mathcal{A}=T,E,B$, define the linear transformations mapping
the primordial scalar and tensor cosmological perturbations to the CMB 
anisotropies as seen by us on the sky today. These functions are plotted 
for two representative values of the multipole number: 
$\ell =2$ (in black) and $\ell =65$ (in red).
\label{Fig:TransferFunctions}}
\end{figure}

This section provides a short theoretical overview
of the extra information provided by polarization data over that of 
temperature alone. (More details can be found in \cite{1994ARA&A..32..319W,Ma:1995ey,bucher2014physics},
and references therein.)
In Sect.~2 
of the \Planck\ 2013 inflation paper \citepalias{planck2013-p17}, we gave an overview of the relation
between the inflationary potential and the three-dimensional
primordial scalar and tensor power spectra, denoted as 
${\cal P}_{\mathcal{R}}(k)$
and ${\cal P}_\mathrm{t}(k),$ 
respectively. (The scalar variable $\mathcal{R}$ is defined precisely in 
Sect.~\ref{sec:methodology}).  We shall not repeat the discussion
there, instead referring the reader to \citetalias{planck2013-p17} and references therein.

Under the assumption of statistical isotropy, which is predicted in all simple models of 
inflation, the two-point correlations of the CMB anisotropies 
are described by the angular power spectra
$C^{TT}_\ell,$
$C^{TE}_\ell,$
$C^{EE}_\ell,$
and 
$C^{BB}_\ell,$
where $\ell $ is the multipole number. (See 
\cite{Kamionkowski:1996ks,Zaldarriaga:1996xe,Seljak:1996gy,Hu:1997hp,Hu:1997mn} and references therein
for early discussions elucidating the role of polarization.)
In principle, one could also envisage
measuring $C^{BT}_\ell$ and $C^{BE}_\ell$, but in theories where parity
symmetry is not explicitly or spontaneously broken, the expectation values for these cross 
spectra (i.e., the theoretical cross spectra) vanish, although the observed realizations of the cross spectra 
are not exactly zero because of cosmic variance.

The CMB angular power spectra are related to the three-dimensional scalar and tensor
power spectra via the transfer functions 
$\Delta _{\ell ,\mathcal{A}}^\mathrm{s}(k)$ 
and  
$\Delta _{\ell ,\mathcal{A}}^\mathrm{t}(k),$ 
so that the contributions from scalar and tensor perturbations are
\begin{equation}
C^{\mathcal{AB},\mathrm{s}}_\ell =\int _0^\infty \frac{\mathrm{d} k}{k}~
\Delta _{\ell ,\mathcal{A}}^\mathrm{s}(k)~
\Delta _{\ell ,\mathcal{B}}^\mathrm{s}(k)~
{\cal P}_{\mathcal{R}}(k)
\end{equation}
and 
\begin{equation}
C^{\mathcal{AB},\mathrm{t}}_\ell =\int _0^\infty \frac{\mathrm{d}k}{k}~
\Delta _{\ell ,\mathcal{A}}^\mathrm{t}(k)~
\Delta _{\ell ,\mathcal{B}}^\mathrm{t}(k)~
{\cal P}_\mathrm{t}(k),
\end{equation}
respectively, where $\mathcal{A,B}=T,E,B.$ The scalar and tensor primordial 
perturbations are uncorrelated in the simplest models, so
the scalar and tensor power spectra add in quadrature, meaning that
\begin{equation}
 C^{\mathcal{AB},\mathrm{tot}}_\ell =
 C^{\mathcal{AB},\mathrm{s}}_\ell 
+C^{\mathcal{AB},\mathrm{t}}_\ell. 
\end{equation}
Roughly speaking, the form of the linear transformations encapsulated in the transfer functions 
$\Delta _{\ell ,\mathcal{A}}^\mathrm{s}(k)$ and  $\Delta _{\ell ,\mathcal{A}}^\mathrm{t}(k)$ 
probe the late time physics,
whereas 
the primordial power spectra 
${\cal P}_{\mathcal{R}}(k)$ 
and 
${\cal P}_\mathrm{t}(k)$ 
are solely determined by the primordial Universe,
perhaps not so far below the Planck scale if large-field inflation turns out to be correct.

To better understand this connection, it is useful to plot and compare the shapes of the 
transfer functions for 
representative values of $\ell $ and characterize their qualitative behavior.
Referring to Fig.~\ref{Fig:TransferFunctions}, we emphasize the following qualitative features:

\begin{enumerate}

\item
For the scalar mode transfer functions, of
which only
$\Delta _{\ell,T}^\mathrm{s}(k)$
and
$\Delta _{\ell,E}^\mathrm{s}(k)$
are non-vanishing (because
to linear order, a three-dimensional scalar mode
cannot contribute to the $B$ mode of the
polarization), both transfer functions start to rise 
at more or less the same 
small values of $k$ (due to the centrifugal barrier
in the Bessel differential equation), but
$\Delta _{\ell,E}^\mathrm{s}(k)$
falls off much faster
at large $k$ and thus smooths sharp features in
$\mathcal{P}_{\mathcal{R}}(k)$ to a lesser extent than
$\Delta _{\ell,T}^\mathrm{s}(k).$ This means that polarization 
is more powerful than temperature for reconstructing possible sharp
features in the scalar primordial power spectrum 
provided that the required signal-to-noise is available. 

\item
For the tensor modes,
$\Delta _{\ell,T}^\mathrm{t}(k)$ starts rising
at about the same small $k$ as $\Delta _{\ell,T}^\mathrm{s}(k)$ and $\Delta _{\ell,E}^\mathrm{s}(k)$
but falls off faster with increasing $k$ than
$\Delta _{\ell,T}^\mathrm{s}(k).$
On the other hand, the polarization components,
$\Delta _{\ell,E}^\mathrm{t}(k)$ and $\Delta _{\ell,B}^\mathrm{t}(k),$ have a shape
completely different from any of the other transfer functions. The shape of
$\Delta _{\ell,E}^\mathrm{t}(k)$ and $\Delta _{\ell,B}^\mathrm{t}(k)$ is much wider in $\ln (k)$ 
than the scalar polarization transfer function, with   
a variance ranging from 0.5 to 1.0 decades. These functions exhibit 
several oscillations with a period smaller than that for scalar transfer functions,
due to the difference between the sound velocity for scalar fluctuations and the light velocity
for gravitational waves \citep{Polarski:1995jg,Lesgourgues:1999eq}.
 
\end{enumerate}

Regarding the scalar primordial cosmological perturbations, the power spectrum
of the $E$-mode polarization provides an important consistency check. As we explore 
in Sects.~\ref{sec:pk} and \ref{sec:oscillations}, to some extent
the fit of the temperature power spectrum can be improved by allowing a complicated
form for the primordial power spectrum (relative to a simple power law), but 
the $C^{TE}_\ell$ and $C^{EE}_\ell$ power spectra provide independent information. Moreover,
in multi-field inflationary models, in which isocurvature modes may have been excited 
(possibly correlated amongst themselves as well as with the adiabatic mode),
polarization information provides a powerful way to break degeneracies
\citep[see, e.g.,][]{IsoPolar}. 

The inability of scalar modes to generate $B$-mode polarization (apart from 
the effects of lensing) has an important consequence.  For the primordial tensor 
modes, polarization information, especially
information concerning the $B$-mode polarization, offers powerful potential for discovery or 
for establishing upper bounds.  \Planck\ 2013 and 
WMAP established upper bounds on a possible tensor mode contribution using
$C^{TT}_\ell$ alone, but
these bounds crucially relied on assuming a simple form for the scalar primordial power
spectrum. For example, as reported in \citetalias{planck2013-p17},
when a simple power law was generalized to allow for running, the bound on the
tensor contribution degraded by approximately a factor of two. The new joint
BICEP2/Keck Array-\Planck\ upper bound (see Sect.~\ref{sec:bicep2}), however, is much more robust and cannot be
avoided by postulating baroque models that alter the scale dependence of the scalar power
spectrum.

%
%
%
%


%% file: section_three.tex

This section describes updates to the 
formalism used to describe cosmological models and the likelihoods used 
with respect to 
the \Planck\ 2013 inflation paper \citepalias{planck2013-p17}.

\subsection{Cosmological model}

The cosmological models that predict observables such as the CMB anisotropies
rely on inputs specifying the conditions
and physics at play during different epochs of the history of the Universe. The
primordial inputs describe the power spectrum of the cosmological
perturbations at a time when all the observable modes were situated outside
the Hubble radius. The inputs from this epoch consist of the primordial power spectra,
which may include scalar curvature perturbations, tensor perturbations, and
possibly also isocurvature modes and their correlations. 
The late time (i.e., $z \lsim 10^4$)
cosmological inputs include parameters such as $\omega _\mathrm{b},$ $\omega _\mathrm{c},$
$\Omega _\Lambda ,$ and $\tau $, which determine the conditions when the primordial
perturbations become imprinted on the CMB and also the evolution of the Universe
between last scattering and today, affecting primarily the angular diameter
distance. Finally, there is a so-called ``nuisance'' component, consisting of parameters
that determine how the measured CMB spectra are contaminated by unsubtracted 
Galactic and extragalactic foreground contamination. 
The focus of this paper is on the primordial inputs and how they are constrained by
the observed CMB anisotropy, but we cannot completely ignore the other 
non-primordial parameters because their presence and uncertainties
must be dealt with in order to correctly extract the primordial
information of interest here.

As in \citetalias{planck2013-p17}, we adopt the minimal six-parameter 
spatially flat base $\Lambda$CDM cosmological model as our baseline 
for the late time cosmology, mainly altering the primordial inputs,
i.e., the simple power-law spectrum parameterized by the scalar amplitude and 
spectral index 
for the adiabatic growing mode, which in this minimal model is the only
late time mode excited. 
This model has 
four free non-primordial cosmological parameters $(\omega_\mathrm{b}, 
\omega_\mathrm{c},\theta_\mathrm{MC},\tau)$ \citep[for a more detailed account of this 
model, we refer the reader to][]{planck2014-a15}.  
On occasion, this assumption will be relaxed in order to consider the impact of 
more complex alternative late time cosmologies on our conclusions about inflation. 
Some of the commonly used cosmological parameters are defined in 
Table~\ref{table:CPDefinitions}.

{
\input CPDefinitions.tex
}


\subsection{Primordial spectra of cosmological fluctuations}

In inflationary models, comoving curvature (${\cal R}$) and tensor ($h$) 
fluctuations are amplified by the nearly exponential expansion from quantum vacuum fluctuations to become
highly squeezed states resembling classical states. Formally, this quantum mechanical phenomenon is most simply
described by 
the evolution in conformal time, $\eta$, of the mode functions for the gauge-invariant inflaton fluctuation, 
$\delta \phi$, and for the tensor fluctuation, $h$:
\begin{equation}
(a y_k)'' +
\left( k^2 - \frac{x''}{x}  \right) a y_k = 0,
\label{fluctuations:Evolution}
\end{equation}
with $(x\,,y)=(a \dot \phi/H \,, \delta \phi)$ for scalars and $(x\,,y)=(a\,, h)$ for tensors. 
Here $a$ is the scale factor, primes indicate 
derivatives with respect to $\eta$, and 
$\dot \phi$ and $H=\dot a/a$ are the proper time derivative of the inflaton and the Hubble parameter,  
respectively.  The curvature fluctuation, ${\cal R}$, and the 
inflaton fluctuation, $\delta \phi$, are related via ${\cal R} = H \delta \phi/\dot \phi$.   
Analytic and numerical calculations of the 
predictions for the primordial spectra of cosmological fluctuations generated during inflation
have reached high standards of precision, which are more than adequate for our purposes, and the largest uncertainty
in testing specific inflationary models arises from our lack of 
knowledge of the history of the Universe between the end of inflation and
the present time, during the so-called ``epoch of entropy generation.''


This paper uses three different methods to compare inflationary 
predictions with \Planck\ data. The first method consists of a phenomenological parameterization 
of the primordial spectra of scalar and tensor 
perturbations according to:
\begin{align}
\mathcal{P}_{\cal R}(k) &= \frac{k^3}{2 \pi^2} |{\cal R}_k|^2 \nonumber \\
&= A_\mathrm{s} \left( \frac{k}{k_*}\right)^{n_\mathrm{s}-1 +
\frac{1}{2} \, \mathrm{d}n_\mathrm{s}/\mathrm{d}\ln k \ln(k/k_*) + \frac{1}{6} \,
\frac{\mathrm{d}^2n_\mathrm{s}}{\mathrm{d}\ln k^2} \left( \ln(k/k_*) \right)^2 + ...}, \label{scalarps}\\
\mathcal{P}_\mathrm{t}(k) &= \frac{k^3}{2 \pi^2} \left( |h^+_k|^2 + |h^\times_k|^2 \right) = 
A_\mathrm{t} \left( \frac{k}{k_*}\right)^{n_\mathrm{t} + \frac{1}{2} \,
\mathrm{d}n_\mathrm{t}/\mathrm{d}\ln k \ln(k/k_*) + ... } ,
\label{tensorps}
\end{align}
where $A_\mathrm{s} \, (A_\mathrm{t})$ is the scalar (tensor) amplitude and $n_\mathrm{s} \,
(n_\mathrm{t})$, $\mathrm{d}n_\mathrm{s}/\mathrm{d}\ln k \, (\mathrm{d}n_\mathrm{t}/\mathrm{d}\ln k)$,
and $\mathrm{d}^2n_\mathrm{s}/\mathrm{d}\ln k^2$ are the scalar (tensor) spectral index,
the running of the scalar (tensor) spectral index, and the running of the running of the scalar spectral index, 
respectively.  
$h^{+,\times}$ denotes the amplitude of 
the two polarization states ($+,\times$) of gravitational waves and $k_*$ is the 
pivot scale.  
Unless otherwise stated, the tensor-to-scalar ratio,
\begin{equation}
r = \frac{\mathcal{P}_{\mathrm t}(k_*)}{\mathcal{P}_{\cal R}(k_*)},
\label{tensortoscalar}
\end{equation}
is fixed to $-8n_\mathrm{t}$,\footnote{When running is considered, we fix 
$n_\mathrm{t}=-r(2-r/8-n_\mathrm{s})/8$ and 
$\mathrm{d}n_\mathrm{t}/\mathrm{d}\ln k = r(r/8+n_\mathrm{s}-1)/8$.} 
which is the relation that holds when inflation is driven by a single slow-rolling scalar field 
with a standard kinetic term. 
We will use a parameterization analogous to Eq.~(\ref{scalarps}) with no running for the power spectra of 
isocurvature modes and their correlations
in Sect.~\ref{sec:iso}.

The second method exploits the analytic dependence of the slow-roll power 
spectra of primordial perturbations in Eqs.~(\ref{scalarps}) and (\ref{tensorps}) on the values of the Hubble 
parameter and the hierarchy of its time derivatives, known as the Hubble flow functions (HFF): 
$\epsilon_1 = - \dot H/H^2$,
$\epsilon_{i+1} \equiv \dot \epsilon_i/(H \epsilon_i)$, with $i \ge 1$. 
We will use the analytic power spectra calculated up to second order
using the Green's function method \citep{Gong:2001he,Leach:2002ar} 
(see \citealt{Habib:2002yi}, \citealt{Martin:2002vn}, and \citealt{Casadio:2006wb} for alternative derivations).
The spectral indices and the relative scale dependence 
in Eqs.~(\ref{scalarps}) and (\ref{tensorps}) are given in terms of the HFFs by: 
\begin{align}
\label{bs}
n_\mathrm{s} - 1 =& - 2 \epsilon_1 - \epsilon_2 - 2
\epsilon_1^2 -\left(2\,C+3\right)\,\epsilon_1\,\epsilon_2 - C \epsilon_2 \epsilon_3, \\
\label{eqn:bs1}
\mathrm{d} n_\mathrm{s}/\mathrm{d} \ln k =& - 2 \epsilon_1 \epsilon_2 - \epsilon_2 \epsilon_3, \\
\label{eqn:bs2}
n_\mathrm{t} =& - 2\epsilon_1 - 2\epsilon_1^2
-2\,\left(C+1\right)\,\epsilon_1\,\epsilon_2 , \\
\label{eqn:bt1}
\mathrm{d} n_\mathrm{t}/\mathrm{d} \ln k =& - 2\epsilon_1\epsilon_2 \,,
\end{align}
where $C \equiv \ln 2+\gamma_\mathrm{E}-2\approx-0.7296$ ($\gamma_\mathrm{E}$
is the Euler-Mascheroni constant). See the Appendix of
\citetalias{planck2013-p17} for more details. Primordial spectra as functions of the $\epsilon_i$ will be employed in 
Sect.~\ref{sec:numerical}, and the expressions generalizing Eqs.~(\ref{bs}) 
to (\ref{eqn:bt1}) 
for a general Lagrangian $p(\phi,X)$, 
where $X \equiv -g^{\mu\nu}\partial_\mu\phi\partial_\nu\phi/2$, 
will be used in Sect.~\ref{sec:putfnl}. The good agreement between the first and second 
method as well as with alternative 
approximations of slow-roll spectra is illustrated in the Appendix of 
\citetalias{planck2013-p17}. 


The third method is fully numerical, suitable for models where the slow-roll conditions are not well 
satisfied and analytical approximations for the primordial fluctuations are not available. 
Two different numerical codes, the inflation module of \citet{Lesgourgues:2007gp}
as implemented in {\tt CLASS}~\citep{Lesgourgues:2011re,Blas:2011rf} and 
{\tt ModeCode}~\citep{Adams:2001vc,2003ApJS..148..213P, 2009PhRvD..79j3519M, 2012PhRvD..85j3533E},
are used in Sects.~\ref{sec:pot} and \ref{sec:putfnl}, 
respectively.~\footnote{\href{http://class-code.net}{http://class-code.net},\href{http://modecode.org}{http://modecode.org}.} 

Conventions for the functions and symbols used to describe inflationary 
physics are defined in Table~\ref{table:InflationDefinitions}.

{
\input InflationDefinitions.tex
}




\subsection{\Planck\ data}

\begin{figure}[!t]
\begin{center}
\includegraphics[angle=0,width=\columnwidth]{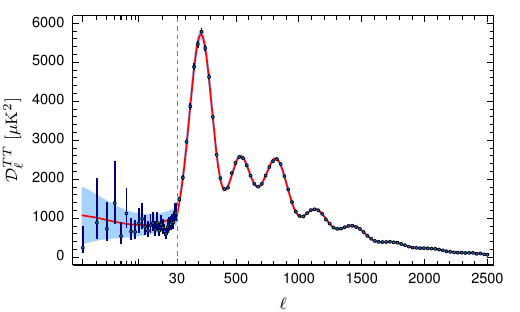}
\includegraphics[angle=0,width=\columnwidth]{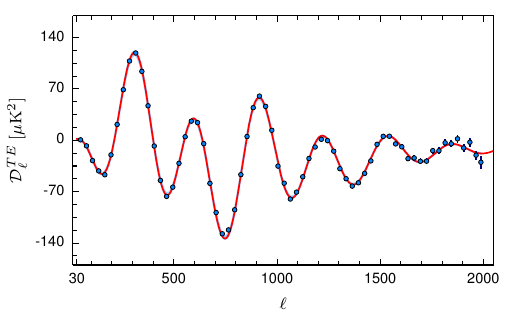}
\includegraphics[angle=0,width=\columnwidth]{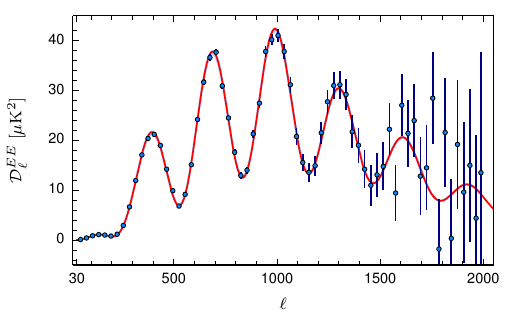}
\end{center}
\caption{\Planck\ $TT$ (top), high-$\ell$ $TE$ (centre), and high-$\ell$ $EE$ (bottom) angular power spectra.  
Here ${\cal D}_\ell \equiv \ell(\ell + 1)C_\ell/(2\pi)$.
\label{fig:cmbdata}}
\end{figure}

The \Planck\ data processing proceeding 
from time-ordered data to maps has been improved for this 2015 release in various aspects 
\citep{planck2014-A03,planck2014-A08}.
We refer the interested reader to \cite{planck2014-A03} and \cite{planck2014-A08} 
for details, and we describe here two of these improvements.
The absolute calibration has been improved using the orbital dipole and more accurate characterization of the \Planck\ beams.
The calibration discrepancy between \Planck\ and WMAP described in 
\cite{planck2013-p01a} for the 2013 release has now been greatly reduced.
At the time of that release, a blind analysis for 
primordial power spectrum reconstruction described a broad feature at $\ell \approx 1800$ 
in the temperature power spectrum, which was most prominent in the 217$\times$217\,GHz auto-spectra \citepalias{planck2013-p17}. 
In work done after the \Planck\ 2013 data release, this feature was shown to be associated with imperfectly subtracted 
systematic effects associated with the 4\,K cooler lines, which were particularly strong in the first survey. 
This systematic effect was shown to potentially lead to $0.5\,\sigma$
shifts in the cosmological parameters, slightly increasing $n_\mathrm{s}$ and $H_0$, 
similarly to the case in which the 217$\times$217 channel was excised 
from the likelihood \citep{planck2013-p08,planck2013-p11}. 
The \Planck\ likelihood \citep{planck2014-a13} is based on the
full mission data and comprises temperature and polarization data (see Fig.~\ref{fig:cmbdata}).

\subsubsection*{\Planck\ low-$\ell$ likelihood}

The \Planck\ low-$\ell$ temperature-polarization likelihood uses 
foreground-cleaned LFI $70$\,GHz polarization maps 
together with the temperature map obtained from the \Planck\ 30 to 353\,GHz channels 
by the {\tt Commander} component separation
algorithm over 94\,\% of the sky (see \cite{planck2014-a11} for further details).
The \Planck\ polarization map uses the  LFI $70$\,GHz (excluding Surveys 2 and 4) 
low-resolution maps of $Q$ and $U$ polarization from which polarized synchrotron and thermal
dust emission components have been removed using 
the LFI $30$\,GHz and HFI
$353$\,GHz maps as templates, respectively. (See 
\cite{planck2014-a13} for more details.) 
The polarization map covers the 46\,\% of the sky outside the
lowP polarization mask.

The low-$\ell$ likelihood 
is pixel-based and treats the temperature and polarization at the 
same resolution of $3\pdeg6$, or {\tt HEALpix} \citep{gorski2005} $N_\mathrm{side}=16$. Its multipole range extends from
$\ell = 2$ to $\ell=29$ in $TT$, $TE$, $EE$, and $BB$.
In the 2015 \Planck\ papers the polarization part of this likelihood is denoted as ``lowP.''\footnote{%
In this paper we use the conventions introduced in \cite{planck2014-a15}. 
We adopt the following labels for likelihoods:
(i) \Planck\ TT denotes the combination of the $TT$ likelihood at multipoles 
$\ell \geq 30$ and a low-$\ell$ temperature-only likelihood based on the CMB map 
recovered with {\tt Commander}; (ii) \Planck\ TT--lowT denotes the $TT$ likelihood at multipoles
$\ell \geq 30$; (iii) \Planck\ TT+lowP further includes the \Planck\ polarization 
data in the low-$\ell$ likelihood, as described in the main text; 
(iv) \Planck\ TE denotes the likelihood at $\ell \geq 30$ using the $TE$ spectrum;
and (v) \Planck\ TT,TE,EE+lowP denotes the combination of the likelihood at $\ell \geq 30$ using 
$TT$, $TE$, and $EE$ spectra and the low-$\ell$ multipole likelihood. 
The label ``$\tau$ prior" denotes the use of a Gaussian prior $\tau=0.07 \pm 0.02$.
The labels ``lowT,P" and ``lowEB" denote the low-$\ell$ multipole likelihood and the $Q,U$ pixel likelihood only, respectively.}
This \Planck\ low-$\ell$ likelihood replaces the \Planck\ temperature low-$\ell$ Gibbs module  
combined with the WMAP 9-year low-$\ell$ polarization module used in the \Planck\ 2013 cosmology papers 
(denoted by WP), which used lower resolution polarization maps at $N_\mathrm{side}=8$ (about $7\pdeg3$). 
With this \Planck-only low-$\ell$ likelihood module, the basic \Planck\ results presented in 
this release are completely independent of external information. 

The \Planck\ low-multipole likelihood alone implies $\tau = 0.067 \pm 0.022$ \citep{planck2014-a13}, 
a value smaller than the value inferred using the WP polarization likelihood, 
$\tau = 0.089 \pm 0.013$, used in the \Planck\ 2013 papers \citep{planck2013-p08}.
See \cite{planck2014-a15} for the important implications of this decrease in $\tau$ for reionization.
However, the LFI $70$\,GHz and WMAP polarization maps are in very good agreement when both
are foreground-cleaned using the HFI $353$\,GHz map as a polarized dust template (see
\cite{planck2014-a13} for further details).
Therefore, it is useful to construct a noise-weighted combination to obtain a joint
\Planck/WMAP low resolution polarization data set, also
described in \cite{planck2014-a13}, using as a polarization mask the union of
the WMAP P06 and \Planck\ lowP polarization
masks and keeping 74\,\% of the sky. 
The polarization part of the combined low multipole likelihood is called lowP+WP.
This combined low multipole likelihood gives $\tau = 0.071^{+0.011}_{-0.013}$ \citep{planck2014-a13}.

\subsubsection*{\Planck\ high-$\ell$ likelihood}

Following \cite{planck2013-p08}, and \cite{planck2014-a13} for polarization, 
we use a Gaussian approximation for  
the high-$\ell$ part of the likelihood ($30<\ell<2500$), so that
\begin{equation}
-\mathrm{log}{\cal L}\Bigl( \hat{C} | C(\boldsymbol{\theta })\Bigr) = 
\frac{1}{2} \Bigl( \hat{C} - C(\boldsymbol{\theta })\Bigr) ^T {\mathcal{M}}^{-1} 
\Bigl(\hat{C} - C(\boldsymbol{\theta })\Bigr),
\label{eq:basic-likelihood}
\end{equation}
where a constant offset has been discarded. Here
$\hat{C}$ is the data vector, 
$C(\boldsymbol{\theta })$ is the model prediction 
for the parameter value vector $\boldsymbol{\theta }$, and $\mathcal{M}$ 
is the covariance matrix.
For the data vector, we use 100\,GHz, 143\,GHz, and 217\,GHz half-mission cross-power spectra, 
avoiding the Galactic plane as well as the brightest point sources and 
the regions where the CO emission is the strongest.  
We retain 66\,\% 
of the sky for 100\,GHz, 57\,\% for 143\,GHz, and 47\,\% for 217\,GHz for the $T$ masks, and respectively 
70\,\%, 50\,\%, and 41\,\% for the
$Q$, $U$ masks. Following \cite{planck2013-pip56}, we do not mask for any other 
Galactic polarized emission. All the spectra are corrected for the beam and pixel window functions
using the same beam for temperature and polarization.
(For details see \cite{planck2014-a13}.)

The model for the cross-spectra can be written as 
\begin{equation}
C_{\mu,\nu}(\theta) = \frac{ C^{\mathrm{cmb}}(\theta) + C^{\mathrm{fg}}_{\mu,\nu}(\theta) }{\sqrt{\mathrm{c}_\mu \mathrm{c}_\nu}},
\end{equation}
where $C^{\mathrm{cmb}}(\theta)$ is the CMB power spectrum, which is independent 
of the frequency, $C^{\mathrm fg}_{\mu,\nu}(\theta)$ is the foreground 
model contribution for the cross-frequency spectrum $\mu\times\nu$, and 
$\mathrm{c}_\mu$ is the calibration factor for the $\mu\times\mu$ spectrum.   
The model for the foreground residuals includes the following components:
Galactic dust, clustered cosmic infrared background (CIB), thermal and kinetic 
Sunyaev-Zeldovich (tSZ and kSZ) effect, tSZ correlations with CIB, and point 
sources, for the  $TT$ foreground modeling; and for polarization, only 
dust is included. All the components are modelled by smooth $C_\ell$ 
templates with free amplitudes, which are determined along with the cosmological 
parameters as the likelihood is explored.  The tSZ and kSZ 
models are the same as in 2013 \citep[see][]{planck2013-p08}, 
although with different priors \citep{planck2014-a13,planck2014-a15}, 
while the CIB and tSZ-CIB correlation models use the updated CIB models described in \citet{planck2013-pip56}. 
The point source contamination is modelled as Poisson noise with an independent 
amplitude for each frequency pair.  Finally, the dust contribution uses an effective 
smooth model measured from high frequency maps. Details of our dust and noise 
modelling can be found in \cite{planck2014-a13}.
The dust is the dominant foreground component for $TT$ at $\ell<500$, while the point source component, 
and for 217$\times$217 also the CIB component, dominate at high $\ell$. The other foreground components
are poorly determined by \Planck.  Finally, our treatment of the calibration 
factors and beam uncertainties and mismatch are described in \cite{planck2014-a13}.

The covariance matrix accounts for the correlation due to the mask and is computed following the equations 
in \cite{planck2013-p08}, extended to polarization in \cite{planck2014-a13} and references therein. 
The fiducial model used to compute the covariance is based on a joint fit of base $\Lambda$CDM and 
nuisance parameters obtained with a previous 
version of the matrix. We iterate the process until the parameters stop changing. 
For more details, see \cite{planck2014-a13}.

The joint unbinned covariance matrix is approximately of size 23\,000$\,\times\,$23\,000. The memory and speed requirements for 
dealing with such a huge matrix are significant, so to reduce its size, we bin the data and the 
covariance matrix to compress the data vector size by a factor of 10. The binning uses varying bin width 
with $\Delta\ell=5$ for $29<\ell<100$, 
$\Delta\ell=9$ for $99<\ell<1504$, $\Delta\ell=17$ for $1503<\ell<2014$, and $\Delta\ell=33$ for $2013<\ell<2509$, and 
a weighting in $\ell(\ell+1)$ to flatten the spectrum. 
Where a higher resolution is desirable, we also use a more finely binned 
version (``bin3", unbinned up to $\ell = 80$ and $\Delta \ell = 3$ beyond that) as well as 
a completely unbinned version (``bin1"). We use odd bin sizes, since for an azimuthally symmetric
mask, the correlation between a multipole and its neighbours is symmetric, oscillating between positive and negative values. 
Using the base $\Lambda$CDM model and single-parameter classical extensions, 
we confirmed that the cosmological and nuisance parameter fits with or without binning are indistinguishable.

As discussed in \citet{planck2014-a13} and \citet{planck2014-a15}, 
the $TE$ and $EE$ high-$\ell$ data are not free of small systematic effects, 
such as leakage from temperature to polarization.
Although the propagated effects of these residual systematics on 
cosmological parameters are small and do not alter the conclusions
of this paper, we mainly refer to \Planck\ TT+lowP in combination with the \Planck\ lensing or additional data
sets as the most reliable results for this release. 

\subsubsection*{\Planck\ CMB bispectrum}

We use measurements of the non-Gaussianity
amplitude $f_\mathrm{NL}$ from the CMB bispectrum presented in~\cite{planck2014-a19}. 
Non-Gaussianity constraints have been obtained using three optimal bispectrum estimators: separable
template fitting (also known as ``KSW''), binned, and modal. The maps analysed are the \Planck\ 2015 full mission sky
maps, both in 
temperature and in $E$ polarization, as cleaned with the four component separation
methods {\tt SMICA}, 
{\tt SEVEM}, {\tt NILC}, and {\tt Commander}. The map is masked to remove the brightest parts of the 
Galaxy as well as the brightest point sources and covers approximately $70\,\%$ of the sky. 
In this paper we mainly exploit the joint constraints on equilateral and 
orthogonal non-Gaussianity (after removing the integrated Sachs-Wolfe effect-lensing bias),
$f_\mathrm{NL}^\mathrm{equil} 
= -16 \pm 70$, $f_\mathrm{NL}^\mathrm{ortho} = -34 \pm 33$ from $T$ only,
and $f_\mathrm{NL}^\mathrm{equil} = -3.7 \pm 43$, $f_\mathrm{NL}^\mathrm{ortho} =
-26 \pm
21$ from $T$ and $E$ (68$\,\%$ CL). For reference, the constraints on local
non-Gaussianity are 
$f_\mathrm{NL}^\mathrm{local}=2.5 \pm 5.7$ from $T$ only, and   
$f_\mathrm{NL}^\mathrm{local}=0.8 \pm 5.0$ from $T$ and $E$ (68$\,\%$ CL).  
Starting from a Gaussian $f_\mathrm{NL}$-likelihood, which is an accurate 
assumption in the regime of small primordial non-Gaussianity, we use these constraints 
to derive limits on the sound speed of the inflaton fluctuations (or other
microscopic parameters 
of inflationary models)~\citep{planck2013-p09a}. The bounds on the sound speed for various models are
then used in combination with \Planck\ power spectrum data.

\subsubsection*{\Planck\ CMB lensing data}

Some of our analysis includes the \Planck\ 2015 lensing likelihood, presented in
\cite{planck2014-a17}, which utilizes the non-Gaussian trispectrum induced by
lensing to estimate the power spectrum of the lensing potential, $C_\ell ^{\phi\phi}$.
This signal is extracted using a full set of temperature- and polarization-based
quadratic lensing estimators \citep{Okamoto:2003zw} applied to the {\tt SMICA} CMB
map over approximately $70\,\%$ of the sky, as described in \cite{planck2014-a11}. 
We
have used the conservative bandpower likelihood, covering multipoles $40 \le \ell \le
400$. This provides a measurement of the lensing potential power at the
$40\,\sigma$ level, giving a $2.5\,\%$-accurate constraint on the overall lensing power in this
multipole range. The measurement of the lensing power spectrum used here is
approximately twice as powerful as the measurement used in our previous 2013 analysis
\citep{planck2013-p17,planck2013-p12}, which used temperature-only data from the
\Planck\ nominal mission data set. 

\subsection{Non-\Planck\ data}

\subsubsection*{BAO data}

Baryon acoustic oscillations (BAO) are the counterpart in the late time matter power
spectrum of the acoustic oscillations seen in the CMB multipole spectrum \citep{Eisenstein:2005su}. 
Both originate from coherent oscillations of the photon-baryon plasma before these two
components become decoupled at recombination. Measuring the position 
of these oscillations in the matter power spectra at different redshifts
constrains the expansion history of the universe after decoupling, thus
removing degeneracies in the interpretation of the CMB anisotropies.

In this paper, we combine 
constraints on $D_V(\bar{z})/r_\mathrm{s}$ (the ratio between the spherically-averaged 
distance scale  $D_V$ to the effective survey redshift, $\bar{z}$, and the sound 
horizon, $r_\mathrm{s}$) inferred from 6dFGRS data \citep{Beutler:2011hx} 
at $\bar{z} = 0.106$, the SDSS-MGS data \citep{Ross:2014qpa} at $\bar{z} = 0.15$, 
and the SDSS-DR11 CMASS and LOWZ data \citep{Anderson:2013zyy} at 
redshifts $\bar{z} = 0.57$ and $0.32$. For details see \cite{planck2014-a15}.

\subsubsection*{Joint BICEP2/Keck Array and \Planck\ constraint on $r$}

Since the \Planck\ temperature constraints on the 
tensor-to-scalar ratio are close to the cosmic variance limit, the 
inclusion of data sets sensitive to the expected $B$-mode signal of primordial 
gravitational waves is particularly useful. In this paper, we provide
results including the joint analysis cross-correlating BICEP2/Keck Array observations 
and \Planck\ \citepalias{pb2015}. Combining 
the more sensitive BICEP2/Keck Array $B$-mode polarization maps in
the approximately 400\,deg$^2$ BICEP2 field with the \Planck\
maps at higher frequencies where dust dominates  allows a statistical 
analysis taking into account foreground contamination. 
Using $BB$ auto- and 
cross-frequency spectra between BICEP2/Keck Array (150\,GHz) and \Planck\ 
(217 and 353\,GHz), \citetalias{pb2015} find a 95\,\% upper limit of 
$r_{0.05} < 0.12$.

\subsection{Parameter estimation and model comparison}
Much of this paper uses a Bayesian approach to parameter 
estimation, and unless otherwise specified, we assign broad top-hat prior 
probability distributions to the cosmological parameters listed in 
Table~\ref{table:CPDefinitions}. We generate posterior probability 
distributions for the parameters using either the Metropolis-Hastings 
algorithm implemented in {\tt CosmoMC}~\mbox{\citep{Lewis:2002ah}} or 
{\tt MontePython}~\mbox{\citep{Audren:2012wb}}, the nested sampling 
algorithm {\tt MultiNest} \citep{Feroz:2007kg,Feroz:2008xx,Feroz:2013hea}, 
or {\tt PolyChord}, which combines nested sampling with slice 
sampling~\citep{Handley:2015fda}. The latter two also compute the Bayesian evidence 
needed for model comparison. Nevertheless, $\chi ^2$ values are often 
provided as well (using \texttt{CosmoMC}'s implementation of the BOBYQA 
algorithm~\citep{bobyqa} for maximizing the likelihood), 
and other parts of the paper employ frequentist methods when appropriate. 


%% file: CPDefinitions.tex
\begin{table*}
\begingroup
\newdimen\tblskip \tblskip=5pt
\caption{Primordial, baseline, and optional late-time cosmological parameters.}
\label{table:CPDefinitions}
\nointerlineskip
\vskip 3mm
\footnotesize
\setbox\tablebox=\vbox{
      \newdimen\digitwidth
      \setbox0=\hbox{\rm 0}
      \digitwidth=\wd0
      \catcode`"=\active
      \def"{\kern\digitwidth}
      \newdimen\signwidth
      \setbox0=\hbox{+}
      \signwidth=\wd0
      \catcode`!=\active
      \def!{\kern\signwidth}
\halign{\tabskip=0pt\hbox to 1.0in{$#$\leaderfil}\tabskip=3em&
 #\hfil\tabskip=0pt\cr
\noalign{\doubleline}
\omit\hfil Parameter\hfil&\omit\hfil Definition\hfil\cr
\noalign{\vskip 3pt\hrule\vskip 5pt}
A_{\mathrm{s}}& Scalar power spectrum amplitude
 (at $k_*=0.05\,{\mathrm{Mpc}}^{-1}$)\cr
n_\mathrm{s}& Scalar spectral index (at $k_*=0.05\,{\mathrm{Mpc}}^{-1}$
 unless otherwise stated)\cr
\mathrm{d}n_\mathrm{s}/\mathrm{d}\ln k& Running of scalar spectral index
 (at $k_*=0.05\,{\mathrm{Mpc}}^{-1}$ unless otherwise stated) \cr
\mathrm{d}^2 n_\mathrm{s}/\mathrm{d} \ln k^2& Running of running of scalar
 spectral index (at $k_*=0.05\,{\mathrm{Mpc}}^{-1}$)\cr
r& Tensor-to-scalar power ratio (at $k_*=0.05\,{\mathrm{Mpc}}^{-1}$ unless
 otherwise stated)\cr
n_\mathrm{t}& Tensor spectrum spectral index (at $k_*=0.05\,{\mathrm{Mpc}}^{-1}$)\cr
\noalign{\hrule\vskip 2pt}
\omega_\mathrm{b} \equiv \Omega_\mathrm{b} \, h^2 & Baryon density today\cr
\omega_\mathrm{c} \equiv \Omega_\mathrm{c} \, h^2  & Cold dark matter density today\cr
\theta_\mathrm{MC}& Approximation to the angular size of sound horizon at
 last scattering\cr
\tau& Thomson scattering optical depth of reionized intergalactic medium\cr
\noalign{\hrule\vskip 2pt}
N_\mathrm{eff}& Effective number of massive and massless neutrinos\cr
\Sigma m_\nu& Sum of neutrino masses\cr 
Y_\mathrm{P}& Fraction of baryonic mass in primordial helium\cr
\Omega_K& Spatial curvature parameter\cr
w_\mathrm{de}& Dark energy equation of state parameter
 (i.e., $p_\mathrm{de}/\rho_\mathrm{de}$) (assumed constant)\cr
\noalign{\vskip 2pt\hrule\vskip 3pt}}}
\endPlancktable
\endgroup
\end{table*}

%% file: InflationDefinitions.tex
\begin{table*}
\begingroup
\newdimen\tblskip \tblskip=5pt
\caption{Conventions and definitions for inflation physics.}
\label{table:InflationDefinitions}
\nointerlineskip
\vskip -3mm
\footnotesize
\setbox\tablebox=\vbox{
      \newdimen\digitwidth
      \setbox0=\hbox{\rm 0}
      \digitwidth=\wd0
      \catcode`"=\active
      \def"{\kern\digitwidth}
      \newdimen\signwidth
      \setbox0=\hbox{+}
      \signwidth=\wd0
      \catcode`!=\active
      \def!{\kern\signwidth}
\halign{\tabskip=0pt\hbox to 1.2in{$#$\leaderfil}\tabskip=3em&
 #\hfil\tabskip=0pt\cr
\noalign{\doubleline}
\omit\hfil Parameter\hfil&\omit\hfil Definition\hfil\cr
\noalign{\vskip 3pt\hrule\vskip 5pt}
\phi& Inflaton\cr
V(\phi)& Inflaton potential\cr
a& Scale factor\cr
t& Cosmic (proper) time\cr
\delta X& Fluctuation of $X$\cr
\dot{X} = dX/dt& Derivative with respect to proper time\cr
X' = dX/d\eta& Derivative with respect to conformal time\cr 
X_\phi  = \partial X /\partial \phi& Partial derivative with respect
 to $\phi$\cr
M_\mathrm{pl}& Reduced Planck mass ($=2.435 \times 10^{18}$\,GeV)\cr
\cal R& Comoving curvature perturbation\cr
h^{+,\times}& Gravitational wave amplitude of $(+,\times)$-polarization
 component \cr 
X_*& $X$ evaluated at Hubble exit during inflation of mode with wavenumber
 $k_*$\cr
X_{\mathrm e}& $X$ evaluated at end of inflation\cr
\epsilon _V={M_\mathrm{pl}^2 V_\phi^2}/(2 V^2)& First slow-roll parameter for
 $V (\phi)$\cr
\eta _V={M_\mathrm{pl}^2 V_{\phi \phi}}/V& Second slow-roll parameter for
 $V(\phi)$\cr
\xi^2_V={M_\mathrm{pl}^4 V_{\phi} V_{\phi \phi \phi}}/{V^2}& Third slow-roll
 parameter for $V (\phi)$\cr
\varpi^3_V={M_\mathrm{pl}^6 V_{\phi}^2 V_{\phi \phi \phi \phi}}/{V^3}&
 Fourth slow-roll parameter for $V (\phi)$\cr      
\epsilon_1 = - {\dot H}/{H^2}& First Hubble hierarchy parameter\cr
\epsilon_{n+1}={\dot \epsilon_n}/(H \epsilon_n)& $(n+1)$st Hubble hierarchy
 parameter (where $n \ge 1$)\cr
N (t) =\int_{t}^{t_{\mathrm e}} \! \mathrm{d}t \; H& Number of $e$-folds to
 end of inflation\cr
\noalign{\vskip 2pt\hrule\vskip 3pt}}}
\endPlancktable
\endgroup
\end{table*}

%% file: section_four.tex
\input macros.tex
\def\ba{\begin{eqnarray}}
\def\ea{\end{eqnarray}}
\def\be{\begin{equation}}
\def\ee{\end{equation}}

One of the most important results of the \Planck\ nominal mission was the determination of the departure
from scale invariance for the spectrum of scalar perturbations at high statistical significance
\citep{planck2013-p11,planck2013-p17}.
We now update these measurements with the \Planck\ full mission data in temperature and polarization.

\subsection{Tilt of the curvature power spectrum}

\input table_LCDM.tex

\begin{figure*}[!ht]
\includegraphics[width=\textwidth,angle=0]{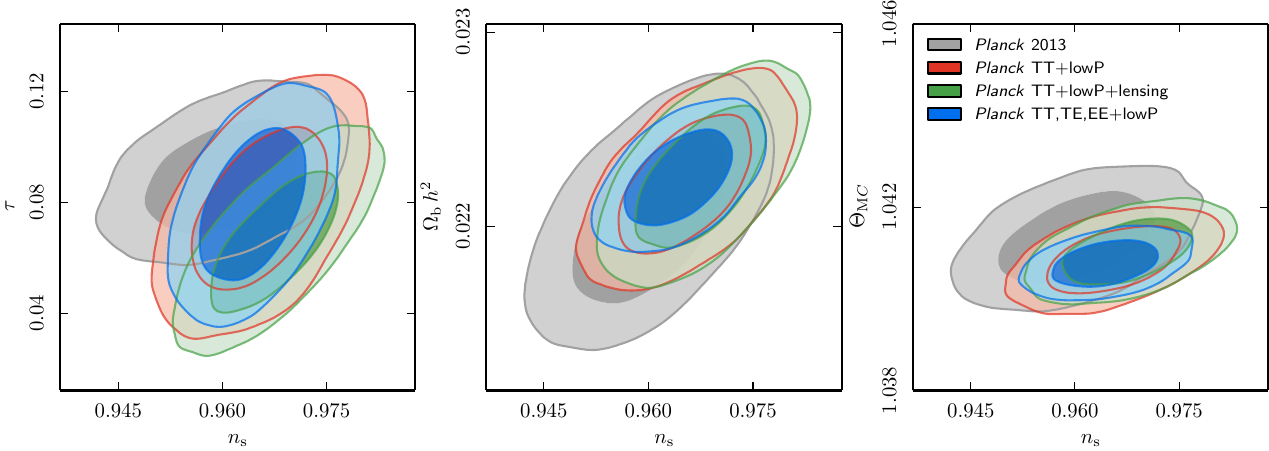}
\caption{Comparison of the marginalized joint 68\,\% and 95\,\% CL constraints on
$(n_{\mathrm s} \,, \tau)$ (left panel), $(n_{\mathrm s} \,, \Omega_\mathrm{b} h^2)$ (middle panel), and
$(n_{\mathrm s} \,, \theta_{\mathrm{MC}})$ (right panel),
for \Planck\ 2013 (grey contours), \Planck\ TT+lowP (red contours), \Planck\ TT+lowP+lensing (green contours),
and \Planck\ TT,TE,EE+lowP (blue contours).}
\label{fig:taunsomegab} 
\end{figure*}

For the base $\Lambda$CDM model with a power-law power spectrum of curvature perturbations, the constraint on
the scalar spectral index, $n_\mathrm{s}$, with
the \Planck\ full mission temperature data is
\be
\label{eq:TT_ns}
n_\mathrm{s} = 0.9655 \pm 0.0062 \, \, (68\,\%\ \text{CL, \Planck\ TT+lowP}) \,.
\ee
This result is compatible with the \Planck\ 2013 constraint, 
$n_\mathrm{s} = 0.9603 \pm 0.0073$ \citep{planck2013-p08,planck2013-p11}.
See Fig.~\ref{fig:taunsomegab} for the accompanying changes in $\tau$, $\Omega_\mathrm{b} h^2$, and $\theta_\mathrm{MC}.$
The shift towards higher values for $n_\mathrm{s}$ with respect to the nominal mission results
is due to several improvements in the data processing and likelihood which are discussed in Sect.~\ref{sec:methodology}, including the removal
of the 4\,K cooler systematics.
For the values of other cosmological parameters in the base $\Lambda$CDM model,
see Table~\ref{tab:lambdaCDM}. We also provide the results for the base $\Lambda$CDM model
and extended models online.\footnote{\url{http://www.cosmos.esa.int/web/planck/pla}}

When the \Planck\ high-$\ell$ polarization is combined with temperature, we obtain
\be
n_\mathrm{s} = 0.9645 \pm 0.0049 \, \, (68\,\%\ \text{CL, \Planck\ TT,TE,EE+lowP}),
\ee
together with $\tau = 0.079 \pm 0.017$ (68\,\% CL), which is consistent with the TT+lowP results.
The \Planck\ high-$\ell$ polarization pulls $\tau$ up to a slightly higher value.
When the \Planck\ lensing measurement is added to the temperature data, we obtain
\be
n_\mathrm{s} = 0.9677 \pm 0.0060 \, \, (68\,\%\ \text{CL, \Planck\ TT+lowP+lensing}),
\ee
with $\tau = 0.066 \pm 0.016$ (68\,\% CL). The shift towards slightly smaller values of the
optical depth is driven by a marginal preference for a smaller primordial amplitude,
$A_\mathrm{s}$, in the \Planck\ lensing data \citep{planck2014-a17}. Given that the
temperature data provide a sharp constraint on the combination $e^{-2 \tau} A_\mathrm{s}$,
a slightly lower $A_\mathrm{s}$ requires a smaller optical depth to reionization.

\subsection{Viability of the Harrison-Zeldovich spectrum}

Even though the estimated scalar spectral index has
risen slightly with respect to the \Planck\ 2013 release,
the assumption of a Harrison-Zeldovich (HZ) scale-invariant
spectrum~\citep{Harrison:1969fb,Peebles:1970ag,Zeldovich:1972zz} continues to be 
disfavoured (with a modest increase in significance, from $5.1\,\sigma $ in 2013 
to $5.6\,\sigma $ today), because the error bar on $n_{\mathrm s}$ has decreased. The value of $n_\mathrm{s}$ inferred from the \Planck\
2015 temperature plus large-scale polarization data lies 5.6 standard deviations
away from unity (with a corresponding $\Delta \chi^2 = 29.9$),
if one assumes the base $\Lambda$CDM late-time cosmological model. If we consider
more general reionization models, parameterized by a principal component analysis
\citep{Mortonson:2007hq} instead of $\tau$
(where reionization is assumed to have occurred instantaneously),
we find $\Delta \chi^2 = 14.9$ for
$n_\mathrm{s} = 1$. Previously, simple one-parameter extensions of the base model, such as $\Lambda$CDM+$N_\mathrm{eff}$
(where $N_\mathrm{eff}$ is the effective number of neutrino flavours)
or
$\Lambda$CDM+$Y_\mathrm{P}$
(where $Y_\mathrm{P}$ is the primordial value of the helium mass fraction),
could nearly reconcile the \Planck\
temperature data with $n_\mathrm{s}=1$. They now lead to $\Delta \chi^2 = 7.6$ and $9.3$, respectively.
For any of the cosmological models that we have considered, the $\Delta \chi^2$ by which the HZ model
is penalized with respect to the tilted model has increased since the 2013 analysis \citepalias{planck2013-p17}
thanks to the constraining power of the full mission temperature data.
Adding \Planck\ high-$\ell$ polarization data further disfavours the HZ model:
in $\Lambda$CDM, the $\chi^2$ increases by $57.8$, for general reionization we obtain $\Delta \chi^2 = 41.3$,
and for $\Lambda$CDM+$N_\mathrm{eff}$
and $\Lambda$CDM+$Y_\mathrm{P}$ we find $\Delta \chi^2 = 22.5$ and $24.0$, respectively.

\subsection{Running of the spectral index}

The running of the scalar spectral index is constrained by the \Planck\ 2015 full mission temperature data to
\be
\frac{d n_\mathrm{s}}{d \ln k} = -0.0084 \pm 0.0082 \, \, (68\,\%\ \text{CL, \Planck\ TT+lowP}) \,.
\ee
The combined constraint including high-$\ell$ polarization is
\be
\frac{d n_\mathrm{s}}{d \ln k} = -0.0057 \pm 0.0071 \, \, (68\,\%\ \text{CL, \Planck\ TT,TE,EE+lowP}) \,.
\ee
Adding the \Planck\ CMB lensing data to the temperature data further reduces the central value for the running,
i.e., $dn_\mathrm{s}/d\ln k = -0.0033 \pm 0.0074$ ($68\,\%$ CL, \Planck\ TT+lowP+lensing).

\begin{figure}[!h]
\includegraphics[height=0.37\textwidth,angle=0]{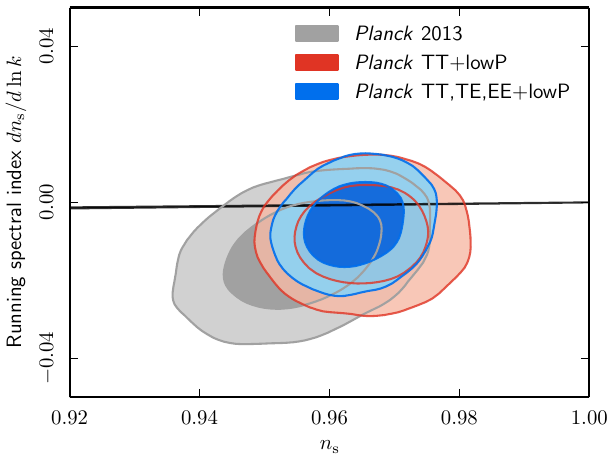}
\caption{Marginalized joint 68\,\% and 95\,\% CL for
$(n_{\mathrm s} \,, \mathrm{d} n_{\mathrm s}/\mathrm{d} \ln k)$
using \Planck\ TT+lowP and \Planck\ TT,TE,EE+lowP. Constraints from the \Planck\ 2013 data release are also shown for comparison. 
For comparison, the thin black stripe shows the
prediction for single-field monomial chaotic inflationary
models with $50 < N_* <60$. \label{fig:nalpha}}
\end{figure}

The central value for the running has decreased in magnitude with respect to the \Planck\ 2013 nominal
mission (\cite{planck2013-p11} found $dn_\mathrm{s}/d\ln k = -0.013 \pm 0.009$; see Fig.~\ref{fig:nalpha}),
and the improvement of the maximum likelihood with respect
to a power-law spectrum is smaller, $\Delta \chi^2 \approx -0.8$.
Among the different effects contributing to the decrease in
the central value of the running with respect to the \Planck\ 2013 result, we mention a change in HFI beams at
$\ell \lsim 200$ \citep{planck2014-a15}.
Nevertheless, the deficit of power at low multipoles in the \Planck\ 2015 temperature power spectrum contributes to
a preference for slightly negative values of the running, but with low statistical significance.

The \Planck\ constraints on $n_\mathrm{s}$ and $dn_\mathrm{s}/d\ln k$
are remarkably stable against the addition of the BAO likelihood. The combination with BAO shifts $n_\mathrm{s}$ to slighly higher values and
shrinks its uncertainty by about $30$\,\% when only high-$\ell$ temperature is considered,
and by only about $15$\,\% when high-$\ell$ temperature and polarization are combined.
In slow-roll inflation, the running of the scalar spectral index is
connected to the third derivative of the potential \citep{Kosowsky:1995aa}.
As was the case for the nominal mission results, values of the
running compatible with the \Planck\ 2015 constraints can be obtained in viable inflationary
models \citep{Kobayashi:2010pz}.

When the running of the running is allowed to float,
the \Planck\ TT+lowP (\Planck\ TT,TE,EE+lowP) data give:
\begin{eqnarray}
&& n_\mathrm{s} = 0.9569\pm0.0077~~(0.9586 \pm 0.0056)~, \nonumber \\
&& d n_\mathrm{s}/d \ln k = 0.011^{+0.014}_{-0.013}~~(0.009 \pm 0.010)~,
~~~~~~(68\,\%~\mathrm{CL})\\
&& d^2 n_\mathrm{s}/d \ln k^2 = 0.029^{+0.015}_{-0.016}~~(0.025 \pm 0.013)~, \nonumber
\end{eqnarray}
at the pivot scale $k_*= 0.05$\,Mpc$^{-1}$.
Allowing for running of the running provides a better fit to the temperature spectrum at low multipoles, such that
$\Delta \chi^2 \approx -4.8$ ($-4.9$) for TT+lowP (TT,TE,EE+lowP), but is not statistically preferred over the simplest 
$\Lambda$CDM model.

Note that the inclusion of small-scale data such as Ly$\alpha$ might further constrain the running of the spectral index and its derivative. The recent analysis of the BOSS one-dimensional Ly$\alpha$ flux power spectrum presented in \cite{Palanque-Delabrouille:2014jca} and \cite{Rossi:2014nea} was optimized for measuring the neutrino mass. It does not include constraints on the spectral index running, which would require new dedicated $N$-body simulations. Hence we do not include  Ly$\alpha$ constraints here.

In Sect.~\ref{sec:pot} on inflaton potential reconstruction we will show that the data cannot accomodate a significant running but are compatible 
with a larger running of the running.

\subsection{Suppression of power on the largest scales \label{LargeScaleSuppression}}

Although not statistically significant, the trend for a negative running or positive running of the running observed
in the last subsection was driven by the lack of power in the \Planck\ temperature power spectrum at low multipoles,
already mentioned in the \Planck\ 2013 release. This deficit could potentially be explained by a primordial
spectrum featuring a depletion of power {\it only} at large wavelengths. Here we investigate two examples of such models.

We first update the analysis (already presented in \citetalias{planck2013-p17})
of a power-law spectrum multiplied by an exponential cutoff:
\begin{equation}
\mathcal{P_R}(k) = \mathcal{P}_{0}(k) \left\{ 1 - \exp \left[- \left( \frac{k}{k_\mathrm{c}}
\right)^{\lambda_\mathrm{c}} \right] \right\}.
\label{eq:cutoff}
\end{equation}
This simple parameterization is motivated by models with
a short inflationary stage in which the onset of the slow-roll phase coincides with the time when the
largest observable scales exited the Hubble radius during inflation.
The curvature spectrum is then strongly suppressed on those scales.
We apply top-hat priors on the parameter $\lambda_\mathrm{c}$, controlling the steepness of the cutoff,
and on the logarithm of the cutoff scale, $k_\mathrm{c}$. We choose prior ranges
$\lambda_\mathrm{c} \in [0,10]$ and $\ln(k_\mathrm{c}/\mathrm{Mpc}^{-1}) \in [-12 \,, -3]$.
For \Planck\ TT+lowP (\Planck\ TT,TE,EE+lowP),
the best-fit model has $\lambda_\mathrm{c} = 0.50$ ($0.53$),
$\ln(k_\mathrm{c}/\mathrm{Mpc}^{-1}) = -7.98$ ($-7.98$), $n_\mathrm{s}=0.9647$ ($0.9649$),
and improves the effective $\chi^2$ by a modest amount, $\Delta \chi^2 \approx -3.4$ ($-3.4$).

As a second model, we consider a broken power-law spectrum for curvature perturbations:
\be
\mathcal{P_R}(k) = \begin{cases}
~~A_\mathrm{low} \left( \frac{k}{k_*} \right)^{n_\mathrm{s}-1 + \delta} & \text{if } k \le
k_\mathrm{b}, \\
~~A_\mathrm{s} \left( \frac{k}{k_*} \right)^{n_\mathrm{s}-1}  & \text{if } k \ge k_\mathrm{b},
\end{cases}
\ee
with $A_\mathrm{low}=A_\mathrm{s} (k_\mathrm{b}/k_*)^{-\delta}$ to ensure continuity at $k=k_\mathrm{b}$. Hence
this model, like the previous one, has two parameters, and also suppresses power at large wavelengths when $\delta > 0$.
We assume top-hat priors $\delta \in [0,2]$ and
$\ln(k_\mathrm{b}/\mathrm{Mpc}^{-1}) \in [-12 \,, -3]$, and standard uniform priors for $\ln(10^{10} A_\mathrm{s})$ and $n_\mathrm{s}$.
The best fit to \Planck\ TT+lowP (\Planck\ TT,TE,EE+lowP) is found for
$n_\mathrm{s} = 0.9658$ ($0.9647$), $\delta = 1.14$ ($1.14$), and
$\ln (k_\mathrm{b}/\mathrm{Mpc}^{-1}) = -7.55$ ($-7.57$), with a very small
$\chi ^2$ improvement of $\Delta \chi^2 \approx -1.9$ ($-1.6$).

We conclude that neither of these two models with two extra parameters is preferred over the base $\Lambda$CDM model.
(See also the discussion of a step inflationary potential in Sect.~\ref{step:potential}.)

%% file: table_LCDM.tex
\begin{table*}
\begin{center}
\begin{tabular}{ccccc}
\noalign{\hrule\vskip 2pt}
\noalign{\hrule\vskip 3pt}
Parameter & TT+lowP & TT+lowP+lensing & TT+lowP+BAO & TT,TE,EE+lowP
\\
\noalign{\vskip 1pt\hrule\vskip 2pt}
$\Omega_{\mathrm{b}}h^2$& $0.02222\pm0.00023$& $0.02226\pm0.00023$
& $0.02226\pm0.00020$ &$0.02225\pm0.00016$
\\
$\Omega_{\mathrm{c}}h^2$& $0.1197\pm0.0022$& $0.1186\pm0.0020$
& $0.1190\pm0.0013$ &$0.1198\pm 0.0015$
\\
$100\theta_{\mathrm{MC}}$& $1.04085\pm0.00047$& $1.04103\pm0.00046$
& $1.04095\pm0.00041$ &$1.04077\pm0.00032$
\\
$\tau$&$0.078\pm0.019$& $0.066\pm0.016$
& $0.080\pm0.017$ &$0.079\pm 0.017$
\\
$\ln(10^{10} A_\mathrm{s})$& $3.089\pm0.036$& $3.062\pm0.029$
& $3.093\pm0.034$ &$3.094\pm0.034$
\\
$n_\mathrm{s}$&$0.9655\pm0.0062$& $0.9677\pm0.0060$
& $0.9673\pm0.0045$ &$0.9645\pm 0.0049$
\\
\noalign{\vskip 1pt\hrule\vskip 3pt}
$H_0$&$67.31\pm0.96\phantom{0}$ & $67.81\pm0.92\phantom{0}$
& $67.63\pm0.57\phantom{0}$ &$67.27\pm0.66\phantom{0}$
\\
$\Omega_{\mathrm{m}}$& $0.315\pm0.013$& $0.308\pm0.012$
& $0.3104\pm0.0076$ &$0.3156\pm0.0091$
\\
\hline
\end{tabular}
\end{center}
\caption{Confidence limits on the parameters of the base $\Lambda$CDM model, for various combinations of \Planck\ 2015 data, at the 68\,\% confidence level.}
\label{tab:lambdaCDM}
\end{table*}

%% file: section_five.tex
\input macros.tex
\def\ba{\begin{eqnarray}}
\def\ea{\end{eqnarray}}

In this section, we focus on the \Planck\ 2015 constraints on tensor perturbations.
Unless otherwise stated, we consider that the tensor spectral index satisfies the
standard inflationary consistency condition to lowest order in slow roll, $n_\mathrm{t} = -r/8$.
We recall that $r$ is defined at the pivot scale $k_*=0.05\,$Mpc$^{-1}$. However, for
comparison with other studies, we also report our bounds in terms of the tensor-to-scalar ratio $r_{0.002}$ at $k_*=0.002\,$Mpc$^{-1}$.

\subsection{\Planck\ 2015 upper bound on $r$}

The constraints on the tensor-to-scalar ratio inferred from the
\Planck\ full mission data for the $\Lambda$CDM+$r$ model are:
\ba
\label{erre_t}
\quad r_{0.002} & < & 0.10 \quad \text{(95\,\% CL, \Planck\ TT+lowP)} \,, \\
\label{erre_tpluslensing}
\quad r_{0.002} & < & 0.11 \quad \text{(95\,\% CL, \Planck\ TT+lowP+lensing)} \,,\\
\quad r_{0.002} & < & 0.11 \quad \text{(95\,\% CL, \Planck\ TT+lowP+BAO)} \,, \\
\label{erre_tplusp}
\quad r_{0.002} & < & 0.10  \quad \text{(95\,\% CL, \Planck\ TT,TE,EE+lowP)} \,.
\ea
Table~\ref{tab:ralphar} also shows the bounds on $n_\mathrm{s}$ in each of these cases.

These results slightly improve over the constraint $r_{0.002} < 0.12$
(95\,\% CL) derived from the \Planck\ 2013 temperature data in combination
with WMAP large-scale polarization data
\citep{planck2013-p11,planck2013-p17}. 
The constraint obtained by \Planck\ temperature and 
polarization on large scales is tighter than the \Planck\ $B$-mode 95\,\% CL upper limit from the 100 and 143\,GHz HFI 
channels, $r < 0.27$ \citep{planck2014-a13}. 
The constraints on $r$ reported
in Table~\ref{tab:ralphar} can be translated into upper bounds on the
energy scale of inflation at the time when the pivot scale exits the
Hubble radius using
\begin{equation}
V_* = \frac{3 \pi^2 A_{\mathrm{s}}}{2} \, r \, M_{\mathrm {pl}}^4
= (1.88 \times 10^{16}~{\mathrm{GeV}} )^4  \frac{r}{0.10}~.
\end{equation}
This gives an upper bound on the Hubble parameter during inflation of 
$H_*/M_{\mathrm {pl}} < 3.6 \times 10^{-5}$ (95\,\% CL) for \Planck\ TT+lowP.

\input Tables/table_rrunning.tex

\begin{figure}[!t]
\includegraphics[height=0.37\textwidth,angle=0]{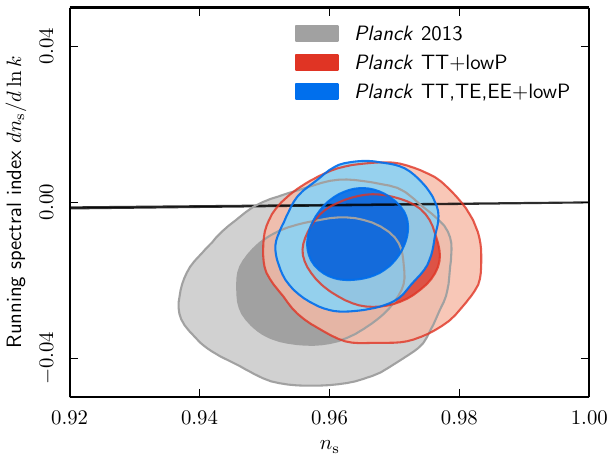}
\caption{Marginalized joint confidence contours for 
$(n_{\mathrm s} \,, \mathrm{d} n_{\mathrm s}/\mathrm{d} \ln k)$, at the
68\,\% and 95\,\% CL, in the presence of a non-zero tensor contribution, and using
\Planck\ TT+lowP or \Planck\ TT,TE,EE+lowP. Constraints from the
\Planck\ 2013 data release are also shown for comparison. The thin black
stripe shows the prediction of single-field monomial inflation models
with $50 < N_* <60$.}
\label{fig:nalphar} 
\end{figure}

\begin{figure}[!t]
\includegraphics[height=0.37\textwidth,angle=0]{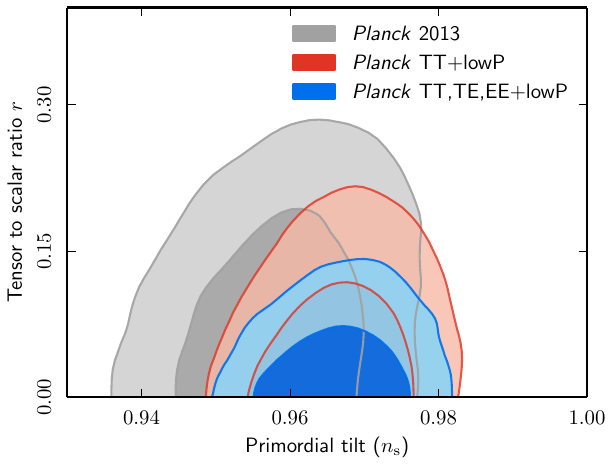}
\caption{Marginalized joint confidence contours for $(n_{\mathrm s} \,, r)$,
at the 68\,\% and 95\,\% CL, in the presence of running of the spectral indices,
and for the same data combinations as in the previous figure.
}
\label{fig:nsrr} 
\end{figure}

These bounds are relaxed when allowing for a scale dependence of the
scalar and tensor spectral indices. In that case, we assume that the
tensor spectral index and its running are fixed by the standard
inflationary consistency condition at second order in slow roll. We
obtain
\begin{align}
\label{erre_run_t}
r_{0.002} & < \,0.18 & \text{(95\,\% CL, \Planck\ TT+lowP),} \\
\label{alpha_run_t}
\frac{d n_\mathrm{s}}{d \ln k} & = \,-0.013^{+0.010}_{-0.009} & \text{(68\,\% CL, \Planck\ TT+lowP),}
\end{align}
with $n_\mathrm{s} = 0.9667 \pm 0.0066$ (68\,\% CL).
At the standard pivot scale, $k_*=0.05\,$Mpc$^{-1}$, the bound is
stronger ($r < 0.17$  at 95\,\% CL), because $k_*$ is closer to the scale
at which $n_\mathrm{s}$ and $r$ decorrelate. The constraint on $r_{0.002}$
in Eq.~(\ref{erre_run_t}) is 21\,\% tighter than the corresponding \Planck\ 2013 constraint.
The mean value of the running in Eq.~(\ref{alpha_run_t}) is
higher (lower in absolute value) than with \Planck\ 2013 by 45\,\%.
Figures~\ref{fig:nalphar} and \ref{fig:nsrr} clearly illustrate this
significant improvement with respect to the previous \Planck\ data release.
Table~\ref{tab:ralphar} shows how bounds on
$(r, \, n_\mathrm{s}, \, {d n_\mathrm{s}}/{d \ln k})$ are affected by
the lensing reconstruction, BAO, or high-$\ell$ polarization data.
The tightest bounds are obtained in combination with polarization:
\ba
r_{0.002} &< & 0.15 \cr
\label{erre_run_tplusp}
&& 
\qquad
\qquad
\text{(95\,\% CL, \Planck\ TT,TE,EE+lowP),} \\
\frac{d n_\mathrm{s}}{d \ln k} &=& -0.009 \pm 0.008 \cr
\label{alpha_run_tplusp}
&&
\qquad  
\qquad  
\text{(68\,\% CL, \Planck\ TT,TE,EE+lowP),}
\ea
with $n_\mathrm{s} = 0.9644 \pm 0.0049$ (68\,\% CL).

Neither the \Planck\ full mission constraints in 
Eqs.~(\ref{erre_t})--(\ref{erre_tplusp}) nor those including a running in
Eqs.~(\ref{erre_run_t}) and (\ref{erre_run_tplusp}) are compatible with
the interpretation of the BICEP2 $B$-mode polarization data in terms
of primordial gravitational waves~\citep{bicep2014A}. Instead they
are in excellent agreement with the results of the BICEP2/Keck Array-\Planck\
cross-correlation analysis, as discussed in Sect.~\ref{sec:bicep2}.

\subsection{Dependence of the $r$ constraints on the low-$\ell$ likelihood}

The constraints on $r$ discussed above are further tightened by adding
WMAP polarization information on large angular scales. The \Planck\ measurement of
CMB polarization on large angular scales at 70\,GHz is consistent with
the WMAP 9-year one, based on the K, Q, and V bands (at 30, 40, and 60\,GHz, respectively),
once the \Planck\ 353\,GHz channel is used to remove the dust contamination,
instead of the theoretical dust model used by the WMAP team \citep{page2007}.
(For a detailed discussion, see \citet{planck2014-a13}.) 
By combining \Planck\ $TT$ data with LFI 70\,GHz and WMAP polarization data
on large angular scales, we obtain a $35\,\%$ reduction of uncertainty, giving 
$\tau=0.074\pm 0.012$ (68\,\% CL) and $n_\mathrm{s} = 0.9660 \pm 0.060$
(68\,\% CL) for the base $\Lambda$CDM model. When tensors are added, the bounds
become
\begin{eqnarray}
r_{0.002} < & 0.09 \,\,\quad\quad\quad\quad & \!\!\text{(95\,\% CL, \Planck\ TT+lowP+WP),} \\
n_\mathrm{s}=& 0.9655 \pm 0.058 & \!\!\text{(68\,\% CL, \Planck\ TT+lowP+WP),} \\
\tau=& 0.073^{+0.011}_{-0.013} \,\,\,\,\quad & \!\!\text{(68\,\% CL, \Planck\ TT+lowP+WP).}
\end{eqnarray}
When tensors and running are both varied, we obtain $r_{0.002} < 0.14$
(95\,\% CL) and $d n_\mathrm{s}/d \ln k = -0.010 \pm 0.008$ (68\,\% CL)
for \Planck\ TT+lowP+WP. These constraints are all tighter
than those based on \Planck\ TT+lowP only.

\subsection{The tensor-to-scalar ratio and the low-$\ell$ deficit in temperature}

As noted previously \citep{planck2013-p08,planck2013-p11,planck2013-p17}, the low-$\ell$ temperature data display a slight lack of 
power compared to the expectation of the best-fit tensor-free base $\Lambda$CDM model.
Since tensor fluctuations add power on small scales, the effect will be exacerbated in models
allowing $r > 0$. 

In order to quantify this tension, we compare the observed constraint on $r$ 
to that inferred from simulated \Planck\ data. In the simulations, we assume the underlying 
fiducial model to be tensor-free, with parameters close to the base $\Lambda$CDM
best-fit values. We limit the simulations to mock temperature power spectra
only and fit these spectra with an exact low-$\ell$ likelihood for
$2 \le \ell \le 29$ \citep[see][]{Perotto:2006rj}, and a high-$\ell$ Gaussian
likelihood for $30 \le \ell \le 2508$  based on the frequency-combined,
foreground-marginalized, unbinned \Planck\ temperature power spectrum
covariance matrix.  Additionally, we impose a Gaussian prior of
\mbox{$\tau = 0.07 \pm 0.02$}.

Based on 100 simulated data sets, we find a
95\,\% CL upper limit on the tensor-to-scalar ratio of $\bar{r}_{2\sigma} \approx 0.260$.
The corresponding constraint from real data (using low-$\ell$
\texttt{Commander} temperature data, the frequency-combined,
foreground-marginalized, unbinned \Planck\ high-$\ell$ $TT$ power spectrum, and the
same prior on $\tau$ as above) reads $r < 0.123$, confirming that the
actual constraint is tighter than what one would have expected.
However, the actual constraint is not excessively unusual: out of the 100
simulations, 4 lead to an even tighter bound, corresponding to a significance
of about 2$\,\sigma$.  Thus, under the hypothesis of the base $\Lambda$CDM cosmology, 
the upper limit on $r$ that we get from the data is not implausible as a chance
fluctuation of the low multipole power.

\begin{figure}[!t]
\begin{center}
\includegraphics[height=0.96\columnwidth,angle=0]{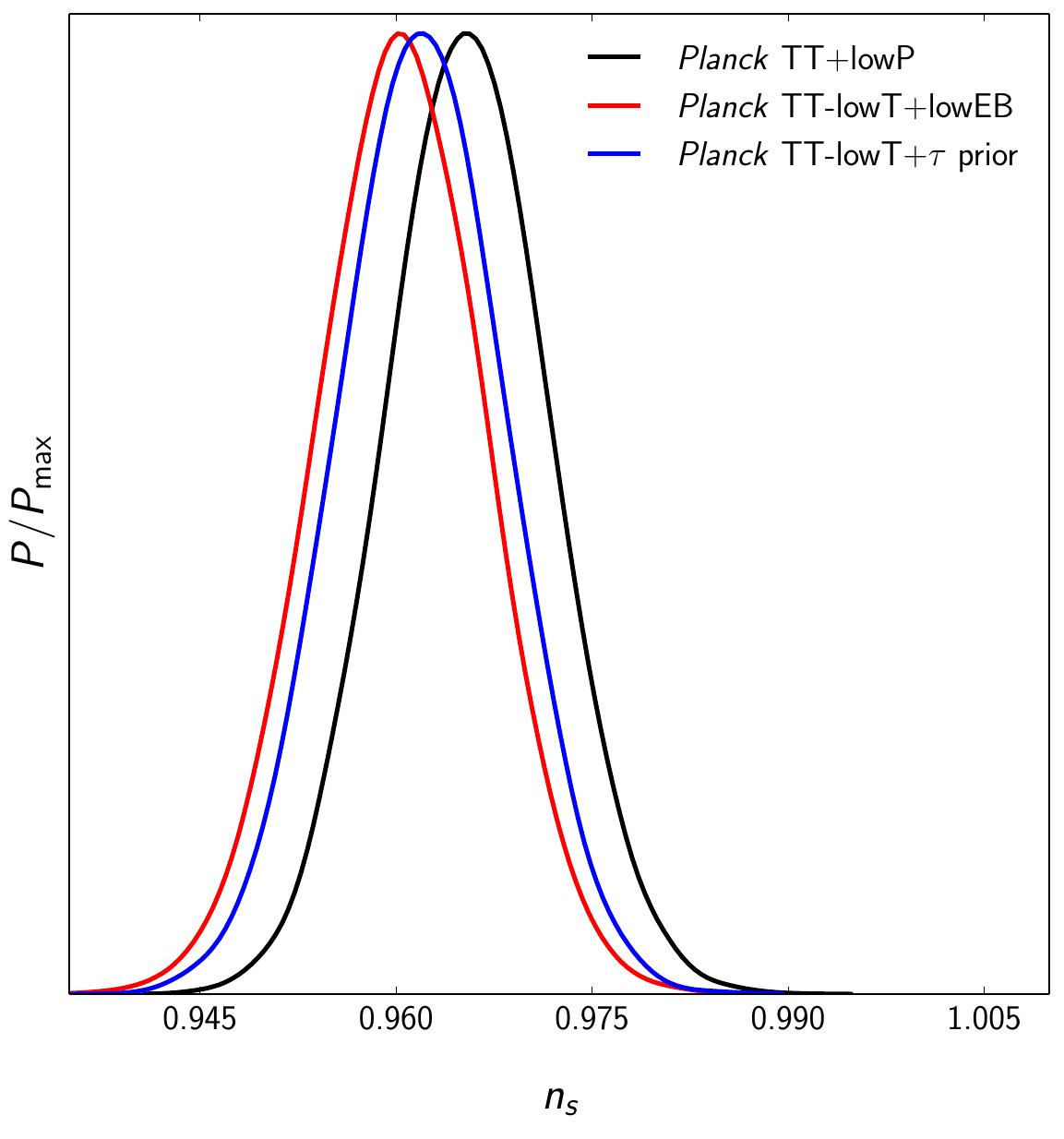} 
\end{center}
\caption{One-dimensional posterior probabilities for $n_\mathrm{s}$ for the
base $\Lambda$CDM model obtained by excluding temperature multipoles
for $\ell < 30$ (``TT$-$lowT''), while either keeping low-$\ell$ polarization data, 
or in addition replacing them with a Gaussian prior on $\tau$.
\label{fig:nsnoTT}}
\end{figure}

To illustrate the contribution of the low-$\ell$ temperature power deficit 
to the estimates of cosmological parameters, we show as an example in Fig.~\ref{fig:nsnoTT} 
how $n_\mathrm{s}$ shifts towards lower values when the $\ell < 30$ temperature information 
is discarded (we will refer to this case as ``\Planck~TT$-$lowT"). The shift in $n_\mathrm{s}$ 
is approximately $-0.005$ (or $-0.003$ when the lowP likelihood is replaced by a Gaussian prior 
$\tau = 0.07 \pm 0.02$). These shifts exceed those found in Sect.~\ref{LargeScaleSuppression}, where 
a primordial power spectrum suppressed on large scales was fitted to the data.

\begin{figure}[!h]
\begin{center}
\includegraphics[height=0.96\columnwidth,angle=0]{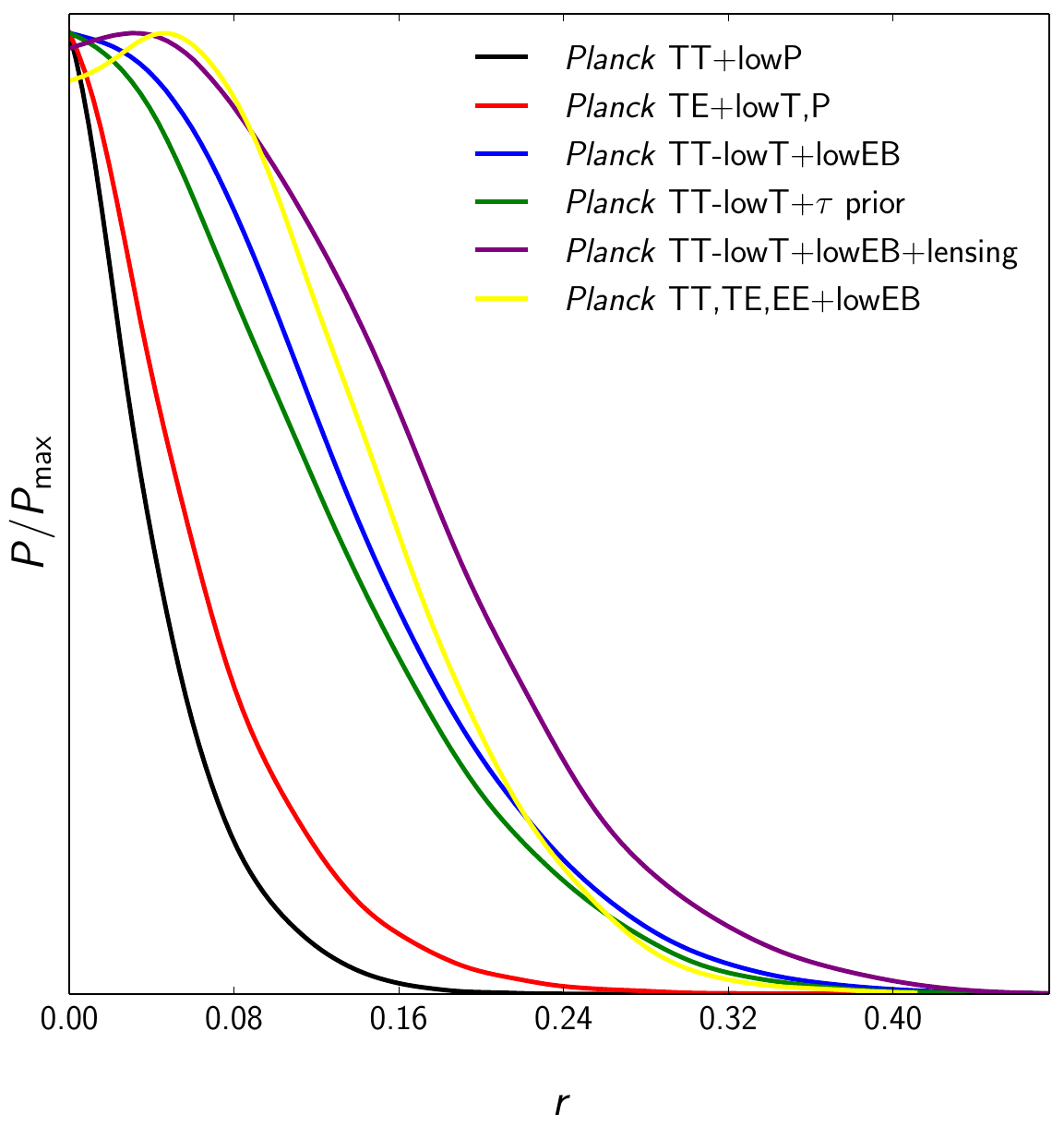}
\end{center}
\caption{One-dimensional posterior probabilities for $r$ for various 
data combinations, either including or not including temperature multipoles for $\ell < 30$,
and compared with the baseline choice (\Planck\ TT+lowP, black curve).}
\label{fig:rnoTT}
\end{figure}

Figure~\ref{fig:rnoTT} displays the posterior probability for $r$ for
various combinations of data sets, some of which exclude the $\ell < 30$
$TT$ data. This leads to the very conservative bounds $r \lsim 0.24$
and $r \lsim 0.23$ at 95\,\% CL when combined with the lowP likelihood 
or with the  Gaussian prior $\tau = 0.07 \pm 0.020$, respectively.

\subsection{Relaxing assumptions on the late-time cosmological evolution}

\begin{figure}[!h]
\begin{center}
\includegraphics[height=0.44\textwidth,angle=0]{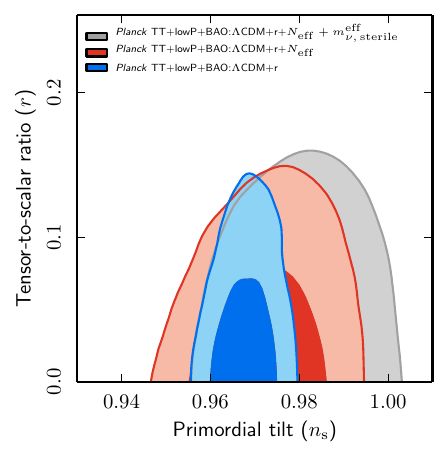} \\
\includegraphics[height=0.44\textwidth,angle=0]{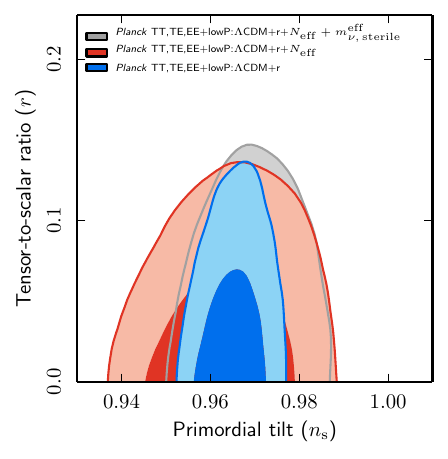}
\end{center}
\caption{Marginalized joint 68\,\% and 95\,\% CL for
$(n_{\mathrm s} \,, r_{0.002})$
using \Planck\ TT+lowP+BAO (upper panel) and \Planck\ TT,TE,EE+lowP (lower panel).
\label{fig:nsrneutrino}}
\end{figure}

\input table_r_extended_rounded.tex

As in the \Planck\ 2013 release \citepalias{planck2013-p17}, we now ask how
robust the \Planck\ results on the tensor-to-scalar ratio are against
assumptions on the late-time cosmological evolution. The results are
summarized in Table~\ref{tab:r_extended}, and some particular cases are
illustrated in Fig.~\ref{fig:nsrneutrino}. Constraints on $r$ turn out to
be remarkably stable for one-parameter extensions of the $\Lambda$CDM+$r$
model, with the only exception the $\Lambda$CDM+$r$+$\Omega_K$
case in the absence of the late time information from \Planck\ lensing or BAO
data. The weak trend towards $\Omega_K<0$, i.e., towards a positively curved
(closed) universe
from the temperature and polarization data alone, and the well-known
degeneracy between $\Omega_K$ and $H_0$/$\Omega_\mathrm{m}$ lead to a
slight suppression of the Sachs-Wolfe plateau in the scalar temperature
spectrum. This leaves more room for a tensor component.

This further degeneracy when $r$ is added builds on the negative values for the curvature
allowed by \Planck\ TT+lowP, \mbox{$\Omega_K = -0.052^{+0.049}_{-0.055}$} at 95\,\% CL \citep{planck2014-a15}.
The exploitation of the information contained in the \Planck\ lensing likelihood
leads to a tighter constraint, \mbox{$\Omega_K = -0.005^{+0.016}_{-0.017}$} at 95\,\% CL,
which improves on the \Planck\ 2013 results (\mbox{$\Omega_K = -0.007^{+0.018}_{-0.019}$} at 95\,\% CL).
However, due to the remaining degeneracies left by the
uncertainties in polarization on large angular scales, a full appreciation
of the improvement
due to the full mission temperature and lensing data can be obtained by using lowP+WP, which leads to
$\Omega_K = -0.003^{+0.012}_{-0.014}$ at 95\,\% CL. 
Note that the negative values allowed
for the curvature are decreased in magnitude when the running is allowed, suggesting that
the low-$\ell$ temperature deficit is contributing to the estimate of the spatial curvature.

The trend found for $\Lambda$CDM+$r$+$\Omega_K$ is even clearer when spatial curvature and the running of
the spectral index are varied at the same time.
In this case, the \Planck\ temperature plus polarization data are compatible with $r$
values as large as 0.19 (95\,\% CL), at the cost of
an almost 4\,$\sigma$ deviation from spatial flatness (which, however,
disappears as soon as lensing or BAO data are considered).

%% file: Tables/table_rrunning.tex
\begin{table*}
\centering
\begin{tabular}{cccccc}
\noalign{\hrule\vskip 2pt}
\noalign{\hrule\vskip 3pt}
Model & Parameter & \Planck\ TT+lowP & \Planck\ TT+lowP+lensing & \Planck\ TT+lowP+BAO & \Planck\ TT,TE,EE+lowP \\
\hline
& $n_{\mathrm{s}}$ & $0.9666 \pm 0.0062$ & $0.9688 \pm 0.0061$ & $0.9680 \pm 0.0045$ & $0.9652 \pm 0.0047$ \phantom{$\Big|$}\\
$\Lambda$CDM+$r$ & $r_{0.002}$ & $< 0.103$ & $< 0.114$ & $< 0.113$ & $< 0.099$ \\
& $-2 \Delta \ln {\cal L}_\mathrm{max}$ & 0 & 0 & 0 & 0 \\
\hline
\multirow{7}{*}{+$\mathrm{d} n_\mathrm{s}/\mathrm{d} \ln k$}& $n_{\mathrm{s}}$ &
$0.9667 \pm 0.0066$ & $0.9690 \pm 0.0063$ & $0.9673 \pm 0.0043$ & $0.9644 \pm 0.0049$ \phantom{$\Big|$}\\
\multirow{2}{*}{$\Lambda$CDM+$r$} & $r_{0.002}$ & $< 0.180$ & $< 0.186$ & $< 0.176$ & $< 0.152$ \\
& $r$ & $< 0.168$ & $< 0.176$ & $< 0.166$ & $< 0.149$ \\
& $\mathrm{d} n_\mathrm{s}/\mathrm{d} \ln k$ &
$-0.0126^{+0.0098}_{-0.0087}$ & $-0.0076^{+0.0092}_{-0.0080}$ & $-0.0125 \pm 0.0091$ & $-0.0085 \pm 0.0076$ \phantom{$\Big|$}\\
& $-2 \Delta \ln {\cal L}_\mathrm{max}$ & $-0.81$ & $-0.08$ & $-0.87$ & $-0.38$ \\
\hline
\end{tabular}
\vspace{.4cm}
\caption{\label{tab:ralphar}
Constraints on the primordial perturbation parameters for
$\Lambda$CDM+$r$ and
$\Lambda$CDM+$r$+$\mathrm{d} n_\mathrm{s}/\mathrm{d} \ln k$ models from {\it Planck}.
Constraints on the spectral index and its dependence on the wavelength are given at the pivot scale of $k_* = 0.05$~Mpc$^{-1}$.}
\end{table*}

%% file: table_r_extended_rounded.tex
\begin{table*}
\begin{center}
\begin{tabular}{ccccc}
\noalign{\hrule\vskip 2pt}
\noalign{\hrule\vskip 3pt}
Extended model, & Parameter & \Planck\ TT+lowP & \Planck\ TT+lowP & \Planck\ TT,TE,EE \\
$\Lambda$CDM+$r$+ & & +lensing & +BAO & +lowP \\
\noalign{\hrule\vskip 2pt}
\multirow{2}{*}{+general reionization} & $r$ & $< 0.11$ & $< 0.10$ & $< 0.10$ \\

& $n_{\mathrm{s}}$ & $0.975 \pm 0.006$ & $0.971 \pm 0.005$ & $0.968 \pm 0.005$ \\
\noalign{\hrule\vskip 2pt}
& $r$ &  $< 0.14$ & $< 0.12$ & $< 0.11$ \\
+$N_\mathrm{eff}$ & $n_{\mathrm{s}}$
& $0.977^{+0.016}_{-0.017}$ & $0.972 \pm 0.009$ & $0.964 \pm 0.010$ \\
\noalign{\vskip 2pt}
& $N_\mathrm{eff}$  &  $3.24^{+0.30}_{-0.35}$ & $3.19 \pm 0.24$ &
$3.02^{+0.20}_{-0.21}$ \\[0.5ex]
\noalign{\hrule\vskip 2pt}
& $r$  & $< 0.14$ & $< 0.12$ & $<0.12$ \\
+$Y_\mathrm{He}$
& $n_{\mathrm{s}}$ & $0.975 \pm 0.007$ & $0.973 \pm 0.009$ & $0.969 \pm 0.008$ \\
& $Y_\mathrm{He}$  & $0.258 \pm 0.022$ & $0.257 \pm 0.022$ & $0.252 \pm 0.014$ \\
\noalign{\hrule\vskip 2pt}
& $r$  &  $< 0.11$ & $< 0.11$ & $< 0.11$\\
+$\sum m_\nu$  & $n_\mathrm{s}$ & $0.963 \pm 0.007$ & $0.967 \pm 0.005$ & $0.962 \pm 0.005$ \\
& $\sum m_\nu$ [eV] & $< 0.67$ & $< 0.21$ &  $< 0.58$ \\
\noalign{\hrule\vskip 2pt}
& $r$ & $< 0.15$  & $< 0.11$ & $< 0.15$\\
+$\Omega_K$ & $n_{\mathrm{s}}$ & $0.971 \pm 0.007$
& $0.971 \pm 0.007$ & $0.969 \pm 0.005$ \\
& $\Omega_K$ & $-0.008^{+0.010}_{-0.008}$  & $-0.001 \pm 0.003$ &
$-0.045^{+0.016}_{-0.020}$ \\[0.5ex]
\noalign{\hrule\vskip 2pt}
& $r$  & $< 0.14$ & $< 0.11$ & $< 0.12$ \\
+$w$&
$n_{\mathrm{s}}$ & $0.969 \pm 0.006$ & $0.967 \pm 0.006$ & $0.966 \pm 0.005$ \\
& $w$ & $-1.46^{+0.20}_{-0.40}$ & $-1.02^{+0.08}_{-0.07}$ & $-1.57^{+0.17}_{-0.37}$ \\[0.5ex]
\noalign{\hrule\vskip 2pt}
\multirow{4}{*}{+$\Omega_K$+$d n_\mathrm{s}/d \ln k$} & $r$ & $< 0.20$  & $< 0.18$ & $< 0.19$ \\
& $n_{\mathrm{s}}$ & $0.971 \pm 0.007 $ & $0.969 \pm 0.007$ & $0.969 \pm 0.005$ \\
& $d n_\mathrm{s}/d \ln k$ & $-0.006 \pm 0.009$ &
$-0.013 \pm 0.009$ & $-0.004 \pm 0.008$ \\
& $\Omega_K$ & $-0.006^{+0.010}_{-0.009}$ & $-0.001 \pm 0.003$ &
$-0.043^{+0.011}_{-0.020}$ \\[0.5ex]
\noalign{\hrule\vskip 2pt}
\multirow{4}{*}{+$N_\mathrm{eff}$+$m^\mathrm{eff}_{\nu \,, \mathrm{sterile}}$}& $r$ & $< 0.14$ & $< 0.13$ & $< 0.12$ \\
& $n_{\mathrm{s}}$ & $0.980^{+0.010}_{-0.014}$ & $0.978^{+0.008}_{-0.011}$ &
$0.968^{+0.006}_{-0.008}$ \\
& $m^\mathrm{eff}_{\nu \,, \mathrm{sterile}}$ [eV] & $< 0.59$ & $<0.55$ & $<0.83$ \\
& $N_\mathrm{eff}$ & $< 3.80$ & $<3.73$ & $<3.47$ \\
\hline
\end{tabular}
\end{center}
\caption{Constraints on extensions of the $\Lambda$CDM+$r$ cosmological
model for \Planck\ TT+lowP+lensing, \Planck\ TT+lowP+BAO, and \Planck\ TT,TE,EE+lowP. 
For each model we quote 68\,\% CL, unless 95\,\% CL upper bounds are reported. 
}
\label{tab:r_extended}
\end{table*}

%% file: section_six.tex
\input macros.tex
\def\ba{\begin{eqnarray}}
\def\ea{\end{eqnarray}}

In this section we study the implications of \Planck\ 2015 constraints on standard slow-roll single-field inflationary 
models.

\subsection{Constraints on slow-roll parameters}

We first present the \Planck\ 2015 constraints on slow-roll parameters obtained through the 
analytic perturbative expansion in terms of the HFFs $\epsilon_i$
for the primordial spectra of cosmological fluctuations during slow-roll inflation \citep{stewart:1993,Gong:2001he,Leach:2002ar}.
When restricting to first order in $\epsilon_i$, we obtain 
\begin{align}
\label{epsilon_1_TT_no}
\epsilon_1 & < \,0.0068  & \text{(95\,\% CL, \Planck\ TT+lowP)} \,, \\
\label{epsilon_2_TT_no}
\epsilon_2 & = \,0.029^{+0.008}_{-0.007}  & \text{(68\,\% CL, \Planck\ TT+lowP)} \,.
\end{align}
When high-$\ell$ polarization 
is included we obtain $\epsilon_1 < 0.0066$ at 95\,\% CL and
$\epsilon_2 = 0.030^{+0.007}_{-0.006}$ at 68\,\% CL.
When second-order contributions in the HFFs are included, we obtain
\ba
\label{epsilon_1_TT}
\epsilon_1 < & 0.012 \,\,\,\quad\quad & \text{(95\,\% CL, \Planck\ TT+lowP)} \,, \\
\label{epsilon_2_TT}
\epsilon_2 = & 0.031^{+0.013}_{-0.011} \,\,\, & \text{(68\,\% CL, \Planck\ TT+lowP)} \,, \\
\label{epsilon_3_TT}
-0.41 < \epsilon_3 < & 1.38 \quad\quad\quad & \text{(95\,\% CL, \Planck\ TT+lowP)} \,.
\ea
When high-$\ell$ polarization
is included we obtain $\epsilon_1 < 0.011$ at 95\,\% CL, 
$\epsilon_2 = 0.032^{+0.011}_{-0.009}$ at 68\,\% CL, and $-0.32 < \epsilon_3 < 0.89$ at 95\,\% CL. 

\begin{figure}[!ht]
\centering
\includegraphics[width=8.8cm]{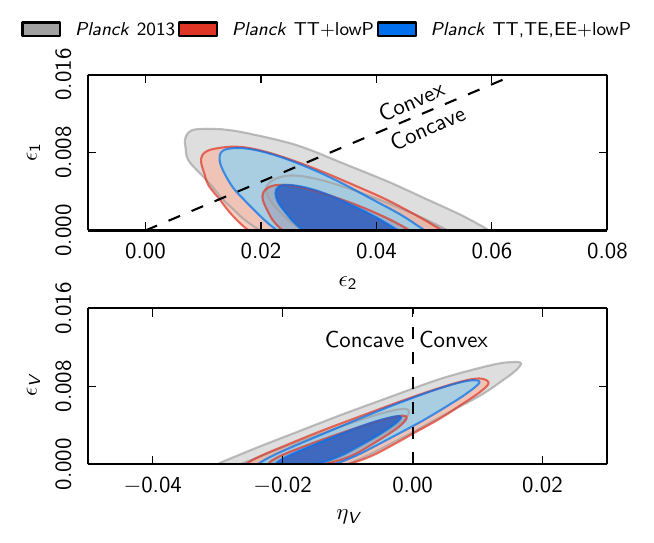}
\caption{Marginalized joint 68\,\% and 95\,\%~CL regions for 
$(\epsilon_1 \,, \epsilon_2)$ (top panel)
and $(\epsilon_V \,, \eta_V)$ (bottom panel) for \Planck\ TT+lowP (red contours), \Planck\ TT,TE,EE+lowP 
(blue contours), and compared with the \Planck\ 2013 results (grey contours).}
\label{fig:epsilon_no3}
\end{figure}

The potential slow-roll parameters are obtained as derived parameters by using their exact expressions 
as function of $\epsilon_i$ \citep{Leach:2002ar,Finelli:2009bs}: 
\begin{align}
\label{epsilon_V}
\epsilon_V &= \,\frac{V_\phi^2 M_\mathrm{pl}^2}{2 V^2} = \epsilon_1 
\frac{\left(1 - \frac{\epsilon_1}{3} +
\frac{\epsilon_2}{6}\right)^2}{\left(1 -
\frac{\epsilon_1}{3}\right)^2} \,, \\
\label{eta_V}
\eta_V &= \,\frac{V_{\phi \phi} M_\mathrm{pl}^2}{V} = \frac{ 2 \epsilon_1 - \frac{\epsilon_2}{2}
- \frac{2\epsilon_1^2}{3} + \frac{5\epsilon_1 \epsilon_2}{6}
-\frac{\epsilon_2^2}{12} - \frac{\epsilon_2 \epsilon_3}{6}}{1 -
\frac{\epsilon_1}{3}} \,, 
\end{align}
\begin{eqnarray}
\label{xi_V}
\xi_V^2 &=& \frac{V_{\phi \phi \phi} V_\phi M_\mathrm{pl}^4}{V^2} =
\frac{1 - \frac{\epsilon_1}{3} + 
\frac{\epsilon_2}{6}}{\left(1-\frac{\epsilon_1}{3}\right)^2} 
\left( 4 \epsilon_1^2 -3 \epsilon_1 \epsilon_2 
+ \frac{\epsilon_2^{\phantom{2}} \epsilon_3}{2} - \epsilon_1 \epsilon_2^2 \right. \nonumber \\
&+& \left.  3 \epsilon_1^2 \epsilon_2 - \frac{4}{3} \epsilon_1^3  - \frac{7}{6} \epsilon_1 \epsilon_2 \epsilon_3  
+\frac{\epsilon_2^2 \epsilon_3}{6}
+ \frac{\epsilon_2 \epsilon_3^2}{6}
+ \frac{\epsilon_2 \epsilon_3 \epsilon_4}{6}
\right)  \,,
\end{eqnarray}
where $V(\phi)$ is the inflaton potential, the subscript $\phi$ denotes the derivative with respect 
to $\phi$, and $M_\mathrm{pl}=(8\pi G)^{-1/2}$ is the reduced Planck mass (see also Table~\ref{table:InflationDefinitions}). 

By using Eqs.~(\ref{epsilon_V}) and (\ref{eta_V}) with $\epsilon_3=0$ and the primordial power spectra to lowest 
order in the HFFs, the derived constraints for the first two slow-roll potential parameters are:
\begin{align}
\label{epsilon_V_TT_no}
\epsilon_V & < \,0.0068 & \text{(95\,\% CL, \Planck\ TT+lowP)} \,, \\
\label{eta_V_TT_no}
\eta_V^2 & = \,-0.010^{+0.005}_{-0.009} & \text{(68\,\% CL, \Planck\ TT+lowP)} \,.
\end{align}
When high-$\ell$ polarization
is included we obtain $\epsilon_V < 0.0067$ at 95\,\% CL and 
$\eta_V = -0.010^{+0.004}_{-0.009}$ at 68\,\% CL. By using Eqs.~(\ref{epsilon_V}), (\ref{eta_V}), and (\ref{xi_V}) 
with $\epsilon_4=0$ and the primordial power spectra to second order in the HFFs, 
the derived constraints for the slow-roll potential parameters are:
\begin{align}
\label{epsilon_V_TT}
\epsilon_V & < \,0.012 & \text{(95\,\% CL, \Planck\ TT+lowP)} \,, \\
\label{eta_V_TT}
\eta_V & = \,-0.0080^{+0.0088}_{-0.0146} & \text{(68\,\% CL, \Planck\ TT+lowP)} \,, \\
\label{xi_V_TT}
\xi_V^2 & = \,0.0070^{+0.0045}_{-0.0069} & \text{(68\,\% CL, \Planck\ TT+lowP)} \,.
\end{align}
When high-$\ell$ polarization is included we obtain $\epsilon_V < 0.011$ at 95\,\% CL, and
$\eta_V = -0.0092^{+0.0074}_{-0.0127}$ and $\xi^2_V = 0.0044^{+0.0037}_{-0.0050}$, both at 68\,\% CL.



\begin{figure*}[!ht]
\centering
\includegraphics[height=11cm,width=16cm]{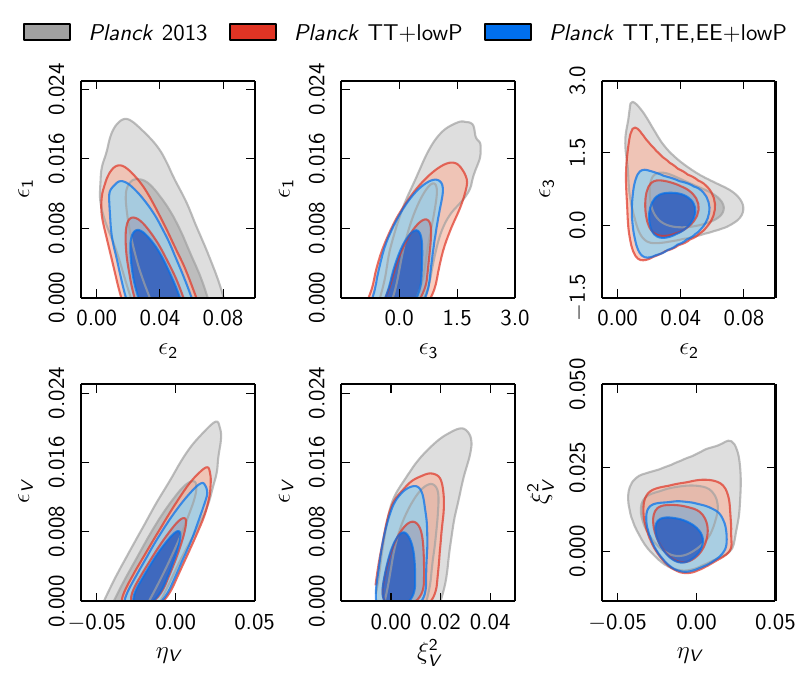}
\caption{Marginalized joint 68\,\% and 95\,\%~CL regions for
$(\epsilon_1 \,, \epsilon_2 \,, \epsilon_3)$ (top panels)
and $(\epsilon_V \,, \eta_V \,, \xi_V^2)$ (bottom panels) for \Planck\ TT+lowP (red contours), \Planck\ TT,TE,EE+lowP
(blue contours), and compared with the \Planck\ 2013 results (grey contours).  
}
\label{fig:epsilon}
\end{figure*}

In Figs.~\ref{fig:epsilon_no3} and \ref{fig:epsilon} we show the 68\,\% CL and 95\,\% CL of the HFFs
and the derived potential slow-roll parameters with and without the high-$\ell$ polarization 
and compare these values with the \Planck\ 2013 results.

\begin{figure*}[!ht]
\includegraphics[width=18cm]{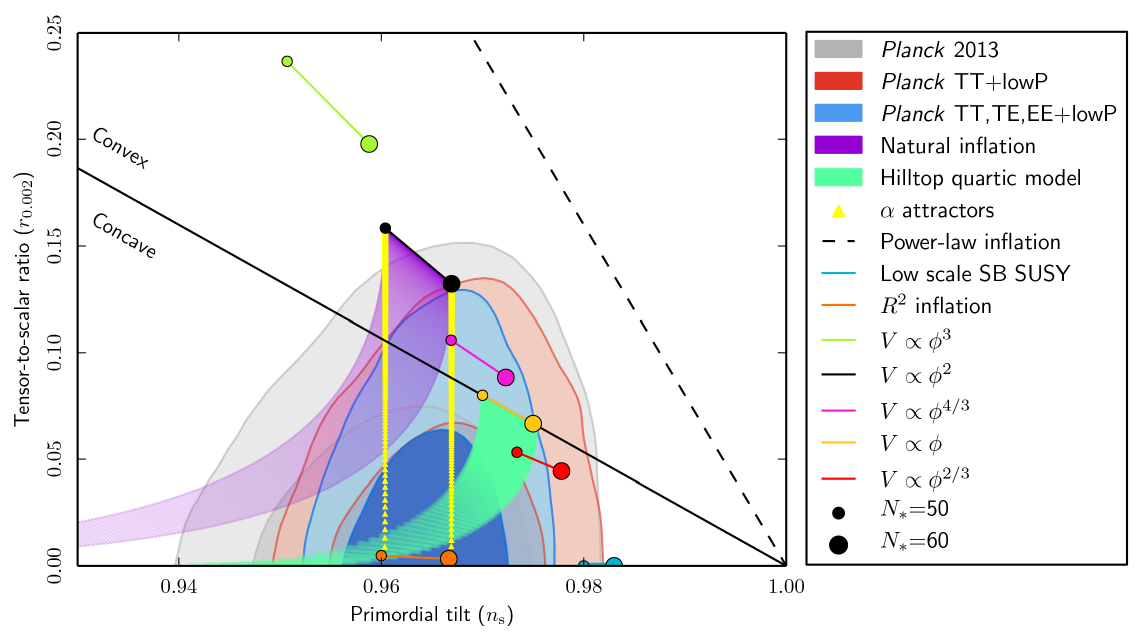}
\caption{Marginalized joint 68\,\% and 95\,\%~CL regions for $n_\mathrm{s}$ and $r$ at $k = 0.002$\,Mpc$^{-1}$ from \Planck\
compared to the theoretical predictions
of selected inflationary models. Note that the marginalized
joint 68\,\% and 95\,\%~CL regions have been obtained by assuming $d n_\mathrm{s}/d \ln k = 0$.
}
\label{fig:nsvsr}
\end{figure*}

\subsection{Implications for selected inflationary models}

The predictions to lowest order in the slow-roll approximation 
for $(n_\mathrm{s},r)$ at $k = 0.002$\,Mpc$^{-1}$ of a few inflationary models with a representative uncertainty for the entropy generation stage
($50 < N_* < 60$) are shown in Fig.~\ref{fig:nsvsr}. Fig.~\ref{fig:nsvsr} updates Fig.~1 of \citetalias{planck2013-p17} 
with the same notation.

In the following we discuss the implications of \Planck\ TT+lowP+BAO data 
for selected slow-roll inflationary models by taking into account the uncertainties in the 
entropy generation stage. We model these uncertainties by two parameters, as in \citetalias{planck2013-p17}: 
the energy scale $\rhorh$ by which the Universe has thermalized, and the parameter $w_\mathrm{int}$ which characterizes 
the effective equation of state between the end of inflation and the energy scale specified by $\rhorh$.
We use the primordial power spectra of cosmological fluctuations generated during slow-roll inflation 
parameterized by the HFFs, $\epsilon_i$, to second order, which can be expressed in 
terms of the number of $e$-folds to the end of inflation, $N_*$, and 
the parameters of the considered inflationary model, using modified routines of 
the public code {\tt ASPIC}\footnote{\url{http://cp3.irmp.ucl.ac.be/~ringeval/aspic.html}} \citep{Martin:2014vha}.
For the number of $e$-folds to the end of inflation \citep{Liddle:2003as,Martin:2010kz} 
we use the expression \citepalias{planck2013-p17}
\be
\begin{aligned}
N_* \approx &  \; 67 - \ln \left(\frac{k_*}{a_0 H_0}\right)
+  \frac{1}{4}\ln{\left( \frac{V_*^2}{\Mpl^4 \rhoend}\right) } \\
&+ \frac{1-3w_\mathrm{int}}{12(1+w_\mathrm{int})}
\ln{\left(\frac{\rhorh}{\rhoend} \right)} - \frac{1}{12} \ln (g_\mathrm{th} ) \; ,
\label{eq:nefolds}
\end{aligned}
\ee
where $\rhoend$ is the energy density at the end of inflation, 
$a_0 H_0$ is the present Hubble scale, 
$V_*$ is the potential energy when $k_*$ left the Hubble radius during inflation, 
$w_\mathrm{int}$ characterizes the effective equation of state between the end of inflation 
and the thermalization energy scale $\rhorh$,
and $g_\mathrm{th}$ is the number of effective bosonic 
degrees of freedom at the energy scale $\rhorh$. 
We consider the pivot scale
$k_*=0.002$\,Mpc$^{-1}$, $g_\mathrm{th}=10^3$, and $\epsilon_\mathrm{end}=1$. We consider the 
uniform priors for the cosmological 
parameters listed in Table~\ref{tab:Bayesian_comparison_priors_six}. 
We also consider a logarithmic prior on $10^{10} A_\mathrm{s}$ 
(over the interval $[(e^{2.5}, e^{3.7}]$) and  
$\rhorh$ (over the interval $[(1\,\mathrm{TeV})^4, \rhoend]$). We consider both the case
in which $w_\mathrm{int}$ is kept fixed at zero and the case in which it is allowed to vary with a uniform prior in the range 
$-1/3 < w_\mathrm{int} < 1/3$.

\input prior_table.tex

We have validated the slow-roll approach by cross-checking 
the Bayes factor computations against the fully numerical inflationary mode equation 
solver {\tt ModeCode} coupled to the {\tt PolyChord} sampler. For each inflationary model we provide in Table~\ref{table:model_compar} and in the
main text the $\Delta \chi^2$ value with respect to the base $\Lambda$CDM model, computed 
with the \texttt{CosmoMC} implementation of the BOBYQA
algorithm for maximizing the likelihood,
and the Bayesian evidence with respect to the $R^2$ inflationary model \citep{Starobinsky:1980te},
computed by \texttt{CosmoMC} connected to \texttt{CAMB}, using \texttt{MultiNest} as the sampler.

\subsubsection*{Power law potentials}

We first investigate the class of inflationary models with a single monomial potential \citep{Linde:1983gd}:
\begin{equation}
V(\phi) = \lambda M_\mathrm{pl}^4 \left( \frac{\phi}{M_\mathrm{pl}} \right)^n \,,
\label{PowerLawPot:Eq}
\end{equation}
in which inflation occurs for large values of the inflaton, $\phi > M_\mathrm{pl}$. 
The predictions for the scalar spectral index and the tensor-to-scalar ratio at first order in the slow-roll 
approximation are $n_\mathrm{s} - 1 \approx - 2 (n+2)/(4 N_* +n)$ and 
$r \approx  16 n/(4 N_* + n)$, respectively.
By assuming a dust equation of state (i.e., $w_\mathrm{int} = 0$) prior to thermalization, 
the cubic and quartic potentials are strongly disfavoured by 
$\ln B = -11.6$ and $\ln B = -23.3$, respectively. The quadratic potential 
is moderately disfavoured by $\ln B = -4.7$. 
Other values, such as $n=4/3$, $1$, and $2/3$, motivated 
by axion monodromy \citep{monodromyI,monodromyII}, are compatible with \Planck\ data with $w_\mathrm{int}=0$.

Small modifications occur when considering the effective equation 
of state parameter, $w_\mathrm{int} = (n-2)/(n+2)$, defined by averaging over the
coherent oscillation regime which follows inflation \citep{1983PhRvD..28.1243T}.
The Bayes factors are slightly modified when 
$w_\mathrm{int}$ is allowed to float, as shown in Table~\ref{table:model_compar}. 

\subsubsection*{Hilltop models}

In hilltop models \citep{Boubekeur:2005zm}, with potential
\begin{equation}
V(\phi) \approx \Lambda^4 \left( 1 - \frac{\phi^{{p}}}{\mu^{{p}}} + ...\right) \,,
\label{newinf}
\end{equation}
the inflaton rolls away from an unstable equilibrium. 
The predictions to first order in the slow-roll approximation are 
$r \approx 8 p^2 (M_\mathrm{pl}/\mu)^2 x^{2p-2}/(1-x^p)^2$ and 
$n_s -1 \approx - 2 p (p-1) (M_\mathrm{pl}/\mu)^2 x^{p-2}/(1-x^p) - 3 r/8$, where $x = \phi_*/\mu$.
As in \citetalias{planck2013-p17}, the ellipsis in Eq.~(\ref{newinf}) and in what follows 
indicates higher-order terms that are negligible during inflation but
ensure positiveness of the potential. 

By sampling $\log_{10} (\mu/M_\mathrm{pl})$ within the prior $[0.30, 4.85]$ for $p=2$, we obtain 
$\log_{10} (\mu/M_\mathrm{pl}) > 1.02$ ($1.05$) at 95\,\% CL and $\ln B = -2.6$ ($-2.4$) for $w_\mathrm{int}=0$ 
(allowing $w_\mathrm{int}$ to float). 

An exact potential which could also apply after inflation, instead of the approximated one in Eq.~(\ref{newinf}), 
might be needed for a better comparison among different models. 
For $\mu/M_\mathrm{pl} \gg 1$, hilltop models as defined in Eq.~(\ref{newinf}) by neglecting the 
additional terms denoted by the ellipsis
lead to $n_\mathrm{s}-1  \approx - 3 r /8$, the same prediction as for
the previously discussed linear potential, $V(\phi) \propto \phi .$ By considering a double well potential, 
$V(\phi) = \Lambda^4 [1-\phi^2/(2 \mu^2)]^2$, instead,  
we obtain a slightly worse Bayes factor than the hilltop $p=2$ model, $\ln B = -3.1$ ($-2.3$) for $w_\mathrm{int}=0$ 
($w_\mathrm{int}$ allowed to vary). This different result can be easily understood.  
Although the double well potential is equal to the hilltop model for \mbox{$\phi \ll \mu$}, 
it approximates $V(\phi) \propto \phi^2$ for $\mu/M_\mathrm{pl} \gg 1$. Since a linear potential 
is a better fit to \Planck\ than $\phi^2$, 
the fit of the double well potential is therefore worse than the hilltop $p=2$ case for $\mu/M_\mathrm{pl} \gg 1$, 
and this partially explains the slightly different Bayes factors obtained.    

In the $p=4$ case, we obtain
$\log_{10} (\mu/M_\mathrm{pl}) > 1.05$ ($1.02$) at 95\,\% CL and $\ln B = -2.8$ ($-2.6$) for $w_\mathrm{int}=0$
(allowing $w_\mathrm{int}$ to float), assuming a prior range $[-2, 2]$ for $\log_{10} (\mu/M_\mathrm{pl})$.

\subsubsection*{Natural inflation}

In {\it natural} inflation \citep{1990PhRvL..65.3233F,Adams:1992bn} a nonperturbative shift symmetry is invoked to 
suppress radiative corrections leading to the periodic potential
\begin{equation}
V(\phi)=\Lambda^4 \left[ 1+\cos \left(\frac{\phi}{f} \right) \right],
\label{NatInf}
\end{equation}
where $f$ is the scale which determines the curvature of the potential. 
We sample $\log_{10} (f/M_\mathrm{pl})$ within the prior $[0.3,2.5]$ as in \citetalias{planck2013-p17}. 
We obtain $\log_{10} (f / M_\mathrm{pl}) > 0.84$ ($> 0.83$) at 95\,\% CL 
and $\ln B = -2.4$ ($-2.3$) for 
$w_\mathrm{int}=0$ (allowing $w_\mathrm{int}$ to vary).

Note that the super-Planckian value for $f$ required by observations is not necessarily a problem for 
this class of models.
When several fields $\phi_i$ with a cosine potential as in Eq.~(\ref{NatInf}) and scales $f_i$ appear in the Lagrangian, 
an effective single-field inflationary trajectory can be found for a suitable choice of parameters \citep{Kim:2004rp}.
In such a setting, the super-Planckian value of the effective scale $f$ 
required by observations can be obtained even if the original scales satisfy $f_i \ll M_\mathrm{pl}$ \citep{Kim:2004rp}.


\subsubsection*{D-brane inflation}

Inflation can arise from physics involving extra dimensions.
If the standard model of particle physics is confined to our 3-dimensional brane, 
the distance between our brane and anti-brane can drive inflation.
We consider the following parameterization for the 
effective potential driving inflation:
\begin{equation}
V(\phi) = \Lambda^4 \left( 1 - \frac{\mu^p}{\phi^p} + ... \right) \,.
\label{Dbrane_infl}
\end{equation}
Sampling $\log_{10}(\mu/M_\mathrm{pl})$ using a uniform prior over $[-6,0.3],$ we consider $p=4$ 
\citep{Kachru:2003sx,Dvali:2001fw} and $p=2$ \citep{GarciaBellido:2001ky}. 
The predictions for $r$ and $n_\mathrm{s}$ can be obtained from the hilltop case with the substitution $p \rightarrow -p$.
These models agree with the \Planck\ data with a Bayes factor of $\ln B = -0.4$ ($-0.6$) and $\ln B = -0.7$ ($-0.9$) 
for $p=4$ and $p=2$, respectively, for $w_\mathrm{int}=0$ (allowing $w_\mathrm{int}$ to vary).

\subsubsection*{Potentials with exponential tails}

Exponential potentials are ubiquitous in inflationary models 
motivated by supergravity and string theory 
\citep{Goncharov:1985yu,Stewart:1994ts,Dvali:1998pa,Burgess:2001vr,Cicoli:2008gp}.
We restrict ourselves to analysing the following class of potentials:
\begin{equation}
V(\phi) = \Lambda^4 \left( 1 - e^{-q \phi/M_\mathrm{pl}} + ... \right) \,.
\label{exp_infl}
\end{equation}
As for the hilltop models described earlier, the ellipsis indicates possible 
higher-order terms that are negligible during inflation but
ensure positiveness of the potential. 
These models predict $r \approx 8 q^2 e^{-2 q \phi/M_\mathrm{pl}}/(1-e^{-q \phi/M_\mathrm{pl}})^2$ and 
$n_s -1 \approx - q^2 e^{-q \phi/M_\mathrm{pl}} (2+e^{-q \phi/M_\mathrm{pl}})/(1-e^{-q \phi/M_\mathrm{pl}})^2$ with 
a slow-roll trajectory characterized by $N \approx f(\phi/M_\mathrm{pl}) - 
f(\phi_\mathrm{end}/M_\mathrm{pl})$, with $f(x) = (e^{q x} - q x)/q^2$. 
By sampling $\log_{10} (q/M_\mathrm{pl})$ with a uniform prior over $[-3,3]$, we obtain 
a Bayes factor of $-0.6$ for $w_\mathrm{int}=0$ ($-0.9$ when $w_\mathrm{int}$ is allowed to vary).



\subsubsection*{Spontaneously broken SUSY}

Hybrid models \citep{Copeland:1994vg,Linde:1993cn} 
predicting $n_\mathrm{s} > 1$ are strongly disfavoured by the \Planck\ data, as for the first cosmological 
release \citepalias{planck2013-p17}.
An example of a hybrid model predicting $n_\mathrm{s} < 1$ is the case in which slow-roll inflation is driven by 
loop corrections in spontaneously broken supersymmetric (SUSY) grand unified theories \citep{Dvali:1994ms} 
described by the potential
\begin{equation}
V(\phi) = \Lambda^4 \left[ 1 + \alpha_h \log\left(\phi/M_\mathrm{pl}\right) \right],
\label{sbsusy}
\end{equation}
where $\alpha_h > 0$ is a dimensionless parameter. 
Note that for $\alpha_h \ll 1$, this model leads to 
the same predictions as the power-law potential for $p \ll 1$ to lowest order in the slow-roll approximation.
By sampling $\log_{10}(\alpha_h)$ on a flat prior $[-2.5,1]$, 
we obtain a Bayes factor of $-2.2$ for $w_\mathrm{int}=0$ ($-1.7$ when $w_\mathrm{int}$ is allowed to vary).

\subsubsection*{$R^2$ inflation}

The first inflationary model proposed \citep{Starobinsky:1980te}, with action
\begin{equation}
S = \int \mathrm{d}^4 x \sqrt{-g} \frac{M^2_\mathrm{pl}}{2} \left( R + \frac{R^2}{6 M^2} \right)\,,
\label{R2}
\end{equation}
still lies within the \Planck\ 68\,\% CL constraints, as for the \Planck\ 2013 
release \citepalias{planck2013-p17}. This model corresponds to the potential
\begin{equation}
V(\phi) = \Lambda^4 \left( 1 - e^{- \sqrt{2/3} \phi/M_\mathrm{pl} }\right)^2
\label{R2_Einsteinframe}
\end{equation}
in the Einstein frame, 
which leads to the slow-roll predictions $n_\mathrm{s}-1 \approx - 2/N$ and $r \approx 12/N^2$ 
\citep{Mukhanov:1981xt,Starobinsky:1983zz}. 

After the \Planck\ 2013 release, several theoretical developments supported the model in Eq.~(\ref{R2}) beyond 
the original motivation of including quantum effects at one-loop \citep{Starobinsky:1980te}. 
No-scale supergravity \citep{Ellis:2013xoa}, the 
large-field regime of superconformal $D$-term inflation \citep{Buchmuller:2013zfa}, or recent developments in 
minimal supergravity \citep{Farakos:2013cqa,Ferrara:2013wka} can lead to a 
generalization of the potential in Eq.~(\ref{R2_Einsteinframe}) 
(see \cite{Ketov:2010qz} for a previous embedding of $R^2$ inflation in $F({\cal R})$ supergravity). 
The potential in Eq.~(\ref{R2_Einsteinframe}) can also be generated by spontaneous breaking of 
conformal symmetry \citep{Kallosh:2013lkr}.
This inflationary model has $\Delta \chi^2 \approx 0.8$ ($0.3$) larger than the base $\Lambda$CDM model
with no tensors for $w_\mathrm{int}=0$ (for $w_\mathrm{int}$ allowed to vary).
We obtain $54 < N_* < 62$ ($53 < N_* < 64$) at 95\,\% CL for $w_\mathrm{int}=0$ 
(for $w_\mathrm{int}$ allowed to vary), compatible with the theoretical prediction, 
$N_* = 54$ \citep{Starobinsky:1980te,Vilenkin:1985md,Gorbunov:2010bn}.

\subsubsection*{$\alpha$ attractors}

We now study two classes of inflationary models motivated by recent developments in conformal symmetry and supergravity 
\citep{Kallosh:2013yoa}. The first class has been motivated by considering a vector rather than a 
chiral multiplet for the inflaton in supergravity \citep{Ferrara:2013rsa} 
and corresponds to the potential \citep{Kallosh:2013yoa}
\begin{equation}
V(\phi) = \Lambda^4 \left( 1 - e^{- \sqrt{2} \phi/\left(\sqrt{3 \alpha} M_\mathrm{pl}\right) }\right)^2 \,.
\label{alpha}
\end{equation}
To lowest order in the slow-roll approximation, 
these models predict $r \approx 64/[3 \alpha (1-e^{\sqrt{2} \phi/(\sqrt{3 \alpha} M_\mathrm{pl})})^2]$ and
$n_\mathrm{s} -1 \approx - 8(1+e^{\sqrt{2} \phi/(\sqrt{3 \alpha} M_\mathrm{pl})})
/[3 \alpha (1-e^{\sqrt{2} \phi/(\sqrt{3 \alpha} M_\mathrm{pl})})^2]$ based on    
an inflationary trajectory characterized by $N \approx g(\phi/M_\mathrm{pl}) - 
g(\phi_\mathrm{end}/M_\mathrm{pl})$ with $g(x) = (3 \alpha^4 e^{\sqrt{2} x/\sqrt{3 \alpha}} 
- \sqrt{6 \alpha} x)/4$. The relation between $N$ and $\phi$ can be inverted through the use of the 
Lambert functions, as carried out for other potentials \citep{Martin:2014vha}. By sampling 
$\log_{10}(\alpha^2)$ with a flat prior over $[0,4]$,
we obtain $\log_{10}(\alpha^2) < 1.7$ ($2.0$) at 95\,\% CL and 
a Bayes factor of $-1.8$ ($-2$) for $w_\mathrm{int}=0$ (for $w_\mathrm{int}$ allowed to vary).

The second class of models has been called super-conformal $\alpha$ 
attractors \citep{Kallosh:2013yoa} and can be understood 
as originating from a different generating function with respect 
to the first class. This second class is described by 
the following potential \citep{Kallosh:2013yoa}:
\begin{equation}
V(\phi) = \Lambda^4 \tanh^{2 m} \left( \frac{\phi}{\sqrt{6 \alpha} M_\mathrm{pl}} \right) \,.
\label{alphaattractors}
\end{equation}
This is the simplest class of models with spontaneous breaking of conformal symmetry, and 
for $\alpha=m=1$ reduces to the original model introduced by \cite{Kallosh:2013lkr}.
The potential in Eq.~(\ref{alphaattractors}) 
leads to the following slow-roll predictions \citep{Kallosh:2013yoa}: 
\begin{align}\label{exact1}
r & \approx {48  \alpha m \over 4 m N^2+ 2 N g(\alpha,m) + 3 \alpha m} \, ,  \\
\notag \\
n_s - 1 & \approx - \frac{8 m N + 6 \alpha m + 2 g(\alpha,m)}{4 m N^2 + 2 N g(\alpha,m) + 3 \alpha m} \, , \label{exact}
\end{align}
where $g(\alpha,m) = \sqrt{3 \alpha (4 m^2 + 3 \alpha)}$.
The predictions of this second class of models interpolate between those of a large-field chaotic model, 
$V(\phi) \propto \phi^{2 m}$, for $\alpha \gg 1$ and the $R^2$ model for $\alpha \ll 1$.

For $\alpha$ we adopt the same priors as for the previous class in Eq.~(\ref{alpha}).
By fixing $m=1$, we obtain $\log_{10}(\alpha^2) < 2.3$ ($2.5$) at 95\,\% CL and
a Bayes factor of $-2.3$ ($-2.2$) for $w_\mathrm{int}=0$ (when $w_\mathrm{int}$ is allowed to vary). 
When $m$ is allowed to vary as well with a flat prior in the range $[0,2]$, we obtain $0.02 < m < 1$  
($m< 1$) at 95\,\% CL for $w_\mathrm{int}=0$  (when $w_\mathrm{int}$ is allowed to vary).

\subsubsection*{Non-minimally coupled inflaton}

Inflationary predictions are quite sensitive to a non-minimal coupling $\xi R \phi^2$ 
of the inflaton to the Ricci scalar.  
One of the most interesting effects due to $\xi \ne 0$ is 
to reconcile the quartic potential $V(\phi) = \lambda \phi^4/4$ with \Planck\ observations, 
even for $\xi \ll 1$. Non-minimal coupling leads as well to attractor behaviour 
towards predictions similar to those in $R^2$ inflation \citep{Kaiser:2013sna,Kallosh:2013maa}.

The Higgs inflation model \citep{Bezrukov:2007ep}, in which inflation occurs with 
$V(\phi) = \lambda (\phi^2-\phi_0^2)^2/4$ and $\xi \gg 1$ for $\phi \gg \phi_0$, 
leads to the same predictions as the $R^2$ model to lowest order in the slow-roll approximation at tree level 
(see \cite{Barvinsky:2008ia} and \cite{Bezrukov:2009db} for the inclusion of loop corrections). 
It is therefore in agreement with the \Planck\ constraints, 
as for the first cosmological data release \citepalias{planck2013-p17}.

\vspace{.5cm}

We summarize below our findings for \Planck\ lowP+BAO.
\begin{itemize}
\item{Monomial potentials with integral $n > 2$ are strongly disfavoured with respect to $R^2$.}
\item{The Bayes factor prefers $R^2$ over chaotic inflation with monomial quadratic potential by odds of 110:1 
under the assumption of a dust equation of state during the entropy generation stage.}

\item{$R^2$ inflation has the strongest evidence (i.e., the greatest Bayes factor)
among the models considered here. 
However, care must be taken not to overinterpret small differences in likelihood lacking statistical significance.}

\item{The models closest to $R^2$ in terms of evidence are brane inflation and exponential 
inflation, which have one more parameter than $R^2$. Both brane inflation considered 
in Eq.~(\ref{Dbrane_infl}) and exponential inflation in Eq.~(\ref{exp_infl}) 
approximate the linear potential for a large portion of parameter space 
(for $\mu/M_\mathrm{pl} \gg 1$ and $q \gg 1$, respectively).  For this reason these models have a higher
Bayesian evidence (although not at a statistically significant level) than those that approximate a quadratic potential, as 
do $\alpha$-attractors, for instance.}

\item{In the models considered here, 
the $\Delta \chi^2$ obtained by allowing $w$ to vary is modest (i.e., less than approximately $1.6$ 
with respect to $w_\mathrm{int}=0$). 
The gain in the logarithm of the Bayesian evidence is even smaller, since an extra parameter is added.}

\end{itemize}

\input TABLE4_rounded.tex

%% file: prior_table.tex
\begin{table}[tb]                 
\begin{center}
  \begingroup
  \newdimen\tblskip \tblskip=5pt
  \nointerlineskip
  \vskip -3mm
  \footnotesize
  \setbox\tablebox=\vbox{
    \newdimen\digitwidth
    \setbox0=\hbox{\rm 0}
    \digitwidth=\wd0
    \catcode`*=\active
    \def*{\kern\digitwidth}
    \newdimen\signwidth
    \setbox0=\hbox{+}
    \signwidth=\wd0
    \catcode`!=\active
    \def!{\kern\signwidth}
    \halign{%
      \hfil#\hfil&
      \hfil#\hfil\cr
      \noalign{\doubleline}
      Parameter range &
      Prior type\cr
      \noalign{\vskip 3pt\hrule\vskip 5pt}
      $0.019< \Omega_\mathrm{b} h^2 <0.025$ &
      uniform
      \cr
      $0.095< \Omega_\mathrm{c} h^2 <0.145$ &
      uniform
      \cr
      $1.03< 100\theta_\mathrm{MC} <1.05$ &
      uniform
      \cr
      $0.01< \tau< 0.4$ &
      uniform
      \cr
      \noalign{\vskip 3pt\hrule\vskip 5pt}}}
    \endPlancktable                    
  \endgroup
\end{center}
\caption{%
Priors for cosmological parameters used in the Bayesian comparison of inflationary models.}
\label{tab:Bayesian_comparison_priors_six}                            
\end{table}                        

%% file: TABLE4_rounded.tex
\begin{table}
\begingroup
\newdimen\tblskip \tblskip=5pt
\caption{Results of the inflationary model comparison.  We provide $\Delta \chi^2$ 
with respect to base $\Lambda$CDM and Bayes factors with respect to $R^2$ inflation.
}
\label{table:model_compar}
\nointerlineskip
\vskip -3mm
\footnotesize 
\setbox\tablebox=\vbox{
\newdimen\digitwidth
\setbox0=\hbox{\rm 0}
\digitwidth=\wd0
\catcode`*=\active
\def*{\kern\digitwidth}
\newdimen\signwidth
\setbox0=\hbox{+}
\signwidth=\wd0
\catcode`!=\active
\def!{\kern\signwidth}
\halign{#\hfil&\hfil#\hfil&\hfil#\hfil&\hfil#\hfil&\hfil#\hfil\cr
\noalign{\doubleline}
Inflationary model & \multispan2\hfil$\Delta \chi^2$\hfil & \multispan2\hfil$\ln B$\hfil\cr
\noalign{\vskip 2pt}
 & \ $w_\mathrm{int}=0\ $ & \ $w_\mathrm{int}\ne0$\ & \ $w_\mathrm{int}=0$\ &
\ $w_\mathrm{int}\ne0$\cr
\noalign{\vskip 5pt\hrule\vskip 5pt}
$R + R^2/(6 M^2)$ & *$+0.8$ & *$+0.3$ & $\dots$ & *$+0.7$ \cr
$n=2/3$ & *$+6.5$  & *$+3.5$ & *$-2.4$ & *$-2.3$ \cr
$n=1$ & *$+6.2$ & *$+5.5$ & *$-2.1$ & *$-1.9$ \cr
$n=4/3$ & *$+6.4$ & *$+5.5$ & *$-2.6$ & *$-2.4$ \cr
$n=2$ & *$+8.6$ & *$+8.1$ & *$-4.7$ & *$-4.6$ \cr
$n=3$ & $+22.8$ & $+21.7$ & $-11.6$ & $-11.4$ \cr
$n=4$ & $+43.3$ & $+41.7$ & $-23.3$ & $-22.7$ \cr
Natural & *$+7.2$ & *$+6.5$ & *$-2.4$ & *$-2.3$  \cr
Hilltop ($p=2$) & *$+4.4$ & *$+3.9$ & *$-2.6$ & *$-2.4$ \cr
Hilltop ($p=4$) & *$+3.7$ & *$+3.3$ & *$-2.8$ & *$-2.6$ \cr
Double well & *$+5.5$ & *$+5.3$ & *$-3.1$ & *$-2.3$ \cr
Brane inflation ($p=2$) & *$+3.0$ & *$+2.3$ & *$-0.7$ & *$-0.9$ \cr
Brane inflation ($p=4$) & *$+2.8$ & *$+2.3$ & *$-0.4$ & *$-0.6$ \cr
Exponential tails & *$+0.8$ & *$+0.3$ & *$-0.7$ & *$-0.9$ \cr
SB SUSY & *$+0.7$ & *$+0.4$ & *$-2.2$ & *$-1.7$ \cr
Supersymmetric $\alpha$-model & *$+0.7$ & *$+0.1$ & *$-1.8$ & *$-2.0$ \cr
Superconformal ($m=1$) & *$+0.9$ & *$+0.8$ & *$-2.3$ & *$-2.2$ \cr
Superconformal ($m\ne1$) & *$+0.7$ & *$+0.5$ & *$-2.4$ & *$-2.6$ \cr
\noalign{\vskip 3pt\hrule}}}
\endPlancktable
\endgroup
\end{table}

%% file: section_seven.tex
\input macros.tex
\subsection{Introdution}

In the previous section, we derived constraints on several types of inflationary potentials assumed to account for the inflaton dynamics between the time at which the largest observable scales crossed the Hubble radius during inflation and the end of inflation. The full shape of the potential was used in order to identify when inflation ends, and thus the field value $\phi_*$ when the pivot scale crosses the Hubble radius.

In section~6 of \citetalias{planck2013-p17}, we explored another approach, consisting of reconstructing the inflationary potential within its observable
range without making any assumptions concerning the inflationary dynamics outside that range. Indeed,
given that the number of $e$-folds between the observable range and the end of inflation can always be adjusted to take a realistic value,
any potential shape giving a primordial spectrum of scalar and tensor perturbations in agreement with observations is a valid candidate.
Inflation can end abruptly by a phase transition,
or can last a long time if the potential becomes very flat after the observable region has been crossed. Moreover, there could be a short inflationary stage
responsible for the origin of observable cosmological perturbations, and another inflationary stage later on (but before
nucleosynthesis), thus contributing to the total $N_*$.

In section~6 of \citetalias{planck2013-p17}, we performed this analysis with a full integration of the inflaton and metric perturbation modes, so that no
slow-roll approximation was made.  The only assumption was that primordial scalar perturbations are generated by the fluctuations of a single inflaton
field with a canonical kinetic term. Since in this approach one is only interested in the potential over a narrow range of observable scales (centred around the field value $\phi_*$ when the pivot scale crosses the Hubble radius), it is reasonable to test relatively simple potential shapes described by a small number of free parameters.

This approach gave very similar results to calculations based on the standard slow-roll analysis.
This agreement can be explained by the fact that
the \Planck\ 2013 data already preferred a primordial spectrum very close to a power law, at least over most of the observable range. Hence
the 2013 data excluded strong deviations from slow-roll inflation, which would either produce a large running of the spectral index or imprint
more complicated features on the primordial spectrum. However, this conclusion did not apply to the largest scales observable by \Planck, for
which cosmic variance and slightly anomalous data points remained compatible with significant deviations from a simple power law spectrum. 
The most striking result in section~6 of \citetalias{planck2013-p17} was that a less restricted functional form for the inflaton
potential gave results compatible with a rather steep potential at the beginning of the observable window, leading to a ``not-so-slow'' roll
stage during the first few observable $e$-folds.
This explains the shape of the potential in figure~14 of \citetalias{planck2013-p17} for a Taylor
expansion at order $n=4$ and in the region where $\phi -\phi_* \leq -0.2$. However, such features were only partially explored because the
method used for potential reconstruction did not allow for an arbitrary value of the inflation velocity $\dot{\phi}$ at the beginning of the
observable window. Instead, our code imposed that the inflaton already tracked the inflationary attractor solution when the largest observable
modes crossed the Hubble scale.

Given that the \Planck\ 2015 data establish even stronger constraints on the primordial power
spectrum than the 2013 results, it is of interest to revisit the reconstruction of the
potential $V(\phi)$. Section~\ref{sec:Vphi_taylor} presents some new results following the same approach as in \citetalias{planck2013-p17}
(explained previously in \cite{Lesgourgues:2007gp} and \cite{Mortonson:2010er}). But in the present work, we also present some more general results,
independent of any assumption concerning the initial field velocity $\dot{\phi}$ when the inflaton enters the observable window. Following
previous studies \citep{Kinney:2002qn,Kinney:2006qm,Peiris:2006ug,Easther:2006tv,Peiris:2006sj,Peiris:2008be,Lesgourgues:2007aa,Powell:2007gu,Hamann:2008pb,Norena:2012rs}, 
we reconstruct the Hubble
function $H(\phi),$ which determines both the potential $V(\phi )$ through
\begin{equation}
V(\phi)= 3 M_\mathrm{Pl}^2 \, H^2(\phi) - 2 M_\mathrm{Pl}^4 \left[H'(\phi)\right]^2~,
\label{eq:VfromH}
\end{equation}
and the solution $\phi (t)$ through
\begin{equation}
\dot \phi = - 2 M_\mathrm{Pl}^2 H'(\phi)~,
\end{equation}
with $H'(\phi) ={\partial H}/{\partial \phi}.$ Note that these two relations are exact. In Sect.~\ref{sec:Hphi_taylor}, we fit $H(\phi)$
directly to the data, implicitly including all canonical single-field models in which the inflaton is rolling not very slowly ($\epsilon$
not much smaller than unity) just before entering the observable window, and the issue of having to start sufficiently early in order to allow the
initial transient to decay is avoided. The only drawback in reconstructing $H(\phi)$ is that one cannot systematically test the simplest 
analytic forms for $V(\phi)$ in the observable range (for instance, polynomials of order $n=1,3,5,\dots$ in $(\phi-\phi_*)$). But our goal in this
section is to explore how much one can deviate from slow-roll inflation in general, independently of the shape of the underlying inflaton
potential.

\begin{figure}[!t]
\includegraphics[width=0.99\columnwidth]{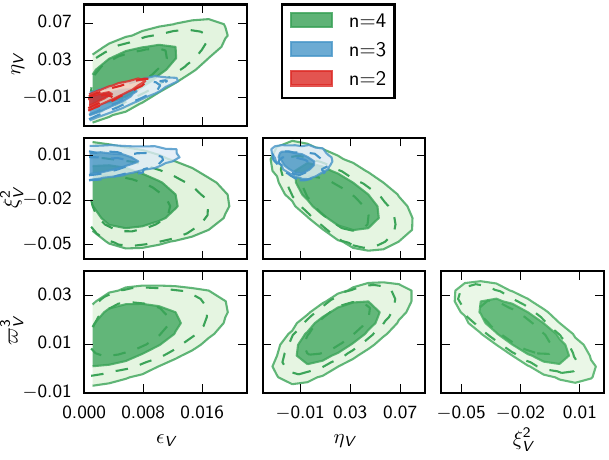}
\caption{Posterior distributions for the first four potential slow-roll parameters when the potential is Taylor 
expanded to $n$th order using
\Planck\ TT+lowP+BAO (filled contours) or TT,TE,EE+lowP (dashed contours). The primordial spectra are computed {\it beyond} the
slow-roll
approximation. \label{fig:PSR}}
\end{figure}

\begin{table}
\begin{center}
\begin{tabular}{c c c c}
\noalign{\hrule\vskip 2pt}
\noalign{\hrule\vskip 3pt}
$n$ & 2 & 3 & 4  \\
\hline
&&&\\[-1.5ex]
$\epsilon_V$  & $<0.0074$ & $<0.010$ & $0.0072_{-0.0069}^{+0.0093}$ \\[1.1ex]
$\eta_V$      & $-0.007_{-0.012}^{+0.014}$ & $-0.020_{-0.018}^{+0.021}$  & $0.021_{-0.042}^{+0.044}$ \\[1.1ex]
$\xi_V^2$      & \dots & $0.006_{-0.010}^{+0.010}$ & $-0.018_{-0.027}^{+0.028}$ \\[1.1ex]
$\varpi_V^3$    & \dots & \dots & $0.015_{-0.017}^{+0.016}$ \\[0.5ex]
\hline
&&&\\[-1.5ex]
$\tau$                  & $0.083_{-0.036}^{+0.036}$ & $0.096_{-0.044}^{+0.046}$ & $0.102_{-0.045}^{+0.046}$ \\[1.1ex]
$n_\mathrm{s}$           & $0.9692_{-0.0093}^{+0.0094}$ & $0.9689_{-0.0097}^{+0.0097}$ & $0.964_{-0.011}^{+0.011}$ \\[1.1ex]
$dn_\mathrm{s}/d\ln k$  & $-0.00034_{-0.00059}^{+0.00055}$ & $-0.013_{-0.019}^{+0.019}$ & $0.003_{-0.026}^{+0.026}$ \\[1.1ex]
$r_{0.002}$              & $<0.11$ & $<0.16$ & $0.11_{-0.11}^{+0.16}$ \\[0.5ex]
\hline
&&&\\[-1.5ex]
$\Delta \chi^2$ & \dots  & $\Delta \chi^2_{3/2}= -1.2$ & $\Delta \chi^2_{4/3}= -2.1$ \\[1.1ex]
$\Delta \ln B$ & \dots & $\Delta \ln B_{3/2} = -4.3$ & $\Delta \ln B_{4/3}=-2.9$ \\[0.5ex]
\hline
\end{tabular}
\end{center}
\caption{Numerical reconstruction of the potential slow-roll parameters {\it beyond} any slow-roll approximation when the potential is
Taylor expanded to $n$th order, using \Planck\ TT+lowP+BAO. We also show the corresponding bounds on the derived parameters (here
$n_\mathrm{s}$, $dn_\mathrm{s}/d\ln k$, and $r_{0.002}$ are derived from the numerically computed primordial spectra). All error bars are at the 95\,\%
CL. The effective $\chi^2$ value and Bayesian evidence logarithm $(\ln B)$ of model $n$ are given relative to the model
of next lowest order $(n-1)$ (assuming flat priors
for $\xi_V^2$ and $\varpi_V^3$ in the range $[-1,1]$).
\label{tab:V_phi}}
\end{table}

\subsection{Reconstruction of a smooth inflaton potential \label{sec:Vphi_taylor}}

Following the approach of \citetalias{planck2013-p17}, we Taylor expand the inflaton potential around $\phi=\phi_*$ to order $n=2, 3, 4$. To
obtain faster-converging Markov chains, instead of imposing flat priors on the Taylor coefficients $\{V, V_\phi,\dots, V_{\phi \phi \phi \phi}\}$,
we sample the potential slow-roll (PSR) parameters $\{ \epsilon_V, \eta_V, \xi_V^2, \varpi_V^3\}$ related to the former as indicated in
Table~\ref{table:InflationDefinitions}. We stress that this is just a choice of prior and does not imply 
any kind of slow-roll
approximation in the calculation of the primordial spectra.

\begin{figure}[t]
\includegraphics[width=0.99\columnwidth]{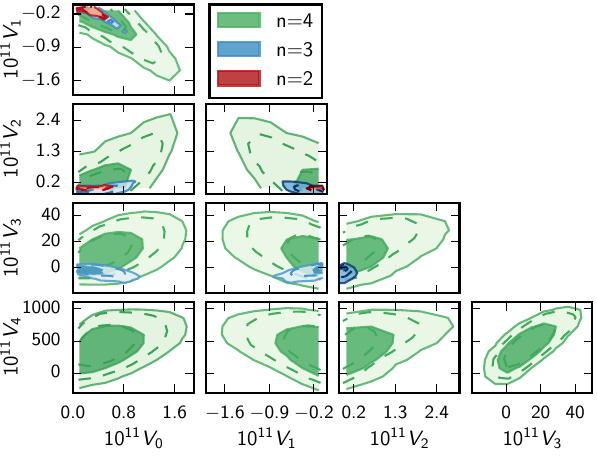}
\caption{Posterior distributions for the coefficients of the inflation potential Taylor expanded to $n$th order 
(in natural units where $\sqrt{8
\pi} M_\mathrm{pl}=1$) reconstructed {\it beyond} the slow-roll approximation 
using \Planck\ TT+lowP+BAO (filled contours) or TT,TE,EE+lowP
(dashed contours). The plot shows only half of the results; the other half is symmetric, with opposite signs for $V_{\phi}$ and
$V_{\phi\phi\phi}$. Note that, unlike Fig.~\ref{fig:PSR}, the parameters shown here do {\it not} have flat priors, since they
are mapped from the slow-roll parameters. \label{fig:V}}
\end{figure}

\begin{figure}[t]
\includegraphics[width=0.99\columnwidth]{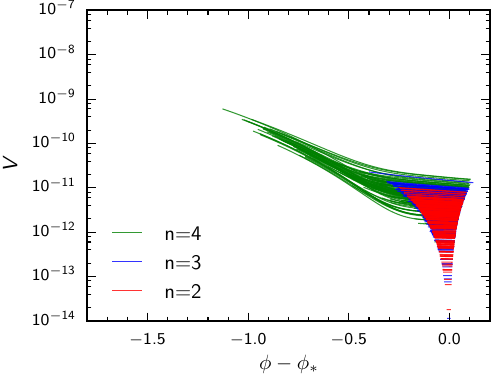}
\caption{Observable range of the best-fit inflaton potentials, when $V(\phi)$ is Taylor expanded to the $n$th order around the pivot value
$\phi_*$ in natural units (where $\sqrt{8 \pi} M_\mathrm{pl}=1)$ 
assuming a flat prior on $\epsilon_V$, $\eta_V$, $\xi^2_V$, and $\varpi_V^3$
and using \Planck\ TT+lowP+BAO. Potentials obtained under the transformation $(\phi-\phi_*)\rightarrow(\phi_*-\phi)$ leave the same
observable signature and are also allowed. The sparsity of potentials with a small $V_0=V(\phi_*)$ is due to our use of a flat prior on $\epsilon_V$
rather than on $\ln(V_0)$; in fact, $V_0$ is unbounded from below in the $n=2$ and $3$ results. The axis ranges are identical to those in Fig.~\ref{fig:H_V_phi}, to make the comparison easier. \label{fig:Vphi}}
\end{figure}

\begin{figure}[t]
\begin{center}
\includegraphics[width=0.99\columnwidth]{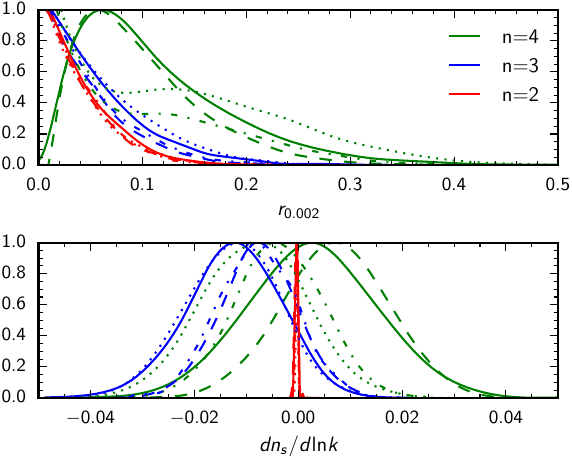}
\end{center}
\caption{Posterior distribution for the tensor-to-scalar ratio (at $k=0.002 \, \mathrm{Mpc}^{-1}$) and for the running parameter $dn_\mathrm{s}/d\ln k$ (at $k=0.05 \, \mathrm{Mpc}^{-1}$), for the potential reconstructions in Sects.~\ref{sec:Vphi_taylor} and \ref{sec:Hphi_taylor}. The $V(\phi)$ reconstruction gives the solid curves for \Planck\ TT+lowP+BAO, or dashed for TT,TE,EE+lowP. The $H(\phi)$ reconstruction gives the dotted curves for \Planck\ TT+lowP+BAO, or dashed-dotted for TT,TE,EE+lowP. The tensor-to-scalar ratio appears as a derived parameter, but by taking a flat prior on either $\epsilon_V$ or $\epsilon_H$, we implicitly also take a nearly flat prior on $r$. The same applies to $dn_\mathrm{s}/d\ln k$. \label{fig:VH_r}}
\end{figure}

The results are given in Table~\ref{tab:V_phi} (for \Planck\ TT+lowP+BAO) and Fig.~\ref{fig:PSR} (for the same data set 
and also for \Planck\
TT,TE,EE+lowP). The second part of Table~\ref{tab:V_phi} shows the corresponding values of the spectral parameters $n_\mathrm{s}$,
$dn_\mathrm{s}/d\ln k$, and $r_{0.002}$ as measured for each numerical primordial spectrum (at the pivot scales $k=0.05 \,
\mathrm{Mpc}^{-1}$ for the scalar and $0.002\, \mathrm{Mpc}^{-1}$ for the tensor spectra), as well as the reionization optical depth. We also show in Fig.~\ref{fig:V} the derived
distribution of each coefficient $V_i$ (with a non-flat prior) and in Fig.~\ref{fig:Vphi} the reconstructed shape of the best-fit inflation
potentials in the observable window. 
Finally, the posterior distribution of the derived parameters $r_{0.002}$ and $dn_\mathrm{s}/d\ln k$ is displayed in
Fig.~\ref{fig:VH_r}.

Figure~\ref{fig:PSR} shows that bounds are very similar when temperature data are combined with either high-$\ell$ polarization data or BAO data. This gives a hint of the robustness of these results. For both data sets, the error bars on the PSR parameters are typically 
smaller by a factor of $1.5$ than in \citetalias{planck2013-p17}.

Since potentials with $n=2$ cannot generate a significant running, the bounds on the scalar spectral index and the tensor-to-scalar 
ratio
and the best-fit
models are very similar to those obtained with the $\Lambda$CDM+$r$ model in Sect.~\ref{sec:modelcomp} and Table~\ref{tab:ralphar}. On the other hand, in
the $n=3$ model, results follow the trend of the previous $\Lambda$CDM+$r$+$dn_\mathrm{s}/d\ln k$ analysis. 
The data prefer potentials with
$V_{\phi}$ and $V_{\phi\phi\phi}$ of the same sign, generating a significant negative running (as can be seen in Fig.~\ref{fig:VH_r}). 
This trend for $V_{\phi\phi\phi}$ occurs because a scalar spectrum with negative running reduces the power on large scales, and provides a
better fit to low-$\ell$ temperature multipoles.
However, such a running also suppresses power on small scales, so $\xi_V^2$ cannot be too large.

\begin{figure}[th]
\begin{center}
\includegraphics[width=0.99\columnwidth]{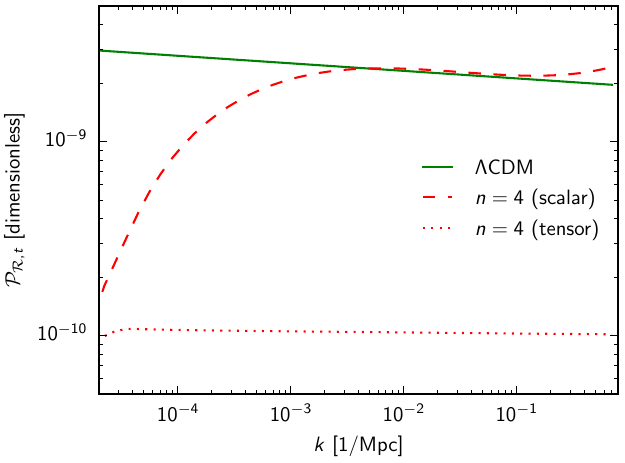}
\end{center}
\caption{Primordial spectra (scalar and tensor) of the best-fit $V(\phi)$ model with $n=4$, for the \Planck\ TT,TE,EE+lowP data set, compared to
the primordial spectrum (scalar only) of the best-fit base $\Lambda$CDM model. The best-fit potential is initially very steep, as can be seen in
Fig.~\ref{fig:Vphi} (note the typical shape of the green curves). The transition from ``marginal slow roll'' ($\epsilon_V(\phi)$ between 0.01 and 1) to ``full
slow roll'' ($\epsilon_V(\phi)$ of order 0.01 or smaller) is responsible for the suppression of the large-scale scalar spectrum.
\label{fig:V4_pk}}
\end{figure}

\begin{figure}[th]
\begin{center}
\includegraphics[width=0.99\columnwidth]{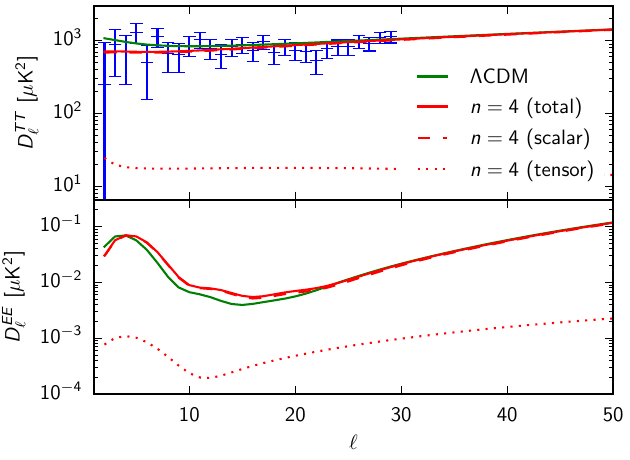}
\end{center}
\caption{Temperature and polarization spectra (total, scalar contribution, tensor contribution) of the best-fit $V(\phi)$ model with $n=4$, for the \Planck\ TT,TE,EE+lowP data set, compared to the spectra (scalar contribution only) of the best-fit base model. We also show the \Planck\ low-$\ell$ temperature data, which is
driving the small differences between the two best-fit models. \label{fig:V4_cl}}
\end{figure}

The $n=4$ case possesses a new feature. The potential has more freedom to generate complicated shapes which would roughly correspond to a running of
the running of the tilt (as studied in Sect.~\ref{sec:updates}). The best-fit models now have $V_{\phi}$ and $V_{\phi\phi\phi}$ of opposite
sign, and a large positive $V_{\phi\phi\phi\phi}$. The preferred combination of these parameters allows for even more suppression of power on
large scales, while leaving small scales nearly unchanged. This can be seen clearly
from the shape of the scalar primordial spectrum
corresponding to the best-fit models, for both data sets \Planck\ TT+lowP+BAO and \Planck\ TT,TE,EE+lowP. These two best-fit models are very
similar, but in Fig.~\ref{fig:V4_pk} we show the one for \Planck\ TT,TE,EE+lowP, for which the trend is even more pronounced. Interestingly,
the preferred models are such that power on large scales is suppressed in the scalar spectrum and balanced by a small tensor contribution, of
roughly $r_{0.002} \sim 0.05$. This particular combination gives the best fit to the low-$\ell$ data, shown in Fig.~\ref{fig:V4_cl}, while
leaving the high-$\ell$ temperature spectrum identical to the best fit base $\Lambda$CDM model. Inflation produces such primordial perturbations with the
family of green potentials displayed in Fig.~\ref{fig:Vphi}. At the beginning of the observable range, the potential is very steep 
[$\epsilon_V(\phi)$
decreases from $O(1)$ to $O(10^{-2})$], and produces a low amplitude of curvature perturbations (allowing a rather large tensor contribution, up
to $r_{0.002} \sim 0.3$). Then there is a transition towards a second region with a much smaller slope, leading to a nearly power-law
curvature spectrum with the usual tilt value $n_\mathrm{s}\approx0.96$. In Fig.~\ref{fig:Vphi}, one can check that the height of the $n=4$ potentials
varies in a definite range, while the $n=2$ and $3$ potentials can have arbitrarily small amplitude at the pivot scale, reflecting the posteriors on
$\epsilon_V$ or $r$.

However, the improvement in $\chi^2$ between the base $\Lambda$CDM and $n=4$ models is only 2.2 (for \Planck\
TT+lowP+BAO) or 4.3 (for \Planck\ TT,TE,EE+lowP). This is marginal and 
offers no statistically significant evidence in favour of these complicated models. This
conclusion is also supported by the calculation of the Bayesian evidence ratios, shown in the last line of Table~\ref{tab:V_phi} (under the
assumption of flat priors in the range [$-1$,1] for $\xi_V^2$ and $\varpi_V^3$): the evidence decreases each time that a new free parameter is
added to the potential. At the 95\,\% CL, $r_{0.002}$ is still compatible with zero, and so are the higher order PSR parameters $\xi_V^2$ and
$\varpi_V^3$. More freedom in the inflaton potential allows fitting the data better, but under the
assumption of a smooth potential in the observable range, a simple quadratic form provides the best explanation of the \Planck\ observations.

With the \Planck\ TT+lowP+BAO and TT,TE,EE+lowP data sets, models with a large running or running of the running can be compatible with an unusually large value of the optical depth, as can be seen in Table~\ref{tab:V_phi}. Including lensing information helps to break the degeneracy between the optical depth and the primordial amplitude of scalar perturbations. Hence the \Planck\ lensing data could be used to strengthen the conclusions of this section.

Since in the $n=4$ model, slow roll is marginally satisfied at the beginning of observable inflation, the reconstruction is very sensitive to the
condition that there is an attractor solution at that time. Hence this case can in principle be investigated in a more conservative way using the
$H(\phi)$ reconstruction method of the next section.

\subsection{Reconstruction of a smooth Hubble function \label{sec:Hphi_taylor}}

In this section, we assume that the shape of the function $H(\phi)$ is well captured within the observable window by a polynomial of order $n$ (corresponding
to a polynomial inflaton potential of order $2n$):
\begin{equation}
H(\phi) = \sum_{i=0}^n H_i \frac{\phi^i}{i!}~.
\end{equation}
We vary $n$ between 2 and 4.
To avoid parameter degeneracies, as in the previous section we assume flat priors
not on the Taylor coefficient
$H_i$, but
on the Hubble slow-roll (HSR) parameters, which are related according to
\begin{align}
\epsilon_H &=  2 M_\mathrm{pl}^2 \left( \frac{H_1}{H_0} \right)^2~,
&\eta_H &= 2 M_\mathrm{pl}^2 \frac{H_2}{H_0}~, \\
\xi_H^2 &= \left(2 M_\mathrm{pl}^2\right)^2 \frac{H_1 H_3}{H_0^2}~,
&\varpi_H^3 &= \left(2 M_\mathrm{pl}^2\right)^3 \frac{H_1^2 H_4}{H_0^3}~.
\end{align}
This is just a choice of prior. This analysis does not rely on the slow-roll approximation.

\begin{table}
\begin{center}
\begin{tabular}{c c c c}
\noalign{\hrule\vskip 2pt}
\noalign{\hrule\vskip 3pt}
$n$ & 2 & 3 & 4  \\
\hline
&&&\\[-1.5ex]
$\epsilon_H$  & $<0.0073$ & $<0.011$ & $<0.020$ \\[1.1ex]
$\eta_H$  & $-0.010_{-0.009}^{+0.011}$ & $-0.012_{-0.013}^{+0.015}$ & $-0.001_{-0.027}^{+0.033}$ \\[1.1ex]
$\xi_H^2$  & \dots & $0.08_{-0.12}^{+0.12}$ & $-0.01_{-0.19}^{+0.19}$ \\[1.1ex]
$\varpi_H^3$  & \dots & \dots & $1.0_{-1.8}^{+2.3}$ \\[0.5ex]
\hline
&&&\\[-1.5ex]
$\tau$                 & $0.082_{-0.036}^{+0.038}$ & $0.096_{-0.043}^{+0.042}$ & $0.096_{-0.042}^{+0.042}$ \\[1.1ex]
$n_\mathrm{s}$          & $0.9693_{-0.0093}^{+0.0094}$ & $0.9680_{-0.0096}^{+0.0096}$ & $0.967_{-0.010}^{+0.010}$ \\[1.1ex]
$10^3dn_\mathrm{s}/d\ln k$ & $-0.251_{-0.41}^{+0.41}$ & $-13_{-19}^{+18}$ & $-8_{-21}^{+21}$ \\[1.1ex]
$r_{0.002}$             & $<0.11$ & $<0.16$ & $<0.32$ \\[0.5ex]
\hline
&&&\\[-1.5ex]
$\Delta \chi^2$ & \dots  & $\Delta \chi^2_{3/2}= -0.6$ & $\Delta \chi^2_{4/3}= -2.3$ \\[0.5ex]
\hline
\end{tabular}
\end{center}
\caption{Numerical reconstruction of the Hubble slow-roll parameters {\it beyond} the slow-roll approximation, using \Planck\ TT+lowP+BAO. We
also show the corresponding bounds on some related parameters (here $n_\mathrm{s}$, $dn_\mathrm{s}/d\ln k$, and $r_{0.002}$ are derived from the
numerically computed primordial spectra).  All error bars are at the 95\,\% CL. The effective $\chi^2$ value of model $n$ is given relative to model
$n-1$.
\label{tab:H_phi}}
\end{table}

\begin{figure}[th]
\includegraphics[width=0.99\columnwidth]{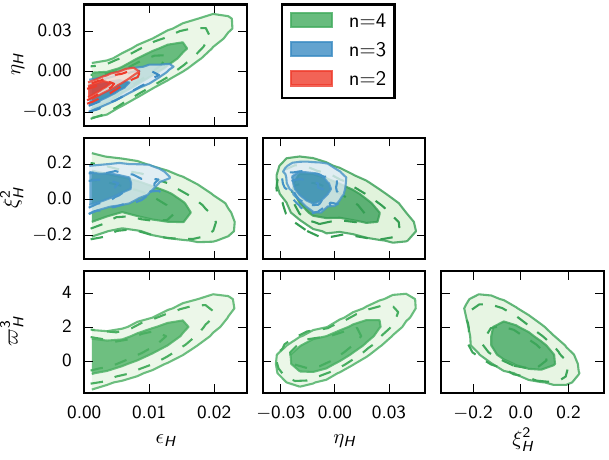}
\caption{Posterior distributions for the first four Hubble slow-roll parameters, when $H(\phi)$ is Taylor expanded to $n$th order, using \Planck\
TT+lowP+BAO (filled contours) or TT,TE,EE+lowP (dashed contours). The primordial spectra are computed {\it beyond} the slow-roll approximation.
\label{fig:HSR}}
\end{figure}

\begin{figure}[th]
\includegraphics[width=0.99\columnwidth]{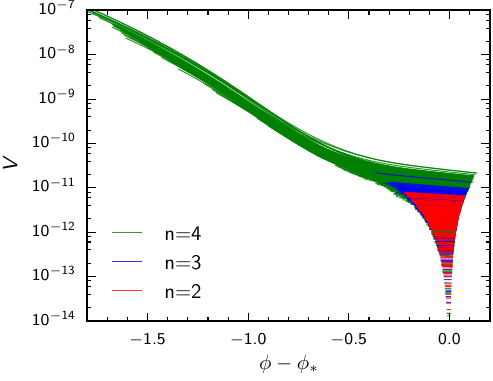}
\caption{Same as Fig.~\ref{fig:Vphi}, when the Taylor expansion to $n$th order is performed on $H(\phi)$ instead of $V(\phi)$, and the
potential is inferred from Eq.~(\ref{eq:VfromH}).
\label{fig:H_V_phi}
}
\end{figure}

Table~\ref{tab:H_phi} and Fig.~\ref{fig:HSR} show our results for the reconstructed HSR parameters. Figure~\ref{fig:H_V_phi} shows a
representative sample of potential shapes $V(\phi-\phi_*)$ derived using Eq.~(\ref{eq:VfromH}), for a sample of models drawn randomly from
the chains, for the three cases $n=2,3,4$.

Most of the discussion of Sect.~\ref{sec:Vphi_taylor} also applies to this section, and so will not be repeated. Results for \Planck\
TT+lowP+BAO and TT,TE,EE+lowP are still very similar. The $n=2$ case still gives results close to $\Lambda$CDM+$r$, and the $n=3$ case to
$\Lambda$CDM+$r$+$dn_\mathrm{s}/d\ln k$. 
The type of potential preferred in the $n=4$ case is very similar to the $n=4$ analysis of the previous
section, for the reasons explained in Sect.~\ref{sec:Vphi_taylor}. There are, however, small differences, because the
range of parametric forms for the potential explored by the two analyses differ. In the $H(\phi)$ reconstruction, the underlying potentials $V(\phi)$
are not polynomials. In the first approximation, they are close to polynomials of order $2n$, but with constraints between the various coefficients.
The main two differences with respect to the results of Sect.~\ref{sec:Vphi_taylor} are as follows:
\begin{itemize}

\item The reconstructed potential shapes for $n=4$ at the beginning of the observable window differ.
Figure~\ref{fig:H_V_phi} shows that even steeper potentials are allowed than for the 
$V(\phi)$ method, with
an even greater excursion of the inflaton field between Hubble crossing for the largest observable wavelengths and the pivot scale. This
is because the $H(\phi)$ reconstruction does not rely on attractor solutions and automatically
explores all valid potentials regardless of their initial field velocity.

\item The best-fit models are different, since they do not explore the same
parametric families of potentials. In particular, for $n=4$, the best-fit models have a negligible tensor contribution, but the distributions still have 
thick tails towards large tensor-to-scalar ratios, so that the upper bound on $r_{0.002}$ is as high as in the previous $n=4$
models, $r_{0.002}<0.32$.

\end{itemize}

Note that $\varpi_H^3$ can be significantly larger than unity for $n=4.$ This does not imply
violation of slow roll within the observable range. By assumption, for all accepted models, $\epsilon_H$ must remain smaller than unity over that
range. In fact, for most of the green potentials visible in Fig.~\ref{fig:H_V_phi}, we checked that $\epsilon_H$ either has a maximum very
close to unity near the beginning of the observable range or starts from unity. So the best-fit models (maximizing the power
suppression at
low multipoles) correspond either to inflation of short duration, or to models nearly violating slow roll just
before the observable window.
However, such peculiar models are not necessary for a good fit. Table~\ref{tab:H_phi} shows that the improvement
in $\chi^2$ as $n$ increases is negligible.

In summary, this section further establishes the robustness of our potential reconstruction and two main conclusions.  
Firstly, under the assumption that the inflaton potential is smooth over the observable range, we showed that the simplest parametric forms
(involving only three free parameters including the amplitude $V(\phi_*)$, no deviation from slow roll, and nearly power law primordial spectra)
are sufficient to explain the data. No high-order derivatives or deviations from slow roll are required. Secondly, if one allows
more freedom in the potential---typically, two more parameters---it is easy to decrease the large-scale primordial spectrum amplitude
with an initial stage of  ``marginal slow roll'' along a steep branch of the potential followed by a transition to a less steep branch.
This type of model can accommodate a large tensor-to-scalar ratio, as high as $r_{0.002} \approx 0.3.$

%% file: section_eight_1.tex
\newcommand{\PR}{\mathcal{P}_\mathcal{R}}
\newcommand{\f}{\mathbf{f}}
\def\be{\begin{equation}}
\def\ee{\end{equation}}
\newcommand{\Lcal}{\mathcal{L}}
\newcommand{\Mcal}{\mathcal{M}}
\newcommand{\BTheta}{\boldsymbol \Theta}

In \citetalias{planck2013-p17} (section~7) we presented the results of a penalized likelihood 
reconstruction, seeking to detect any possible deviations from a homogeneous power-law form (i.e., ${\cal 
P}_{\cal R}(k)\propto k^{n_\mathrm{s}-1}$) for the primordial power spectrum (PPS) for various values of a smoothing 
parameter, $\lambda .$ (For an extensive set of
references to the prior literature concerning
the methodology for reconstructing the power spectrum, 
see \citetalias{planck2013-p17}.) In the initial March 2013 preprint version of that paper, we reported evidence for
a feature at 
moderate statistical significance around $k\approx 0.15$\,Mpc$^{-1}$. However, in the November 2013 revision 
we retracted this finding, because subsequent tests indicated that the feature was no longer statistically 
significant when more aggressive cuts were made to exclude sky survey rings where contamination from 
electromagnetic interference from the 4\,K cooler was largest, as indicated in the November 2013 ``Note Added.'' 

In this section we report on results using the 2015 $C^{TT}_\ell$ likelihood (Sect.~\ref{PenalizedUpdate}) using 
essentially the same methodology as described in \citetalias{planck2013-p17}. 
(See \cite{Gauthier} and references therein for more technical details.) This 
method is also extended to include the $EE$ and $TE$ likelihoods
in Sect.~\ref{PenalizedLikeliPolar}. 
As part of this 2015 release, we include the results of two other methods 
(see Sects.~\ref{sec:PR_Bayes} and \ref{sec:csrecon})
to search for features. We
find that all three methods yield broadly consistent reconstructions and reach the following 
main conclusion: there is no statistically significant evidence for any features departing from a 
simple power-law (i.e., $\mathcal{P}_{\cal R}(k)\propto k^{n_\mathrm{s}-1}$) PPS. Given the 
substantial differences between these methods, it is satisfying to observe this convergence.

\subsection{Method I: penalized likelihood \label{PenalizedUpdate}}

\subsubsection{Update with 2015 temperature likelihood}

We repeated the same maximum likelihood analysis used to reconstruct the PPS in 
\citetalias{planck2013-p17} using the updated \Planck\ TT+lowP likelihood. 
Since we are interested in deviations 
from the nearly scale-invariant model currently favoured by the parametric approach, we replaced the true 
PPS $\PR(k)$ by a fiducial power-law spectrum $\PR^{(0)}(k) = A_{\mathrm s} (k/k_{\ast})^{n_\mathrm{s}-1}$, modulated by 
a small deviation function $f(k)$:
\be
\PR(k) = \PR^{(0)}(k) \exp \left[ f(k) \right] .
\ee
The deviation function $f(k)$\footnote{The definition of $f(k)$ used here differs from 
that of \citetalias{planck2013-p17}
in that $\exp (f)$ is used in place of $1+f$  
to ensure that the reconstructed primordial power spectrum is always 
non-negative.} was represented by $B$-spline basis functions parameterized by $n_\mathrm{knot}$ control points 
$\f = \{f_{i}\}_{i=1}^{n_\mathrm{knot}}$, which are the values of $f(k)$ along a grid of knot points 
$\kappa_{i} = \ln k_{i}$.

Naively maximizing the \Planck\ TT+lowP likelihood with respect to $\f$ results in over-fitting to cosmic 
variance and noise in the data. Furthermore, due to the limited range of scales over which \Planck\ measures 
the anisotropy power spectrum, the likelihood is very weakly dependent on $f(k)$ at extremely small and large scales. To 
address these issues, the following two penalty functions were added to the \Planck\ likelihood:
\be
\begin{aligned}
&
\mathbf{f}^\mathrm{T}
\mathbf{R}(\lambda,\alpha)
\mathbf{f} \equiv
 \lambda  \int
\mathrm{d}\kappa~
\left(
\frac{
\partial^2f(\kappa)
}{
\partial \kappa^2
}
\right) ^2
 \cr
&\qquad 
+
\alpha \int _{-\infty }^{\kappa _{\mathrm{min}}}
\mathrm{d}\kappa~f^2(\kappa)
+
\alpha \int ^{+\infty }_{\kappa _{\mathrm{max}}} \mathrm{d}\kappa~f^2(\kappa).
\label{Priors}
\end{aligned}
\ee

The first term on the right-hand side of Eq.~(\ref{Priors}) is a roughness penalty, which disfavours 
$f(\kappa)$ that ``wiggle'' too much. The last two terms drive $f(\kappa)$ to zero 
for scales below $\kappa_\mathrm{min}$ and above $\kappa_\mathrm{max}$. The values of $\lambda$ and $\alpha$ 
represent the strengths of the respective penalties. The exact value of $\alpha$ is unimportant as 
long as it is large enough to drive $f(\kappa)$ close to zero on scales outside $[\kappa_\mathrm{min}, \kappa_\mathrm{max}]$. 
However, the magnitude of the roughness penalty $\lambda$ controls the smoothness of the reconstruction.

Since the anisotropy spectrum depends linearly on the PPS, the Newton-Raphson method is well suited 
to optimizing with respect to $\f$. However, a maximum likelihood analysis 
also has to take into account the cosmological parameters, $\boldsymbol \Theta \equiv \{H_{0}, \Omega_\mathrm{b} 
h^{2}, \Omega_\mathrm{c} h^{2}\} .$\footnote{Due to the high 
correlation between $\tau$ and $A_{\mathrm{s}}$, $\tau$ is not included as a free 
parameter. Any change in $\tau$ can be almost exactly compensated for by a change in $A_{\mathrm{s}}$. We 
fix $\tau$ to its best-fit fiducial model value.} These additional parameters are not easy to 
include in the Newton-Raphson method since it is difficult to evaluate the derivatives
${\partial C_{\ell}}/{\partial \boldsymbol \Theta}$, 
${\partial^{2} C_{\ell}}/{\partial \boldsymbol \Theta^{2}}$, 
etc., to the accuracy required by the method. Therefore a non-derivative method, such as 
the downhill simplex algorithm, is best suited to optimization over these parameters. 
Unfortunately the downhill simplex method is inefficient given the large number of control points in 
our parameter space. Since each method  has its drawbacks, we combined the two methods to draw on 
their respective strengths. We define the function $\Mcal$ as
\begin{equation}
\mathcal{M}(\boldsymbol \Theta) =\underset{f_{i} \in [-1,1]}{\textrm{min}} \left\{-2 \ln 
\mathcal{L}(\boldsymbol \Theta,\mathbf{f}) + \mathbf{f}^{T} \mathbf{R}(\lambda,\alpha) 
\mathbf{f}\right\}.
\end{equation}
Given a set of non-PPS cosmological parameters $\BTheta$, $\Mcal$ is the value of the penalized log 
likelihood, minimized with respect to $\f$ using the Newton-Raphson method. The function $\Mcal$ is 
in turn minimized with respect to $\BTheta$ using the downhill simplex method. In contrast to the 
analysis done in \citetalias{planck2013-p17}, the \Planck\ low-$\ell$ likelihood has been modified so that it 
can be included in the Newton-Raphson minimization. Thus the reconstructions presented here extend to 
larger scales than were considered in 2013.

\begin{figure*}[!t]
\begin{center}
\includegraphics[scale=.9]%
{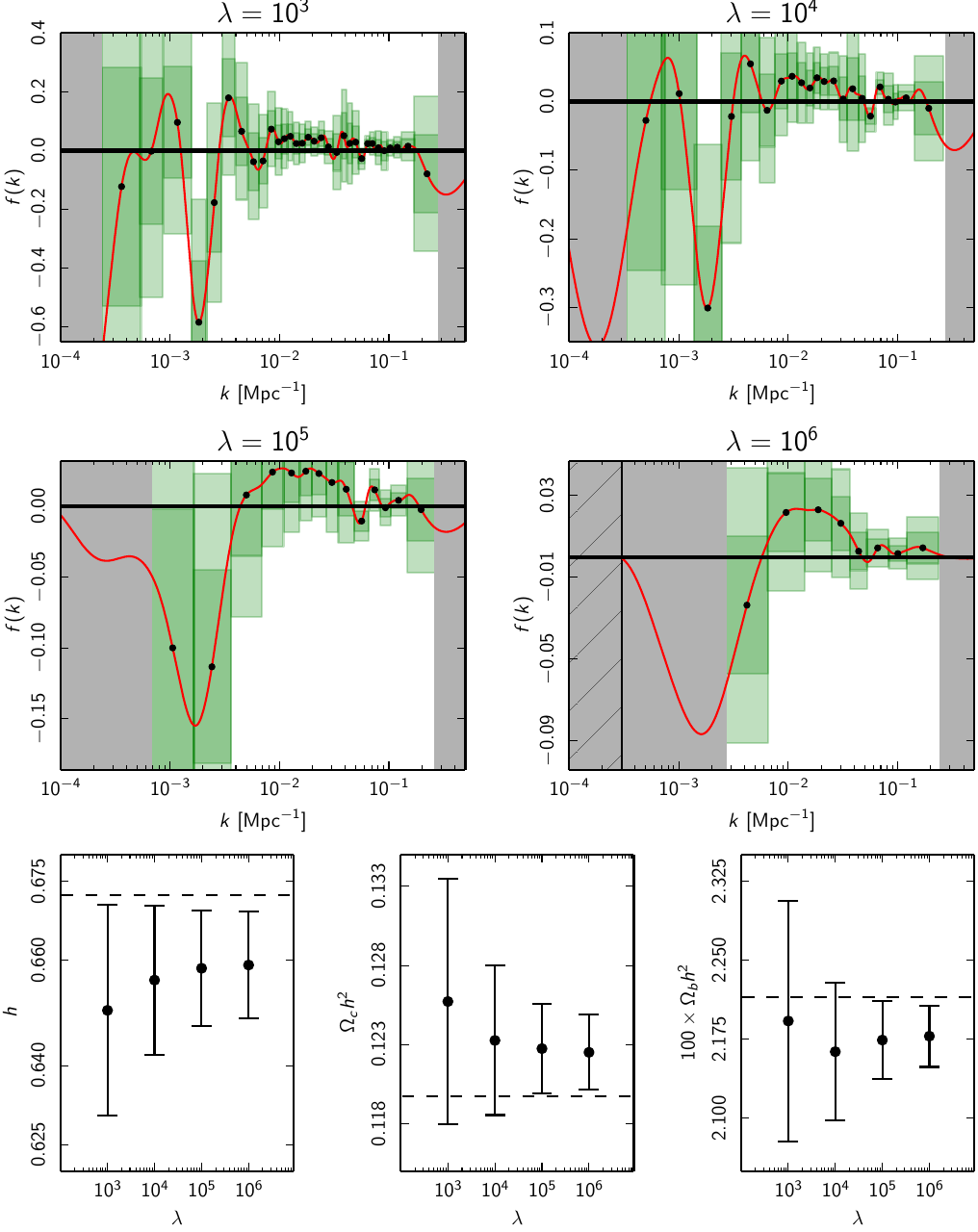}
\end{center}
\caption{
\Planck\ TT+lowP likelihood primordial power spectrum (PPS) reconstruction results. {\it Top four panels}:
Reconstruction of the deviation $f(k)$ using four different roughness penalties. The red curves
represent the best-fit deviation $f(k)$ using the \Planck\ TT+lowP likelihood. $f(k)=0$ would represent a
perfectly featureless spectrum with respect to the fiducial PPS model, which is obtained from the
best-fit base $\Lambda$CDM model with a power-law PPS. The vertical extent of the dark and light green
error bars indicates the $\pm1\,\sigma$ and $\pm2\,\sigma$ errors, respectively. The width of the error bars represents the
minimum reconstructible width (the minimum width for a Gaussian feature so that the mean square
deviation of the reconstruction is less than 10\,\%). The grey regions
indicate where the minimum reconstructible width is undefined, indicating that the reconstruction in
these regions is untrustworthy. The hatched region in the $\lambda = 10^{6}$ plot shows where the fixing
penalty has been applied. These hashed regions are not visible in the other three reconstructions,
for which $\kappa_\mathrm{min}$ lies outside the range shown in the plots. For all
values of the roughness penalty, all data points are within 2\,$\sigma$ of the fiducial PPS except for the
deviations around $k \approx 0.002$\,Mpc$^{-1}$ in the $\lambda = 10^{3}$ and $\lambda = 10^{4}$
reconstructions.
{\it Lower three panels:} $\pm1\,\sigma$ error bars of the
three non-PPS cosmological parameters included in the maximum likelihood reconstruction. All values
are consistent with their respective best-fit fiducial model values indicated by the dashed
lines.}
\label{pps_cp_TT}
\end{figure*}

\begin{figure*}[!t]
\begin{center}
\includegraphics[scale=.9]%
{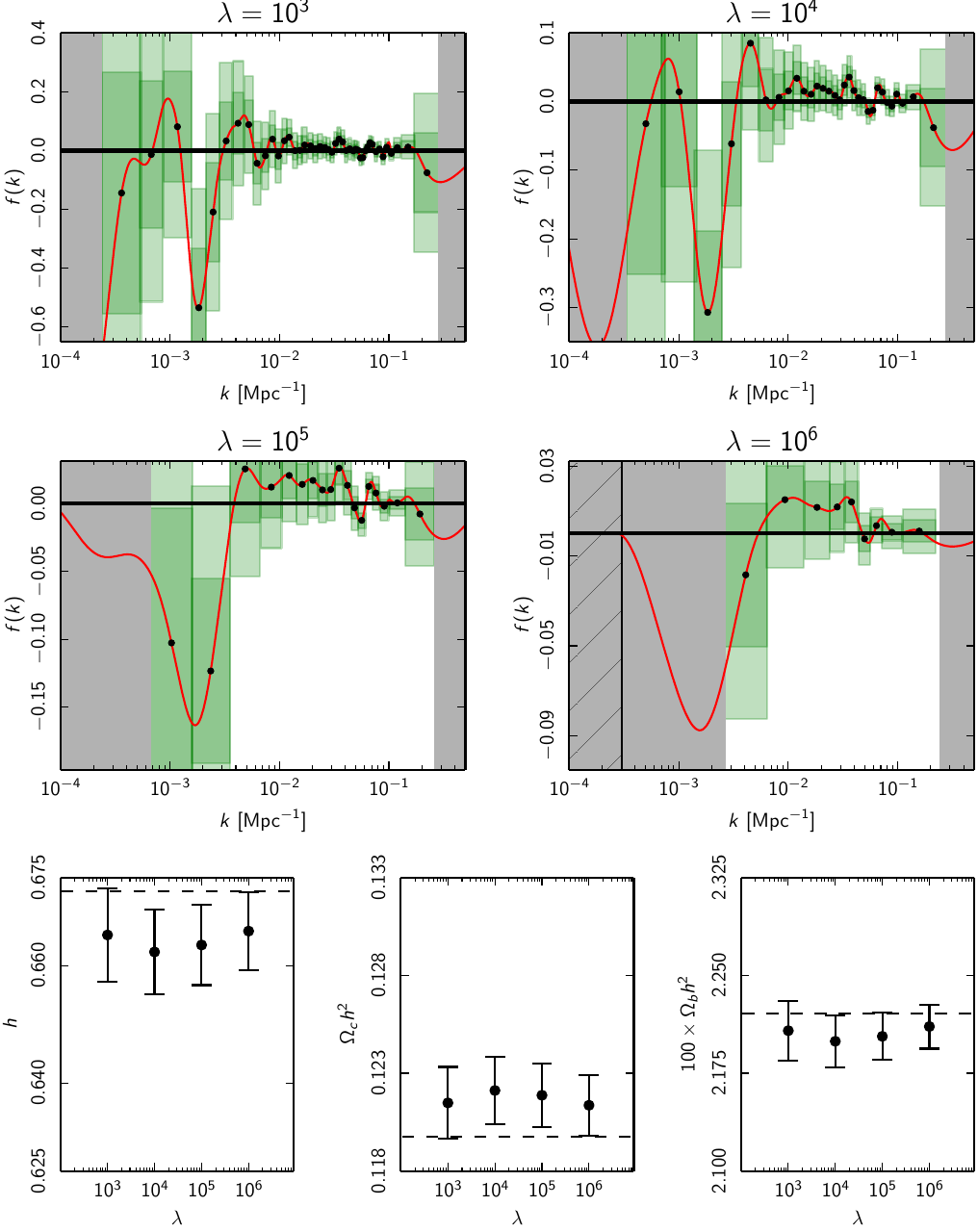}
\end{center}
\caption{
\Planck\ TT,TE,EE+lowP likelihood primordial power spectrum reconstruction results. {\it Top
four panels}: Reconstruction of the deviation $f(k)$ using four different roughness penalties. As in
Fig.~\ref{pps_cp_TT}, the red curves represent the best-fit deviation $f(k)$ and the height and
width of the green error bars represent the error and minimum reconstructible width, respectively. For all
values of the roughness penalty, the deviations are consistent with a featureless spectrum.
{\it Lower three panels:} $\pm1\,\sigma$ error bars of the three non-PPS cosmological
parameters included in the maximum likelihood reconstruction. All values are consistent with their
respective best-fit fiducial model values (indicated by the dashed lines).}
\label{pps_cp_TTTEEE}
\end{figure*}

Figure~\ref{pps_cp_TT} shows the best-fit PPS reconstruction using 
the \Planck\ TT+lowP likelihood. The penalties in Eq.~(\ref{Priors}) introduce a
bias in the reconstruction by smoothing and otherwise 
deforming potential features in the power spectrum. To assess this bias,
we define the ``minimum reconstructible width'' (MRW) to be the minimum width of a
Gaussian feature needed so that the integrated squared difference between the feature 
and its reconstruction is less than 1\,\% 
of the integrated square of the input Gaussian,
which is equivalent to 10\,\% rms. Due to the 
combination of the roughness and fixing penalties, it is impossible to satisfy the MRW 
criterion too close to $\kappa_\mathrm{min}$ and $\kappa_\mathrm{max}$. Wherever the MRW is 
undefined, the reconstruction is substantially biased and therefore 
suspect. 
An MRW cannot be defined too close to the endpoints $\kappa_\mathrm{min}$ and $\kappa_\mathrm{max}$ for two reasons:
(1) lack of data; and (2) if a feature is too close to where the fixing penalty has been applied,
the fixing penalty distorts the reconstruction. Consequently a larger roughness penalty decreases 
the range over which an MRW is well defined.  The grey shaded areas in Fig.~\ref{pps_cp_TT}
show where the MRW is undefined and thus the reconstruction cannot be trusted. 
The cutoffs $\kappa_\mathrm{min}$ and $\kappa_\mathrm{max}$ have been chosen to maximize the range 
over which an MRW is defined for a  given value of $\lambda .$
The 1\,$\sigma$ and 2\,$\sigma$ error bars in Fig.~\ref{pps_cp_TT} are estimated using the Hessian of the 
log-likelihood evaluated at the best-fit PPS reconstruction. More detailed MC investigations suggest that
the nonlinear corrections to these error bars are small.

For the $\lambda =10^5$ and $10^6$ cases of the $TT$ reconstruction,
no deviation exceeds $2\,\sigma ,$ so we do not comment on the probability of obtaining a
worse fit. For the other cases, we use the maximum of the deviation, expressed
in $\sigma $, of the plotted points as a metric of the quality of fit. Then using Monte Carlo
simulations we compute the $p$-value, or the probability to obtain a worse fit, according
to this metric. For $\lambda =10^3$ and $10^4,$ 
we obtain $p$-values
of $0.304$ and $0.142,$ respectively, corresponding to $1.03$
and $1.47\,\sigma$.  We thus conclude that the observed deviations are not
statistically significant. 

\subsubsection{Penalized likelihood results with polarization}
\label{PenalizedLikeliPolar}

In Fig.~\ref{pps_cp_TTTEEE} the best-fit reconstruction of the PPS from the \Planck\ TT,TE,EE+lowP likelihood is shown. 
We observe that the reconstruction including polarization broadly agrees with the reconstruction
obtained using temperature only.
For the \Planck\ TT,TE,EE+lowP likelihood, we
obtain for $\lambda =10^3$, $10^4$, and $10^5$ the $p$-values
$0.166,$ $0.107,$ and $0.045,$ respectively, corresponding to
$1.38,$ $1.61,$ and $2.00\,\sigma ,$ 
and likewise conclude the
absence of any statistically significant evidence for deviations
from a simple power-law scalar primordial power spectrum.

%% file: section_eight_2.tex
\input macros.tex
\newcommand{\PR}{\mathcal{P}_\mathcal{R}}
\newcommand{\alphamink}{\alpha_\mathrm{min}^{(k)}}
\newcommand{\alphamaxk}{\alpha_\mathrm{max}^{(k)}}
\newcommand{\Pknotj}[1]{\mathcal{P}_{#1}}
\newcommand{\Pknot}{\mathcal{P}}
\newcommand{\As}{A_\mathrm{s}}
\newcommand{\Asj}[1]{A_\mathrm{s}^{(#1)}}
\newcommand{\PolyChord}{{\tt PolyChord}}
\newcommand{\Nint}{N_\mathrm{int}}
\newcommand{\Nknots}{N_\mathrm{knots}}

\subsection{Method II: Bayesian model comparison}
\label{sec:PR_Bayes}

\input wl_fig.tex

In this section we model the PPS $\PR(k)$ using a nested
family of models where $\PR(k)$ is piecewise linear
in the $\ln (\mathcal{P})$-$\ln (k)$ plane
between a number of knots, $\Nknots$,
that is allowed to vary. The question arises as to how many knots one should use,
and we address this question using Bayesian model comparison.
A family of priors
is chosen where both the horizontal and vertical positions of the knots are allowed
to vary. We examine the ``Bayes factor'' or ``Bayesian evidence'' as a function of
$\Nknots$ to decide how many knots are statistically justified. 
A similar analysis has been performed by \cite{vazquez_knots} and \cite{knottedsky1}.
In addition, we marginalize over all possible numbers of knots to obtain 
an averaged reconstruction weighted according to the Bayesian evidence.

The generic prescription is illustrated in Fig.~\ref{fig:linear_spline_reconstruction}. $\Nknots$ knots
$\{(k_i,\Pknotj{i})\,$: $i=1,\ldots,\Nknots\}$ are placed in the $(k,\PR)$ plane and the function $\PR(k)$ is
constructed by logarithmic interpolation (a linear interpolation in $\log$-$\log$ space) between adjacent points.
Outside the horizontal range $[k_1,k_N]$ the function is extrapolated using the outermost interval.

Within this framework, base $\Lambda$CDM arises when ${\Nknots=2}$---in other words, when there are two boundary knots
and no internal knots, and the parameters $\Pknotj{1}$ and $\Pknotj{2}$ (in place of $A_\mathrm{s}$ and $n_\mathrm{s}$) parameterize
the simple power-law PPS. There are also, of course,
the four standard cosmological parameters $(\Omega_{\mathrm{b}} h^2$, $\Omega_{\mathrm{c}} h^2$, $100\theta_{\mathrm{MC}}$, and 
$\tau$), as well as the numerous foreground parameters associated with the \Planck\ high-$\ell$ likelihood, all of which are unrelated to the PPS.
This simplest model can be extended iteratively by successively inserting an additional internal knot, thus requiring with each iteration
two more variables to parameterize the new knot position.

\input wl_table.tex

We run models for a variety of numbers of internal knots, $\Nint =\Nknots -2$, evaluating the evidence for $\Nint$.
Under the assumption that the prior is justified, the most likely, or preferred, model is the one with the highest
evidence.  Evidences are evaluated using the \PolyChord\ sampler \citep{Handley:2015fda} 
in \CAMB\ and \CosmoMC. The use of \PolyChord\ is
essential, as the posteriors in this parameterization are often multi-modal. Also, the ordered log-uniform priors on
the $k_i$ are easy to implement within the \PolyChord\ framework. All runs were performed with $1000$ live points, 
oversampling the semi-slow and fast parameters by a factor of $5$ and $100$, respectively.

\begin{figure*}[t!]
\begin{center}
  \includegraphics[width=\textwidth]{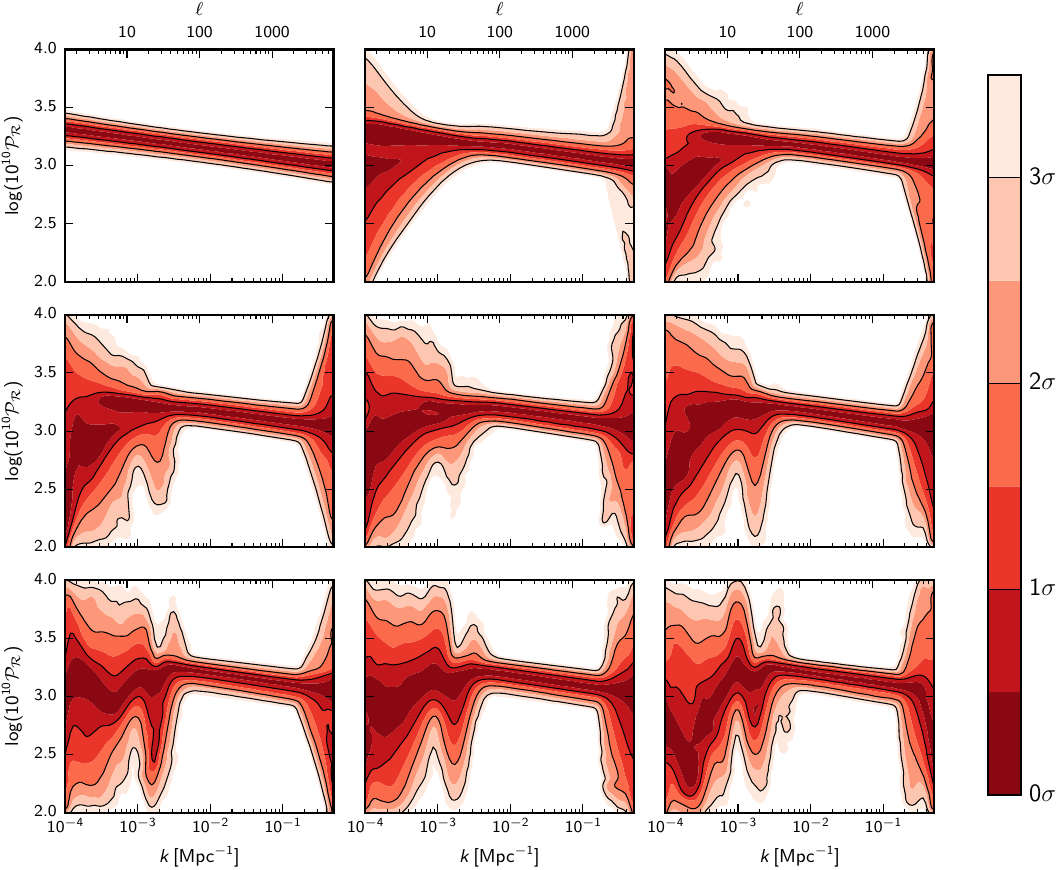}
\end{center}
\caption{Bayesian movable knot reconstructions of the primordial power spectrum $\PR(k)$ using \Planck\ TT data.
The plots indicate our knowledge of the PPS $P(\PR(k)|k,\Nint)$ for a given number of knots.
The number of internal knots $\Nint$ increases (left to right and top to bottom) from $0$ to $8$.
For each $k$-slice, equal colours have equal probabilities. The colour scale is chosen so that darker regions
correspond to lower-$\sigma$ confidence intervals.
$1\,\sigma $ and $2\,\sigma $ confidence intervals are also indicated (black curves).
The upper horizontal axes give the approximate corresponding multipoles via $\ell \approx kD_\mathrm{rec}$,
where $D_\mathrm{rec}$ is the comoving distance to recombination.
    \label{fig:Pkr0}}
\end{figure*}

\begin{figure}[hb!]
\begin{center}
\includegraphics[width=1.05\columnwidth]{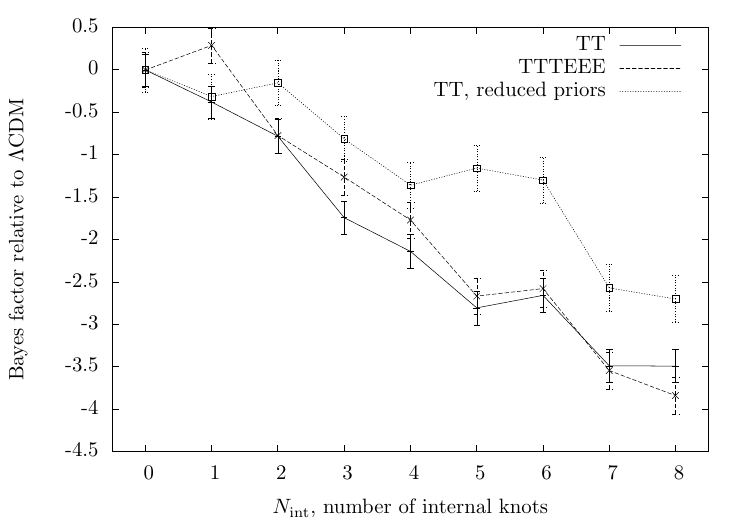}
\end{center}
  \caption{
    Bayes factor (relative to the base $\Lambda$CDM model) as a function of the number of knots
for three separate runs. Solid line: \Planck\ TT. Dashed line: \Planck\ TT,TE,EE. Dotted line:
\Planck\ TT, with priors on the $\mathcal{P}_i$ parameters reduced in width by a factor of 2 ($2.5<\ln(10^{10}\mathcal{P}_i)<3.5$).
\label{fig:Bayes_Factors}
}
\end{figure}

\begin{figure}[hb!]
\begin{center}
  \includegraphics[width=\columnwidth]{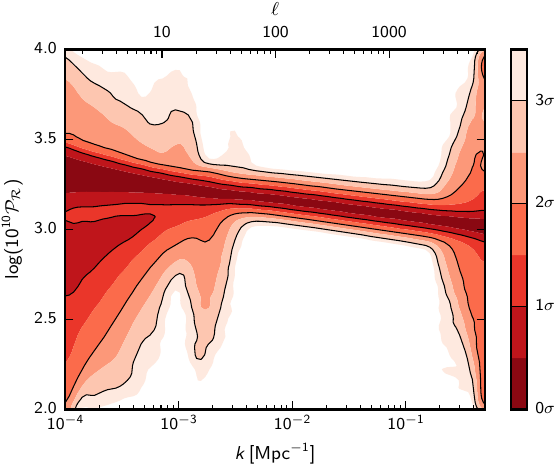}
\end{center}
\caption{
Bayesian reconstruction of the primordial power spectrum averaged over different values of $N_\mathrm{int}$
(as shown in Fig.~\protect\ref{fig:Pkr0}), weighted according to the Bayesian evidence.
The region ${30<\ell<2300}$ is highly constrained, but the resolution is lacking to say anything precise 
about higher $\ell$. At lower $\ell$, cosmic variance reduces our knowledge of $\PR(k).$
The weights assigned to the lower $\Nint$ models outweigh those of the higher models, so no oscillatory 
features are visible here.
\label{fig:full_bayes_knots}}
\end{figure}

Priors for the reconstruction parameters are detailed in Table~\ref{tab:P_k_priors}.
We report evidence ratios with respect to the base $\Lambda$CDM case. The cosmological priors remain the
same for all models, and this part of the prior has almost no impact on the evidence ratios.
The choice of prior on the reconstruction parameters
$\{\Pknotj{i}\}$ does affect the Bayes factor. \CosmoMC , however, puts an implicit prior on all models by excluding
parameter choices that render the internal computational approximations in {\tt CAMB} invalid.
The baseline prior for the vertical position of the knots includes all
of the range allowed by \CosmoMC, so slighly increasing this prior range will not affect the evidence ratios. If
one were to reduce the prior widths significantly, the evidence ratios would be increased.
The allowed horizontal range includes all $k$-scales accessible to \Planck. Thus, altering this
width would be unphysical.

After completion of an evidence calculation, \PolyChord\ generates a representative set of samples of the posterior for each model, 
$P(\Theta) \equiv P(\Theta|\mathrm{data},\Nint)$. We may use this to calculate a marginalized probability distribution
for the PPS:
\begin{equation}
  P(\log\PR|k,\Nint) = \int \delta\left(\log\PR - \log\PR(k;\Theta)\right)P(\Theta)\:d\Theta.
  \label{eqn:margPR}
\end{equation}
This expression encapsulates our knowledge of $\PR$ at each value of $k$ for a given number of knots.
Plots of this PPS posterior are shown in Fig.~\ref{fig:Pkr0} using \Planck\ TT data.

If one considers the Bayesian evidence of each model, Fig.~\ref{fig:Bayes_Factors} shows that although no model is 
preferred over base $\Lambda$CDM, the case $\Nint=1$
is competitive. This model is analogous to the broken-power-law spectrum of
Sect.~\ref{LargeScaleSuppression},
although the models differ significantly in terms of the priors used. In this case, the
additional freedom of one knot allows a reconstruction of the suppression of power at low $\ell$. Adding polarization data does not alter 
the evidences significantly, although $\Nint=1$ is strengthened. We also plot a \Planck\ TT run,
but with the reduced vertical priors ${2.5<\ln\left(10^{10}\mathcal{P}_i\right)<3.5}$. 
As expected, this increases the evidence ratios, but does not alter the above conclusion.

For increasing numbers of internal knots, the Bayesian evidence monotonically decreases. Occam's razor dictates,
therefore, that these models should not be preferred, due to their higher complexity. However, there is an
intriguing stable oscillatory feature, at $20\lsim\ell\lsim50$, that appears once there are enough knots to reconstruct it.
This is a qualitative feature predicted by several inflationary models (discussed in Sect.~\ref{sec:oscillations}),
and a possible hint of new physics, although
its statistical significance is not compelling.

A full Bayesian analysis marginalizes over all models weighted according to the normalized evidence $Z_{\Nint},$
so that
\begin{equation}
  P(\log\PR|k) = \sum\limits_{\Nint} P(\log\PR|k,\Nint) Z_{\Nint},
  \label{eqn:margPR_full}
\end{equation}
as indicated in Fig.~\ref{fig:full_bayes_knots}.
This reconstruction is sensitive to how model complexity is penalized in the prior distribution. 

%% file: wl_fig.tex
\begin{figure}[b!]
  \resizebox{\columnwidth} {!} {%

    \begin{tikzpicture}
      \def\xwidth{7}
      \def\ywidth{4}
      \def\xmn{0.5}
      \def\ymn{2}
      \def\xstart{2}
      \def\ystart{3}
      \def\xmid{3}
      \def\ymid{1}
      \def\xend{5.5}
      \def\yend{3}
      \def\xmx{6.5}
      \def\ymx{1.5}

      \def\croslen{0.4}

      \newcommand{\movablecross}[1]{%
        \draw[->](#1) -- ++(0:\croslen);
        \draw[->](#1) -- ++(90:\croslen);
        \draw[->](#1) -- ++(180:\croslen);
        \draw[->](#1) -- ++(270:\croslen);
        \fill[red!70!black] (#1) circle (2pt);
      }

      \newcommand{\movablevert}[1]{%
        \draw[->](#1) -- ++(90:\croslen);
        \draw[->](#1) -- ++(270:\croslen);
        \fill[red!70!black] (#1) circle (2pt);
      }

      \draw [<->,thick] (0,\ywidth) node (yaxis) [above] {$\log\PR(k)$}
      |- (\xwidth,0) node (xaxis) [right] {$\log k$};

      \draw (\xmn,\ymn) coordinate (mn) -- (\xstart,\ystart) coordinate (start) -- (\xmid,\ymid) coordinate (mid) --  (\xend,\yend) coordinate(end) -- (\xmx,\ymx) coordinate(mx);

      \draw (mn) node[below right]    {$(k_1,\Pknotj{1})$};
      \draw (start) node[above right] {$(k_2,\Pknotj{2})$};
      \draw (mid) node[below right]   {$(k_3,\Pknotj{3})$};
      \draw (end) node[above right]   {$(k_4,\Pknotj{4})$};
      \draw (mx) node[below left]     {$(k_{\Nknots},\Pknotj{{\Nknots}})$};


      \movablevert{mn}
      \movablecross{start}
      \movablecross{mid}
      \movablecross{end}
      \movablevert{mx}


    \end{tikzpicture}
  }
  \caption{%
Linear spline reconstruction. The primordial power spectrum is reconstructed using $\Nknots$ interpolation points
${\{(k_i,\Pknotj{i}):i=1,2,\ldots {\Nknots}\}}$. The end knots are fixed in $k$ but allowed to
vary in ${\Pknot}$, whereas the internal knots can vary subject to the constraint that ${k_1<k_2<\cdots<k_{\Nknots}}$.
The function $\mathcal{P}_\mathcal{R}(k)$ is constructed within the range $[k_1,k_{\Nknots}]$
by interpolating logarithmically between adjacent knots (i.e., linearly in $\log$-$\log$ space). Outside this range the function is extrapolated logarithmically.
The function $\mathcal{P}_\mathcal{R}(k;\{k_i,\Pknotj{i}\})$ thus has $2\Nknots-2$ parameters.
\label{fig:linear_spline_reconstruction}
}
\end{figure}

%% file: wl_table.tex
\begin{table}
\begin{center}
  \begingroup
  \newdimen\tblskip \tblskip=5pt
  \nointerlineskip
  \vskip -3mm
  \footnotesize
  \setbox\tablebox=\vbox{
    \newdimen\digitwidth
    \setbox0=\hbox{\rm 0}
    \digitwidth=\wd0
    \catcode`*=\active
    \def*{\kern\digitwidth}
    \newdimen\signwidth
    \setbox0=\hbox{+}
    \signwidth=\wd0
    \catcode`!=\active
    \def!{\kern\signwidth}
    \halign{%
      \hfil#\hfil&
      \hfil#\hfil\cr
      \noalign{\doubleline}
      Parameter range &
      Prior type
      \cr
      \noalign{\vskip 3pt\hrule\vskip 5pt}
      $10^{-4}\,\mathrm{Mpc}^{-1} = k_1< k_2 < \ldots < k_{\Nknots} = 0.3\,\mathrm{Mpc}^{-1}$ &
      log uniform (sorted)
      \cr
      $ 2 < \ln\left(10^{10}\Pknotj{1}\right), \ldots ,\ln\left(10^{10}\Pknotj{{\Nknots}}\right) < 4 $  &
      log uniform
      \cr
      $2\le \Nknots \le 10 $ &
      integer uniform
      \cr
      \noalign{\vskip 5pt\hrule\vskip 3pt}}}
    \endPlancktable                    
  \endgroup
\end{center}
\caption{%
Prior for moveable knot positions.
The $\PR$ positions are distributed in a log-uniform manner across a wide range.
The $k$ positions are also log-uniformly distributed
across the entire range needed by {\tt CosmoMC} and are sorted so that ${k_1<\ldots<k_{\Nknots}}$. 
When we marginalize over the number of knots, $\Nknots$, we assume a uniform prior between 2 and 10. }
\label{tab:P_k_priors}                            
\end{table}                        

%% file: section_eight_3.tex
\def\gtorder{\mathrel{\raise.3ex\hbox{$>$}\mkern-14mu \lower0.6ex\hbox{$\sim$}}}
\def\ltorder{\mathrel{\raise.3ex\hbox{$<$}\mkern-14mu \lower0.6ex\hbox{$\sim$}}}

\subsection{Method III: cubic spline reconstruction \label{sec:csrecon}}

In this section we investigate another reconstruction algorithm based on cubic splines 
in the $\ln (k)$-$\ln \mathcal{P}_{\cal R}$ plane, where (unlike for the approach of the previous subsection)
the horizontal positions of the knots are uniformly spaced in $\ln (k)$ and fixed.
A prior on the vertical positions (described in detail
below) is chosen and the reconstructed power spectrum is calculated using {\tt CosmoMC} for various numbers of knots.
This method differs from the method in Sect.~\ref{PenalizedUpdate} 
in that the smoothness is controlled by the number of discrete knots rather than by a continuous parameter of a 
statistical model having a well-defined continuum limit.
With respect to the Bayesian model comparison of Sect.~\ref{sec:PR_Bayes}, 
the assessment of model complexity differs because here the knots are not movable.

\begin{figure}[th]
\begin{center}
  \includegraphics[width= 0.4\textwidth]{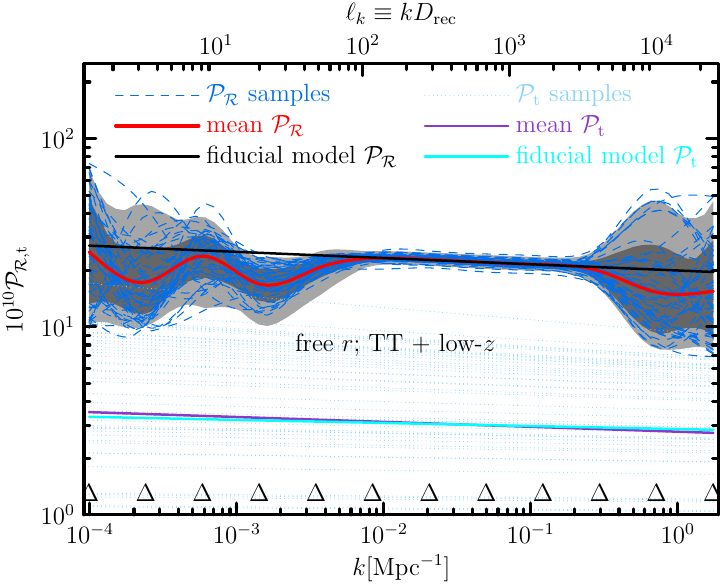}
  \includegraphics[width= 0.4\textwidth]{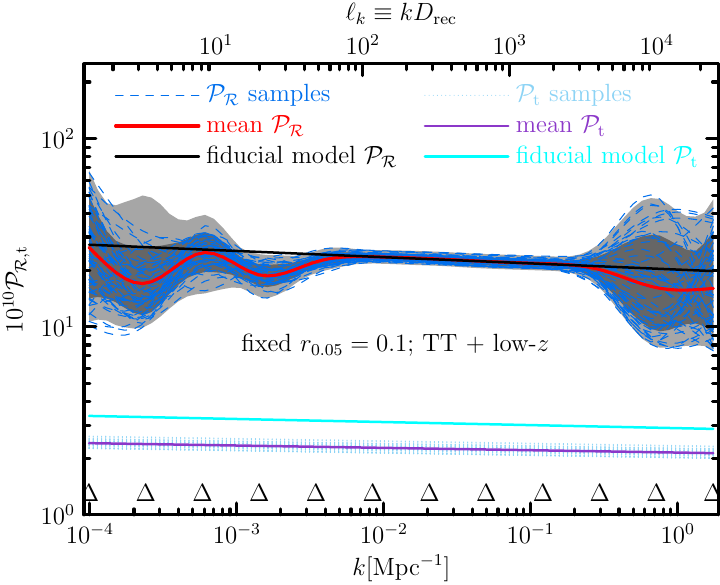}
  \includegraphics[width= 0.4\textwidth]{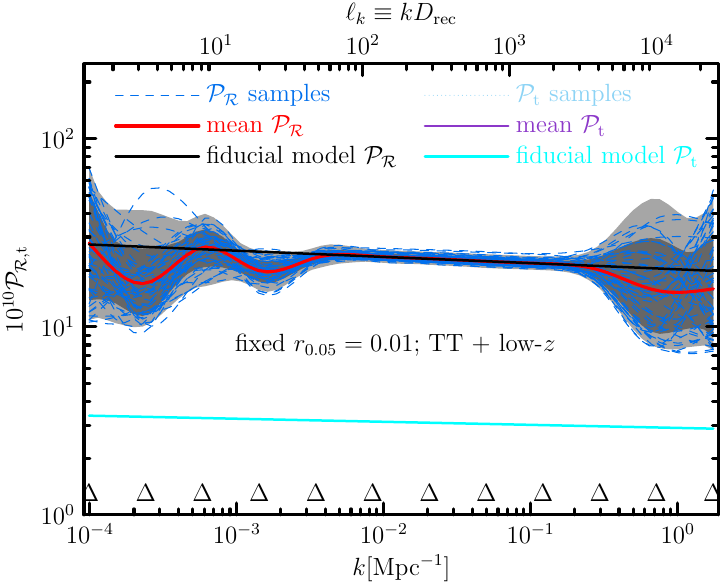}
\end{center}
  \caption{Reconstructed power spectra applied to the \Planck\ 2015 data using 12 knots 
(with positions marked as $\Delta$ at the bottom of each panel) with cubic spline interpolation.
Mean spectra as well as sample trajectories are shown for scalars and tensors, and
$\pm1\,\sigma$ and $\pm2\,\sigma$ limits are shown for the scalars.
The fiducial tensor spectrum corresponds, arbitrarily, to $r = 0.13$.  
\emph{Top:} uniform prior, $0\le r\le 1$. \emph{Middle:} fixed, $r = 0.1$. \emph{Bottom:} fixed, $r=0.01$. 
Data sets: \Planck\ TT+lowP+BAO+SN+HST+$z_{\mathrm{re}}$$>$6 prior.
$D_{\mathrm{rec}}$ is the comoving distance to recombination.
\label{fig:traj_power_spline}}
\end{figure}

\begin{figure*}[t!]
\begin{center}
 \includegraphics[width= 0.4\textwidth]{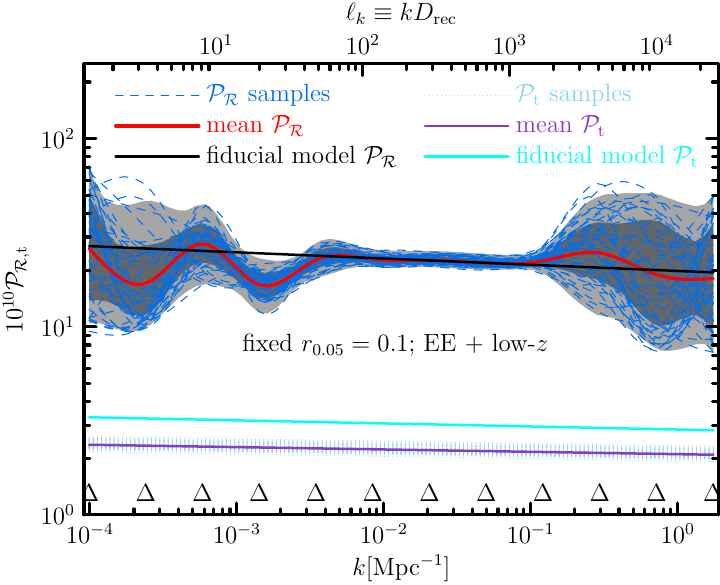}
 \includegraphics[width= 0.4\textwidth]{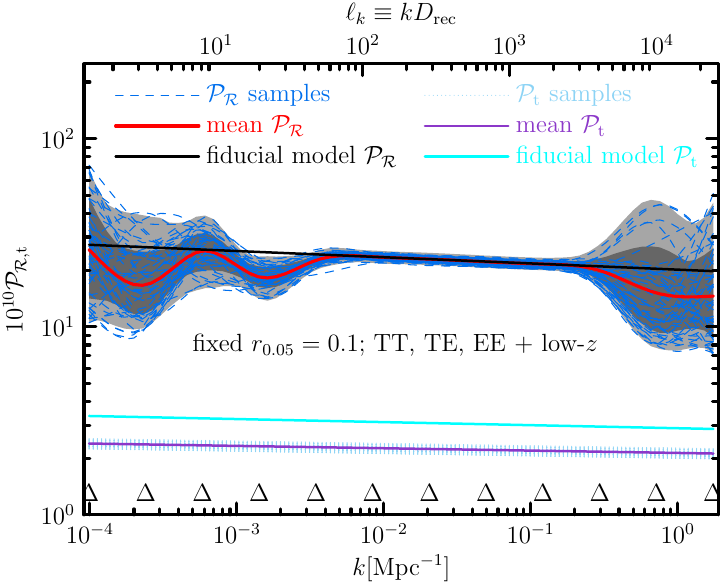}\\
 \includegraphics[width= 0.4\textwidth]{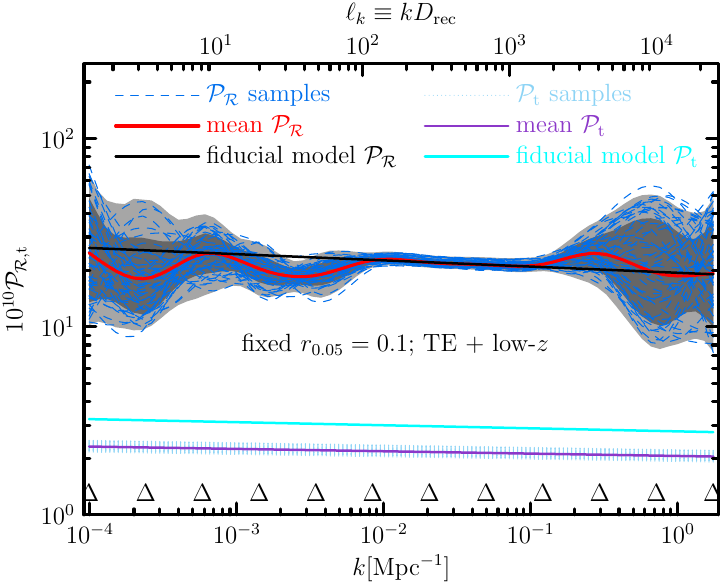}
 \includegraphics[width= 0.4\textwidth]{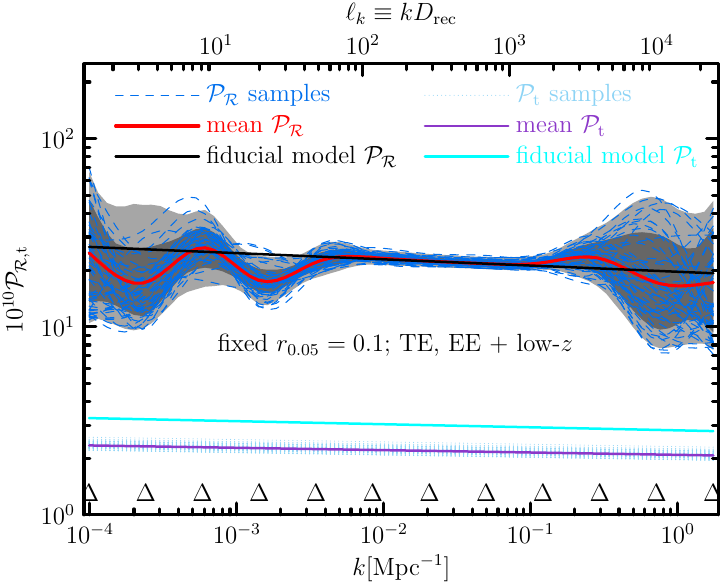}
\end{center}
  \caption{ Reconstructed 12-knot power spectra with polarization included. Data sets in common:
lowT+lowP+BAO+SN+HST+$z_{\mathrm{re}}$$>$6 prior.
\label{fig:traj_compare_datasets}}
\end{figure*}

Let the horizontal positions of the $n$ knots be given by $k_b$, where $b=1,\ldots , n$, spaced 
so that $k_{b+1}/k_b$ is independent of $b.$ We single out a ``pivot knot'' $b\!=\!p$, so that
$k_p=k_*=0.05\,\mathrm{Mpc}^{-1},$ which is the standard scalar power spectrum pivot scale.
For a given number of knots $n$ we choose $k_1$ and $k_n$ so that the interval of relevant cosmological scales, taken to
extend from $10^{-4}$\,Mpc$^{-1}$ to $O(1)$\,Mpc$^{-1},$  is included. 
We now define the prior on the vertical knot coordinates. For the pivot point, 
we define $\ln A_\mathrm{s} = \ln \mathcal{P}_{\cal R} (k_*)$, where $\ln A_\mathrm{s}$ has 
a uniformly distributed prior, and for the other points with $b\ne p,$
we define the derived variable
\begin{equation}
q_b \equiv \ln \left(
\frac{\mathcal{P}_{\cal R} (k_b)}{\mathcal{P}_{\mathrm{{\cal R},fid}}(k_b)}
\right),
\end{equation}
where 
$\mathcal{P}_{{\cal R},\mathrm{fid}}(k) \equiv A_\mathrm{s}(k/k_*)^{n_{\mathrm{s}, \mathrm{fid}}-1}.$ Here
the spectral index $n_{\mathrm{s}, \mathrm{fid}}$ is fixed. A uniform prior is imposed on each variable $q_b$ $(b\ne p)$
and the constraint $-1\le q_b\le 1$ is also imposed to force the reconstruction to behave reasonably
near the endpoints, where
it is hardly constrained by the data.  The quantity $\ln \mathcal{P}_{\cal R} (k)$ is interpolated between the knots using
cubic splines with natural boundary conditions (i.e., the second derivatives vanish at the first and the last knots). 
Outside $[k_1,k_n]$ we set
$\mathcal{P}_{\cal R}(k) = e^{q_1} \mathcal{P}_{{\cal R},\mathrm{fid}}(k)$ (for $k<k_1$) and 
$\mathcal{P}_{\cal R}(k) = e^{q_n} \mathcal{P}_{{\cal R},\mathrm{fid}}(k)$ (for $k>k_n$). 
For most knots, we found that the 
upper and lower bounds of the $q_b$ prior hardly affect the reconstruction, 
since the data sharpen the allowed range significantly. 
However, for super-Hubble scales (i.e., $k\lsim 10^{-4}$\,Mpc$^{-1}$) and very small scales 
(i.e., $k\gsim 0.2$\,Mpc$^{-1}$), which are only weakly constrained by the cosmological data, the prior dominates the 
reconstruction.
For the results here, a fiducial spectral index $n_{\mathrm{s}, \mathrm{fid}} = 0.967$ for 
$\mathcal{P}_{{\cal R} , \mathrm{fid}}$ was chosen, which is close to the estimate from 
\Planck\ TT+lowP+BAO.  
A different choice of $n_{\mathrm{s}, \mathrm{fid}}$ leads to a trivial linear shift in the 
$q_b.$

The possible presence of tensor modes (see Sect.~\ref{sec:modelcomp}) has the potential to bias and introduce
additional uncertainty in the reconstruction of the primordial scalar power spectrum as parameterized
above. Obviously, in the absence of a detection of tensors at high statistical significance, it is not
sensible to model a possible tensor contribution with more than a few degrees of freedom. A complicated
model would lead to prior dominated results. We therefore use the power law parameterization, 
$\mathcal{P}_\mathrm{t}(k) = r A_\mathrm{s}(k/k_*)^{n_\mathrm{t}}$, where the consistency relation $n_\mathrm{t} = -r/8$ is
enforced as a constraint.

\begin{figure}
\begin{center}
 \includegraphics[width= \columnwidth]{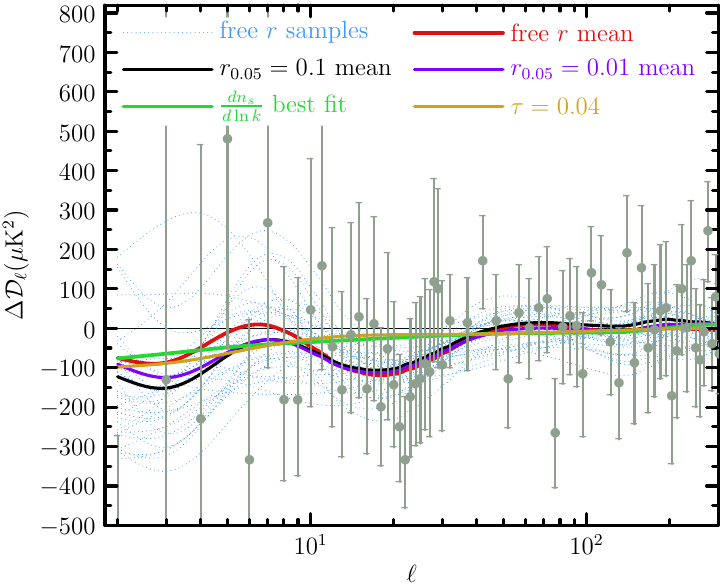}
\end{center}
  \caption{Reconstructed $\mathcal{D}_\ell^{TT}$ power spectra with the base $\Lambda$CDM best fit subtracted.  
The mean spectra shown are for the floating $r$ and the two fixed $r$ cases with 12 cubic spline knots.  These
should be contrasted with the running best-fit mean (green) and the similar looking uniform $n_\mathrm{s}$ case
in which $\tau$ has been lowered from its best-fit base $\Lambda$CDM value to 0.04.
Data points are the \Planck\ 2015 {\tt Commander} ($\ell < 30$) and {\tt Plik} ($\ell \ge 30$) 
temperature power spectrum.
\label{fig:traj_dcl}}
\end{figure}

Primordial tensor fluctuations contribute to CMB temperature and polarization angular power spectra, in particular 
at spatial scales larger than the recombination Hubble length, $k\lsim (aH)_{\mathrm{rec}} \approx 0.005\,\mathrm{Mpc}^{-1}.$
If a large number of knots in $\ln \mathcal{P}_{\cal R}(k)$ 
is included over that range, then a modified $\mathcal{P}_{\cal R}$ can mimic a tensor contribution, leading to a 
near-degeneracy.  This can lead to large uncertainty in the tensor amplitude, $r.$ Once $r$ is 
measured or tightly constrained in $B$-mode experiments, this near degeneracy will be broken. As examples here, we do allow 
$r$ to float, but also show what happens when $r$ is constrained to take the values $r=0.1$ and $r=0.01$ 
in the reconstruction.

Figure~\ref{fig:traj_power_spline} shows the reconstruction obtained using the 2015 \Planck\ TT+lowP
likelihood, BAO, SNIa, HST, and a $z_\mathrm{re}>6$ prior.
Including these ancillary likelihoods improves the constraint on the PPS by
helping to fix the cosmological parameters (e.g., $H_0$, $\tau$, and the late-time expansion history), 
which in this context may be regarded as nuisance parameters.
These results were obtained by modifying 
{\tt CosmoMC} to 
incorporate the $n$-knot parameterization of the PPS. 
Here 12 knots were used and the mean reconstruction as well as the $1\,\sigma $ and $2\,\sigma $ limits are shown.
Some $1\,\sigma$ sample trajectories (dashed curves) are also shown to illustrate the degree of correlation or smoothing of the reconstruction. 
The tensor trajectories are also shown, but, as explained above, have been constrained to be straight lines.
In the top panel $r$ is allowed to freely float, and  
a wide range of $r$ is allowed because of the near-degeneracy with the low-$k$ scalar power. Two illustrative
values of fixed $r$ (i.e., $r=0.1$ and $r=0.01$) are also shown to give an idea of how much the reconstruction 
is sensitive to variations in $r$ within the range of presently plausible values.

The reconstructions using the 2013 \Planck\ likelihood in place of the 2015 likelihood are broadly 
consistent with the reconstruction shown in Fig.~\ref{fig:traj_power_spline}. To demonstrate robustness with respect
to the interpolation scheme we tried using linear interpolation instead of cubic splines and found that the reconstruction
was consistent provided enough knots (i.e., $n_\mathrm{knot}\approx 14$) were used.
At intermediate $k$ the reconstruction is consistent with a simple power law, corresponding to a straight line in Fig.~\ref{fig:traj_power_spline}. 
We observe that once $k$ drops, so that the effective multipole being probed is below about $60$, deviations from a power law appear, but the dispersion in 
allowed trajectories also rises as a consequence of cosmic variance. The power deficit at 
$k\approx 0.002\,\mathrm{Mpc}^{-1}$ (i.e., $\ell_k \equiv k D_{\mathrm{rec}} \approx 30$, 
where $D_{\mathrm{rec}}$ is the comoving distance to recombination) is largely driven by the power spectrum 
anomaly in the $\ell \approx 20$--$30$ range that has been evident since the early spectra from WMAP \citep{bennett2010}, and verified by \Planck.

We also explore the impact of including the \Planck\ polarization likelihood in the reconstruction.
Figure~\ref{fig:traj_compare_datasets} shows the reconstructed power spectra using various combinations of the polarization 
and temperature data. The $\ell < 30$ treatments are the same in all cases, so this is mainly a test 
of the higher $k$ region. What is seen is that, except at high $k$, the $EE$ polarization data also enforce a nearly 
uniform $n_\mathrm{s}$, consistent with that from $TT$, over a broad $k$-range. When $TE$ is used alone, or $TE$ and $EE$ are used in 
combination, the result is also very similar. The upper right panel shows the constraints from all three spectra together, and the errors on the 
reconstruction are now better than those from $TT$ alone.

It is interesting to examine how the $TT$ power spectrum obtained
using the above reconstructions compares to the CMB data, in particular around
the range $\ell \approx 20$--$30$, corresponding roughly to 
$k_4\approx 1.5 \times 10^{-3}\,\textrm{Mpc}^{-1}.$ In Fig.~\ref{fig:traj_dcl}
the differences in ${\cal D}_\ell ^{TT}$ from the best-fit simple power-law model
are plotted for various assumptions concerning $r$.  We see that a better
fit than the power-law model can apparently be obtained around $\ell \approx 20$--$30$.  
We quantify this improvement below.

\begin{figure}
\begin{center}
  \includegraphics[width=\columnwidth]{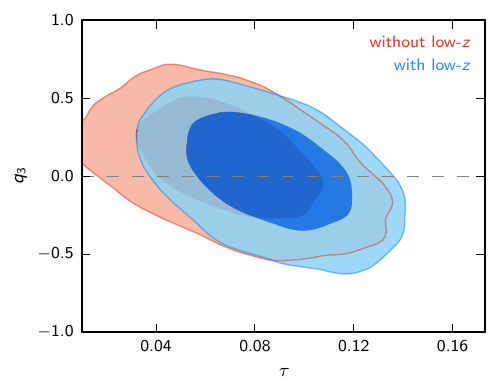}
  \includegraphics[width=\columnwidth]{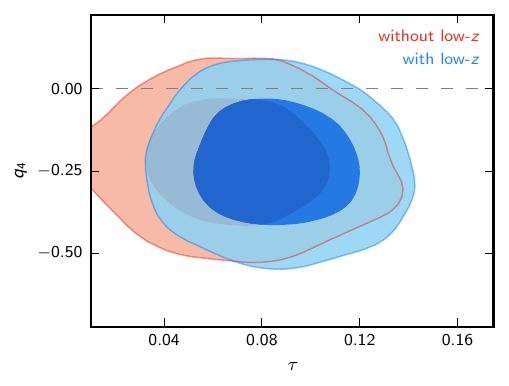}
\end{center}
  \caption{The degeneracy between $\tau$ and the knot variables $q_3$ and $q_4$ in the 12-knot case shown in  Fig.~\ref{fig:traj_power_spline}. \label{fig:taudeg}}
\end{figure}

Due to the degeneracy of scalar and tensor contributions to ${\cal D}_{\ell}^{TT}$, the 
significance of the low-$\ell$ anomaly depends on the tensor prior and whether polarization data are used. 
For $k < 10^{-3}\,\textrm{Mpc}^{-1}$, once more degeneracy appears: the shape of ${\cal D}^{TT}_\ell $ also depends 
on the reionization optical depth, $\tau$. In Fig.~\ref{fig:traj_dcl} we also show 
the effect of replacing the best-fit $\tau$ for tilted base $\Lambda$CDM with a low value, 
while keeping $A_\mathrm{s}e^{-2\tau}$ unchanged. A low $\tau$ bends 
${\cal D}_\ell^{TT}$ downward at $\ell \lsim 10$. For the 12-knot (or similar) runs, if $\tau$ is allowed to run into the 
(nonphysically) small values $\tau \lsim 0.04,$ 
a slight rise in $\mathcal{P}_{\cal R}(k)$ at $k \approx 3 \times 10^{-4}\,\textrm{Mpc}^{-1}$ is 
preferred to compensate the low-$\tau$ effect. 
This degeneracy can be broken to a certain extent using low-redshift data: $z_{\mathrm{re}}>6$ from quasar observations 
\citep{becker2001}, BAO (SDSS), Supernova (JLA), and HST.

It is evident that allowing $n_\mathrm{s}$ to run is not what the ${\cal D}_\ell^{TT}$ data prefer. The best-fit running is also shown 
in Fig.~\ref{fig:traj_dcl}. 
The $k$-space $\mathcal{P}_{\cal R}(k)$-response in Fig.~\ref{fig:traj_power_spline} shows that running does not capture the 
shape of the low-$\ell$ residuals.

\begin{figure}
\begin{center}
\includegraphics[width= 0.95\columnwidth]{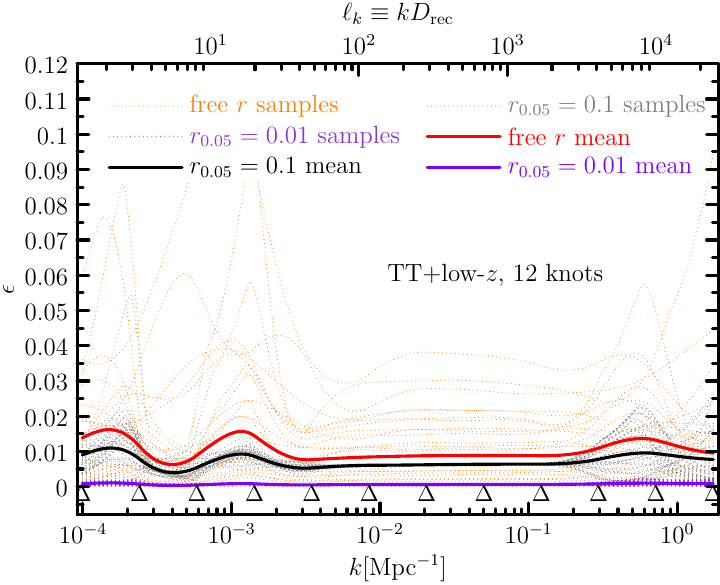}
\end{center}
\caption{Slow-roll parameter $\epsilon$ for reconstructed trajectories using 12 knots (marked as $\Delta$ at the bottom of the figure) 
with cubic spline interpolation. The mean values are shown for floating $r$ and $r$ fixed to be 0.1 and 0.01. 
Sample $1\,\sigma$ trajectories shown for the floating $r$ case show wide variability, which is significantly 
diminished if $r$ is fixed to $r=0.1$, as shown. \label{fig:traj_eps_spline}}
\label{SlowRollBond}
\end{figure}

\begin{figure}
\begin{center}
  \includegraphics[width= 0.95\columnwidth]{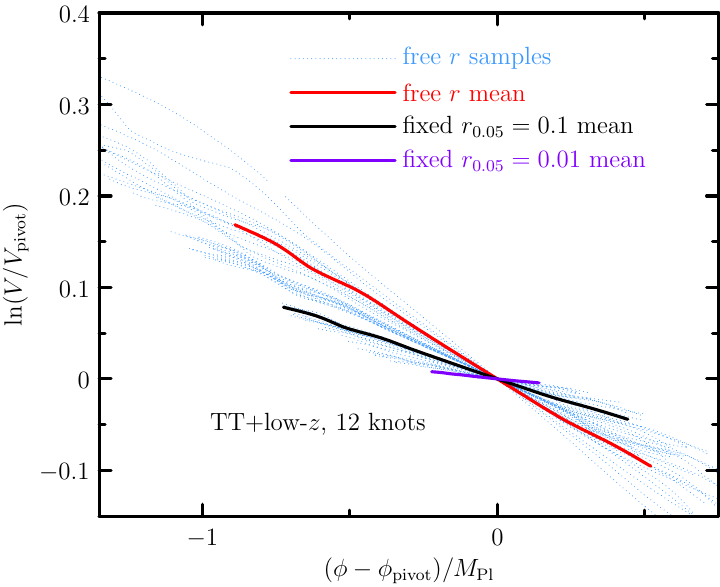}
\end{center}
  \caption{ Reconstructed single-field inflaton potentials from the cubic spline power spectra mode expansion using 
12 knots. }
\label{fig:traj_pot_spline}
\end{figure}

\input{Tables/table_ps_chisq.tex}

We have shown that the cubic spline reconstruction studied in this section consistently
produces a dip in $q_4,$ corresponding to $k\approx 1.5\times 10^{-3}\,\textrm{Mpc}^{-1}.$
We now turn to the question of whether this result is real or simply the result of cosmic variance.
To assess the statistical significance of the departures of the mean reconstruction
from a simple power law, we calculate the low-$k$ and high-$k$ reduced $\chi ^2$
for the five $q_b$ values for scales below and six $q_b$ values ($b\ne p$) for scales above $50/D_{\mathrm{rec}}$, respectively, indicating
the corresponding $p$-values (i.e., probability to exceed), for various data combinations,
in Table~\ref{tab:ps_chisq}. 
The high-$k$  fit is better than expected for reasons that we
do not understand, but we attribute this situation to chance. 
The low-$k$ region shows
a poor fit, but in no case does the $p$-value fall below 10\,\%. Therefore, even though
the low-$k$ dip is robust against the various choices made for the reconstruction, we
conclude that it is not statistically significant. The plot for the knot position of
the dip (corresponding to $q_4$) in Fig.~\ref{fig:taudeg} does not contradict this conclusion.

Because of the $r$ degeneracy associated with the scalar power, it is best when quoting statistics to use the fixed 
$r$ cases, although for completeness we show the floating $r$ case as well. There is also a smaller effect 
associated with the $\tau$ degeneracy, and the values quoted have restricted the redshift of 
reionization to exceed 6. The value $z_{\mathrm{re}} = 6.5$ was used in \citet{planck2014-a15}. The significance of the 
low-$k$ anomaly is meaningful only if an explicit $r$ prior and low-redshift constraint on $\tau$ 
have been applied.

Finally, we relate the reconstructed ${\cal P}_{\cal R}(k)$ calculated above to the
trajectories of the slow-roll parameter $\epsilon = -\dot H/H^2\vert_{k=aH}$
plotted as a function of $k$ (see Fig.~\ref{fig:traj_eps_spline}). 
We also plot in Fig.~\ref{fig:traj_pot_spline} the reconstructed inflationary potential in the region
over which the inflationary potential is constrained by the data.
Here canonical single-field inflation is assumed, and the value of $r$ enters
solely to fix the height of the potential at the pivot scale. This is not
entirely self-consistent, but justified by the lack of constraining power on
the tensors at present.

%% file: Tables/table_ps_chisq.tex
\newcommand{\pz}{\phantom{0}}
\newcommand{\pd}{\phantom{-}}

\begin{table*}[t]
\centering
\caption{Reduced $\chi^2$ and $p$-values for low-$k$ knots (5 knots) and high-$k$ knots (6 knots, pivot knot excluded), with 
the null hypothesis being the best-fit power law spectrum. 
Low-$z$ data refers to BAO+SN+HST+$z_{\mathrm{re}}$$>$6 prior. In all cases lowP data are used.\label{tab:ps_chisq}}
\begin{tabular}{ccccccccc}
\noalign{\hrule\vskip 2pt}
\noalign{\hrule\vskip 3pt}
$r$ prior & low-$z$ data & \Planck\ data & low-$k$ $\chi^2_{\mathrm{reduced}}$ & low-$k$ $p$-value & high-$k$ $\chi^2_{\mathrm{reduced}}$ & high-$k$ $p$-value & $q_3$ constraint & $q_4$ constraint\\
\noalign{\vskip 1pt\hrule\vskip 3pt}
$0\le r \le 1$ & used & TT  & 0.95  &    0.45   &  0.17   &    0.98\pz\pz  & $-0.07\pm  0.28$ & $-0.39\pm 0.20$ \\
$ r = 0.01$  & used  & TT  & 1.13 &     0.34    &  0.09 &    0.997\pz & $\pd0.01\pm 0.24$ & $-0.23 \pm 0.12$  \\
$ r = 0.01$ & not used & TT  & 0.89 &     0.49  &    0.36 &  0.90\pz\pz & $\pd0.10 \pm 0.24$ & $-0.23 \pm 0.12$ \\
$ r = 0.1$ & used & TT  & 1.70 &   0.13 &    0.12 & 0.994\pz & $-0.04 \pm 0.26$ & $-0.28 \pm 0.13$ \\
$ r = 0.1$ & not used & TT  & 1.46 &    0.20 &    0.38 &  0.89\pz\pz & $\pd0.05 \pm 0.27$ & $-0.28 \pm 0.13$ \\
$r = 0.1$ & used &   TT,TE,EE  &  1.71 &   0.13    &    0.17 &  0.985\pz &$-0.02\pm 0.25$ & $-0.30 \pm 0.12$ \\
$r = 0.1$ & used & TE,EE  & 1.72 &    0.13    &      0.38 &    0.89\pz\pz  & $\pd0.06\pm 0.25$ & $-0.32 \pm 0.15$ \\
$r = 0.1$ & used & TE  & 1.80   &   0.11 &    0.26 & 0.95\pz\pz & $-0.02\pm 0.27$ & $-0.17 \pm 0.16$ \\
$r = 0.1$ & used & EE  & 1.78  &   0.11   &   0.18 &   0.98\pz\pz & $\pd0.09 \pm 0.25$ & $-0.39 \pm 0.16$ \\
$r = 0.1$ & used & TT+lensing  &  1.54 &   0.17 &    0.05 & 0.9995 & $\pd0.05 \pm 0.25$ & $-0.27 \pm 0.13$ \\
\hline
\end{tabular}
\vspace{.4cm}
\end{table*}

%% file: section_eight_4.tex
\subsection{Power spectrum reconstruction summary}

The three non-parametric methods for reconstructing the primordial
power spectrum explored here support the following two conclusions:

\begin{enumerate}

\item 
Except possibly at low $k,$ over the range of $k$
where the CMB data best constrain the form
of the primordial power spectrum, none of
the three methods finds any 
statistically significant evidence for
deviations from a simple power-law form.
The fluctuations seen in this regime are entirely
consistent with the expectations from cosmic variance and noise. 

\item 
At low $k$, all three methods reconstruct
a power deficit at $k\approx 1.5$--$2.0\times 10^{-3}\,\textrm{Mpc}^{-1},$
which can be linked to the dip in the ${TT}$
angular power spectrum
at $\ell \approx 20$--$30.$ This agreement 
suggests that the reconstruction of this
``anomaly'' is not an artefact of any of the methods,
but rather inherent in the CMB data themselves. 
However, the evidence for this feature is marginal since it is in
a region of the spectrum where the fluctuations from cosmic variance are large.

\item 

We have verified that the power deficit at $\ell = 20$--30 is not
substantially modified (a) by removing from the CMB pattern the hottest and
coldest peaks selected by the Kolmogorov-Smirnov test studied in 
Sects.~4.5.3 and 4.5.4 of \citet{planck2014-a18} or (b) by substituting the anomalously
cold region around the Cold Spot with Gaussian constrained realizations.

\end{enumerate}

%% file: section_nine.tex
\input macros.tex
\def\ba{\begin{eqnarray}}
\def\ea{\end{eqnarray}}
\newcommand{\pz}{\phantom{0}}

In this section, we explore the possibility of a radical departure from the near-scale-invariant power-law spectrum
\mbox{$\mathcal{P}_\mathcal{R}^0(k) = A_\mathrm{s} (k/k_*)^{n_\mathrm{s}-1}$} of the standard slow-roll 
scenario for a selection of theoretically motivated parameterizations of the spectrum 
(see~\cite{Chluba:2015bqa} for a recent review). 

\subsection{Models}

\subsubsection{Step in the inflaton potential}
\label{step:potential}

A sudden, step-like feature in the inflaton potential \citep{Adams:2001vc} or the sound speed \citep{Achucarro:2010da} 
leads to a localized oscillatory burst in the scalar primordial power spectrum.
A general parameterization describing both a $\tanh$-step in the potential and in the warp term of a DBI model 
was proposed in \citet{Miranda:2013wxa}:
\begin{equation}
\label{eq:step1}
\ln \mathcal{P}_\mathcal{R}^\mathrm{s}(k) = \exp \left[ \ln \mathcal{P}_\mathcal{R}^0(k) +  \mathcal{I}_0(k) + \ln \left(1 + \mathcal{I}_1^2(k)\right) \right],
\end{equation}
where the first- and second-order terms are given by
\begin{align}
\mathcal{I}_0 &= \left[ \mathcal{A}_\mathrm{s} \, \mathcal{W}_1^{(0)}(k/k_\mathrm{s}) + \mathcal{A}_2 \, \mathcal{W}_2^{(0)}(k/k_\mathrm{s}) \right. \cr
              & \left. + \; \mathcal{A}_3 \, \mathcal{W}_3^{(0)}(k/k_\mathrm{s})  \right] \, \mathcal{D}\left( \frac{k/k_\mathrm{s}}{x_\mathrm{s}} \right), \\
\mathcal{I}_1 &= \frac{1}{\sqrt{2}} \left\{ \frac{\pi}{2} (1 - n_\mathrm{s}) + 
\left[ \mathcal{A}_\mathrm{s} \, \mathcal{W}_1^{(1)}(k/k_\mathrm{s}) \right. \vphantom{\mathcal{D}\left( \frac{k/k_\mathrm{s}}{x_\mathrm{s}} \right)} \right. \cr
 & \left. \left.  + \; \mathcal{A}_2 \, \mathcal{W}_2^{(1)}(k/k_\mathrm{s}) + \mathcal{A}_3 \, 
 \mathcal{W}_3^{(1)}(k/k_\mathrm{s}) \right]\, \mathcal{D}\left( \frac{k/k_\mathrm{s}}{x_\mathrm{s}} \right) \right\},
\end{align}
with window functions
\begin{align}
\mathcal{W}_1^{(0)}(x) &= \frac{1}{2 x^3} \left[ \left( 18 x - 6 x^3 \right) \cos 2x + \left( 15 x^2 - 9 \right) \sin 2x \right],\\
\mathcal{W}_2^{(0)}(x) &= \frac{3}{2 x^3} \left[ \sin (2x) - 2x \cos (2x) - x^2 \sin (2x) \right],\\
\mathcal{W}_3^{(0)}(x) &= \frac{1}{x^3} \left[ 6x \cos (2x) + (4x^2 -3) \sin (2x) \right],\\
\mathcal{W}_1^{(1)}(x) &= -\frac{1}{x^3} \left\{ 3(x \cos x - \sin x) \left[3 x \cos x + \left(2 x^2 -3\right) \sin x\right] \right\},\\
\mathcal{W}_2^{(1)}(x) &= \frac{3}{x^3} \left( \sin x - x \cos x \right)^2, \\
\mathcal{W}_3^{(1)}(x) &= -\frac{1}{x^3} \left[ 3 + 2x^2 - \left(3-4x^2\right) \cos (2x) - 6x \sin (2x) \right],
\end{align}
and damping function
\begin{equation}
\label{eq:step2}
\mathcal{D}(x) = \frac{x}{\sinh x}.
\end{equation}
Due to the high complexity of this model, we focus on the limiting case of a step in the potential 
($\mathcal{A}_2 = \mathcal{A}_3 = 0$).

\subsubsection{Logarithmic oscillations}

Logarithmic modulations of the primordial power spectrum generically appear, for example, in models with non-Bunch-Davies 
initial conditions \citep{Martin:2000xs,Danielsson:2002kx,Bozza:2003pr}, or, approximately, in the axion monodromy model, 
explored in more detail in Sect.~\ref{sec:putfnl}.  We assume a constant modulation amplitude and use
\begin{equation}
\mathcal{P}_\mathcal{R}^\mathrm{log}(k) =  \mathcal{P}_\mathcal{R}^0(k)  \left\{ 1 + \mathcal{A}_\mathrm{log} \cos \left[ \omega_\mathrm{log} \ln \left( \frac{k}{k_*} \right) + \varphi_\mathrm{log} \right] \right\}.
\end{equation}

\subsubsection{Linear oscillations}

A modulation linear in $k$ can be obtained, for example, in boundary effective field theory models \citep{Jackson:2013vka}, 
and is typically accompanied by a scale-dependent modulation amplitude.  We adopt the parameterization used in 
\cite{Meerburg:2013dla}, which allows for a strong scale dependence of the modulation amplitude:
\begin{equation}
\mathcal{P}_\mathcal{R}^\mathrm{lin}(k) =  \mathcal{P}_\mathcal{R}^0(k)  \left[ 1 + \mathcal{A}_\mathrm{lin} \left(\frac{k}{k_*}\right)^{n_\mathrm{lin}} \cos \left( \omega_\mathrm{lin} \frac{k}{k_*} + \varphi_\mathrm{lin} \right) \right].
\end{equation}

\subsubsection{Cutoff model}
If today's largest observable scales exited the Hubble radius before the inflaton field reached the slow-roll attractor, the amplitude 
of the primordial power spectrum is typically strongly suppressed at low $k$.
As an example of such a model, we consider a scenario in which
slow roll is preceded by a stage of kinetic energy domination.  The resulting power spectrum was derived by \cite{Contaldi:2003zv} and can
be expressed as
\begin{equation}
\ln \mathcal{P}_\mathcal{R}^\mathrm{c}(k) = \ln \mathcal{P}_\mathcal{R}^0(k) + \ln \left( \frac{\pi}{16} \, \frac{k}{k_\mathrm{c}} \left| C_\mathrm{c} - D_\mathrm{c} \right|^2 \right),
\end{equation}
with
\begin{align}
C_\mathrm{c} &= \exp\left(\frac{-i k}{k_\mathrm{c}}\right) \left[H_0^{(2)}\left(\frac{k}{2 k_\mathrm{c}}\right) - \left( \frac{k_\mathrm{c}}{k} + i \right) H_1^{(2)}\left(\frac{k}{2 k_\mathrm{c}}\right) \right],\\
D_\mathrm{c} &= \exp\left(\frac{i k}{k_\mathrm{c}}\right) \left[H_0^{(2)}\left(\frac{k}{2 k_\mathrm{c}}\right) - \left( \frac{k_\mathrm{c}}{k} - i \right) H_1^{(2)}\left(\frac{k}{2 k_\mathrm{c}}\right) \right],
\end{align}
where $H_n^{(2)}$ denotes the Hankel function of the second kind.  The power spectrum in this model is 
exponentially suppressed for wavenumbers smaller than the cutoff scale $k_\mathrm{c}$ and converges to a standard power-law spectrum for 
$k \gg k_\mathrm{c}$, with an oscillatory transition region for $k \gsim k_\mathrm{c}$.

\begin{table}
\centering
\begin{tabular}{ccc}
\noalign{\hrule\vskip 2pt}
\noalign{\hrule\vskip 3pt}
Model & Parameter & Prior range \\
\noalign{\vskip 1pt\hrule\vskip 2pt}
\multirow{3}{*}{Step}   &   $\mathcal{A}_\mathrm{s}$     & $[0,2]$ \\
                &    $ \log_{10} \left(k_\mathrm{s}/\mathrm{Mpc}^{-1}\right)$            &  $[-5,0]$ \\
                &    $\ln x_\mathrm{s}$          &  $[-1,5]$ \\
\noalign{\vskip 1pt\hrule\vskip 2pt}
\multirow{3}{*}{Log osc.}   &   $\mathcal{A}_\mathrm{log}$   &  $[0,0.5]$ \\
                &   $\log_{10} \omega_\mathrm{log}$      & $[0,2.1]$ \\
                &   $\varphi_\mathrm{log}$       & $[0,2\pi]$ \\
\noalign{\vskip 1pt\hrule\vskip 2pt}
\multirow{4}{*}{Linear osc.}    &   $\mathcal{A}_\mathrm{lin}$   & $[0,0.5]$ \\
                &   $\log_{10} \omega_\mathrm{lin}$      & $[0,2]$ \\
                &   $n_\mathrm{lin}$         & $[-1,1]$ \\
                &   $\varphi_\mathrm{lin}$       & $[0,2\pi]$ \\
\noalign{\vskip 1pt\hrule\vskip 2pt}
\multirow{1}{*}{Cutoff}     &   $\log_{10} \left( k_\mathrm{c}/\mathrm{Mpc}^{-1} \right)$    & $[-5,-2]$ \\
\hline
\end{tabular}
\vspace{.4cm}
\caption{\label{tab:features_priors} Parameters and prior ranges.}
\end{table}

\subsection{Analysis and results}

\begin{table}[t]
\centering
\begin{tabular}{cccccc}
\noalign{\hrule\vskip 2pt}
\noalign{\hrule\vskip 3pt}
\multirow{2}{*}{Model} & \multicolumn{2}{c}{\Planck\ TT+lowP} & \multicolumn{2}{c}{\Planck\ TT,TE,EE+lowP} & \multirow{2}{*}{PTE}\\
                                & \,\,\,\,$\Delta \chi^2$ & $\ln B$ & \quad\quad$\Delta \chi^2$ & $\ln B$ &  \\
\noalign{\vskip 3pt\hrule\vskip 3pt}
Step    & \pz$-8.6$ & $-0.3$ & \quad\pz$-7.3$ & $-0.6$ & 0.09 \\
Log osc.     & $-10.6$ & $-1.9$ & \quad$-10.1$ & $-1.5$ & 0.24 \\
Linear osc. & \pz$-8.9$ & $-1.9$ & \quad$-10.9$ & $-1.3$ & 0.50 \\
Cutoff  & \pz$-2.0$ & $-0.4$ & \quad\pz$-2.2$ & $-0.6$ & 0.12 \\
\hline
\end{tabular}
\vspace{.4cm}
\caption{\label{tab:features_results} Improvement in fit and Bayes factors with respect to power-law base $\Lambda$CDM for 
\Planck\ TT+lowP and \Planck\ TT,TE,EE+lowP data, as well as approximate probability to exceed the observed
$\Delta \chi^2$ ($p$-value), constructed from simulated \Planck\ TT+lowP data.  
Negative Bayes factors indicate a preference for the power-law model.}
\end{table}

\begin{table*}
\centering
\begin{tabular}{lccccc}
\noalign{\hrule\vskip 2pt}
\noalign{\hrule\vskip 3pt}
Parameter & Step & Log osc. & Linear osc. & Cutoff & Power law \\
\noalign{\hrule\vskip 3pt}
$100 \, \omega_\mathrm{b}$ & $2.23 \pm 0.02$ & $2.22 \pm 0.02$ & $2.23 \pm 0.02$ & $2.23 \pm 0.02$ & $2.23 \pm 0.02$ \\
$10 \, \omega_\mathrm{c}$ & $1.20 \pm 0.02$ & $1.20 \pm 0.02$ & $1.20 \pm 0.02$ & $1.19 \pm 0.02$ & $1.19 \pm 0.02$ \\
$100 \, \theta_\mathrm{MC}$ & $1.0409 \pm 0.0004$ & $1.0409 \pm 0.0004$ & $1.0409 \pm 0.0004$ & $1.0410 \pm 0.0005$ & $1.0409 \pm 0.0005$ \\
$\tau$ & $0.083 \pm 0.015$ & $0.082 \pm 0.015$ & $0.084 \pm 0.014$ & $0.086 \pm 0.017$ & $0.085 \pm 0.016$ \\
$\ln \left( 10^{10} A_\mathrm{s} \right)$ & $3.10 \pm 0.03$ & $3.10 \pm 0.03$ & $3.10 \pm 0.03$ & $3.11 \pm 0.03$ & $3.10 \pm 0.03$ \\
$n_{\mathrm s}$ & $0.966 \pm 0.005$ & $0.970 \pm 0.007$ & $0.967 \pm 0.004$ & $0.968 \pm 0.005$ & $0.968 \pm 0.005$  \\
\noalign{\hrule\vskip 2pt}
$\mathcal{A}_{\mathrm s}$ & 0.374 & \dots & \dots & \dots & \dots \\
$\log_{10} \left(k_{\mathrm s}/\mathrm{Mpc}^{-1} \right)$ & $-3.10$ & \dots  & \dots & \dots & \dots \\
$\ln x_{\mathrm s}$ & 0.342 & \dots  & \dots & \dots & \dots \\
\noalign{\hrule\vskip 2pt}
$\mathcal{A}_\mathrm{log}$ & \dots & 0.0278 & \dots & \dots & \dots \\
$\log_{10} \omega_\mathrm{log}$ & \dots & 1.51 & \dots & \dots & \dots \\
$\varphi_\mathrm{log}/2\pi$ & \dots & 0.634 & \dots & \dots & \dots \\
\noalign{\hrule\vskip 2pt}
$\mathcal{A}_\mathrm{lin}$       & \dots & \dots & 0.0292 & \dots & \dots \\
$\log_{10} \omega_\mathrm{lin}$ & \dots & \dots &  1.73 & \dots & \dots \\
$n_\mathrm{lin}$  & \dots & \dots & 0.662 & \dots & \dots \\
$\varphi_\mathrm{lin}/2\pi$  & \dots & \dots & 0.554 & \dots & \dots \\
\noalign{\hrule\vskip 2pt}
$\log_{10} \left( k_\mathrm{c}/\mathrm{Mpc}^{-1} \right)$ & \dots & \dots & \dots & $-3.44$ & \dots \\
\noalign{\vskip 2pt\hrule}
\end{tabular}
\vspace{.4cm}
\caption{\label{tab:features_parameters}
Best-fit features parameters
and parameter constraints on the remaining cosmological parameters for the four features models for \Planck\ TT+lowP data. 
Note that the foreground parameters have been fixed to their power-law base $\Lambda$CDM best-fit values.}
\end{table*}

\begin{figure}[t]
\begin{center}
\includegraphics[angle=270,width=88mm]{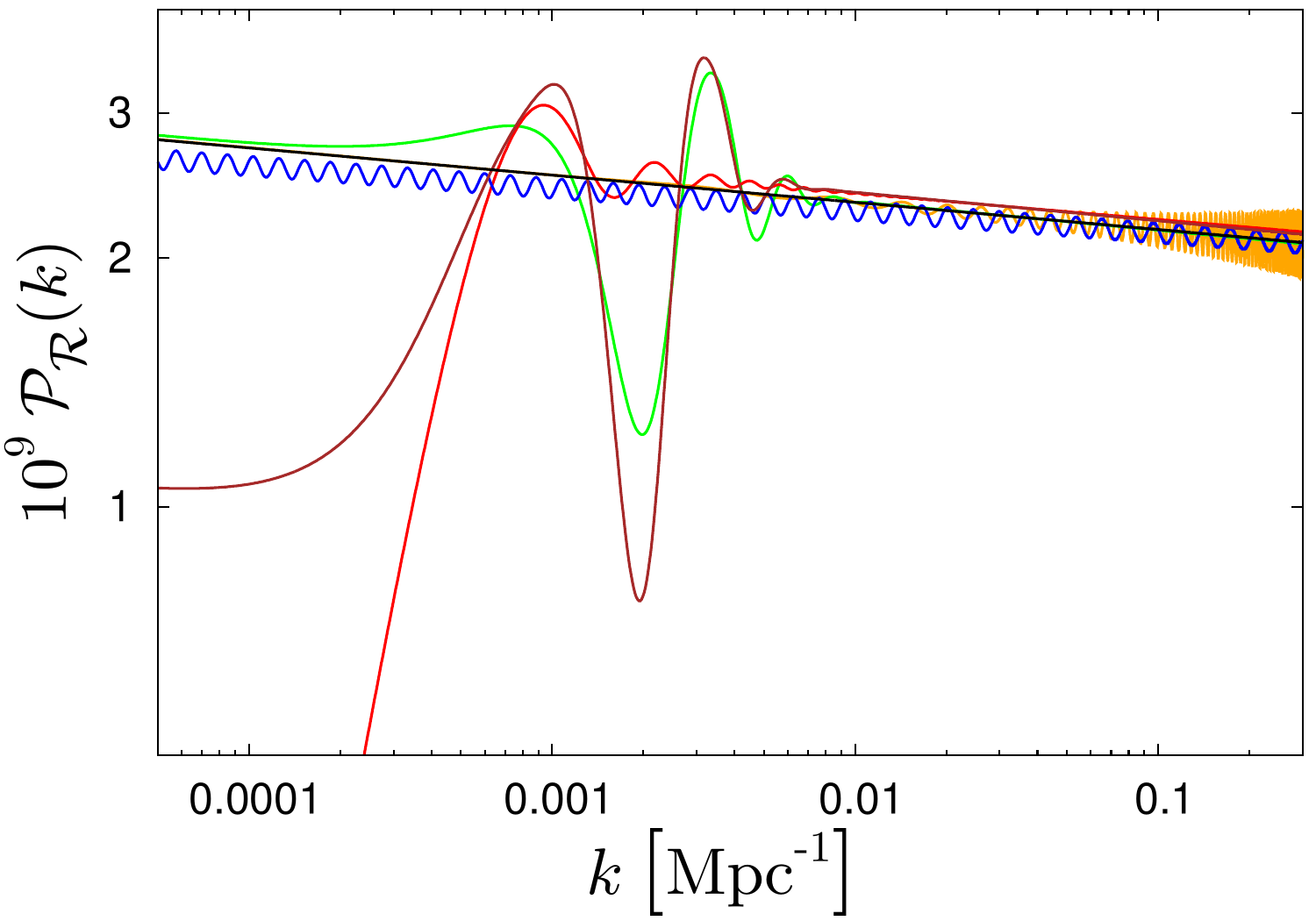}\\
\end{center}
\caption{Best-fit power spectra for the power-law (black curve), step (green), logarithmic oscillation (blue),
linear oscillation (orange), and cutoff (red) models using \Planck\ TT+lowP data.
The brown curve is the best fit for a model with a step in the warp {\it and} potential (Eqs.~\ref{eq:step1}--\ref{eq:step2}).
\label{fig:featuresspectra}}
\end{figure}

\begin{figure}[t]
\begin{center}
\includegraphics[angle=270,width=44mm]{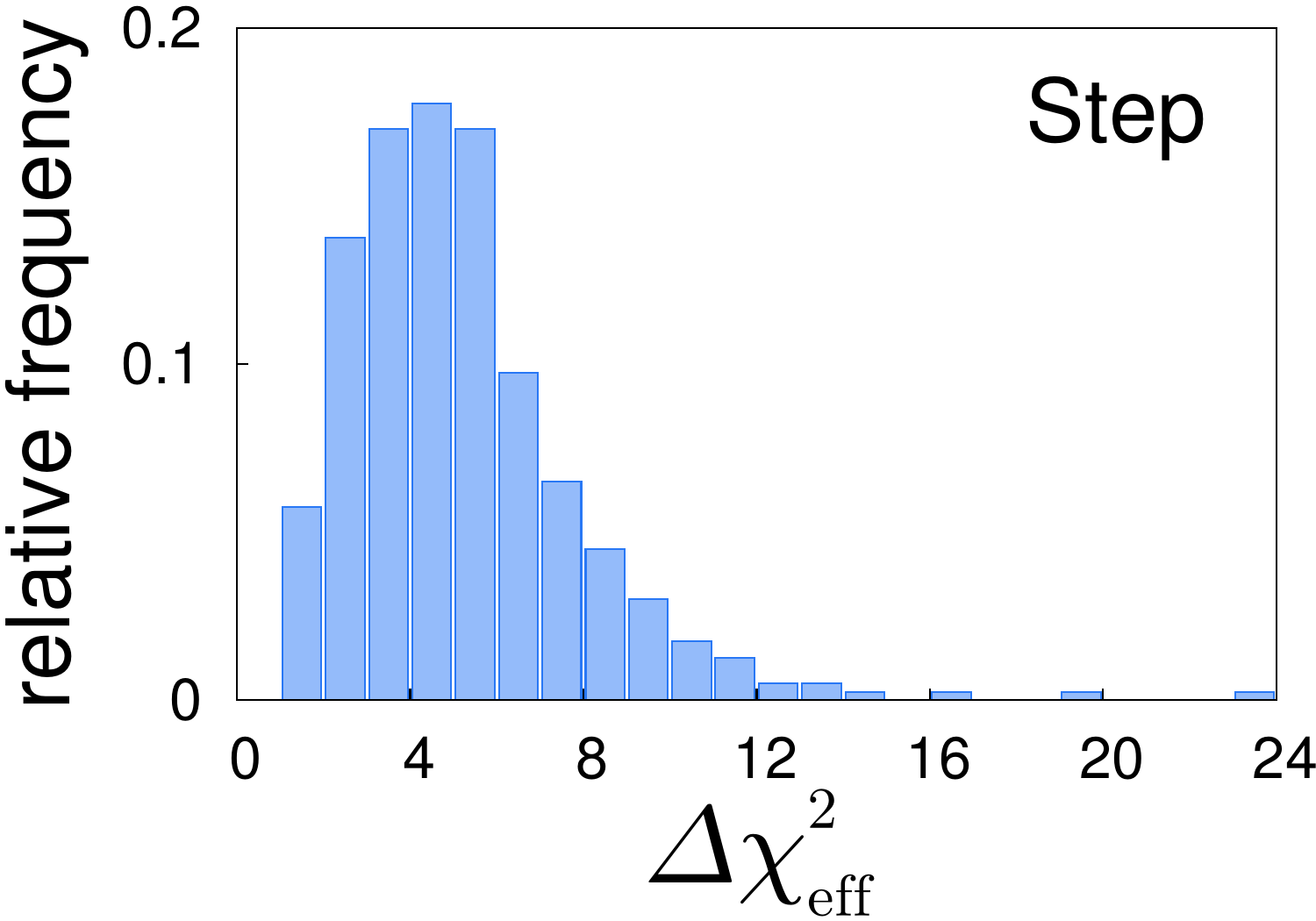}
\includegraphics[angle=270,width=44mm]{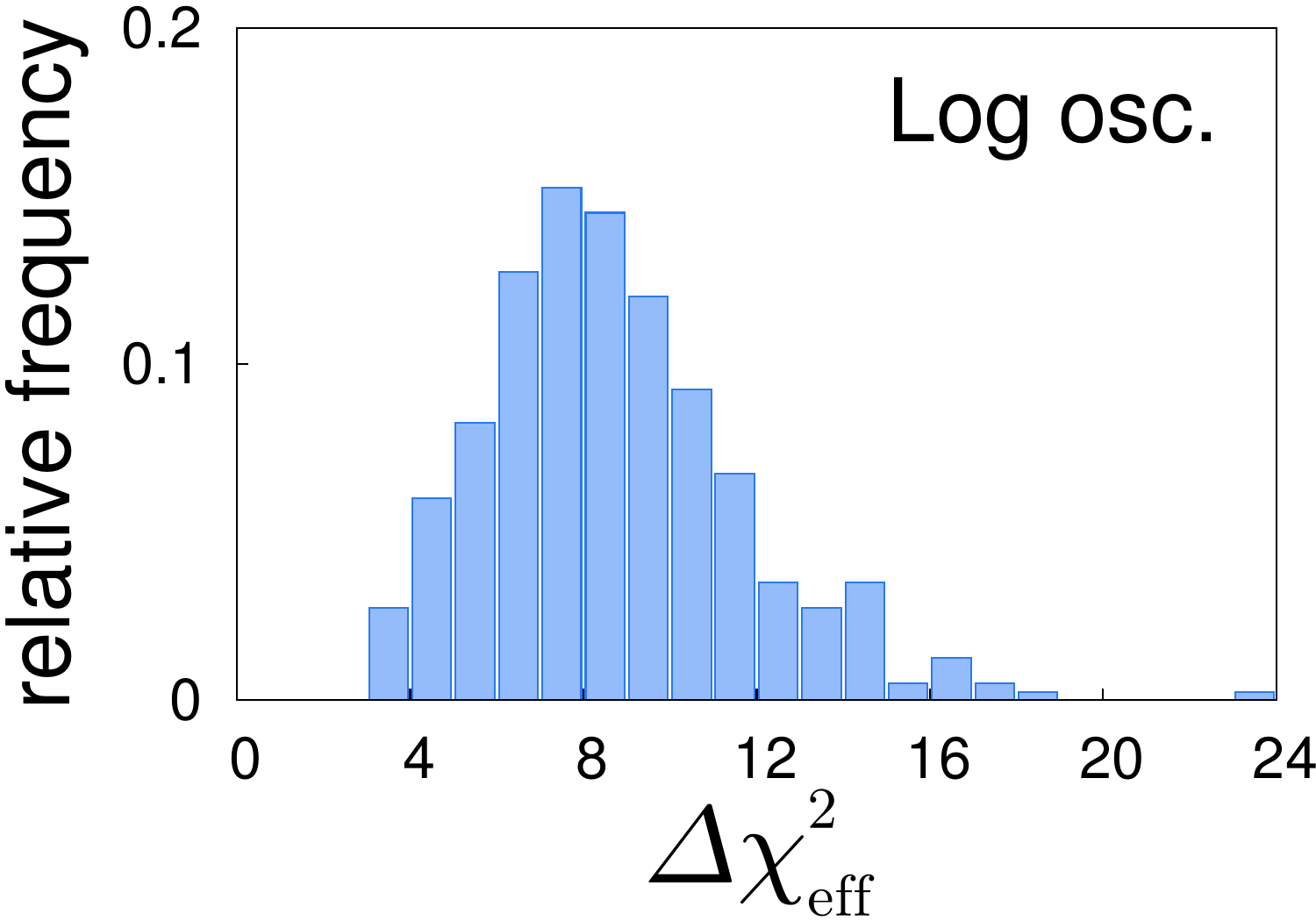}\\
\includegraphics[angle=270,width=44mm]{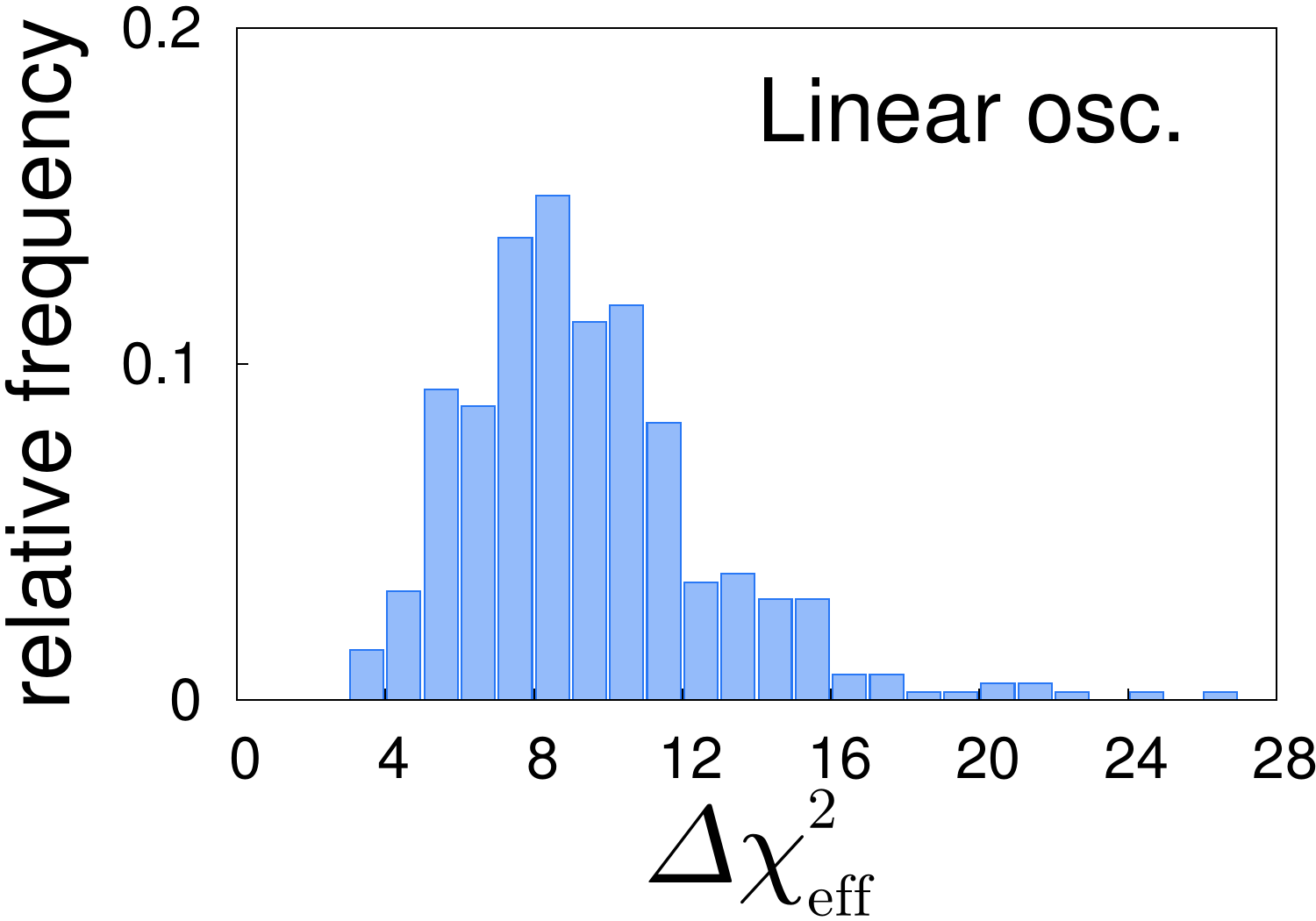}
\includegraphics[angle=270,width=44mm]{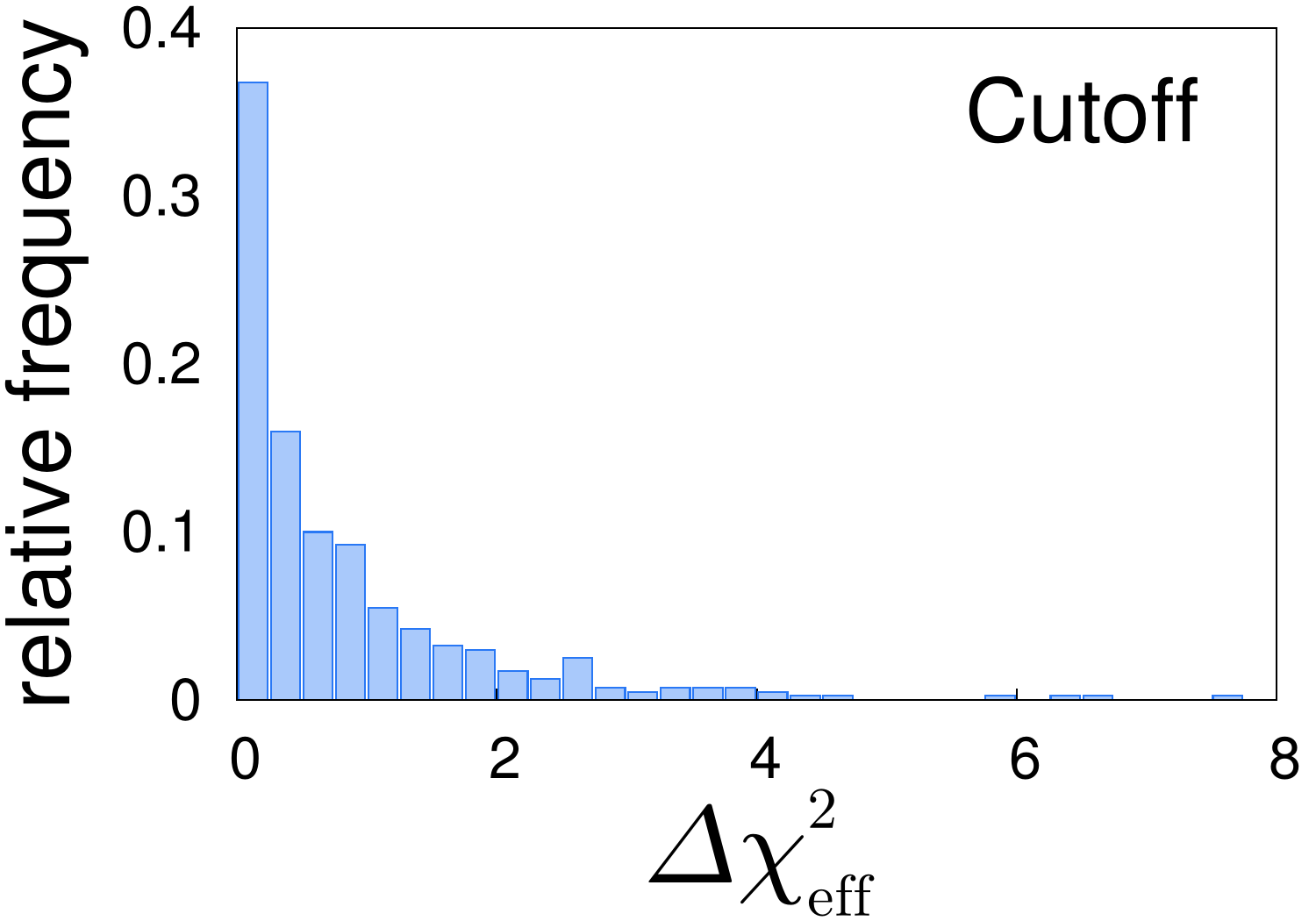}
\end{center}
\caption{Distribution of $\Delta \chi^2$ from 400 simulated \Planck\ TT+lowP data sets.
\label{figure:featureshisto}}
\end{figure}


We use \texttt{MultiNest} to evaluate the Bayesian 
evidence for the models, establish parameter constraints, and roughly identify the global maximum likelihood 
region of parameter space.  The features model best-fit parameters and $\ln \mathcal{L}$ are then obtained 
with the help of the \texttt{CosmoMC} minimization algorithm 
taking narrow priors around the \texttt{MultiNest} best fit. 
We assign flat prior probabilities to the parameters of the features models with prior ranges 
listed in Table~\ref{tab:features_priors}.  
Note that throughout this section for the sake of maximizing sensitivity to very sharp features, 
the unbinned (``bin1") versions of the 
high-$\ell$ TT and TT,TE,EE likelihoods are used instead of the standard binned versions.

Since the features considered here can lead to broad distortions of the CMB angular power spectrum degenerate 
with the late time cosmological parameters \citep{Miranda:2013wxa},
in all cases we simultaneously vary primordial parameters and all the $\Lambda$CDM parameters,
but keep the foreground parameters fixed to their 
best-fit values for the power-law base $\Lambda$CDM model. 

We present the Bayes factors with respect to the power-law base $\Lambda$CDM model 
and the improvement in the effective $\chi^2$ over the power-law model 
in Table~\ref{tab:features_results}.  For our choice of priors, none of the features models is 
preferred over a power-law spectrum.  
The best-fit power spectra are plotted in Fig.~\ref{fig:featuresspectra}.  While the cutoff and step model best fits reproduce the
large-scale suppression at $\ell \approx 20$--30 also obtained by direct power spectrum reconstruction in Sect.~\ref{sec:pk},
the oscillation models prefer relatively high frequencies beyond the resolution of the reconstruction methods.

In addition to the four features models we also show in Fig.~\ref{fig:featuresspectra} 
the best fit of a model allowing for steps in both inflaton potential and warp (brown line).  
Note the strong resemblance to the reconstructed features of the previous section.  
The effective $\Delta \chi^2$ for this model is $-12.1$ ($-11.5$) for \Planck\ TT+lowP (\Planck\ TT,TE,EE+lowP) data at the cost
of adding five new parameters, resulting in a ln-Bayes factor of $-0.8$ ($-0.4$).  
A similar phenomenology can be also be found for a model with a sudden change in the slope of the inflaton potential
\mbox{\citep{Starobinsky:1992ts,Choe:2004zg}}, which yields a best-fit $\Delta \chi^2 = -4.5$ ($-4.9$) for two extra parameters.

As shown in Table~\ref{tab:features_parameters}, constraints on the remaining cosmological parameters are not
significantly affected when allowing for the presence of features.

For the cutoff and step models, the inclusion of \Planck\ small-scale polarization data does not add much in 
terms of direct sensitivity. The best fits lie in the same 
parameter region as for \Planck\ TT+lowP data, and the $\Delta \chi^2$ and Bayes factors
are not subject to major changes.  The two oscillation models' \Planck\ TT+lowP best fits, 
on the other hand, also predict a non-negligible
signature in the polarization spectra at high $\ell$.  Therefore, if the features were real, one would 
expect an additional improvement in  
$\Delta \chi^2$ for \Planck\ TT,TE,EE+lowP.  This is not the case here.  
Though the linear oscillation model's maximum $\Delta \chi^2$ does 
increase, the local $\Delta \chi^2$ in the \Planck\ TT+lowP best-fit regions is in fact reduced for both models, and the global likelihood maxima
occur at different frequencies ($\log_{10} \omega_\mathrm{log} = 1.25$ and $\log_{10} \omega_\mathrm{lin} = 1.02$) compared to their \Planck\ TT+lowP
counterparts.

In addition to the Bayesian model comparison analysis, we also approach the matter of the statistical relevance of the features 
models from a frequentist statistics perspective in order to give the 
$\Delta \chi^2$ numbers a quantitative interpretation.  Assuming that the underlying 
$\mathcal{P_R}(k)$ was actually a featureless power law, we can ask how large an improvement to $\ln \mathcal{L}$ 
the different features models would yield on average just by overfitting scatter from cosmic variance and noise.   For this purpose, we simulate 
\Planck\ power spectrum data sets consisting of temperature and polarization up to $\ell = 29$ and unbinned temperature for 
$30 \leq \ell \leq 2508$, taking as input fiducial spectra the power-law base $\Lambda$CDM model's best-fit spectra.

For each of these simulated \Planck\ data sets, we perform the following procedure: 
(i)~find the power-law $\Lambda$CDM model's best-fit parameters with \texttt{CosmoMC}'s minimization algorithm, 
(ii)~fix the non-primordial parameters ($\omega_\mathrm{b},\omega_\mathrm{c},\theta_\mathrm{MC},\tau$) to their respective best-fit values, 
(iii)~using \texttt{MultiNest}, find the best fit of the features 
models,\footnote{Due to the multimodal nature of the posterior, usual minimization routines perform poorly here.} 
and (iv) extract the effective $\Delta \chi^2$ between power-law and features models.

The resulting distributions are shown in Fig.~\ref{figure:featureshisto}.  Compared to the real data $\Delta \chi^2$ values 
from Table~\ref{tab:features_results}, they are biased towards lower values, since we do not vary the late-time 
cosmological parameters in the analysis of the simulated data.  Nonetheless, the observed improvements in the fit do not appear to be
extraordinarily large, with the respective (conservative) $p$-values ranging between 0.09 and 0.50.

These observations lead to the conclusion that even though some of the peculiarities seen in the 
residuals of the \Planck\ data with respect to a power-law primordial spectrum may be explained in terms 
of primordial features, none of the simple model templates considered here is required by \Planck\ data.  
The simplicity of the power-law spectrum continues to give it an edge over more complicated initial spectra and the 
most plausible explanation for the apparent features in the data remains that we are just observing 
fluctuations due to cosmic variance at large scales and noise at small scales.

%% file: section_ten.tex
The combination of power spectrum constraints and primordial non-Gaussianity (NG) constraints, such as the 
\Planck\ upper bound on the NG amplitude $f_\mathrm{NL}$ \citep{planck2014-a19}, can be exploited to limit 
extensions to the simplest standard single-field models of slow-roll inflation. The next subsection considers 
inflationary models with a non-standard kinetic term \citep{1999PhLB..458..219G}, where the inflaton Lagrangian is 
a general function of the scalar inflaton field and its first derivative, i.e., $\mathfrak{L}=P(\phi,X)$, where $X=- 
g^{\mu \nu} \partial_\mu \phi \partial_\nu \phi/2$ \citep{1999PhLB..458..219G,2007JCAP...01..002C}. 
Sect.~\ref{Ginflation} focuses on a specific example of a single-field model of inflation with more general 
higher-derivative operators, the so-called ``Galileon inflation.'' Sect.~\ref{axion} presents constraints on axion 
monodromy inflation. See \cite{planck2014-a19} for the analysis of other interesting non-standard inflationary 
models, including warm inflation \citep{Berera:1995ie}, whose $f_\mathrm{NL}$ predictions can be constrained by 
\Planck.

\vspace{.3cm}
 
\subsection{Inflation with a non-standard kinetic term}

This class of models includes $k$-inflation \citep{1999PhLB..458..209A,1999PhLB..458..219G} and Dirac-Born-Infield 
(DBI) models introduced in the context of brane inflation 
\citep{2004PhRvD..70j3505S,2004PhRvD..70l3505A,2005PhRvD..71f3506C,2005JHEP...08..045C}. In these models inflation 
can take place despite a steep potential or may be driven by the kinetic term.

Moreover, one of the main predictions of inflationary models with a non-standard kinetic term is that the inflaton 
perturbations can propagate with a sound speed $c_\mathrm{s} <1$. We show how the \Planck\ combined measurement of 
the power spectrum and the nonlinearity parameter $f_\mathrm{NL}$ \citep{planck2014-a19} improves constraints on 
this class of models by breaking degeneracies between the parameters determining the observable power spectra. 
Such degeneracies (see, e.g., \citet{2007PhRvD..76j3517P}, \citet{2009JCAP...04..019P}, 
\citet{2008PhRvD..78h3513L}, \citet{2009PhRvD..79b3503A}, and \citet{2015JCAP...01..016B}) are evident from the 
expressions for the power spectra. We adopt the same notation as~\cite{planck2013-p09a}.  At leading order in the 
slow-roll parameters the scalar power spectrum depends additionally on the sound speed $c_\mathrm{s}$ via 
\citep{1999PhLB..458..219G}
\begin{equation}
\label{sps}
A_\mathrm{s} \approx \frac{1}{8 \pi^2 M^2_\mathrm{pl}} \frac{H^2}{c_\mathrm{s} \epsilon_1}\, ,
\end{equation}
which is evaluated at $k c_\mathrm{s}=aH$. Correspondingly, the scalar spectral index 
\begin{equation}
\label{nnskt}
n_\mathrm{s}-1=-2 \epsilon_1-\epsilon_2-s\,
\end{equation} 
depends on an additional slow-roll parameter $s = \dot{c_\mathrm{s}}/(c_\mathrm{s} H)$, which describes the 
running of the sound speed. The usual consistency relation holding for the standard single-field models of 
slow-roll inflation ($r= -8 n_\mathrm{t}$) is modified to $r \approx -8 n_\mathrm{t} c_\mathrm{s}$, with 
$n_{\mathrm t}=-2 \epsilon_1$ as usual \citep{1999PhLB..458..219G}, potentially allowing models which otherwise would predict a large 
tensor-to-scalar ratio for the Klein-Gordon case \citep{Unnikrishnan:2012zu}.\footnote{We use the more accurate 
relation 
\begin{equation}
r = 16 \epsilon_1 c_\mathrm{s}^{(1+\epsilon_1)/(1-\epsilon_1)} \,,
\label{c1}
\end{equation}
accounting for different epochs of freeze-out for the scalar fluctuations (at sound
horizon crossing, 
$k c_\mathrm{s}=a H$) and tensor perturbations (at Hubble radius crossing, $k=a H$) 
\citep{2007PhRvD..76j3517P,2009JCAP...04..019P,2008PhRvD..78h3513L,2009PhRvD..79b3503A,2015JCAP...01..016B}.
}

At lowest order in the slow-roll parameters, there are strong degeneracies between the parameters 
($A_\mathrm{s},c_\mathrm{s},\epsilon_1, \epsilon_2,s$). This makes the constraints on these parameters from the 
power spectrum alone not very stringent, and for parameters like $\epsilon_1$ and $\epsilon_2$ less stringent 
compared with the standard case. However, combining the constraints on the power spectra observables with those on 
$f_{\mathrm{NL}}$ can also result in a stringent test for this class of inflationary models. Models where the 
inflaton field has a non-standard kinetic term predict a high level of primordial NG of the scalar perturbations 
for $c_\mathrm{s} \ll 1$, \citep[see, e.g.,][]{2007JCAP...01..002C}. Primordial NG is generated by the 
higher-derivative interaction terms arising from the expansion of the kinetic part of the Lagrangian, $P(\phi,X)$. 
There are two main contributions to the amplitude of the NG (i.e., to the nonlinearity parameter 
$f_{\mathrm{NL}}$), coming from the inflaton field interaction terms $\dot{\delta \phi}\, (\nabla \delta \phi)^2$ 
and $(\dot{\delta \phi})^3$ \citep{2007JCAP...01..002C,2010JCAP...01..028S}. The NG from the first term scales as 
$c_\mathrm{s}^{-2}$, while the NG arising from the other term is determined by a second parameter, $\tilde{c}_3$ 
\citep[following the notation of][]{2010JCAP...01..028S}. Each of these two interactions produces bispectrum shapes 
similar to the so-called {\it equilateral} shape~(\citealt{2004JCAP...08..009B}) for which the signal peaks for 
equilateral triangles with $k_1=k_2=k_3.$ \citep[These two shapes are called, respectively, ``EFT1'' and ``EFT2'' in][]
{planck2014-a19}. However, the difference between the two shapes is such that the total signal is a linear 
combination of the two, leading to an ``orthogonal'' bispectral template \citep{2010JCAP...01..028S}.

The equilateral and orthogonal NG amplitudes can be expressed in terms of the two ``microscopic'' parameters, 
$c_\mathrm{s}$ and $\tilde{c}_3$ (for more details see \cite{planck2014-a19}), according to
\begin{eqnarray}
\label{meanfNL}
f^{\mathrm{equil}}_{\mathrm{NL}} &=&\frac{1-c_\mathrm{s}^2}{c_\mathrm{s}^2} \left [-0.275 - 0.0780 c_s^2 -  (2/3) \times 0.780\,  \tilde{c}_3 \right], \\
f^{\mathrm{ortho}}_{\mathrm{NL}} &=&\frac{1-c_\mathrm{s}^2}{c_\mathrm{s}^2} \left[ 0.0159 - 0.0167 c_s^2 - (2/3) \times 0.0167\,  \tilde{c}_3\right] \, .
\end{eqnarray}
Thus the measurements of $f^{\mathrm{equil}}_{\mathrm{NL}}$ and $f^{\mathrm{ortho}}_{\mathrm{NL}}$ obtained in the 
companion paper \citep{planck2014-a19} provide a constraint on the sound speed, $c_\mathrm{s}$, of the inflaton 
field. Such constraints allow us to combine the NG information with the analyses of the power spectra, since the 
sound speed is the NG parameter also affecting the power spectra.

In this subsection we consider three cases. In the first case we perform a general analysis as described above 
(focusing on the simplest case of a constant sound speed, $s=0$), improving on~\citetalias{planck2013-p17} 
and~\cite{planck2013-p09a} by exploiting the full mission temperature and polarization data. The \Planck\ 
constraints on primordial NG in general single-field models of inflation provide the most stringent bound on the 
inflaton sound speed \citep{planck2014-a19}:\footnote{This section uses results based on $f_{\mathrm{NL}}$ 
constraints from $T$ and $E$ \citep{planck2014-a19}. In \cite{planck2014-a19} it is shown that, although conservatively 
considered as preliminary, the $f_{\mathrm{NL}}$ constraints from $T$ and $E$ are robust, since they pass an extensive 
battery of validation tests and are in full agreement with $T$-only constraints.}
\begin{equation}
c_{\mathrm{s}} \geq 0.024\quad (95\,\%~\mathrm{CL}).
\end{equation}

\begin{figure}[ht]
\includegraphics[width=8.8cm]{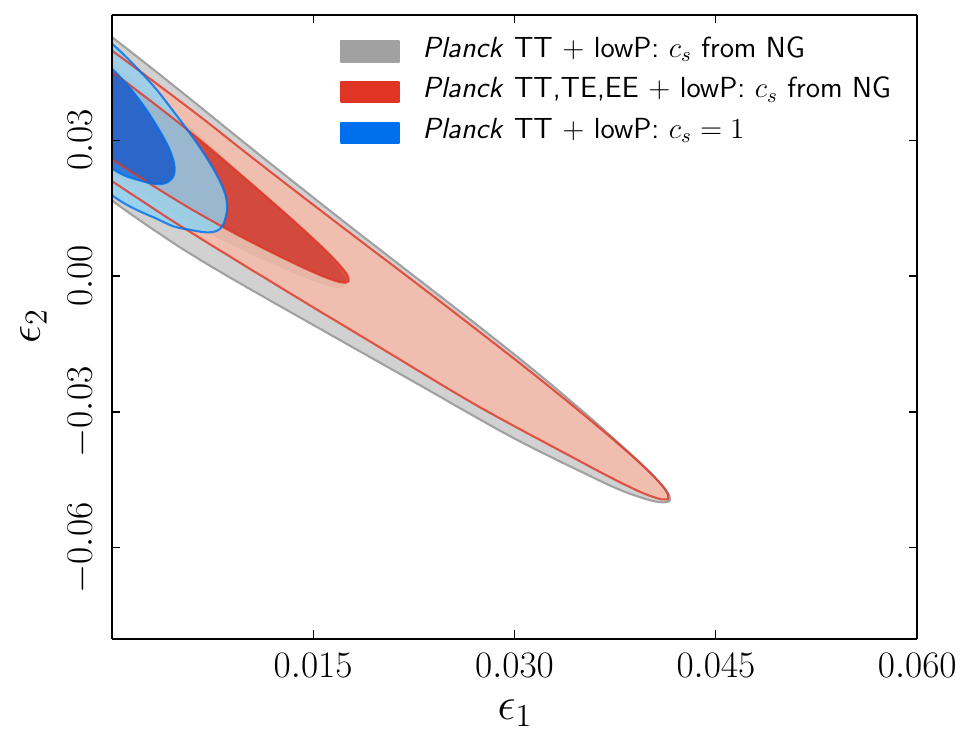}
\caption{
$(\epsilon_1\,, \epsilon_2)$ 68\,\%\ and 95\,\% CL constraints
for \Planck\ data comparing the canonical Lagrangian case with $c_\mathrm{s}=1$ to the
case of varying $c_\mathrm{s}$ with a uniform prior $0.024< c_\mathrm{s} < 1$ derived from the
\Planck\ NG measurements.
}
\label{canonvscs}
\end{figure}

We then use this information on $c_\mathrm{s}$ as a uniform prior $0.024 \leq c_{\mathrm{s}} \leq 1$ in 
Eq.~(\ref{c1}) within the HFF formalism, as in \citetalias{planck2013-p17}. Fig.~\ref{canonvscs} shows the joint 
constraints on $\epsilon_1$ and $\epsilon_2.$ \Planck\ TT+lowP yields $\epsilon_1 < 0.031$ at 95\,\%~CL. No 
improvement in the upper bound on $\epsilon_1$ results when using \Planck\ TT,TE,EE+lowP. This constraint improves 
the previous analysis in \citetalias{planck2013-p17} and can be compared with the restricted case of 
$c_\mathrm{s}=1$, also shown in Fig.~\ref{canonvscs}, with $\epsilon_1 < 0.0068$ at 95\,\% CL. The limits on the 
sound speed from the constraints on primordial NG are crucial for deriving an upper limit on $\epsilon_1$, because 
the relation between the tensor-to-scalar ratio and $\epsilon_1$ also involves the sound speed (see, e.g., 
Eq.~\ref{c1}). This breaks the degeneracy in the scalar spectral index.

The other two cases analysed involve DBI models. The degeneracy between the different slow-roll parameters can be 
broken for $s=0$ or in the case where $s \propto \epsilon_2$. We first consider models defined by an action of the 
DBI form
\begin{equation}
P(\phi,X)=- f(\phi)^{-1} \sqrt{1-2f(\phi) X}+f(\phi)^{-1}-V(\phi)\, ,
\end{equation}
where $V(\phi)$ is the potential and $f(\phi)$ describes the warp factor determined by the geometry of the extra 
dimensions. We follow an analogous procedure to exploit the NG limits derived in~\cite{planck2014-a19} on 
$c_\mathrm{s}$ in the case of DBI models: $c_\mathrm{s} \geq 0.087$ (at 95\,\% CL). Assuming a uniform prior, 
$0.087 \leq c_\mathrm{s} \leq 1$, and $s=0$, \Planck\ TT+lowP gives $\epsilon_1 < 0.024$ at 95\,\% CL, a $43\,\%$ 
improvement with respect to~\citetalias{planck2013-p17}. The addition of high-$\ell$ $TE$ and $EE$ does not 
improve the upper bound on $\epsilon_1$ for this DBI case.

Next we update the constraints on the particularly interesting case of infrared DBI models 
(\citealt{2005PhRvD..71f3506C,2005JHEP...08..045C}), where $f(\phi) \approx \lambda/\phi^4.$ (For details, see 
\cite{2004PhRvD..70j3505S}, \cite{2004PhRvD..70l3505A}, \cite{2007JCAP...01..002C}, and references therein). In 
these models the inflaton field moves from the IR to the UV side with an inflaton potential
\begin{equation}
V(\phi)=V_0-\frac{1}{2} \beta H^2 \phi^2\, .
\end{equation}
From a theoretical point of view a wide range of values for $\beta$ is allowed: \mbox{$0.1 < \beta < 10^9$} 
\citep{2008PhRvD..77b3527B}.  \citetalias{planck2013-p17} dramatically restricted the allowed parameter space of 
these models in the limit where stringy effects can be neglected and the usual field theory computation of the 
primordial curvature perturbation holds (see \citet{2005JHEP...08..045C,2005PhRvD..72l3518C} and 
\citet{2008PhRvD..77b3527B} for more details). In this limit of the IR DBI model, one finds 
\citep{2005PhRvD..72l3518C,2007JCAP...01..002C} $c_\mathrm{s} \approx (\beta N_*/3)^{-1}$, 
$n_\mathrm{s}-1=-4/N_*$, and $dn_\mathrm{s}/d\ln k=-4/N_*^2.$ (In this model one can verify that $s\approx 
1/N_*\approx \epsilon_2/3$.) Combining the uniform prior on $c_\mathrm{s}$ with \Planck\ TT+lowP, we obtain
\begin{equation}
\beta \leq  0.31 \quad (95\,\%~\mathrm{CL}),
\end{equation}
and a preference for a high number of $e$-folds: $78 < N_* < 157$ at 95\,\% CL. 

We now constrain the general case of the IR DBI model, including the ``stringy'' regime, which occurs when the 
inflaton extends back in time towards the IR side \citep{2008PhRvD..77b3527B}. The stringy phase transition is 
characterized by an interesting phenomenology altering the predictions for cosmological perturbations. A 
parameterization of the power spectrum of curvature perturbations interpolating between the two regimes 
is~(\cite{2008PhRvD..77b3527B}; see also~\cite{2013PhRvD..87j3516M})
\begin{equation}
\mathcal{P}_{\cal R}(k)=  \frac{A_\mathrm{s}}{\left(N_e^{\mathrm{DBI}}\right)^4}
\left[1-\frac{1}{(1+x)^2} \right],
\label{eq:DBIps}
\end{equation}
where $A_\mathrm{s}=324 \pi^2/(n_{\mathrm{B}} \, \beta^4)$ 
is the amplitude of the perturbations which depends on various 
microscopic parameters ($n_\mathrm{B}$ is the number of branes at the B-throat; 
see~\cite{2008PhRvD..77b3527B} for more details), while $x=(N_e^{\mathrm{DBI}}/N_c)^8$ 
sets the stringy phase transition taking place at the critical 
$e$-fold $N_c.$ (Here $N_e^{\mathrm{DBI}}$ is the number of $e$-folds to the end of IR DBI inflation.)  
The spectral index and its running are
\begin{align}
n_\mathrm{s}-1 &= \frac{4}{N_e^{\mathrm{DBI}}} \frac{x^{2}+3x-2}{(x+1)(x+2)}\, ,
\\
\frac{d n_\mathrm{s}}{d \ln k} &=\frac{4}{{\left(N_e^{\mathrm{DBI}}\right)^2}}
\frac{x^4+6x^3-55x^{2}-96x-4}{(x+1)^{2}(x+2)^{2}} \, .
\end{align}
A prediction for the primordial NG in the stringy regime is not available. We assume the standard field-theoretic 
result for a primordial bispectrum of the equilateral type with an amplitude 
$f^\mathrm{DBI}_\mathrm{NL}=-(35/108)\, [(\beta^2\, (N_e^{\mathrm{DBI}})^2/9)-1]$. By considering the same uniform 
prior on $c_\mathrm{s}$, we obtain $\beta < 0.77$, $66 < N_e^{\mathrm{DBI}} < 72$, and $x < 0.41$ at 95\,\% CL, 
which severely limits the general IR DBI model and strongly restricts the allowed parameter space.

 \subsection{Galileon inflation} 
 \label{Ginflation}

As a further example of the implications of the NG constraints on (non-standard) inflationary models we consider 
Galileon inflation \citep{2011JCAP...01..014B} (see also \cite{Kobayashi:2010cm}, \citet{2010PhRvD..82j3518M}, and 
\citet{2012JCAP...10..035O}).  This represents a well-defined and well-motivated model of inflation with more 
general higher derivatives of the inflaton field compared to the non-standard kinetic term case analysed above. 
The Galileon models of inflation are based on the so-called ``Galilean symmetry'' \citep{2009PhRvD..79f4036N}, and 
enjoy some well understood stability properties (absence of ghost instabilities and protection from large quantum 
corrections). This makes the theory also very predictive, since observable quantities (scalar and tensor power 
spectra and higher-order correlators) depend on a finite number of parameters. From this point of view this class 
of models shares some of the same properties as the DBI inflationary models 
\citep{2004PhRvD..70j3505S,2004PhRvD..70l3505A}. The Galileon field arises naturally within fundamental physics 
constructions (e.g., \citealt{2010PhLB..693..334D,2010PhRvD..82d4020D}). These models also offer an interesting 
example of large-scale modifications to Einstein gravity.
 
The Galileon model is based on the action \citep{2009PhRvD..80f4015D,2009PhRvD..79h4003D}
 \begin{equation}
S=\int d^4x\sqrt{-g}\left(\frac{M_\mathrm{pl}^2}{2}R+\sum_{n=0}^{3}\mathcal{L}_{n}\right),
\end{equation}
where
\begin{eqnarray}
\mathcal{L}_0 & = & c_2X, \\
\mathcal{L}_1 & = & -2\left(c_3/\Lambda^3\right)X \Box\phi,\\
\mathcal{L}_2 & = &2 \left(c_4/\Lambda^6\right)X \left[\left(\Box\phi\right)^{2}-\left(\nabla_{\mu}\nabla_{\nu}\phi\right)^{2}\right]+\left(c_4/\Lambda^6\right) X^2\,R,\\
\mathcal{L}_3 & = & -2 \left(c_5/\Lambda^9\right)X \left[\left(\Box\phi\right)^{3}-3\Box\phi\left(\nabla_{\mu}\nabla_{\nu}\phi\right)^{2}  +2 \left(\nabla_{\mu}\nabla_{\nu}\phi\right)^{3}\right] \cr 
&+& 6 \left(c_5/\Lambda^9\right)X^2 G_{\mu\nu}\nabla^{\mu}\nabla^{\nu}\phi\, \, .
\end{eqnarray}
Here $X=-\nabla_\mu\phi\nabla^\mu\phi/2$, $(\nabla_\mu\nabla_\nu\phi)^2=\nabla_\mu\nabla_\nu\phi\nabla^\mu\nabla^\nu\phi$, and $(\nabla_\mu\nabla_\nu\phi)^3=
\nabla_\mu\nabla_\nu\phi\nabla^\mu\nabla^\rho\phi\nabla^\nu\nabla_\rho\phi$. 
The coupling coefficients $c_i$ are dimensionless and $\Lambda$ is the cutoff of the theory. 
The case of interest includes a potential term $V(\phi)=V_0+\lambda \phi+(1/2) m^2 \phi^2+\dots$ to drive inflation.

The predicted scalar power spectrum at leading order is~(\cite{2012JCAP...10..035O,2011JCAP...01..014B,2013PhRvD..88b3529T}; see also~\cite{2011PThPh.126..511K} and \cite{2011JCAP...12..019G})\footnote{For 
the following expressions it is convenient to define the quantities
\begin{equation}
A=c_2/2+6\bar c_3+27 \bar c_4+60 \bar c_5, \quad
B=-c_2/2-4\bar c_3-13 \bar c_4-24 \bar c_5,\quad
\end{equation}
where $\bar{c}_i=c_iZ^{i-2}$ for $i=2$ to $5$, with $Z=H \dot{\phi}_0/\Lambda^3$.  
In order to have a viable model we require $A>0$ (no ghosts) and $B<0$ (no gradient instabilities).
}
\begin{equation}
\label{PSG}
\mathcal{P}_\mathcal{R}=\frac{H^2}{8\pi^2M_{\mathrm{Pl}}^2\epsilon_\mathrm{s} F c_\mathrm{s}}\Bigg |_{c_\mathrm{s}k=aH}=\frac{H^4}{8\pi^2 A (\dot{\phi_0})^2 c_\mathrm{s}^3},
\end{equation}
where $F=1+\bar c_4(\dot\phi_0)^2/(2H^2M_{\mathrm{Pl}}^2)$ and $c_\mathrm{s}^2=-B/A$ is the sound speed of the Galileon field. 
$\epsilon_\mathrm{s}$ is different from the usual slow-roll parameter $\epsilon_1$ and at leading order related according to 
$\epsilon_\mathrm{s} = -2 B/(1+6 \bar c_3+18\bar c_4+30 \bar c_5) \epsilon_1 .$ 
The scalar spectral index 
\begin{equation}
\label{nsG}
n_\mathrm{s}-1=-2\epsilon_1-\eta_\mathrm{s}-s
\end{equation}
depends on the slow-roll parameters $\epsilon_1$, $\eta_{\mathrm{s}}=\dot\epsilon_\mathrm{s}/(H\epsilon_\mathrm{s})$, 
and $s=\dot c_\mathrm{s}/(Hc_\mathrm{s})$. As usual the slow-roll parameter $s$ describes the running of the sound speed. 
In the following we restrict ourselves to the case of a constant sound speed with $s=0$. 
The tensor-to-scalar ratio is 
\begin{equation}
\label{cG}
r=16\epsilon_s c_{\mathrm{s}}=16 \epsilon_1 \bar c_{\mathrm{s}}\, ,
\end{equation}
where we have introduced the parameter $\bar{c}_\mathrm{s}=-[2 B /(1+6 \bar c_3+18\bar c_4+30 \bar c_5)] c_\mathrm{s} $, 
which is related to the Galileon sound speed. The parameter $\bar c_{\mathrm{s}}$ can be either positive or negative.
In the negative branch a blue spectral tilt for the primordial gravitational waves is allowed, contrary to the situation for standard slow-roll models 
of inflation.
We introduce such a quantity so that the consistency relation takes the form 
$r \approx -8 n_\mathrm{t} \bar c_\mathrm{s}$, with $n_{\mathrm t}=-2 \epsilon_1$,
analogous to Eq.~(\ref{c1}). The measurements of primordial NG constrain $\bar{c}_\mathrm{s},$ which in turn
constrains $\epsilon_1$ and $\eta_\mathrm{s}$ in Eq.~(\ref{nsG}).  This is analogous to the constraints on 
$\epsilon_1$ and $\eta$ of Eq.~(\ref{nnskt}) in the previous subsection. 

\begin{figure}[t]
\includegraphics[width=\columnwidth]{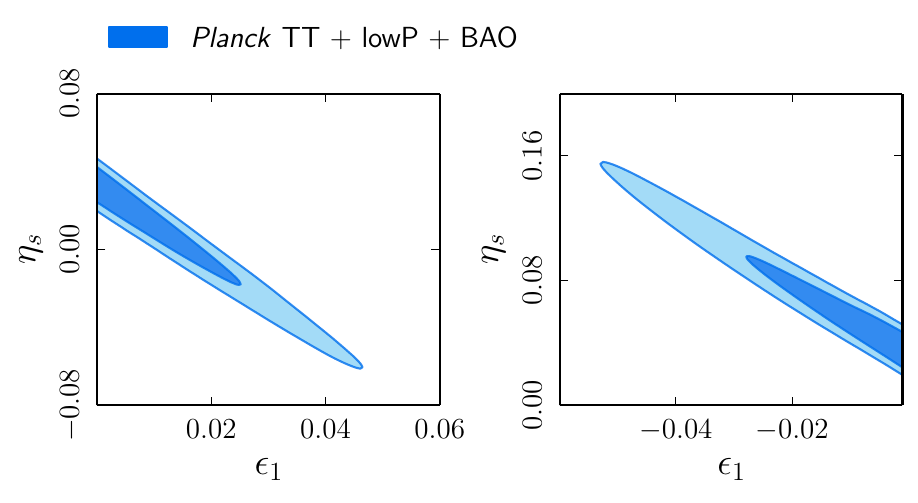}
\caption{Marginalized joint 68\,\% and 95\,\% CL for 
the Galileon parameters $(\epsilon_1\,, \eta_s)$ for $n_{\mathrm t}<0$ (left panel) and $n_{\mathrm t}>0$ (right panel).}
\label{Gplus}
\end{figure}

Galileon models of inflation predict interesting NG signatures 
\citep{2011JCAP...01..014B,2013PhRvD..88b3529T}.\footnote{See also \cite{2010PhRvD..82j3518M}, \cite{2011JCAP...12..019G}, 
\cite{2011PhRvD..83j3524K}, \cite{2013JCAP...03..030D}, and \cite{2014arXiv1411.4501R}.} 
We have verified \citep[see also][]{2011JCAP...02..006C} that bispectra can be generated with the 
same shapes as the ``EFT1'' and ``EFT2'' \citep{2010JCAP...01..028S,2007JCAP...01..002C} constrained 
in the companion paper \citep{planck2014-a19}, which usually arise in models of inflation with non-standard kinetic terms, with 
\begin{eqnarray}
\label{fnlG}
f_\mathrm{NL}^\mathrm{EFT1}&=&\frac{17}{972}\left(-\frac{5}{c_{\mathrm{s}}^4}+\frac{30}{c_{\mathrm{s}}^2}-\frac{40}{c_{\mathrm{s}} \bar c_{\mathrm{s}}}+15 \right),\\
\label{fnlG2}
f_\mathrm{NL}^\mathrm{EFT2}&=&\frac{1}{243} \left(\frac{5}{c_{\mathrm{s}}^4}+
\frac{30/A-55}{c_{\mathrm{s}}^2}+\frac{40}{c_{\mathrm{s}} \bar c_{\mathrm{s}}}-320 \frac{c_{\mathrm{s}}}{\bar c_{\mathrm{s}}}-\frac{30}{A}+275 
\right. \nonumber\\
&-& \left.225 c_{\mathrm{s}}^2+280 \frac{{c_{\mathrm{s}}^3}}{\bar c_{\mathrm{s}}}\right)\, .
\end{eqnarray}
As explained in the previous subsection, the linear combinations of these two bispectra produce both equilateral and orthogonal bispectrum templates.
Given Eqs.~(\ref{PSG})--(\ref{fnlG2}), we can proceed as in the previous section to exploit 
the limits on primordial NG in a combined analysis with the power spectra analysis. 
In~\cite{planck2014-a19} the constraint $c_{\mathrm{s}} \geq 0.23$ ($95\,\%$ CL) is obtained based on 
the constraints on $f^{\mathrm{equil}}_{\mathrm{NL}}$ and $f^{\mathrm{ortho}}_{\mathrm{NL}}$.  
One can proceed as described in~\cite{planck2014-a19} to constrain the parameter 
$\bar{c}_{\mathrm{s}}$ modifying the consistency relation, Eq.~(\ref{cG}). Adopting a log-uniform prior 
on $A$ in the interval $10^{-4} \leq A \leq 10^4$ and a uniform prior $10^{-4} \leq c_{\mathrm s} \leq 1$, 
the \Planck\ measurements on $f^{\mathrm{equil}}_{\mathrm{NL}}$ and $f^{\mathrm{ortho}}_{\mathrm{NL}}$ 
constrain $\bar c_{\mathrm{s}}$ to be $0.038 \leq \bar c_{\mathrm{s}}\leq  100$ ($95\,\%$ CL) \citep{planck2014-a19}. 
We also explore the possibility of the negative branch (corresponding to a blue tensor spectral index), 
finding $-100 \leq \bar c_{\mathrm{s}} \leq  -0.034$ ($95\,\%$ CL) \citep{planck2014-a19}. 
By allowing a logarithmic prior on $\bar c_{\mathrm{s}}$ based on the $f_{\mathrm{NL}}$ measurements, 
Fig.~\ref{Gplus} shows the joint constraints on $\epsilon_1$ and
$\eta_{\mathrm s}$ for the $n_{\mathrm t}<0$ branch and for the $n_{\mathrm t}>0$ branch.
\Planck\ TT+lowP+BAO and the NG bounds on $\bar c_{\mathrm{s}}$ constrain $\epsilon_1 < 0.036$
at $95\,\%$ CL for $n_{\mathrm t}<0$ (and $|\epsilon_1| < 0.041$ for $n_{\mathrm t}>0$).

\subsection{Axion monodromy inflation 
\label{axion}}

\subsubsection{Introduction}

The mechanism of monodromy inflation \citep{monodromyI, monodromyII, Kaloper:2011jz,Flauger:2014ana} in string theory motivates
a broad class of inflationary potentials of the form
\begin{equation}
V(\phi)=\mu^{4-p} \phi^p +\Lambda_0^4\, e^{- C_0 \left(\frac{\phi}{\phi_0}\right)^{p_\Lambda}}
\cos\left[\gamma_0 +\frac{\phi_0}{f_0} \left(\frac{\phi}{\phi_0}\right)^{p_f+1}\right] \, .
\label{powersform}
\end{equation}
Here $\mu$, $\Lambda_0$, $f_0$, and $\phi_0$ are constants with the dimension of mass and
$C_0$, $p$, $p_\Lambda$, $p_f$, and  $\gamma_0$  are dimensionless.

In simpler parameterizations used in prior analyses of oscillations from axion monodromy inflation 
\citep{Peiris:2013opa, planck2013-p17, Easther:2013kla, Jackson:2013mka, 
Meerburg:2013cla,Meerburg:2013dla,Meerburg:2014kna,Meerburg:2014bpa}, one assumes $p_\Lambda=p_f=0$, corresponding 
to a sinusoidal term with constant amplitude throughout inflation taken to be a periodic function of the 
canonically-normalized inflaton $\phi$. Taking $p_\Lambda \neq 0$ and $p_f \neq 0$ allows the {\it{magnitude}} and 
{\it{frequency}}, respectively, of the modulation to depend on $\phi$.  For example, the frequency is always a 
periodic function of an underlying angular axion field, but its relation to the canonically normalized inflaton 
field is model-dependent.

The microphysical motivation for $p_\Lambda \neq 0$ and $p_f \neq 0$ is that in string theory additional scalar 
fields, known as ``moduli,'' evolve during inflation. The inflationary potential depends on a subset of these 
fields. Because the magnitude and frequency of modulations are determined by the vacuum expectation values of 
moduli, both quantities are then naturally functions of $\phi$. The case $p_\Lambda=p_f=0$ corresponds to when 
these fields are approximately fixed, stabilized strongly by additional terms in the scalar potential.  But in 
other cases, the axion potential that drives inflation also provides a leading term stabilizing the moduli. The 
exponential dependence of the magnitude in the potential of Eq.~(\ref{powersform}) arises because the modulations 
are generated non-perturbatively, e.g., by instantons. For this reason, the modulations can be undetectably small 
in this framework, although there are interesting regimes where they could be visible.

Specific examples studied thus far yield exponents $p$, $p_\Lambda$, and $p_f$ that are rational numbers of modest 
size. For example, models with $p=3$, $2$, $4/3$, $1$, and $2/3$ have been constructed 
\citep{monodromyI,monodromyII,powers}, or in another case $p=4/3$, $p_\Lambda=-1/3$, and $p_f=-1/3$. Following 
\cite{Flauger:2014ana}, we investigate the effect of a drift in frequency arising from $p_f$, neglecting a 
possible drift in the modulation amplitude by setting $p_\Lambda = C_0 = 0$. Even in this restricted model, a 
parameter exploration using a fully numerical computation of the primordial power spectrum following the 
methodology of \cite{Peiris:2013opa} is prohibitive, so we follow \cite{Flauger:2014ana} to study two templates 
capturing the features of the primordial spectra generated by this potential.

The first template, which we call the ``semi-analytic'' template, is given by
\begin{equation}
\label{eq:semianalytic_mono}
{\cal P}_{\cal R} (k)  = {\cal P}_{\cal R} (k_*) \left( \frac{k}{k_*} \right)^{n_\mathrm{s} -1}
\left\{1 + \delta n_\mathrm{s}  \cos\left[ \frac{\phi_0}{f}\left(\frac{\phi_k}{\phi_0}\right)^{p_f+1} + \Delta \phi \right] \right\}.
\end{equation}
The parameter $f$ is higher than 
the underlying axion decay constant 
$f_0$ 
of the potential
by a few percent, but this difference will be neglected in this analysis. The quantity $\phi_0$ is some
fiducial value for the scalar field, and $\phi_k$ is the value of the scalar field at the
time when the mode with comoving momentum $k$ exits the Hubble radius. At leading order in the
slow-roll expansion, in units where the reduced Planck mass $M_\mathrm{Pl}=1$,
$\phi_k = \sqrt{2p\,(N_0 - \ln(k/k_*))}$, where $N_0 = N_* + \phi_\mathrm{end}^2/(2p)$,
and $\phi_\mathrm{end}$ is the value of the scalar field at the end of inflation.

The second ``analytic'' template was derived by \cite{Flauger:2014ana} by expanding
the argument of the trigonometric function in Eq.~(\ref{eq:semianalytic_mono}) in $\ln(k/k_*)$, leading to
\begin{eqnarray}
\label{eq:analytic_mono}
&&{\cal P}_{\cal R}(k) = {\cal P}_{\cal R} (k_*) \left( \frac{k}{k_*} \right)^{n_\mathrm{s} -1} \\ \nonumber
& \times & \left\{1 + \delta n_\mathrm{s}
\cos\left[\Delta \phi  + \alpha\left(\ln\left(\frac{k}{k_*}\right)
+ \sum_{n=1}^{2} \frac{c_n}{N_*^n} \ln^{n+1}\left(\frac{k}{k_*}\right)\right)\right] \right\}.
\end{eqnarray}
The relation between the empirical parameters in the templates
and the potential parameters are approximated by $\delta n_\mathrm{s} = 3b\sqrt{2\pi/\alpha}$, where
\begin{equation}
\label{eq:alpha}
\alpha = (1+p_f)\frac{\phi_0}{2 f N_0} \left(\frac{\sqrt{2 p N_0}}{\phi_0}\right)^{1+p_f},
\end{equation}
and $b$ is the monotonicity parameter defined in \cite{Flauger:2014ana},
providing relations converting
bounds on $c_n$ into bounds on the microphysical parameters of the potential.
However, the analytic template can describe more general shapes of primordial spectra than just axion monodromy.

As discussed by \cite{Flauger:2014ana}, there is a degeneracy between $p$
(or alternatively $n_\mathrm{s}$) and $f$. For both templates we fix $p=4/3$ and also fix 
the tensor power spectrum to its form in the absence of oscillations.
This is an excellent approximation because tensor oscillations are suppressed relative to the scalar oscillations by a
factor $\alpha(f/M_\mathrm{Pl})^2\ll1$. A uniform prior $-\pi < \Delta\phi < \pi$ is adopted for the
phase parameter of both templates as well as a prior $0 < \delta n_\mathrm{s} < 0.7$ for the modulation amplitude parameter.

\begin{figure*}
\begin{center}
\includegraphics[height=0.255\textwidth]{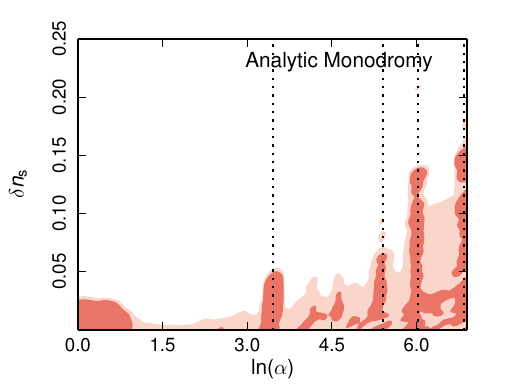}%
\includegraphics[height=0.255\textwidth]{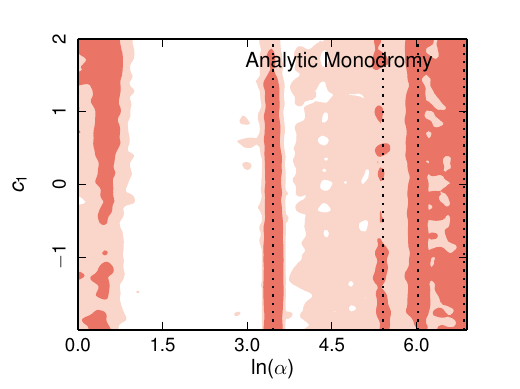}%
\includegraphics[height=0.255\textwidth]{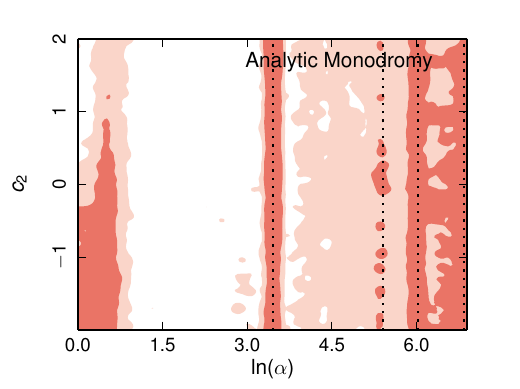} \\
\end{center}
\caption{Constraints on  the parameters of the analytic template, showing joint 68\,\% and 95\,\%~CL. The dotted lines correspond to the frequencies showing the highest-likelihood improvements (see text).
\label{fig:mono_analytic}}
\end{figure*}

\begin{figure*}
\begin{center}
\includegraphics[height=0.255\textwidth]{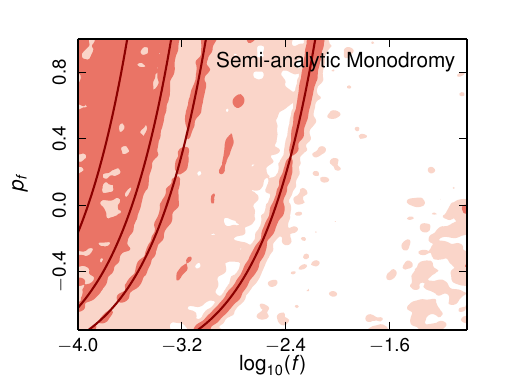}%
\includegraphics[height=0.255\textwidth]{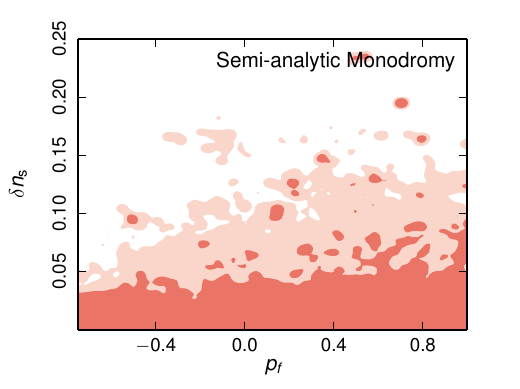}%
\includegraphics[height=0.255\textwidth]{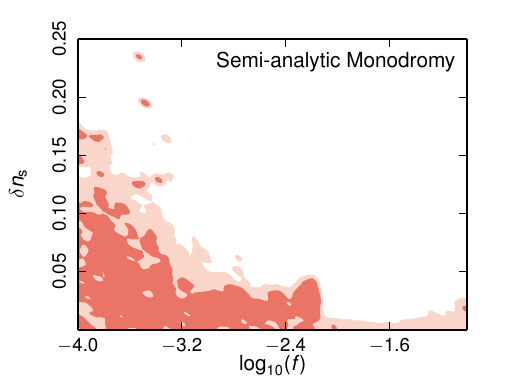} \\
\end{center}
\caption{Constraints on  the parameters of the semi-analytic template 
showing joint 68\,\% and 95\,\%~CL. The solid lines on the left-hand panel 
mark the frequencies showing the highest-likelihood improvements (see text).
\label{fig:mono_semianalytic}}
\end{figure*}

In order to specify the semi-analytic template, 
we assume instantaneous reheating, which for $p=4/3$ corresponds to 
$N_* \approx 57.5$ for $k_* = 0.05$\,Mpc$^{-1}$. 
We set $\phi_0 = 12.38M_\mathrm{Pl}$ with $\phi_\mathrm{end} = 0.59M_\mathrm{Pl}$. We adopt uniform priors 
$-4 < \log_{10} (f/M_\mathrm{Pl}) < -1$ and $-0.75<p_f < 1$ for the remaining parameters. The priors $0< \ln (\alpha) < 6.9$ and $-2 < c_{1,2} < 2$ specify the {\it 
analytic} template. The single-field effective field theory becomes strongly coupled for $\alpha > 200$. However, in principle the string construction 
remains valid in this regime.

\subsubsection{Power spectrum constraints on monodromy inflation} 

We carry out a Bayesian analysis of axion monodromy inflation using a high-resolution version of {\tt CAMB} 
coupled to the {\tt PolyChord} sampler (see Sect.~\ref{sec:PR_Bayes}). For our baseline analysis we conservatively 
adopt \Planck\ TT+lowP, using the ``bin1'' high-$\ell$ $TT$ likelihood.
In addition to the primordial template priors specified above, we marginalize over the standard priors for the cosmological 
parameters, the primordial amplitude, and foreground parameters.

\begin{figure}
\includegraphics[width=\columnwidth]{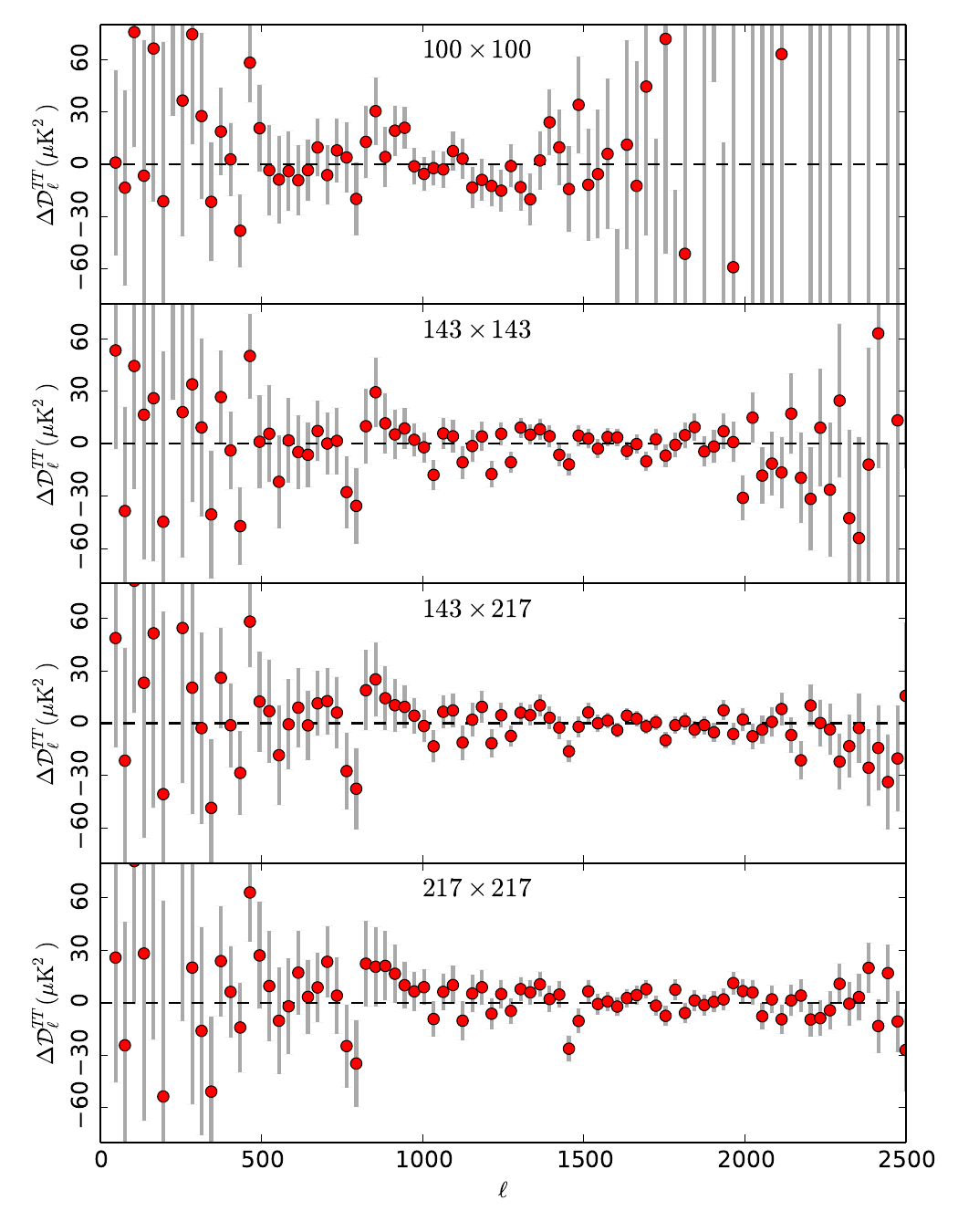} 
\caption{Frequency residuals for the $\ln(\alpha) \approx 3.5$ likelihood peak, binned at $\Delta \ell = 30$. The $\pm1\,\sigma$ errors are given by the square root of the diagonal elements of the covariance matrix.
\label{fig:mono_residuals}}
\end{figure}

The marginalized joint posterior constraints on pairs of primordial parameters for the analytic and semi-analytic 
templates are shown in Figs.~\ref{fig:mono_analytic} and \ref{fig:mono_semianalytic}, respectively.

The complex structures seen in these plots arise due to degeneracies in the likelihood frequency ``beating'' 
between underlying modulations in the data and the model \citep{Easther:2004vq}. Parameter combinations where 
``beating'' occurs over the largest $k$ ranges lead to discrete local maxima in the likelihood. Fortuitous 
correlations in the observed realization of the $C_\ell$ can give the same effect.

The four frequencies picked out by these structures, $\ln(\alpha) \approx \left\{3.5, 5.4, 6.0, 6.8 \right\}$, 
show improvements of $\Delta \chi^2 \approx \left\{-9.7, -7.1, -12.2, -12.5 \right\}$ relative to 
$\Lambda$CDM, respectively. These frequencies are marked by dotted lines in Fig.~\ref{fig:mono_analytic}, and by 
solid lines in Fig.~\ref{fig:mono_semianalytic} using Eq.~(\ref{eq:alpha}). The semi-analytic and analytic 
templates lead to self-consistent results as expected, with analogous structures being picked out by the 
likelihood in each template. There is no evidence for a drifting frequency, $p_f \neq 0$ or $c_n \neq 0$. Thus, 
these parameters serve to smooth out structures in the marginalized posterior.

The improvement in $\chi^2$ is not compelling enough to suggest a primordial origin. Fitting a modulated model to 
simulations with a smooth spectrum can give rise to $\Delta \chi^2 \sim -10$ improvements 
\citep{Flauger:2014ana}. Furthermore, as the monodromy model contains only a single frequency, at least three of 
these structures must correspond to spurious fits to the noise.  Considering the two models defined by the two 
templates and the parameter priors specified above, the Bayes factors calculated using {\tt PolyChord} favours 
base $\Lambda$CDM over both templates by odds of roughly 8:1.

Compared to previous analyses of the linear ($p=1$) axion monodromy model for WMAP9 \citep{Peiris:2013opa} and the 
2013 \Planck\ data \citep{planck2013-p17,Easther:2013kla} the common frequencies are shifted slightly upward. The 
lower frequency in common appears shifted by a factor of order $\sqrt{p}$ from $\alpha \approx 28.9$ to $31.8$ and 
the higher frequency in common from $\alpha \approx 210$ to $223$. \cite{Flauger:2014ana} suggest that the lower 
frequency (which had $\Delta \chi^2 = -9$ in \citetalias{planck2013-p17}) was associated with the 4\,K 
cooler line systematic effects in the 2013 \Planck\ likelihood. However, its presence at similar significance in 
the 2015 likelihood with improved handling of the cooler line systematics suggests that this explanation is not 
correct. The second frequency, which appeared with $\Delta \chi^2 \approx -20$ in WMAP9 
\citep{Peiris:2013opa} is still present but with much reduced significance, suggesting that the high multipoles do 
not give evidence for this frequency. Additionally, two higher frequencies are present, which if interpreted as 
being of primordial origin, correspond to a regime well beyond the validity of the single-field effective field 
theory. If one of these frequencies were to be confirmed as primordial, a significantly improved understanding of 
the underlying string construction would need to be undertaken.

In order to check whether the improvement in fit at these four modulation frequencies is responding to residual 
foregrounds or other systematics, we examine the frequency residuals. Figure~\ref{fig:mono_residuals} shows the 
residuals of the data minus the model (including the best-fit foreground model) for the four {\tt PLIK} frequency 
combinations binned at $\Delta \ell = 30$ for the lowest modulation frequency, $\ln(\alpha) \approx 3.5$. This 
plot shows no significant frequency dependence, and thus there is no indication that the fit is responding to 
frequency dependent systematics. Furthermore, the plot does not show evidence that the improvement for this 
modulation frequency comes from the feature at $\ell \approx 800$, as suggested by \cite{Easther:2013kla}. This 
feature and another at $\ell \approx 1500$ are apparent at all frequency combinations. Similar plots for the three 
other modulation frequencies also do not show indications of frequency dependence.

\begin{figure}
\includegraphics[height=0.4\textwidth]{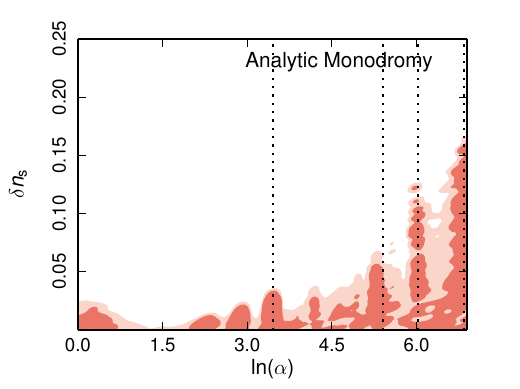} 
\includegraphics[height=0.4\textwidth]{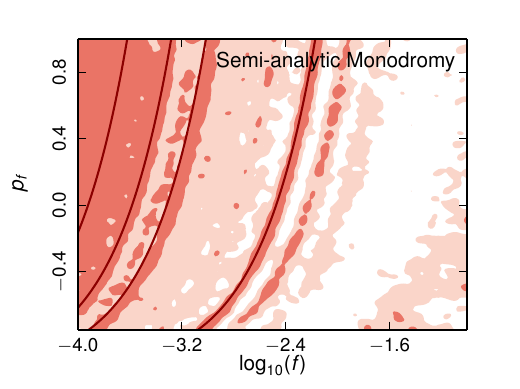} 
\caption{Constraints on  the parameters of the analytic (top) 
and semi-analytic (bottom) templates with the addition of high-$\ell$ 
polarization data in the likelihood, showing joint 68\,\% and 95\,\%~CL. 
The lines mark the frequencies showing the highest-likelihood improvements identified in the baseline 
temperature-only analysis.
\label{fig:mono_pol}}
\end{figure}

\begin{figure}
\includegraphics[height=0.4\textwidth]{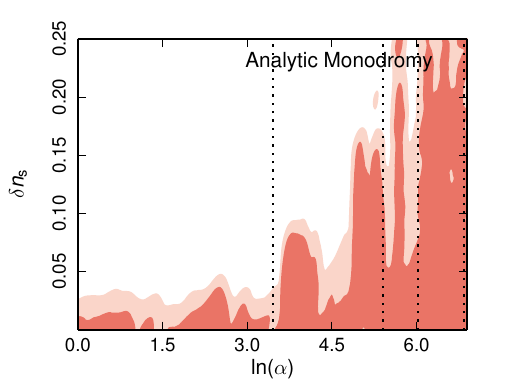}
\includegraphics[height=0.4\textwidth]{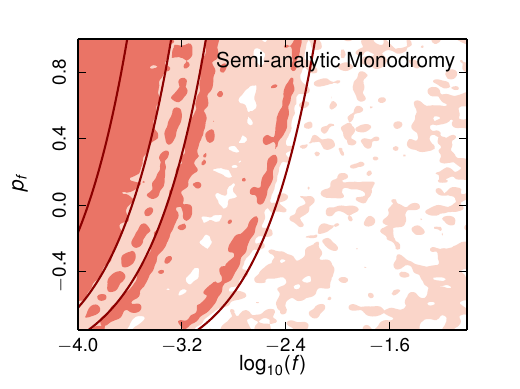}
\caption{Constraints on  the parameters of the analytic (top) and semi-analytic
(bottom) templates with $EE$-only high-$\ell$ polarization data plus low-$\ell$
temperature and polarization data, showing joint 68\,\% and 95\,\%~CL. The lines
mark the frequencies showing the highest-likelihood improvements identified in the
baseline temperature-only analysis.
\label{fig:mono_pol_only}}
\end{figure}

In order to confirm whether any of the frequencies picked out here is of primordial origin, one can exploit 
independent information in the polarization data to perform a cross-check of the temperature prediction, thus 
minimizing the ``look-elsewhere'' effect \citep{2009PhRvD..79j3519M}. Leaving a complete analysis of the 
independent information in the polarization for future work, we now check whether the temperature-only result 
remains stable when high-$\ell$ polarization is added in the likelihood. In Fig.~\ref{fig:mono_pol} we show a 
preliminary analysis using the {\tt PLIK} temperature and polarization ($TT$, $TE$, and $EE$) ``bin1'' likelihood plus 
low-$\ell$ polarization data. A comparison with the left-hand panels of Figs.~\ref{fig:mono_analytic} and
\ref{fig:mono_semianalytic} indicates slight differences from the $T$-only analysis.
However, all the four frequencies identified in the temperature are present when high-$\ell$ polarization is added. There is a maximum 
$\Delta \chi^2 \approx -8.0$ improvement over $\Lambda$CDM. We also repeat the analysis using only 
the $EE$ polarization ``bin1'' likelihood plus low-$\ell$ temperature and polarization data. These results 
are presented in Fig.~\ref{fig:mono_pol_only}. The $EE$-only frequencies are offset with respect to the 
temperature-only frequencies: the best-fit $EE$-only frequencies are at $\ln(\alpha) \approx \left\{3.8, 5.0, 5.4, 
5.8, 6.2\right\}$. The maximum improvement over $\Lambda$CDM for this case is $\Delta \chi^2  \approx 
-12.5$.

\subsubsection{Predictions for resonant non-Gaussianity} 

\begin{figure*} [!htb]
\begin{center}
\includegraphics[height=0.255\textwidth]{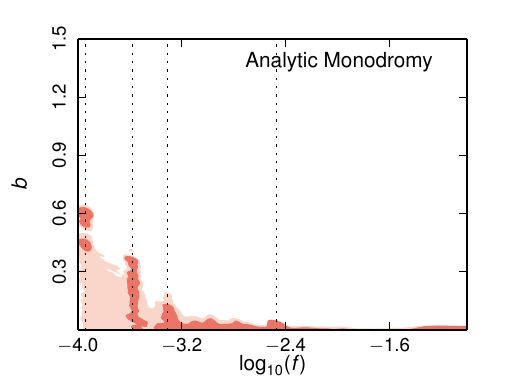}%
\includegraphics[height=0.255\textwidth]{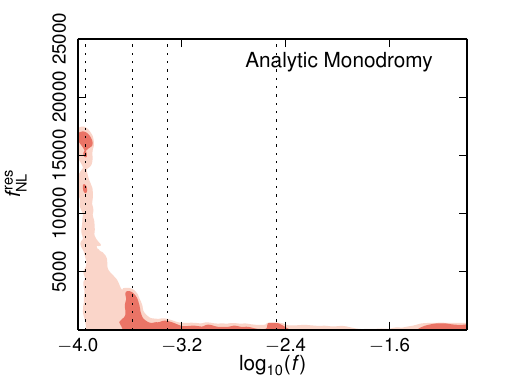}%
\includegraphics[height=0.255\textwidth]{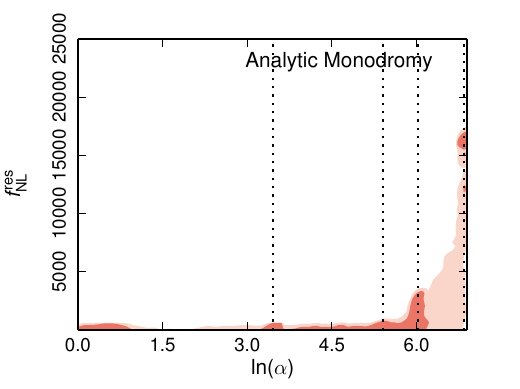} \\
\end{center}
\caption{Derived constraints on  the parameters of the potential, Eq.~(\ref{powersform}), as well as the predicted resonant NG, $f_\mathrm{NL}^\mathrm{res}$, using the analytic template, showing joint 68\,\% and 95\,\%~CL. The dotted lines mark the frequencies showing the highest-likelihood improvements (see text).
\label{fig:mono_derived}}
\end{figure*}

The left-hand panel of Fig.~\ref{fig:mono_derived} presents derived constraints on the 
parameters of the potential in Eq.~(\ref{powersform}) 
calculated using the analytic template. Another cross-check of primordial origin is available since the monodromy model predicts 
resonant NG, generating a bispectrum whose properties would be strongly correlated with that of the power spectrum 
\citep{Chen:2008wn,Flauger:2010ja}. Using the mapping
\begin{equation}
f_\mathrm{NL}^\mathrm{res} = \frac{\delta n_\mathrm{s}}{8} \alpha^2 ,
\end{equation}
we use the analytic template to derive the posterior probability for the resonant NG signal predicted by constraints from the power 
spectrum, presented in the middle and right panels of Fig.~\ref{fig:mono_derived}.

\cite{planck2014-a19} use an improved modal estimator to scan for resonant NG. The resolution of this 
scan is currently limited to $\ln(\alpha) < 3.9$, which potentially can probe the lowest frequency picked out 
in the power spectrum search. However, the modal estimator's sensitivity (imposed by cosmic variance) of $\Delta 
f_\mathrm{NL}^\mathrm{res} \approx 80$ is significantly greater than the predicted value for this frequency from 
fits to the power spectrum, $f_\mathrm{NL}^\mathrm{res} \sim 10$. Efforts to increase the resolution of the modal 
estimator are ongoing and may allow consistency tests of the significantly higher levels of resonant 
NG predicted by the higher frequencies in the future.

\subsubsection{Power spectrum and bispectrum constraints on axion inflation with a gauge field coupling  \label{sec:Model}}

We now consider the case where the axion field is coupled to a gauge field. Such a scenario is physically well 
motivated. From an effective field theory point of view the derivative coupling is natural and must be included 
since it respects the same shift symmetry that leads to axion models of inflation 
\citep{2010PhRvD..81d3534A,2011PhRvL.106r1301B,2013CQGra..30u4002P}. This type of coupling is also ubiquitous in 
string theory (see, e.g., \cite{2012PhRvD..85b3525B} and \cite{2013PhRvD..87j3506L}). The coupling term in the 
action is \citep{2010PhRvD..81d3534A,2011PhRvL.106r1301B,2011JCAP...04..009B}
\begin{equation}
S\supset\int d^4x\sqrt{-g}\left(-\frac{\alpha}{4f}\phi F^{\mu\nu}\tilde F_{\mu\nu}\right),
\end{equation}
where $F_{\mu\nu}=\partial_\mu A_\nu-\partial_\nu A_\mu$, its dual is $\tilde 
F^{\mu\nu}=\epsilon^{\mu\nu\alpha\beta} F_{\alpha\beta}/2$, and $\alpha$ is a dimensionless constant which, from 
an effective field theory perspective, is expected to be of order one. For the potential of the axion field, we 
will not investigate further the consequences of the oscillatory part of the potential, focusing on the coupling 
of the axion field to the U(1) gauge field (effectively setting $\Lambda_0=0$).

The coupling of a pseudo-scalar axion with the gauge field has interesting phenomenological consequences, both for 
density perturbations and primordial gravitational waves 
\citep{2011PhRvL.106r1301B,2011JCAP...06..003S,2011JCAP...04..009B,2012PhRvD..85b3525B,2013JCAP...02..017M,2014JHEP...12..139F}. 
Gauge field quanta source the axion field via an inverse decay process $\delta A+\delta A \rightarrow \delta 
\varphi$, modifying the usual predictions already at the power spectrum level. Additionally, the inverse decay can 
generate a high level of primordial NG.

The parameter \begin{equation} \xi=\frac{\alpha|\dot\phi|}{2fH} \end{equation} characterizes the strength of the 
inverse decay effects. If $\xi < 1$ the coupling is too small to produce any modifications to the usual 
predictions of the uncoupled model. For previous constraints on $\xi$ see 
\cite{2011JCAP...04..009B,2012PhRvD..85b3525B} and \cite{2013JCAP...02..017M}. Using the slow-roll approximation 
and neglecting the small oscillatory part of the potential, one can express
\begin{equation} 
\xi=M_\mathrm{Pl}\frac{\alpha}{f}\sqrt{\frac{p}{8N+2p}},\label{xi} 
\end{equation} 
where $N$ is, as usual, the number of $e$-folds to the end of inflation. The scalar power spectrum of curvature 
perturbations is given by
\begin{eqnarray} \label{p1new} \mathcal{P}_{\mathcal {R}}(k)=\mathcal{P}_* \left( \frac{k}{k_*} \right)^{n_\mathrm{s}-1} \left[1+ 
\mathcal{P}_* \left( \frac{k}{k_*} \right)^{n_\mathrm{s}-1} f_2(\xi(k))\, e^{4 \pi \xi_*} 
\left( \frac{k}{k_*}\right)^{2 \pi \xi_* \epsilon_2} \right]\, , 
\nonumber\\ 
\end{eqnarray} 
where \citep{2013JCAP...02..017M} 
\begin{equation} \label{xik} \xi(k)=\xi_* \left[1+\frac{\epsilon_2}{2} 
\ln \left( \frac{k}{k_*} \right) \right]\, . \end{equation} 
Here an asterisk indicates evaluation at the pivot scale, $k_*=0.05$\,Mpc$^{-1}$, and $\mathcal{P}_*=H_*^4/(4 \pi^2 
\dot \phi_*^2)$ and $n_\mathrm{s}-1=-2 \epsilon_1-\epsilon_2$ are the amplitude and spectral index, respectively, of the standard 
slow-roll power spectrum of vacuum-mode curvature perturbations (the usual power spectrum in the absence of the 
gauge-coupling). By numerically evaluating the function $f_2(\xi)$ \citep[defined in equation~(3.27) of][]
{2011JCAP...04..009B}, we created an analytical fit to this function accurate to better than $2\,\%$ for 
$0.1 < \xi_* <7$.\footnote{The fitting function used is $\exp\{-a-b\,\ln(\xi)-c\,[\ln(\xi)]^2+d\, [\ln(\xi)]^3+ 
e\, [\ln(\xi)]^4\}$, where the coefficients are $a=10.8$, $b=4.58$, $c=0.51$, $d=0.01$, and $e=0.02$.} In the 
following, unless stated otherwise, we fix $p=4/3$ as in the previous subsection and assume instantaneous 
reheating so that $N_* \approx 57.5$ and the slow-roll parameters $\epsilon_1$ and $\epsilon_2$ are fixed.  For 
the tensor power spectrum we adopt the approximation \citep{2011JCAP...04..009B}
\begin{equation}
\mathcal{P}_{\mathrm t}(k)=
\mathcal{P}_{\mathrm t}\left(\frac{k}{k_*}\right)^{n_\mathrm{t}} 
\left[1+\frac{\pi^2}{2} \mathcal{P}_{\mathrm t} f_{\mathrm t,L}(\xi(k)) 
e^{4 \pi \xi_*} \left(\frac{k}{k_*}\right)^{n_\mathrm{t}+2 \pi \epsilon_2 \xi_*}\right]\, ,
\end{equation}
where
\begin{equation}
\label{fTL}
f_{\mathrm t,L}(\xi(k))=2.6 \times 10^{-7} \xi^{-5.7}(k)\, .
\end{equation}
Here $\mathcal{P}_{\mathrm t}=2 H_*^2/(\pi^2 M_\mathrm{Pl}^2)$ and $n_\mathrm{t}=-2 \epsilon_1$ are the ``usual'' 
expressions for the tensor amplitude and tensor tilt in standard slow-roll inflation.

The total bispectrum is \citep{2012PhRvD..85b3525B}  
\begin{subequations}
\begin{align}
  B(k_i) &= B_{\text{inv.dec.}}(k_i) + B_{\mathrm{res}}(k_i)  \\
&= f_\mathrm{NL}^{\text{inv.dec.}}(\xi) \, F_{\mathrm{\text{inv.dec.}}}(k_i)
+B_{\mathrm{res}}(k_i),\label{bispectrum}
\end{align}
\end{subequations}
where the explicit expression for $F_{\mathrm{inv.dec.}}(k_i)$ (\cite{2011JCAP...04..009B}; see also 
\cite{2013JCAP...02..017M}) is reported in~\cite{planck2014-a19}. This shows that the inverse decay effects and 
the resonant effects (which arise from the oscillatory part of the potential) simply ``add up" in the bispectrum. 
The nonlinearity parameter is

\begin{equation}
  f_\mathrm{NL}^{\mathrm{inv.dec.}} = \frac{f_3(\xi_*) \mathcal{P}_*^3 e^{6\pi\xi_*}}{\mathcal{P}_{\mathcal{R}}^2(k_*)} \, .\label{fNLequil}
\end{equation}
The function $f_3(\xi_*)$ corresponds to the quantity $f_3(\xi_*;1,1)$ defined in equation~(3.29) of 
~\cite{2011JCAP...04..009B}. We have computed $f_3(\xi_*)$ numerically and used a fit with an accuracy of better 
than $2\,\%$.\footnote{The fit has the same expression as the one for $f_2(\xi)$ with coefficients $a=17.0048$, 
$b=6.6578$, $c=0.96479$, $d=0.0506098$, and $e=0.039139$.} We use the observational constraint 
$f_\mathrm{NL}^\mathrm{inv.dec.}=22.7 \pm 25.5$ ($68\,\%$ CL) obtained in~\cite{planck2014-a19} from an analysis 
where only the inverse decay type NG is assumed present. We omit the explicit expression for the 
resonant bispectrum $B_{\mathrm{res}}$, since it will not be used here.

We carried out an MCMC analysis of constraints on the (scalar and tensor) power spectra predicted by this model 
with the \Planck\ TT+lowP likelihood, 
marginalizing over standard 
priors for the cosmological parameters and foreground parameters with the uniform priors $2.5 \leq \ln [10^{10} 
\mathcal{P}_* ] \leq 3.7$ and $0.1\leq \xi_* \leq 7.0$.

The power spectrum constraint gives  
\begin{equation}
0.1 \leq \xi_* \leq 2.3 \quad (95\,\%~\mathrm{CL}).
\end{equation}
Given that $f_\mathrm{NL}^{\mathrm{inv.dec.}}$ is exponentially sensitive to $\xi$, this translates into the 
prediction (using Eq.~\ref{fNLequil}) $f_\mathrm{NL}^{\mathrm{inv.dec.}} \leq 1.2$, which is significantly 
tighter than the current bispectrum constraint from \cite{planck2014-a19}. Indeed, importance sampling with the 
likelihood for $f_\mathrm{NL}^{\mathrm{inv.dec.}}$, taken to be a Gaussian centred on the NG estimate 
$f_\mathrm{NL}^\mathrm{inv.dec.}=22.7 \pm 25.5$ ($68\,\%$ CL) \citep{planck2014-a19}, changes the limit on $\xi_*$ 
only at the second decimal place.

We now derive constraints on model parameters using only the observational constraint on 
$f_\mathrm{NL}^\mathrm{inv.dec}.$ The constraints thus derived are applicable for generic $p$ and also to the 
axion monodromy model discussed in Sect.~\ref{axion}, even in the case $\Lambda_0 \neq 0$. We follow the procedure 
described in section~11 of~\cite{planck2014-a19}. The likelihood for $f_\mathrm{NL}^{\mathrm{inv.dec.}}$ is taken 
to be a Gaussian centred on the NG estimate $f_\mathrm{NL}^\mathrm{inv.dec.}=22.7 \pm 25.5$ ($68\,\%$ CL) 
\citep{planck2014-a19}. We use the expression of Eq.~(\ref{fNLequil}), where $f_3(\xi_*)$ is numerically 
evaluated. To find the posterior distribution for the parameter $\xi_*$ we choose uniform priors in the intervals 
$1.5\times 10^{-9}\leq\mathcal{P}_* \leq3.0\times 10^{-9}$ and $0.1\leq\xi_*\leq 7.0$. This yields $95\,\%$ CL 
constraints for $\xi_*$ (for any value of $p$) of
\begin{equation}
\xi_*\leq2.5\quad (95\,\%~\mathrm{CL}).
\end{equation}
If we choose a log-constant prior on $\xi_*$ we find 
\begin{equation}
\xi_*\leq2.2 \quad (95\,\%~\mathrm{CL}).
\end{equation}
For both cases the results are insensitive to the upper limit chosen for the prior on $\xi_*$ since the likelihood 
quickly goes to zero for $\xi _*>3$. As the likelihood for $\xi_*$ is fairly flat, the tighter constraint seen for 
the log-constant case is mildly prior driven. The constraints from the bispectrum are consistent with, and slightly 
worse than, the result from the power spectrum alone.

Using a similar procedure and Eq.~(\ref{xi}) one can also obtain a constraint on $\alpha/f$. Adopting a 
log-constant prior $2\leq\alpha/f\leq100$\footnote{We give only the results for a log-constant prior on 
$\alpha/f$, which is well-motivated since it corresponds to a log-constant prior on the axion decay constant for 
some fixed $\alpha$.} and uniform priors $50\leq N_*\leq70$ and $1.5\times 10^{-9}\leq\mathcal{P}_*\leq3.0\times 
10^{-9}$ we obtain the $95\,\%$ CL constraints
\begin{equation}
\alpha/f\leq48M^{-1}_\mathrm{Pl}\,\,\,\mathrm{for}\,\,\, p=1,\qquad\alpha/f\leq35M^{-1}_\mathrm{Pl} \,\,\,\mathrm{for}\,\,\, p=2, 
\end{equation}
and 
\begin{equation}
\alpha/f\leq 42M^{-1}_\mathrm{Pl} \,\,\,\mathrm{for}\,\,\, p=4/3\, .
\end{equation}
For example, for a linear potential, $p=1$, if $\alpha \sim 1$ as suggested by effective field theory, 
then the axion decay constant $f$ is constrained to be
\begin{equation}
f\geq 0.020M_\mathrm{Pl} \quad (95\,\%~\mathrm{CL}) \, ,
\end{equation}
while for a potential with $p=4/3$ we find
\begin{equation}
f\geq 0.023M_\mathrm{Pl}\quad (95\,\%~\mathrm{CL}).
\end{equation}
These limits are complementary to those derived in Sect.~\ref{axion}. 

%% file: section_eleven.tex
\begin{figure}
\begin{center}
\includegraphics{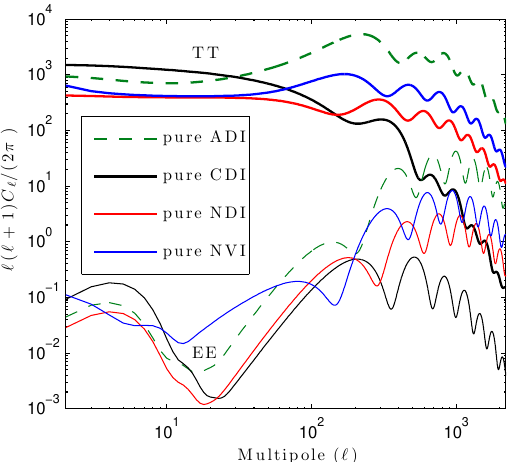}
\end{center}
\caption{Angular power spectra for the scale-invariant (i.e., $n_{\cal R \cal R} = 1$) 
pure adiabatic mode (ADI, green 
dashed curves) and for the scale invariant ($n_{\cal I \cal I} = 1$) 
pure isocurvature (CDI, NDI, or NVI) modes, with 
equal primordial perturbation amplitudes. The thick lines represent the temperature auto-correlation ($TT$) and the 
thin lines the $E$-mode polarization auto-correlation ($EE$).
\label{fig:JVmodecomparison}
}
\end{figure}

In \citetalias{planck2013-p17}, we presented constraints on a number of simple models featuring a mixture of the 
adiabatic (ADI) mode and one type of isocurvature mode. We covered the cases of CDM density isocurvature (CDI), 
neutrino density isocurvature (NDI), and neutrino velocity isocurvature (NVI) modes
\citep{Bucher:1999re},
with different assumptions 
concerning the correlation 
\citep{Langlois:1999dw,Amendola:2001ni}
between the primordial adiabatic and isocurvature perturbations. 
Isocurvature modes, possibly correlated among themselves and with the adiabatic mode, can be generated in multi-field
models of inflation; however, at present a mechanism for exciting the neutrino velocity isocurvature mode
is lacking.
Section~\ref{IsoUpdate} shows how these constraints have evolved with the new \Planck\ TT+lowP likelihoods, how 
much including the \Planck\ lensing likelihood changes the results, and what extra information the \Planck\ high-$\ell$ 
polarization contributes. A pure isocurvature mode as a sole source of perturbations has been ruled out 
\citep{Enqvist:2001fu}, since, as can be seen from Fig.~\ref{fig:JVmodecomparison}, any of the isocurvature modes 
leads to an
acoustic peak structure for the temperature angular power 
very different from 
the adiabatic mode, which 
fits the data very well. The different phases and tilts of the various modes also occur in the polarization spectra, as shown in  
Fig.~\ref{fig:JVmodecomparison} for the $E$ mode.\footnote{The transfer function mapping the 
primordial CDI mode to $C_\ell^{TT}$ is suppressed by a factor $(k/k_{\mathrm{eq}})^{-2} \sim 
(\ell/\ell_{\mathrm{eq}})^{-2}$ relative to the ADI mode, where $k_{\mathrm{eq}}$ is the 
wavenumber of matter-radiation equality.
As seen in Fig.~\ref{fig:JVmodecomparison}, there is a similar damping for the $E$ mode in the CDI 
versus the ADI case. Therefore, to be observable at high $\ell$, a CDI mode should be (highly) blue tilted. So, if 
the data favoured as small as possible a disturbance by CDI over all scales, then the CDI should have a spectral 
index, $n_{\cal I \cal I}$, of roughly three. In practice, the lowest-$\ell$ part of the data has very little 
weight due to cosmic variance, and thus we expect that the data should favour $n_{\cal I \cal I}$ less than three, 
but significantly larger than one. This should be kept in mind when interpreting the results in the CDI case, 
i.e., one cannot expect strong constraints on the \emph{primordial} CDI fraction at small scales, even if the data 
are purely adiabatic. The imprint of the baryon density isocurvature (BDI) mode in the CMB, at least at linear order, 
is indistinguishable 
from the CDI case, and hence we do not consider it separately as it can be described by 
${\cal I}_{{\mathrm{CDI}}}^\mathrm{effective}={\cal I}_{{\mathrm{CDI}}}+(\Omega_{\mathrm b} / \Omega_{{\mathrm{c}}}){\cal 
I}_{{\mathrm{BDI}}}$. The trispectrum, however, can in principle be used to
distinguish the BDI and CDI modes \citep{Grin:2013uya}. \label{foot:CDIshape}}

In Sect.~\ref{tensorsWithCDI:sect} we add one extra degree of freedom to the generally-correlated ADI+CDI model by 
allowing primordial tensor perturbations (assuming the inflationary consistency relation for the tilt of the 
tensor power spectrum and its running). Our main goal is to explore a possible degeneracy between tensor modes and 
negatively-correlated CDI modes, tending to tilt the large-scale temperature spectrum in opposite directions. In 
Sect.~\ref{sec:specialisoc}, we update the constraints on three special cases motivated by axion or curvaton 
scenarios.

The goal of this analysis is to test the hypothesis of adiabaticity and establish the robustness of the base 
$\Lambda$CDM model against different assumptions concerning initial conditions (Sect.~\ref{robustness:sect}). 
Adiabaticity is also an important probe of the inflationary paradigm, since any significant detection of 
isocurvature modes would exclude the possibility that all perturbations in the Universe emerged from quantum 
fluctuations of a single inflaton field, which can excite only one degree of freedom, the curvature (i.e., 
adiabatic) perturbation.\footnote{However, conversely, if no isocurvature was detected, the fluctuations could 
have been seeded either by single- or multi-field inflation, since later processes easily wash out inflationary 
isocurvature perturbations \citep{Mollerach:1989hu,Weinberg:2004kf,Beltran:2005gr}. An example is the curvaton 
model, in which perturbations can be purely isocurvature at Hubble exit during inflation, but are later converted 
to ADI if the curvaton or curvaton particles \citep{Linde:2005yw} dominate the energy density at the curvaton's 
decay. For a summary of various curvaton scenarios, see, e.g., \citet{Gordon:2002gv}.}

In this section, theoretical predictions were obtained with a modified version of the {\tt CAMB} code (version 
{\tt Jul14}) while parameter exploration was performed with the {\tt MultiNest} nested sampling algorithm.


\begin{figure*}
\begin{center}
(a)\hspace{130mm}$\phantom{1}$\\
\vspace{-6mm}\includegraphics{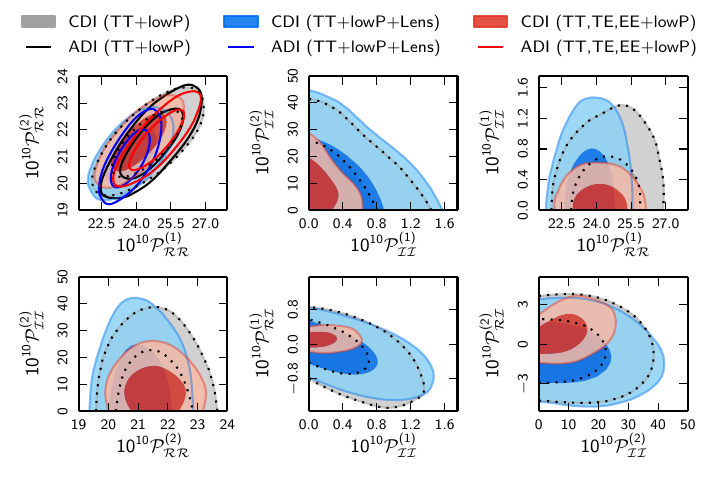}\\
(b)\hspace{130mm}$\phantom{1}$\\
\vspace{-6mm}\includegraphics{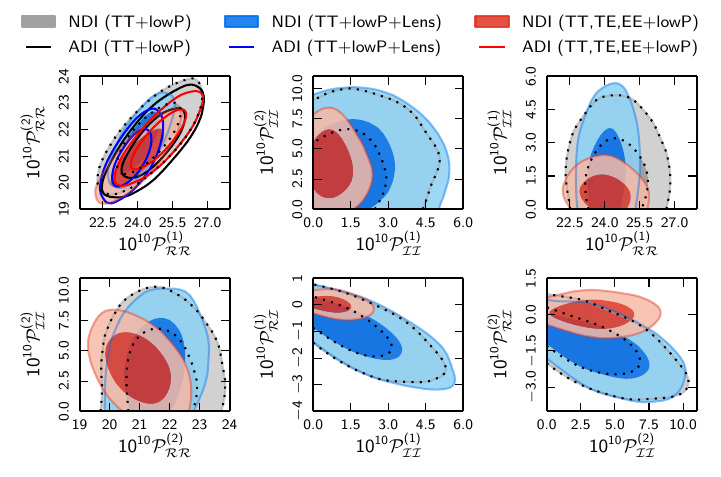}\\
(c)\hspace{130mm}$\phantom{1}$\\
\vspace{-6mm}\includegraphics{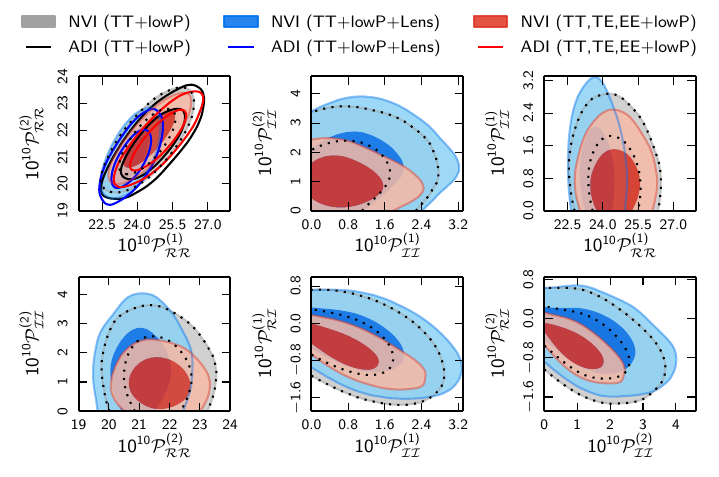}\\
\vspace{-7mm}
\end{center}
\caption{68\,\% and 95\,\% CL constraints on the primordial perturbation power in general mixed ADI+CDI (a), 
ADI+NDI (b), and ADI+NVI (c) models at two scales, $k_1=0.002\,\mathrm{Mpc}^{-1}$ (1) and 
$k_2=0.100\,\mathrm{Mpc}^{-1}$ (2), for \Planck\ TT+lowP (grey regions highlighted by dotted contours), \Planck\ 
TT+lowP+lensing (blue), and \Planck\ TT,TE,EE+lowP (red). In the first panels, we also show contours for the pure 
adiabatic base $\Lambda$CDM model with the corresponding colours of solid lines.
\label{fig:JVprimordialPowers}
}
\end{figure*}

\begin{figure*}
\begin{center}
\includegraphics{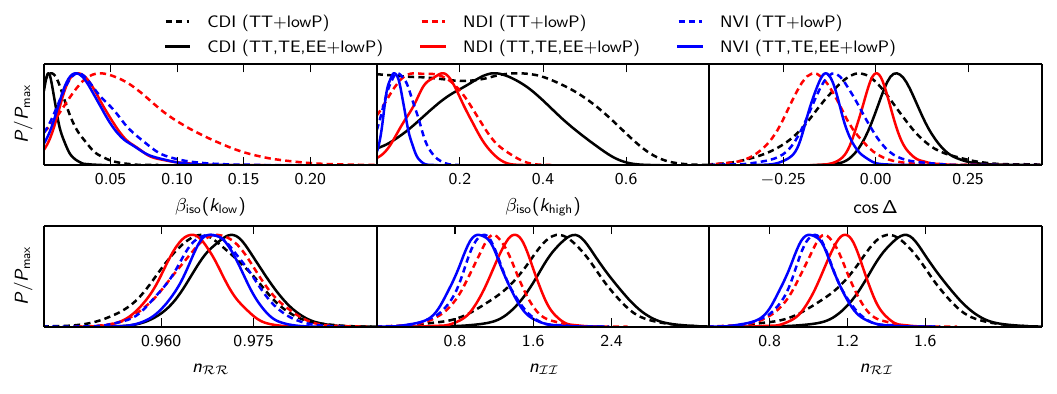}
\end{center}
\caption{Constraints on the primordial isocurvature fraction, $\beta_{\mathrm{iso}}$, at 
$k_{\mathrm{low}}=0.002\,\mathrm{Mpc}^{-1}$ and $k_{\mathrm{high}}=0.100\,\mathrm{Mpc}^{-1}$, the primordial 
correlation fraction, $\cos\Delta$, the adiabatic spectral index, $n_{\cal R \cal R}$, the isocurvature spectral 
index, $n_{\cal I \cal I}$, and the correlation spectral index, $n_{\cal R \cal I} = (n_{\cal R \cal R} + n_{\cal 
I \cal I}) / 2$, with \Planck\ TT+lowP data (dashed curves) and TT,TE,EE+lowP data (solid curves), for the 
generally-correlated mixed ADI+CDI (black), ADI+NDI (red), and ADI+NVI (blue) models. All these parameters are 
derived, and the distributions shown here result from a uniform prior on the primary 
parameters, as detailed in Eqs.~(\ref{eq:RRprior})--(\ref{eq:RIprior}). However, the effect of the non-flat 
derived-parameter priors is negligible for all parameters except for $n_{\cal I \cal I}$ (and $n_{\cal R \cal I}$) 
where the prior biases the distribution toward one. With TT+lowP, the flatness of 
$\beta_{\mathrm{iso}}(k_{\mathrm{high}})$ in the CDI case up to a ``threshold'' value of about 0.5 is a 
consequence of the $(k/k_{\mathrm{eq}})^{-2}$ damping of its transfer function as explained in 
Footnote~\ref{foot:CDIshape}.
\label{fig:JVprimordialFractions}
}
\end{figure*}

\begin{figure*}
\begin{center}
\includegraphics{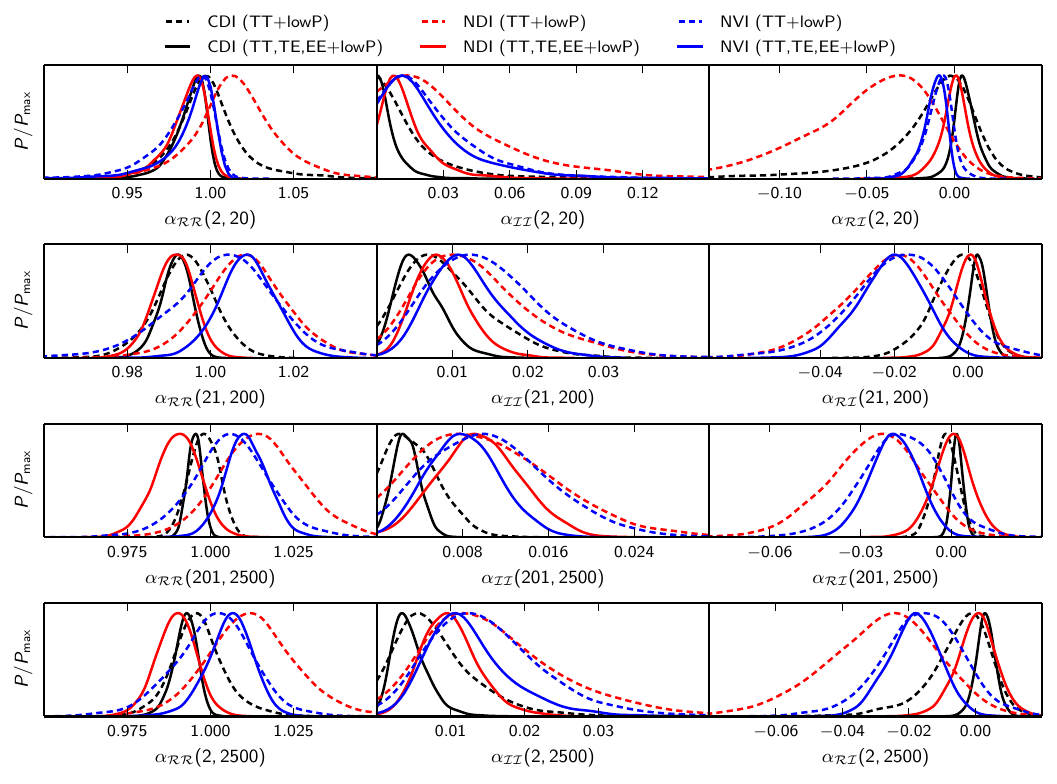}
\end{center}
\caption{Constraints on the fractional contribution of the adiabatic ($\cal R \cal R$), isocurvature ($\cal I \cal 
I$), and correlation ($\cal R \cal I$) components to the CMB temperature variance in various multipole ranges, 
as defined in Eq.~(\ref{eq:FracDef}), with \Planck\ TT+lowP data (dashed curves) and with \Planck\ TT,TE,EE+lowP data (solid 
curves).  These are shown for the generally-correlated mixed ADI+CDI (black), ADI+NDI (red), or ADI+NVI (blue) models.
\label{fig:JVobservedFractions}
}
\end{figure*}

\subsection{Parameterization and notation}
\label{sec:isocurvatureparametrization}

A general mixture of the adiabatic mode and one isocurvature mode
is described by the 
 three functions
${\cal P}_{\cal R\cal R}(k),$
${\cal P}_{\cal I\cal I}(k),$
and ${\cal P}_{\cal R\cal I}(k)$
describing the curvature, isocurvature, and cross-correlation power spectra, respectively. Our sign conventions are 
such that positive values for ${\cal P}_{\cal RI}$ correspond to a positive contribution of the cross-correlation 
term to the Sachs-Wolfe component of the total temperature spectrum.

As in \citetalias{planck2013-p17}, we specify the amplitudes at two 
scales $k_1<k_2$ and assume power-law behaviour, so that
\begin{align}
\begin{split}
{\cal P}_{\mathrm{ab}}(k)=&\exp
\Biggl[
\left(
\frac%
{\ln (k  )-\ln (k_2)}
{\ln (k_1)-\ln (k_2)}
\right)
\ln\left( {\cal P}^{(1)}_{\mathrm{ab}}\right) \\
&\qquad
+
\left(
\frac%
{\ln (k  )-\ln (k_1)}
{\ln (k_2)-\ln (k_1)}
\right)
\ln\left( {\cal P}^{(2)}_{\mathrm{ab}}\right)
\Biggr] ,
\label{matInter}
\end{split}
\end{align}
where ${\mathrm{a,b}}={\cal I}, {\cal R}$ and
${\cal I}={\cal I}_{{\mathrm{CDI}}}$, ${\cal I}_{{\mathrm{NDI}}},$ or ${\cal I}_{{\mathrm{NVI}}}$.
We set \mbox{$k_1 = 0.002\,\textrm{Mpc}^{-1}$}
and $k_2 = 0.100\,\textrm{Mpc}^{-1},$ so that
$[k_1, k_2]$ spans most of the range constrained by the \Planck\ data.
The positive definiteness of the initial condition matrix imposes a constraint on its elements at
any value of $k$:
\begin{equation}
\left[\mathcal{P}_{\mathrm{ab}}(k)\right]^2 \le \mathcal{P}_{\mathrm{aa}}(k) \mathcal{P}_{\mathrm{bb}}(k)\,.
\label{PD:Constraint}
\end{equation}
We take uniform priors on the positive amplitudes,
\begin{align}
\mathcal{P}_{\cal R\cal R}^{(1)},
\mathcal{P}_{\cal R\cal R}^{(2)} & \in (10^{-9},\, 10^{-8})\,, \label{eq:RRprior}\\
\mathcal{P}_{\cal I\cal I}^{(1)},
\mathcal{P}_{\cal I\cal I}^{(2)} & \in (0,\, 10^{-8})\,.
\end{align}
The correlation spectrum can be positive or negative. For $\mathrm{a} \neq \mathrm{b}$ we apply a uniform 
prior at large scales (at $k_1$):
\begin{equation}
\mathcal{P}_{\mathrm{ab}}^{(1)}  \in  (-10^{-8},\, 10^{-8})\,,\label{eq:RIprior}
\end{equation}
but reject all parameter combinations violating the constraint in Eq.~(\ref{PD:Constraint}).
To ensure that Eq.~(\ref{PD:Constraint}) holds for all $k$, we restrict ourselves to a 
scale-independent correlation fraction:
\begin{equation}
\cos\Delta_{\mathrm{ab}} \equiv \frac{\mathcal{P}_{\mathrm{ab}}}{\left( \mathcal{P}_{\mathrm{aa}} \mathcal{P}_{\mathrm{bb}} \right)^{1/2}} \in (-1,1)\,.
\label{eq:defCosDelta}
\end{equation}
Thus ${\cal P}_{\mathrm{ab}}^{(2)}$ is a derived parameter\footnote{
Given our ansatz of power-law primordial spectra,
if we treated $\mathcal{P}_{\mathrm{ab}}^{(2)}$ as an independent parameter as we do with $\mathcal{P}_{\mathrm{ab}}^{(1)}$,
Eq.~(\ref{PD:Constraint}) would always be violated somewhere outside $[k_1,k_2]$. In \citetalias{planck2013-p17}, we dealt with this by assuming that when maximal (anti-)correlation is reached at some scale, the correlation remains at ($-$)100\,\% beyond this scale. This introduced a kink in the cross-correlation spectrum, located at a different wavenumber for each model.
Even though the range $[k_1, k_2]$ was chosen to span most of the observable scales, this kink tended to impact 
the smallest (or largest) multipole values used in the analysis.
In particular, the kink helped fit the dip in the temperature angular power in the multipole range $\ell \approx 
10$--$40$.
\label{foot:kink}
}
given by
\begin{equation}
\mathcal{P}_{\mathrm{ab}}^{(2)} =
\mathcal{P}_{\mathrm{ab}}^{(1)}
\frac{\left( \mathcal{P}_{\mathrm{aa}}^{(2)} \mathcal{P}_{\mathrm{bb}}^{(2)} \right)^{1/2} }
{\left( \mathcal{P}_{\mathrm{aa}}^{(1)} \mathcal{P}_{\mathrm{bb}}^{(1)} \right)^{1/2}},
\label{eq:Pab2}
\end{equation}
which in terms of spectral indices is equivalent to 
\begin{equation}
n_{\mathrm{ab}} = \frac{1}{2}(n_{\mathrm{aa}} + n_{\mathrm{bb}})\,.
\label{eq:ncor}
\end{equation}



The conservative baseline likelihood is \Planck\ TT+lowP. The results 
obtained with \Planck\ TT,TE,EE+lowP should be interpreted with caution
because the data used in the 2015 release are known to contain some low
level systematics, in particular arising from $T$-to-$E$ leakage, and it is possible 
that such systematics may be fit by the isocurvature auto-correlation and
cross-correlation templates. (See \citet{planck2014-a15} for a detailed discussion.)

In what follows, we quote our results in terms of derived parameters identical to those in 
\citetalias{planck2013-p17}. We define the primordial isocurvature fraction as
\begin{equation}
\beta _\mathrm{iso}(k)=\frac{\mathcal{P}_\mathcal{II}(k)}{\mathcal{P}_\mathcal{RR}(k)+\mathcal{P}_\mathcal{II}(k)}\,.
\label{PrimFrac}
\end{equation}
Unlike the primordial correlation fraction $\cos\Delta$ defined in Eq.~(\ref{eq:defCosDelta}), $\beta 
_\mathrm{iso}$ is scale-dependent in the general case. We present bounds on this quantity at $k_{\mathrm{low}} 
=k_1$, $k_{\mathrm{mid}}=0.050\,\mathrm{Mpc}^{-1}$, and $k_{\mathrm{high}} =k_2$.

We report constraints on the relative adiabatic ($\mathrm{ab} = \cal R \cal R$), 
isocurvature ($\mathrm{ab} =\cal I \cal I$), and correlation ($\mathrm{ab} = \cal R \cal I$) 
components according to their contribution to the 
observed CMB temperature variance in various multipole ranges:
\begin{align}
\alpha _\mathrm{ab}(\ell _\mathrm{min},\ell _\mathrm{max})
\equiv&\frac%
{(\Delta T)^2_\mathrm{ab}(\ell _\mathrm{min},\ell _\mathrm{max})}
{(\Delta T)^2_\mathrm{tot}(\ell _\mathrm{min},\ell _\mathrm{max})},
\label{eq:FracDef}
\end{align}
where
\begin{equation}
{(\Delta T)^2_\mathrm{ab}(\ell _\mathrm{min},\ell _\mathrm{max})}=
\sum _{\ell =\ell _\mathrm{min}}^{\ell _\mathrm{max}}
(2\ell +1)C_{\mathrm{ab}, \ell }^{TT}.
\end{equation}
The ranges considered are $(\ell _\mathrm{min},\ell _\mathrm{max}) = (2,20)$, $(21,200),$ $(201,2500)$, and 
$(2,2500)$, where the last range describes the total contribution to the observed CMB temperature variance. 
Here $\alpha _\mathcal{RR}$ measures the adiabaticity of the temperature angular power 
spectrum, a value of unity meaning ``fully adiabatic initial conditions.'' Values less than unity mean that some of 
the observed power comes from the isocurvature or correlation spectrum, while values larger than unity mean that 
some of the power is ``cancelled'' by a negatively-correlated isocurvature contribution. The relative non-adiabatic 
contribution can be expressed as $\alpha _{\text{non-adi}} \equiv 1 - \alpha _\mathcal{RR} = \alpha _\mathcal{II} + 
\alpha_\mathcal{RI}$.

\begin{figure}
\includegraphics{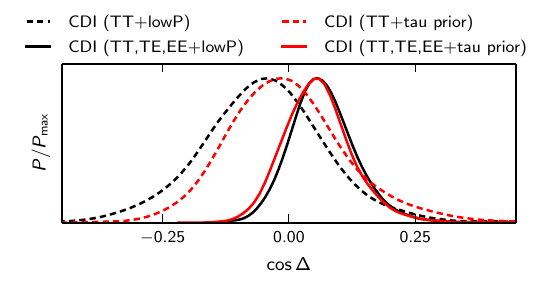}
\caption{Constraints on the primordial correlation fraction, $\cos\Delta$, in the mixed ADI+CDI model with \Planck\ 
TT+lowP data (dashed black curve) compared to the case where \Planck\ lowP data are not used, but replaced by a 
Gaussian prior $\tau=0.078\pm0.019$ (dashed red curve). The same exercise is repeated with \Planck\ TT,TE,EE data 
(solid curves) demonstrating that to a great extent the preferred value of $\cos\Delta$ is driven by the 
high-$\ell$ data.}
\label{fig:JVcosDeltaWithTau}
\end{figure}


\subsection{Results for generally-correlated adiabatic and one isocurvature mode (CDI, NDI, or NVI)}
\label{IsoUpdate}

Results are reported as 2D and 1D marginalized posterior probability 
distributions. Numerical 95\,\% CL intervals or upper bounds are tabulated 
in 
Table~\ref{tab:how_much_ic}.

Figure~\ref{fig:JVprimordialPowers} shows the \Planck\ 68\,\% and 95\,\% CL contours for various 2D combinations 
of the primordial adiabatic and isocurvature amplitude parameters at large scales (\mbox{$k_1 = 
0.002\,\textrm{Mpc}^{-1}$}) and small scales ($k_2 = 0.100\,\textrm{Mpc}^{-1}$) for (a) the generally-correlated 
ADI+CDI, (b) ADI+NDI, and (c) ADI+NVI models.  Overall, the results using \Planck\ TT+lowP are consistent with the 
nominal mission results in \citetalias{planck2013-p17}, but slightly tighter. In the first panels of 
Figs.~\ref{fig:JVprimordialPowers}~(a), (b), and (c) we also show the constraints on the curvature perturbation 
power in the pure adiabatic case. Comparing the generally-correlated isocurvature case to the pure adiabatic case 
with the same data combination summarizes neatly what the data tell us about the initial conditions. If the 
contours in the $\mathcal{P}_{\cal R\cal R}^{(1)}$-$\mathcal{P}_{\cal R\cal R}^{(2)}$ plane were shifted 
significantly relative to the pure adiabatic case, the missing power could come either from the isocurvature 
and postive correlation contributions, or the extra adiabatic power could be cancelled by a negative correlation 
contribution. We can see that these shifts are small. The low-$\ell$ temperature data continue to mildly 
favour a negative correlation (see in particular the bottom middle panel for each of the three models), since 
compared to the prediction of the best-fit adiabatic base $\Lambda$CDM model, the $TT$ angular power at multipoles 
$\ell \lsim 40$ is somewhat low. But the dotted grey shaded contours in the three middle top panels show that 
for \Planck\ TT+lowP, the posterior peaks at values ($\mathcal{P}_{\cal I\cal I}^{(1)}$,$\,\mathcal{P}_{\cal I\cal 
I}^{(2)}$) entirely consistent with $(0,\,0)$, i.e., the pure adiabatic case is preferred. The best-fit values of 
($\mathcal{P}_{\cal I\cal I}^{(1)}$,$\,\mathcal{P}_{\cal I\cal I}^{(2)}$) are 
($1.4\times10^{-11}$,$\,4.7\times10^{-13}$) for CDI, ($1.2\times10^{-12}$,$\,4.6\times10^{-10}$) for NDI, and 
($1.6\times10^{-12}$,$\,2.3\times10^{-10}$) for NVI, while ($\mathcal{P}_{\cal R\cal 
R}^{(1)}$,$\,\mathcal{P}_{\cal R\cal R}^{(2)}$) $\approx$ ($2.4\times10^{-9}$,$\,2.1\times10^{-9}$). It 
may appear from the bottom-centre panels of Fig.~\ref{fig:JVprimordialPowers} that there is nonzero posterior 
probability for $\mathcal{P}_{\cal R\cal I}^{(1)} \ne 0$ when $\mathcal{P}_{\cal I\cal I}^{(1)} = 0$, which would 
violate the positivity constraint, Eq.~(\ref{PD:Constraint}).  However, the leftmost pixels of the plots are 
actually evaluated at values of $\mathcal{P}_{\cal I\cal I}^{(1)}$ large enough 
that the constraint is satisfied.

Including the \Planck\ lensing likelihood does not significantly affect the 
non-adiabatic primordial powers, except for  tightening the constraints on the adiabatic 
power (see the blue versus black contours in the first panels of Figs.~\ref{fig:JVprimordialPowers}~(a), (b), 
and (c)). Including the lensing ($C_\ell^{\phi\phi}$) likelihood constrains the optical depth $\tau$ more tightly 
than the high-$\ell$ temperature and low-$\ell$ polarization alone \citep{planck2014-a15}. As there is a strong 
degeneracy between $\tau$ and the primordial (adiabatic) perturbation power $\mathcal{P}_{\cal R\cal R}$ (denoted 
in the other sections of this paper by $A_\mathrm{s}$), it is natural that adding the lensing data leads to 
stronger constraints on $\mathcal{P}_{\cal R\cal R}$. Moreover, replacing the low-$\ell$ likelihood \Planck\ lowP by 
\Planck\ lowP+WP constrains $\tau$ better \citep{planck2014-a15}. In the ADI+CDI case the effect of 
this replacement was very similar to adding the \Planck\ lensing data (see also 
Table~\ref{tab:how_much_ic}). Although the \Planck\ lensing data do not directly constrain the isocurvature 
contribution,\footnote{This is expected, since already with \Planck\ TT+lowP, the allowed 
isocurvature fraction is so small that it hardly affects the lensing potential spectrum, $C_\ell^{\phi\phi}$.} they can 
shift and tighten the constraints on some derived isocurvature parameters by affecting the favoured values of the 
standard parameters (present even in the pure adiabatic model). However this effect is small as confirmed in 
Table~\ref{tab:how_much_ic}. Therefore, in the figures we do not show 1D posteriors of the derived isocurvature 
parameters for \Planck\ TT+lowP+lensing, since they would be (almost) indistinguishable from \Planck\ TT+lowP, as 
we see in Fig.~\ref{fig:JVprimordialPowers} for the primary non-adiabatic parameters.

In contrast, the high-$\ell$ \emph{polarization} data significantly tighten the bounds on isocurvature and 
cross-correlation parameters, as seen by comparing the dotted grey and red contours in 
Fig.~\ref{fig:JVprimordialPowers}. The significant negative correlation previously allowed by the temperature data 
in the ADI+CDI and ADI+NDI models is now disfavoured. This is also clearly visible in the 1D posteriors of 
primordial and observable isocurvature and cross-correlation fractions shown, respectively, in 
Fig.~\ref{fig:JVprimordialFractions} and \ref{fig:JVobservedFractions}. Note how the $\cos \Delta$ and 
$\alpha_{\cal R \cal I}$ parameters are driven towards zero by the inclusion of the high-$\ell$ TE,EE data (from 
the dashed to the solid lines) in the ADI+CDI and ADI+NDI cases. We also observed 
that when the lowP data are 
replaced by a simple Gaussian prior on the reionization optical depth ($\tau=0.078\pm0.019$), the trend is 
similar. The high-$\ell$ ($\ell\ge30$) \Planck\ TT data allow a large negative correlation, while the high-$\ell$ 
\Planck\ TE,EE data prefer positive correlation. This is clearly seen in Fig.~\ref{fig:JVcosDeltaWithTau} for the 
ADI+CDI case. The best-fit values show an even more dramatic effect. We find $\cos\Delta = -0.55$ with TT+lowP, 
and $+0.15$ with TT,TE,EE+lowP.

Hence there is a competition between the temperature and polarization data that balances out and yields almost 
symmetric results about zero correlation (except in the ADI+NVI case). The isocurvature auto-correlation amplitude 
is also strongly reduced, especially in the ADI+CDI case. The best-fit values are slightly offset from 
$(\mathcal{P}_{\cal I\cal I}^{(1)},\,\mathcal{P}_{\cal I\cal I}^{(2)}) = (0,0)$, but the pure adiabatic model 
still lies inside the 68\,\% CL (for ADI+CDI and ADI+NDI) or 95\,\% CL (for ADI+NVI) regions.  In summary, the 
high-$\ell$ polarization data exhibit a strong preference for adiabaticity, although one should keep in mind the 
possibility of unaccounted systematic effects in the polarization data, possibly leading to artificially 
strong constraints.  For example, the tendency for polarization to shift the constraints towards positive 
correlation may be due to particular systematic effects that mimic modified acoustic peak structure, as we 
discussed in Sect.~\ref{sec:isocurvatureparametrization}.


\begin{figure*}
\begin{center}
\includegraphics{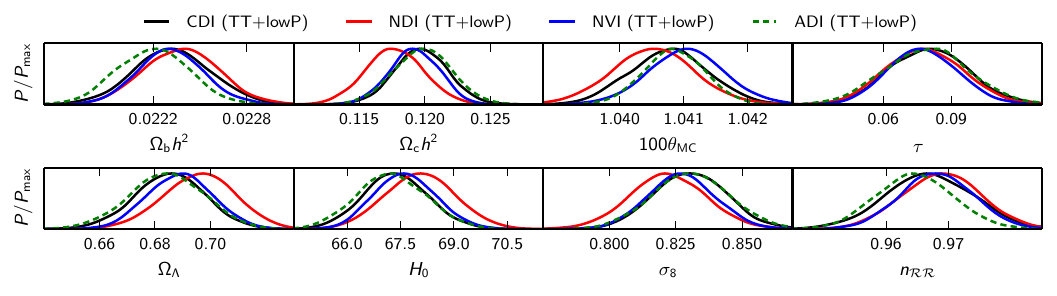}
\end{center}
\caption{Constraints on selected ``standard'' cosmological parameters with \Planck\ TT+lowP data
for the generally-correlated ADI+CDI (black), ADI+NDI (red), and ADI+NVI (blue) models compared to the pure 
adiabatic case (ADI, green dashed curves).
\label{fig:JVadiparams}
}
\end{figure*}
\begin{figure}
\includegraphics{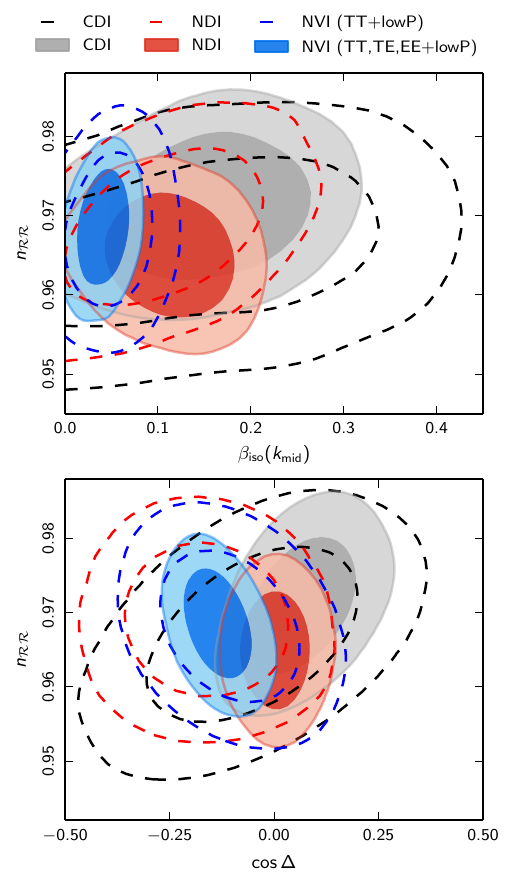}
\caption{Dependence of the determination of the adiabatic spectral index $n_{\cal R \cal R}$ (called $n_\mathrm{s}$ in the other sections of this paper) on the primordial isocurvature fraction $\beta_{\mathrm{iso}}$ and correlation fraction $\cos\Delta$, with \Planck\ TT+lowP data (dashed contours) and with \Planck\ TT,TE,EE+lowP data (shaded regions).
\label{fig:JVisocvsns}
}
\end{figure}

We also performed a parameter extraction with the \Planck\ TT,TE,EE+lowP+lensing data, but this combination did 
not provide interesting new constraints. We found 
only a tightening of bounds on the standard adiabatic parameters as in 
the \Planck\ TT+lowP+lensing case.

We provide 95\,\% CL upper limits or ranges for $\beta_{\mathrm{iso}}$, $\cos\Delta$, and $\alpha_{\cal R \cal R}$ 
in Table~\ref{tab:how_much_ic}. With \Planck\ TT+lowP, the constraints on the non-adiabatic contribution to the 
temperature variance, $1 - \alpha_{\cal R \cal R}(2,2500)$, are ($-1.5$\,\%,\,1.9\,\%), ($-4.0$\,\%,\,1.4\,\%), 
and ($-2.3$\,\%,\,2.4\,\%) in the ADI+CDI, ADI+NDI, and ADI+NVI cases, respectively.\footnote{These numbers 
can be positive even if the correlation contribution is negative. This happens whenever 
$\alpha_{\cal I \cal I} > |\alpha_{\cal R \cal I}|$. Thus in the observational non-adiabaticity estimator $1 - 
\alpha_{\cal R \cal R}(2,2500)$, the negative numbers are not as pronounced as in the primordial correlation 
fraction $\cos\Delta$.} With \Planck\ TT,TE,EE+lowP these tighten to (0.1\,\%,\,1.5\,\%), ($-0.1$\,\%,\,2.2\,\%), 
and ($-2.0$\,\%,\,0.8\,\%). In the ADI+CDI case, zero is not in the 95\,\% CL interval, but this should not be 
considered a detection of non-adiabaticity. For example, as mentioned above, $(\mathcal{P}_{\cal I\cal 
I}^{(1)},\,\mathcal{P}_{\cal I\cal I}^{(2)}) = (0,0)$ is in the 68\,\% CL region, and the best-fit values are 
($\mathcal{P}_{\cal I\cal I}^{(1)}$,$\,\mathcal{P}_{\cal I\cal I}^{(2)}$) = 
($1.0\times10^{-13}$,$\,3.5\times10^{-9}$). Moreover, the improvement in $\chi^2$ with respect to the adiabatic 
model is only 5.3 with 3 extra parameters, so this is not a significant improvement of fit. Indeed, for all 
generally-correlated mixed models the improvement in $\chi^2$ is very small. In particular, with \Planck\ TT+lowP 
it does not even exceed the number of extra degrees of freedom, which is three (see Table~\ref{tab:how_much_ic}).

Finally, we checked whether there is any Bayesian evidence for the presence of generally-correlated adiabatic and 
isocurvature modes. In all cases and with all data combinations studied, the Bayesian model comparison supports 
the null hypothesis, i.e., adiabaticity. Indeed, the logarithm of the evidence ratio is $\ln B = 
\ln(P_{\mathrm{ISO}}/P_{\mathrm{ADI}}) < -5$ (i.e., odds of much greater than 150:1 in 
favour of pure adiabaticity within \Planck's accuracy and given the parameterization and prior ranges used in our 
analysis), except for ADI+NDI with \Planck\ TT+lowP+lensing, for which the evidence ratio is slightly larger, 
$-4.6$, corresponding to odds of 1:100 for the ADI+NDI model compared to the pure adiabatic model.

\subsection{Robustness of the determination of standard cosmological parameters}
\label{robustness:sect}

Another outcome of our analysis is the robustness of the determination of the standard cosmological 
parameters against assumptions on initial conditions. Figure~\ref{fig:JVadiparams} shows the 1D marginalized 
posteriors for several cosmological parameters (not all independent of each other) with the \Planck\ TT+lowP data 
alone. For the first time, we observe that in the presence of one generally-correlated isocurvature mode (CDI, 
NDI, or NVI), predictions for these parameters remain very stable with respect to the pure adiabatic case. Except 
for the ADI+NDI case, the posteriors neither broaden nor shift significantly. A small broadening is only observed 
in the sound horizon angle $\theta_\mathrm{MC}$, which is naturally the most sensitive parameter to 
tiny disturbances of the acoustic peak structure. In the ADI+NDI case, the peak of the posterior distribution for 
some parameters shifts slightly, but the largest shift (for $\Omega_\mathrm{c} h^2$) is less than 1\,$\sigma$.

It is striking that a scale-invariant adiabatic spectrum ($n_{\cal R \cal R}=1$) is 
excluded at many $\sigma$ even when isocurvature modes are allowed: at 4.7$\,\sigma$ (ADI+CDI), 5.0$\,\sigma$ 
(ADI+NDI), and 5.4$\,\sigma$ (ADI+NVI). This illustrates how much the 
constraining power of the CMB has improved. With WMAP data, there was still a strong degeneracy between, 
for example, the 
primordial isocurvature fraction and the adiabatic spectral index \citep{Valiviita:2009bp,Savelainen:2013iwa}. 
This degeneracy nearly disappears with \Planck\ TT+lowP, and even more so with \Planck\ TT,TE,EE+lowP, as shown in 
the upper panel of Fig.~\ref{fig:JVisocvsns}. Contours in the ($n_{\cal R \cal R}$, $\cos\Delta$) space also 
shrink considerably, with some correlation remaining between these parameters in the ADI+CDI and ADI+NVI cases 
(Fig.~\ref{fig:JVisocvsns}, lower panel).

\subsection{CDI and primordial tensor perturbations}
\label{tensorsWithCDI:sect}

A primordial tensor contribution adds extra temperature angular power at low multipoles, where the adiabatic base 
$\Lambda$CDM model predicts slightly more power than seen in the data. Hence allowing for a nonzero 
tensor-to-scalar ratio $r$ might tighten the constraints on positively-correlated isocurvature, but degrade them 
in negatively-correlated models. We test how treating $r$ as a free parameter affects the constraints on 
isocurvature and how allowing for the generally-correlated CDI mode affects the constraints 
on $r$. These cases are denoted as ``CDI+$r$.'' For comparison, we examine the pure adiabatic case in the same 
parameterization, and call it ``ADI+$r$.'' We also consider another approach where we fix $r = 0.1$. These cases are 
named ``CDI+$r$$=$0.1'' and ``ADI+$r$$=$0.1.''


In the pure adiabatic case (where the curvature and tensor perturbations stay constant on super-Hubble scales), the 
primordial $r$ is the same as the tensor-to-scalar ratio at the Hubble radius exit of perturbations during 
inflation, which we call $\tilde r$. However, in the presence of an isocurvature component, ${\cal P}_{\cal R \cal 
R}$ is not constant in time even on super-Hubble scales \citep{GarciaBellido:1995qq}. Instead, the isocurvature component may source ${\cal 
P}_{\cal R \cal R}$, for example if the background trajectory in the field space is curved between Hubble exit 
and the end of inflation
\citep{Langlois:1999dw,Langlois:2000ar,Gordon:2000hv,Amendola:2001ni}.
As a result, we will have at the primordial time ${\cal P}_{\cal R \cal R} = 
\tilde{\cal P}_{\cal R \cal R} / (1-\cos^2\Delta)$, where $\tilde{\cal P}_{\cal R \cal R}$ is the curvature power 
at Hubble exit.  That is, by the primordial time the curvature perturbation power is larger than at the Hubble 
radius exit time \citep{Bartolo:2001rt,Wands:2002bn,Byrnes:2006fr}. Thus we find a relation 
\citep{Savelainen:2013iwa,Valiviita:2012ub,Kawasaki:2007mb}:
\begin{equation}
r  =  \left(1-\cos^2\!\Delta\right) \tilde r\,,
\label{eq:rvstilder}
\end{equation}
i.e., the tensor-to-scalar ratio at the primordial time ($r$) is smaller than the ratio at the Hubble radius exit 
time ($\tilde r$).

The derivation of Eq.~(\ref{eq:rvstilder}) assumes that the adiabatic and isocurvature perturbations are 
uncorrelated at Hubble radius exit ($\cos\tilde\Delta = 0$), and that all the possible primordial correlation 
($\cos\Delta \neq 0$) appears from the evolution of super-Hubble perturbations between Hubble exit and the 
primordial time. This is true to leading order in the slow-roll parameters, but inflationary models that break 
slow roll might produce perturbations that are strongly correlated already at the Hubble radius exit time.
In these cases the correlation would depend on the details of the particular model, such as the detailed shape of 
the potential and the interactions of the fields. However, a generic prediction of slow-roll inflation is that, at 
Hubble radius exit, the cross-correlation $\tilde{\cal P}_{\cal R \cal I}$ is very weak, and indeed is of the 
order of the slow-roll parameters compared to the auto-correlations $\tilde{\cal P}_{\cal R \cal R}$ and 
$\tilde{\cal P}_{\cal I \cal I}$ \citep[see, e.g.,][]{Byrnes:2006fr}. Thus, for slow-roll models, 
$|\cos\tilde\Delta| = {\cal O}\mbox{(slow-roll parameters)}\ll 1$.

In our analysis, we fix the tensor spectral index by the leading-order inflationary consistency relation, which 
now reads \citep{Wands:2002bn}
\begin{equation}
n_\mathrm{t} = -\frac{\tilde r}{8} =  -\frac{r}{8\left(1-\cos^2\Delta\right)}\,. 
\end{equation}
Assuming a uniform prior for $r$ would lead to huge negative $n_\mathrm{t}$ whenever $\cos^2\Delta$ was close to one. 
Therefore, when studying the CDI+$r$ case we assume a uniform prior on $\tilde r$ at $k=0.05\,$Mpc$^{-1}$ \citep[for 
details, see][]{Savelainen:2013iwa}.

\begin{figure}
\includegraphics{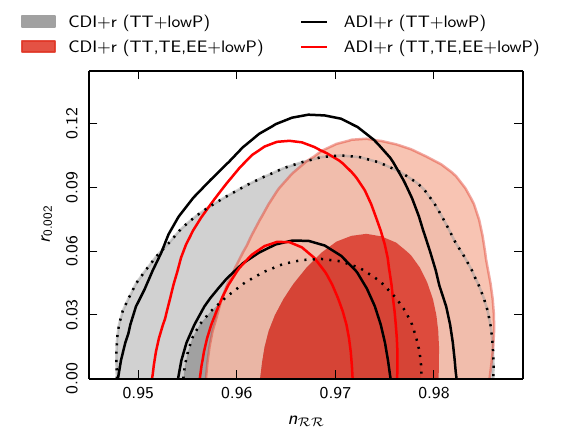}
\caption{ 68\,\% and 95\,\% CL constraints on the primordial adiabatic spectral index $n_{\cal R \cal R}$ and the primordial tensor-to-scalar ratio $r$ (more accurately, in the CDI+$r$ model, the primordial tensor-to-curvature power ratio) at $k=0.002\,$Mpc$^{-1}$. Filled contours are for generally-correlated ADI+CDI and solid contours for the pure adiabatic model.
\label{fig:JVnadir0}
}
\end{figure}

Surprisingly, allowing for a generally-correlated CDI mode (i.e., three extra parameters) hardly changes the 
constraints on $r$ from those obtained in the pure adiabatic model. In Fig.~\ref{fig:JVnadir0} we demonstrate this 
in a ``standard'' plot of $r_{0.002}$ versus adiabatic spectral index.

From Table~\ref{tab:how_much_ic} we notice that, with \Planck\ TT+lowP and TT,TE,EE+lowP, fixing $r$ to 0.1 
tightens constraints on the primordial isocurvature fraction at large scales. This is as we expected, since both 
tensor and isocurvature perturbations add power at low $\ell$, and the data do not prefer this. 
However, the shapes  
of the tensor spectrum and correlation spectrum are such that negative correlation cannot efficiently cancel the 
unwanted extra power over all scales produced by tensor perturbations (at $\ell \lsim 70$). Therefore, the 
correlation fraction $\cos\Delta$ is almost unaffected. However, when we allow $r$ to vary, the cancelation 
mechanism works to some degree when using \Planck\ TT+lowP data, leading to more negative $\cos\Delta$ than 
without $r$: with varying $r$ we have $\cos\Delta$ in the range ($-0.43$,\,0.20), while without $r$ it is in 
($-0.30$,\,0.20), at 95\,\% CL. As there is now some cancellation of power at large scales, the constraint on 
$\beta_{\mathrm{iso}}(k_{\mathrm{low}})$ weakens slightly from 0.041 without $r$ to 0.043 with $r$. On the other 
hand, the high-$\ell$ polarization data constrain the correlation to be so close to zero that with \Planck\ 
TT,TE,EE+lowP the results for $\cos\Delta$ with and without $r$ are almost identical.

The mean value of $\cos\Delta$ in the CDI+$r$ cases is $-0.071$ (TT+lowP) and $-0.076$ (TT,TE,EE+lowP). Therefore, 
$1-\cos^2\Delta \approx 0.99$, and so we do not expect a large difference between the primordial $r$ and the 
Hubble radius exit value $\tilde r$. The smallness of the difference is evident in Table~\ref{table:isocr}. To 
summarize, CDI hardly affects the determination of $r$ from the \Planck\ data, and allowing for tensor 
perturbations hardly affects the determination of the non-adiabaticity parameters.

{
\input{Tables/table_isocurvaturer.tex}
}

{
\input{special_CDI_cases.tex}
}

{
\input{Tables/table_isocurvature.tex}
}


%% file: Tables/table_isocurvaturer.tex
\begin{table}
\begin{center}
\begin{tabular}{lccc}
\noalign{\hrule\vskip 2pt}
\noalign{\hrule\vskip 3pt}
Model (and data) & $r_{0.05}$ & $\tilde r_{0.05}$ &
$C_{10}^{\mathrm{tens}} / C_{10}^{\mathrm{scal}}$ \cr
\noalign{\vskip 1pt\hrule\vskip 4pt}
CDI+$r$ (TT+lowP)     & 0.086     & 0.089 & 0.041  \cr 
ADI+$r$ (TT+lowP)     & 0.101     & 0.101 & 0.048  \cr 
CDI+$r$ (TT,TE,EE+lowP) & 0.092     & 0.092 &0.043 \cr 
ADI+$r$ (TT,TE,EE+lowP) & 0.094     & 0.094 & 0.044 \cr 
\hline
\end{tabular}
\end{center}
\caption{95\,\% CL upper bounds on the tensor-to-scalar ratio (actually
  the tensor-to-curvature power ratio) at the primordial time, $r$, and
  earlier, at the Hubble radius exit time during inflation, $\tilde r$, at
  $k=0.05\,$Mpc$^{-1}$. (In the pure adiabatic case $r$ and $\tilde r$
  are equal.) In the last column $C_{10}^{\mathrm{tens}} /
  C_{10}^{\mathrm{scal}}$ indicates the tensor contribution to the
  temperature angular power at $\ell = 10$ relative to the temperature
  power from scalar perturbations ($C_{10}^{\mathrm{scal}} =
  C_{10}^{\cal R \cal R} + C_{10}^{\cal R \cal I} + C_{10}^{\cal I
    \cal R} + C_{10}^{\cal I \cal I} $).
\label{table:isocr}}
\end{table}

%% file: special_CDI_cases.tex
\subsection{Special CDI cases}
\label{sec:specialisoc}

\begin{figure}
\includegraphics{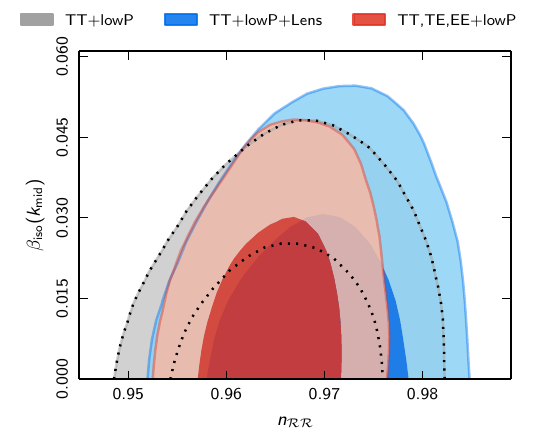}
\caption{ Uncorrelated ADI+CDI with $n_{\cal I \cal I} = 1$ (``axion'').}
\label{fig:JVaxion}
\end{figure}

Next we study three one-parameter CDI extensions to the adiabatic model. In all these extensions the isocurvature 
mode modifies only the largest angular scales, since we either fix $n_{\cal I \cal I}$ to unity (``axion'') or to 
the adiabatic spectral index (``curvaton I/II''). As can be seen from Fig.~\ref{fig:JVmodecomparison}, the 
polarization $E$ mode at multipoles $\ell \gtrsim 200$ will not be significantly affected by this type of CDI 
mode. Therefore, these models are much less sensitive to residual systematic effects in the high-$\ell$ 
polarization data than the generally correlated models.

\subsubsection{Uncorrelated ADI+CDI (``Axion'')}

We start with an uncorrelated mixture of adiabatic and CDI modes ($\mathcal{P}_{\cal R\cal I}=0$), and make the 
additional assumption that $\mathcal{P}_{\cal I\cal I}^{(2)} = \mathcal{P}_{\cal I\cal I}^{(1)}$, i.e., we assume 
unit isocurvature spectral index, $n_{\cal I \cal I} = 1$. Constraints in the ($n_{\cal R \cal 
R}$,$\,\beta_{\mathrm{iso}}$) plane are presented in Fig.~\ref{fig:JVaxion}. This model is the only case for which 
our new results do not improve over bounds from \citetalias{planck2013-p17}. At 
$k_\mathrm{mid}=0.050\,\mathrm{Mpc}^{-1}$, we find $\beta_{\mathrm{iso}}<0.038$ (95\,\% CL, TT,TE,EE+lowP; see 
Table~\ref{tab:how_much_ic}), compared with $\beta_{\mathrm{iso}}< 0.039$ using \Planck\ 2013 and low-$\ell$ WMAP 
data. This is not surprising, since fixing $n_{\cal I \cal I}$ to unity implies that bounds are dominated by 
measurements on very large angular scales, $\ell\lesssim30$, as can easily be understood from 
Fig.~\ref{fig:JVmodecomparison}. Hence the results are insensitive to the addition of better high-$\ell$ 
temperature data, or new high-$\ell$ polarization data.

We summarized in \citetalias{planck2013-p17} why an uncorrelated CDI mode with $n_{\cal I \cal I} \approx 1$ can 
be produced in axion models, under a number of restrictive assumptions: the Peccei-Quinn symmetry should be broken 
before inflation; it should not be restored by quantum fluctuations of the inflaton or by thermal fluctuations 
when the Universe reheats; and axions produced through the misalignement angle should contribute to a sizable 
fraction (or all) of the dark matter. Under all of these assumptions, limits on $\beta_{\mathrm{iso}}$ can be used 
to infer a bound on the energy scale of inflation, using equation~(73) of \citetalias{planck2013-p17}. This bound is 
strongest when all the dark matter is assumed to be in the form of axions. In that case, the limit on 
$\beta_{\mathrm{iso}}(k_\mathrm{mid})$ for \Planck\ TT,TE,EE+lowP gives
\begin{equation}
H_\mathrm{inf} < 0.86 \times 10^7\,\mathrm{GeV}
\left( \frac{f_a}{10^{11}\,\mathrm{GeV}} \right)^{0.408}
\quad \mbox{ (95\,\% CL)}\,,
\end{equation}
where $H_\mathrm{inf}$ is the expansion rate at Hubble radius exit of the scale corresponding to 
$k_\mathrm{mid}=0.050\,\mathrm{Mpc}^{-1}$ and $f_a$ is the Peccei-Quinn symmetry breaking energy scale.

\subsubsection{Fully correlated ADI+CDI (``Curvaton I'')}

Another interesting special case of mixed adiabatic and CDI (or BDI) perturbations is a model where these 
perturbations are primordially fully correlated and their power spectra have the same shape. These cases are 
obtained by setting $\mathcal{P}_{\cal R\cal I}^{(1)} = ( \mathcal{P}_{\cal R\cal R}^{(1)} \mathcal{P}_{\cal I\cal 
I}^{(1)} )^{1/2}$, which, by condition~(\ref{eq:Pab2}), implies that the corresponding statement holds at scale 
$k_2$ and indeed at any scale. In addition, we set $\mathcal{P}_{\cal I\cal I}^{(2)} = (\mathcal{P}_{\cal R\cal 
R}^{(2)} / \mathcal{P}_{\cal R \cal R}^{(1)}) \mathcal{P}_{\cal I\cal I}^{(1)}$, i.e., $n_{\cal I \cal I} = 
n_{\cal R \cal R}$. From this it follows that $\beta_{\mathrm{iso}}$ is scale independent. Therefore, this model 
has only one primary non-adiabaticity parameter, $\mathcal{P}_{\cal I\cal I}^{(1)}$.

A physically motivated example of this type of model is the curvaton model 
\citep{Mollerach:1989hu,Linde:1996gt,Enqvist:2001zp,Moroi:2001ct,Lyth:2001nq,Lyth:2002my} with the following assumptions. (1) 
The average curvaton field value, $\bar\chi_\ast$, is sufficiently below the Planck mass when 
cosmologically interesting scales exit the Hubble radius during
inflation. (2) At Hubble radius exit, the curvature perturbation from the 
inflaton is negligible compared to the perturbation caused by the curvaton. (3) The same is true 
for any inflaton decay products after reheating. This means that, after reheating, the Universe is homogeneous, 
except for the spatially varying entropy (i.e., isocurvature perturbation) due to the curvaton field 
perturbations. (4) Later, CDM is created from the curvaton decay and baryon number after curvaton decay. This 
corresponds to case 4 presented in \citet{Gordon:2002gv}. (5) The curvaton contributes a significant amount to the energy density of the 
Universe at the time of the curvaton's decay to CDM, i.e., the curvaton decays late enough. (6) The energy density of 
curvaton particles possibly produced during reheating should be sufficiently low 
\citep{Bartolo:2002vf,Linde:2005yw}. (7) The small-scale variance of curvaton perturbations, $\Delta_s^2 = \langle 
\delta\chi^2 \rangle_s / \bar\chi^2$, is negligible, so that it does not significantly contribute to the average energy density 
on CMB scales; see equation~(102) in \citet{Sasaki:2006kq}. The last two conditions are necessary in order to have an almost-Gaussian curvature 
perturbation, as required by the \Planck\ observations. Namely, if they are not valid, a large $f_{\mathrm{NL}}^\mathrm{local}$ follows, as discussed below.
Indeed, the conditions (6) and (7) are related, since curvaton particles would add a homogeneous component to
the average energy density on large scales, and hence we can describe their effect by $\Delta_s^2 = \rho_{\chi,\,\mathrm{particles}} / \rho_{\bar\chi,\,\mathrm{field}}$, where
$\rho_{\bar\chi,\,\mathrm{field}}$ is the average energy density of the classical curvaton field on large scales; see  equation~(98) in \citet{Sasaki:2006kq}.
Then the total energy density carried by the curvaton will be $\bar\rho_\chi = \rho_{\bar\chi,\,\mathrm{field}} + \rho_{\chi,\,\mathrm{particles}}$.

The amount of isocurvature and non-Gaussianity present after curvaton decay depends on the ``curvaton decay fraction'',
\begin{equation}
r_D = \frac{3\bar\rho_\chi}{3\bar\rho_\chi + 4\bar\rho_{\mathrm{radiation}}},
\label{eq:curvatondecayfraction}
\end{equation}
evaluated at curvaton decay time. If conditions (6) and (7) do not hold, then the isocurvature perturbation
disappears.\footnote{Indeed, if curvaton particles are produced during reheating, 
they can be expected to survive and outweigh other particles
at the moment of curvaton decay, but by how much depends on the details of the model.
As the curvaton field (during its oscillations) and the curvaton particles have the same equation of
state and they decay simultaneously, no isocurvature perturbations are produced.}

\begin{figure}
\includegraphics{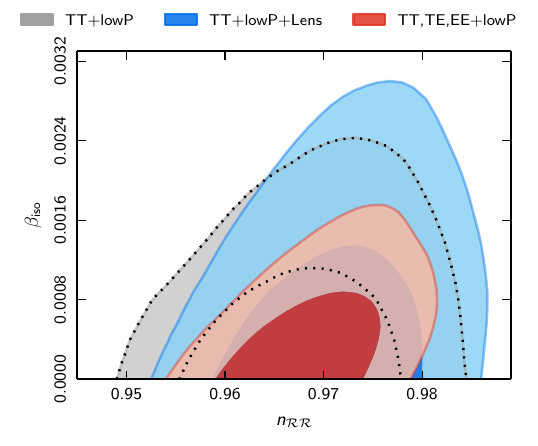}
\caption{ Fully correlated ADI+CDI with $n_{\cal I \cal I} = n_{\cal R \cal R}$ (``curvaton I''). Since the spectral 
indices are equal, the primordial isocurvature fraction $\beta_{\mathrm{iso}}$ is scale independent.
\label{fig:JVcurvaton}
}
\end{figure}

\begin{figure}
\includegraphics{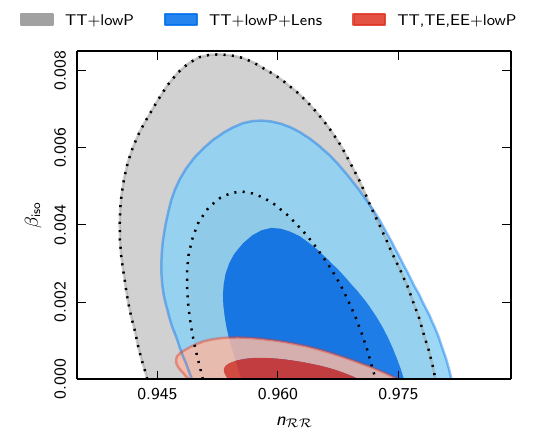}
\caption{ Fully anticorrelated ADI+CDI with $n_{\cal I \cal I} = n_{\cal R \cal R}$ (``curvaton II'').
\label{fig:JVanticorr}
}
\end{figure}

The curvaton scenario presented here is one of the simplest to test against observations. It should be noted that 
at least the conditions (1)--(5) listed at the beginning of this subsection should be satisfied simultaneously. 
Indeed, if we relax some of these conditions, almost any type of correlation can be produced. For example, the 
relative correlation fraction can be written as
$\cos\Delta = \sqrt{\lambda / (1+\lambda)}$,
where $\lambda = (8/9)r_D^2 \epsilon_\ast (M_\mathrm{Pl} / 
\bar\chi_\ast)^2$. Therefore, the model is fully correlated only if $\lambda \gg 1$. If the slow-roll parameter 
$\epsilon_\ast$ is very close to zero or the curvaton field value $\bar\chi_\ast$ is large compared to the Planck 
mass, this model leads to almost uncorrelated perturbations.

As seen in Fig.~\ref{fig:JVcurvaton} and Table~\ref{tab:how_much_ic}, the upper bound on the primordial 
isocurvature fraction in the fully correlated ADI+CDI model weakens slightly when we add the \Planck\ lensing data 
to \Planck\ TT+lowP, whereas adding high-$\ell$ TE,EE tightens the upper bound moderately. With all of these three 
data combinations, the pure adiabatic model gives an equally good best-fit $\chi^2$ as the fully-correlated 
ADI+CDI model. Bayesian model comparison strengthens the conclusion that the data disfavour this model with 
respect to the pure adiabatic model.

The isocurvature fraction is connected to the curvaton decay fraction, Eq.~(\ref{eq:curvatondecayfraction}), by
\begin{equation}
  \beta_{\mathrm{iso}} \approx \frac{9(1-r_D)^2}{r_D^2 + 9(1-r_D)^2}
\label{eq:betaisocurvatonI}
\end{equation}
\citep[see case 4 in][]{Gordon:2002gv}.
Reading the constraints on $\beta_{\mathrm{iso}}$ from Table~\ref{tab:how_much_ic}, we can convert them into 
constraints on $r_D$ and further into the non-Gaussianity parameter assuming a quadratic potential for the 
curvaton and instantaneous decay\footnote{It should be noted that, in particular, in the older curvaton literature
a formula $f_{\mathrm{NL}}^\mathrm{local} = \frac{5}{4 r_D}$ is often quoted or utilized. This result, which follows from considering only squares of first order perturbations, is valid when $r_D$ is close to zero, i.e., when $f_{\mathrm{NL}}^\mathrm{local}$ is very large. However, when $r_D$ is close to unity or $f_{\mathrm{NL}}^\mathrm{local} \lesssim 10$, which is the case with the \Planck\ measurements, the second and third terms in Eq.~(\ref{eq:fnlcurvaton}) are vitally important. These follow from genuine second order perturbation theory calculations. Coincidentally, if one erroneously uses the expression
$\frac{5}{4 r_D}$ in the limit $r_D \rightarrow 1$, one obtains a result $+5/4$, whereas the correct formula (\ref{eq:fnlcurvaton}), with $\Delta_s^2=0$, leads to $-5/4$, when $r_D \rightarrow 1$.}
\citep{Sasaki:2006kq}:
\begin{equation}
f_{\mathrm{NL}}^\mathrm{local}=(1+\Delta_s^2)\frac{5}{4 r_D} - \frac{5}{3} - \frac{5 r_D}{6}.
\label{eq:fnlcurvaton}
\end{equation}
If conditions (6) and (7) hold, i.e., $\Delta_s^2 = 0$, as implicitly assumed, e.g., in \citet{Bartolo:2004ty,Bartolo:2003jx}, then the smallest possible value of 
$f_{\mathrm{NL}}^\mathrm{local}$ is $-5/4$, which is obtained when $r_D = 1$, and
Eqs.~(\ref{eq:betaisocurvatonI}) and (\ref{eq:fnlcurvaton}) yield for the various 
\Planck\ data sets (at 95\,\% CL):\footnote{However, if  $\Delta_s^2$ was non-negligible, then all the 
constraints on $f_{\mathrm{NL}}^\mathrm{local}$ would shift upward. For example, with $\Delta_s^2 = 1$, 
our constraints on  $\beta_{\mathrm{iso}}$ would translate to $0 \le f_{\mathrm{NL}}^\mathrm{local} \lesssim 0.03$.
On the other hand, the \Planck\ constraint of $f_{\mathrm{NL}}^\mathrm{local}$ can be converted to an upper bound
 $\Delta_s^2 = \rho_{\chi,\,\mathrm{particles}}  / \rho_{\bar\chi,\,\mathrm{field}} <  8.5$ (95\,\% CL from $T$ only) as shown in \citet{planck2014-a19}.}
\begin{eqnarray}
\mbox{TT+lowP: }   & \beta_{\mathrm{iso}} < 0.0018 \Rightarrow 0.9860 < r_D \le 1  \nonumber        \\    & \Rightarrow -1.250 \le f_{\mathrm{NL}}^{\mathrm{local}} < -1.220\,,    \\
\mbox{TT+lowP+lensing: } &  \beta_{\mathrm{iso}} < 0.0022  \Rightarrow  0.9845 < r_D \le 1 \nonumber\\    & \Rightarrow -1.250 \le f_{\mathrm{NL}}^{\mathrm{local}} < -1.217\,,     \\
\mbox{TT,TE,EE+lowP: }   & \beta_{\mathrm{iso}} < 0.0013 \Rightarrow 0.9882 < r_D \le 1 \nonumber   \\    & \Rightarrow -1.250 \le f_{\mathrm{NL}}^{\mathrm{local}} < -1.225\,.
\end{eqnarray}
Thus the results for the simplest curvaton model remain unchanged from those presented 
in \citetalias{planck2013-p17}, i.e., in order to produce almost purely adiabatic perturbations, the 
curvaton should decay when it dominates the energy density of the Universe ($r_D > 0.98$), and the 
non-Gaussianity parameter is constrained to close to its smallest possible value 
($-5/4 < f_{\mathrm{NL}}^{\mathrm{local}}< -1.21$), which is consistent with the result 
$f_\mathrm{NL}^\mathrm{local}=2.5 \pm 5.7$ (68\,\% CL, from $T$ only) found in \citet{planck2014-a19}.

\subsubsection{Fully anticorrelated ADI+CDI (``Curvaton II'')}

The curvaton scenario or some other mechanism could also produce 100\,\% anticorrelated perturbations, with 
$n_{\cal I \cal I} = n_{\cal R \cal R}$. The constraints in the ($n_{\cal R \cal R},\beta_{\mathrm{iso}}$) plane 
are presented in Fig.~\ref{fig:JVanticorr}.  Examples of this kind of model are provided by cases 2, 3, and 6 in 
\citet{Gordon:2002gv}. These lead to a fixed, large amount of isocurvature, e.g., in case 2 to 
$\beta_{\mathrm{iso}} = 9/10$, and are hence excluded by the data at very high significance. However, case 9 in 
\citet{Gordon:2002gv}, with a suitable $r_D$ (i.e., $r_D > R_\mathrm{c}$, where $R_\mathrm{c} = \rho_\mathrm{c} / 
(\rho_\mathrm{c}+\rho_\mathrm{b})$), leads to fully anticorrelated perturbations and might provide a good fit to 
the data. In this case CDM is produced by curvaton decay while baryons are created earlier from inflaton decay products
 and do not carry a curvature perturbation. We obtain a very 
similar expression to Eq.~(\ref{eq:betaisocurvatonI}), namely
\begin{equation}
\beta_{\mathrm{iso}} \approx \frac{9(1-r_D/R_\mathrm{c})^2}{r_D^2 + 9(1-r_D/R_\mathrm{c})^2}\,.
\label{eq:betaisocurvatonII} 
\end{equation}
We convert this approximately to a constraint on $r_D$ by fixing $R_\mathrm{c}$ to its best-fit value, $R_\mathrm{c} = 0.8437$ (\Planck\ TT+lowP), within this model. The results for the various \Planck\ data sets are:
\begin{eqnarray}
\mbox{TT+lowP: }   & \beta_{\mathrm{iso}} < 0.0064  \Rightarrow  0.8437 < r_D < 0.8632  \nonumber        \\    & \Rightarrow  -0.9379 < f_{\mathrm{NL}}^{\mathrm{local}} < -0.8882\,,    \\
\mbox{TT+lowP+lensing: } &  \beta_{\mathrm{iso}} < 0.0052  \Rightarrow   0.8437< r_D < 0.8612 \nonumber   \\ & \Rightarrow  -0.9329 < f_{\mathrm{NL}}^{\mathrm{local}}< -0.8882\,,     \\
\mbox{TT,TE,EE+lowP: }   & \beta_{\mathrm{iso}} < 0.0008 \Rightarrow  0.8437 < r_D < 0.8505 \nonumber     \\    & \Rightarrow -0.9056 < f_{\mathrm{NL}}^{\mathrm{local}} < -0.8882\,. 
\end{eqnarray}

After all the tests conducted in this section, both for the generally correlated CDI, NDI, and NVI cases as well 
as for the special CDI cases, we conclude that, within the spatially flat base $\Lambda$CDM model, the initial 
conditions of perturbations are consistent with the hypothesis of pure adiabaticity, a conclusion that is also 
overwhelmingly supported by the Bayesian model comparison.  Moreover, \citet{planck2014-a19} report a null 
detection of \emph{isocurvature non-Gaussianity}, with polarization improving constraints significantly.

%% file: Tables/table_isocurvature.tex
\newcommand{\pz}{\phantom{0}}
\newcommand{\pzz}{\phantom{00}}
\newcommand{\pd}{\phantom{-}}

\begin{table*}
\begin{center}
\footnotesize
\begin{tabular}{lcccccccc}
\hline
\hline
Model (and data)\phantom{$\Biggr( $} &
$100\beta_\mathrm{iso}(k_{\mathrm{low}})$ &
$100\beta_\mathrm{iso}(k_{\mathrm{mid}})$ &
$100\beta_\mathrm{iso}(k_{\mathrm{high}})$ &
$100\cos\Delta$ &
$100\alpha _{\cal R\cal R}{(2,2500)}$ &
$\Delta n$ &
$\Delta\chi^2$&
$\ln B$ \\
\noalign{\hrule\vskip 3pt}
General models:
& & & & & & & \cr
$\quad $ CDI (TT+lowP)
& \pz4.1\pz & 35.4 & 56.9&[$-30$:20]
&[98.1:101.5] 
& 3 & $-2.1$ & $-8.8$ \cr
$\quad $ CDI (TT+lowP+WP)
& \pz4.2\pz & 35.5 & 57.2&[$-31$:23]
&[97.9:101.4] 
& 3 & $-1.8$  & $-9.1$\cr
$\quad $ CDI (TT,TE,EE+lowP)
& \pz2.0\pz & [3.4:28.1]& [3.1:51.8] & [\pz$-6$:20]
& [98.5:\pz99.9] 
&3 & $-5.3$  & $-8.8$\cr
$\quad $ CDI (TT,TE,EE+lowP+WP)
& \pz2.1\pz & [2.3:28.4]& [2.6:52.1] & [\pz$-7$:21]
& [98.5:\pz99.9] 
&3 & $-5.5$  & $-8.2$\cr
$\quad $ CDI (TT+lowP+lensing)
 &\pz4.5\pz &37.9& 59.4& [$-28$:17]
& [98.1:101.1]
& 3&$-1.2$ & $-8.8$ \cr
\noalign{\hrule\vskip 2pt}
$\quad $ NDI (TT+lowP) &14.3\pz &22.4 &  27.4& [$-33$:\pz1]
&[98.6:104.0]
 & 3&$-2.0$ & $-5.3$ \cr
$\quad $ NDI (TT,TE,EE+lowP)
& \pz7.3\pz  &  [3.4:19.3]& [3.5:26.7] & [\pz$-9$:10]
& [97.8:100.1] 
& 3 & $-5.5$ & $-5.5$\cr
$\quad $ NDI (TT+lowP+lensing) &15.8\pz &[1.4:24.1] &[0.3:28.4] &[$-32$:\pz0]
 &[98.6:104.0]
  &3 &$-2.8$ & $-4.6$ \cr
\noalign{\hrule\vskip 2pt}
$\quad $ NVI (TT+lowP) &  \pz8.3\pz &[0.1:10.2]& 11.9 & [$-26$:$\phantom{-}6$]
 & [97.6:102.3]
 &3 &$-2.8$ & $-6.3$ \cr
$\quad $ NVI (TT,TE,EE+lowP)
& \pz7.4\pz &[0.9:\pz7.4]& [0.4:\pz8.8]&[$-22$:$-4$]
& [99.2:102.0] 
& 3 & $-6.2$ & $-6.5$\cr
$\quad $ NVI (TT+lowP+lensing) &\pz9.7\pz & [0.4:11.6]&13.1 &[$-23$:$\phantom{-}7$]
 &[97.1:102.0]
  & 3 &  $-2.5$ & $-6.5$ \cr
\noalign{\hrule\vskip 2pt}
General models + $r$:
& & & & & & & \cr
$\quad $ CDI+r=0.1 (TT+lowP) & \pz3.4\pz & 38.7&63.9 & [$-33$:24]
& [98.1:101.4]
& 3 & $-5.4$ & $-8.9$ \cr
$\quad $ CDI+r=0.1 (TT,TE,EE+lowP) & \pz1.6\pz &[4.4:31.7] &[6.9:59.2] &[\pz$-6$:22]
 &[98.6:\pz99.9]
  & 3 & $-6.3$ & $-8.1$\cr
$\quad $ CDI+r (TT+lowP) & \pz4.3\pz & 34.9 &  56.2 &[$-43$:20]
&  [97.9:102.4]
  & 3 & $-3.3$ & $-7.7$ \cr
$\quad $ CDI+r (TT,TE,EE+lowP) & \pz1.7\pz & [3.9:29.0] & [5.8:53.8]& [\pz$-5$:21]
&[98.6:\pz99.9]
 & 3 & $-5.1$ & $-7.2$\cr
\noalign{\hrule\vskip 2pt}
Special CDI cases:
& & & & & & &  \cr
$\quad $ Uncorrelated, $n_\mathcal{II}=1$ & & & & & & & \cr
$\quad\quad$ ``axion'' (TT+lowP)
& \pz3.3\pz & 3.7\pz &3.8\pz & $\pd\pzz0$
&[98.5:100]
& 1 & $\pd0.0$ & $-5.2$ \cr
$\quad\quad$ ``axion'' (TT,TE,EE+lowP)
&  \pz3.5\pz &3.8\pz &3.9\pz & $\pd\pzz0$
& [98.4:100]
& 1 & $-0.2$ & $-4.9$ \cr
$\quad\quad$ ``axion'' (TT+lowP+lensing)
& \pz3.9\pz &4.3\pz &4.4\pz & $\pd\pzz0$
& [98.3:100]
& 1 & $\pd0.0$ & $-5.0$ \cr
$\quad $ Fully correlated, $n_\mathcal{II}=n_\mathcal{RR}$  & & & & &
& & \cr
$\quad\quad$ ``curvaton I'' (TT+lowP)
&\pz0.18 & 0.18 &0.18 & $\pd100$
& [97.5:100.0]
& 1 & $-0.1$ & $-8.1$ \cr
$\quad\quad$ ``curvaton I'' (TT,TE,EE+lowP)
& \pz0.13& 0.13 & 0.13 & $\pd100$
& [97.8:\pz99.9]
& 1 & $\pd0.0$ & $-7.8$\cr
$\quad\quad$ ``curvaton I'' (TT+lowP+lensing)
&\pz0.22& 0.22 & 0.22 & $\pd100$
& [97.3:\pz99.7]
& 1 & $\pd0.0$ & $-8.5$ \cr
$\quad $ Fully anti-correlated, $n_\mathcal{II}=n_\mathcal{RR}$ & & &
& & & & \cr
$\quad\quad$ ``curvaton II'' (TT+lowP)
& \pz0.64 &  0.64 &  0.64 & $-100$
& [100.5:105.1]
& 1 & $-1.1$ & $-5.4$ \cr
$\quad\quad$  ``curvaton II'' (TT,TE,EE+lowP)
& \pz0.08 & 0.08 & 0.08 & $-100$
& [100.1:101.8]
& 1 & $\pd0.0$ & $-8.9$ \cr
$\quad\quad$  ``curvaton II'' (TT+lowP+lensing)
&\pz0.52 & 0.52 &0.52 & $-100$
& [100.4:104.4]
& 1 & $-0.6$ & $-6.3$ \cr
\noalign{\vskip 2pt\hrule}
\end{tabular}
\end{center}
\caption{\label{tab:how_much_ic} Constraints on mixed adiabatic
  and isocurvature models. For each mixed model,
we report 95\% CL bounds on the fractional primordial contribution of
isocurvature modes at three comoving wavenumbers ($k_{\mathrm{low}}= 0.002~$Mpc$^{-1}$,
$k_{\mathrm{mid}}= 0.050~$Mpc$^{-1} ,$ and $k_{\mathrm{high}}=
0.100~$Mpc$^{-1}$), as well as the scale-independent primordial
correlation fraction, $\cos\Delta$.
The fractional adiabatic contribution to the observed temperature
variance is denoted by $\alpha_{\cal R \cal R}(2,2500)$, and from this
the nonadiabatic contribution can be calculated as
$\alpha_{\text{non-adi}} = 1 - \alpha_{\cal R \cal R}(2,2500)$. The number of
extra parameters compared with the
corresponding pure adiabatic model is denoted by $\Delta n$, and
$\Delta \chi^2$ is the difference between the $\chi^2$ of the
best-fitting mixed and pure adiabatic models. (A negative $\Delta \chi^2$
means that the mixed model is a better fit to the data.) In the last
column we give the difference between the logarithm of Bayesian
evidences. (A negative $\ln B = \ln(P_{\mathrm{ISO}}/P_{\mathrm{ADI}})$
means that Bayesian model comparison disfavours the mixed model.
With our settings of {\tt MultiNest} the uncertainty in these numbers is about $\pm$0.5.)
}
\end{table*}

%% file: section_twelve.tex

\def\gtorder{\mathrel{\raise.3ex\hbox{$>$}\mkern-14mu
             \lower0.6ex\hbox{$\sim$}}}
\def\ltorder{\mathrel{\raise.3ex\hbox{$<$}\mkern-14mu
             \lower0.6ex\hbox{$\sim$}}}

\noindent

A key prediction of standard inflation, which in the present context includes
all single field models of inflation as well as many multi-field models, is that
the stochastic process generating the primordial cosmological
perturbations is completely characterized
by its power spectrum, constrained by statistical isotropy to depend only
on the multipole number $\ell .$  This statement applies at least to the accuracy
that can be probed using the CMB given the limitations imposed by cosmic
variance, since all models exhibit {\em some} level of non-Gaussianity.
Nevertheless, more general
Gaussian stochastic processes can be envisaged for which one or more special
directions on the sky are singled out, so that the expectation values for the
temperature multipoles take the form
\begin{equation}
\left< a_{\ell m}^T ~\left({a_{\ell'm'}^T}\right)^* \right> =C_{\ell m; \ell'm'}^{TT},
\label{eq:gencovariance}
\end{equation}
rather than the very special form
\begin{equation}
\left< a_{\ell m}^T ~\left(a_{\ell'm'}^T\right)^* \right>
=C_\ell ^{TT}\, \delta _{\ell , \ell '}\, \delta _{m, m'},
\end{equation}
which is the only possibility consistent with statistical isotropy.

The most general form for a Gaussian stochastic process on the sphere violating
the hypothesis of statistical isotropy in Eq.~(\ref{eq:gencovariance}) is too broad
to be useful, given that we have only one sky to analyse. For $\ell < \ell _{\mathrm{max}},$
there are $O(\ell _{\mathrm{max}}^2)$ multipole expansion coefficients, 
compared with $O(\ell _{\mathrm{max}}^4)$
model parameters. Therefore, in order to make some progress on testing the hypothesis
of statistical isotropy, we must restrict ourselves to examining only the simplest
models violating statistical isotropy, for which the available data can establish 
meaningful constraints and for
which one can hope to find a simple theoretical motivation.

\subsection{Asymmetry: observations versus model building}

In one simple class of statistically anisotropic models, we start with
a map produced by a process respecting statistical isotropy, which becomes modulated by another
field in the following manner to produce the observed sky map:
\begin{equation}
\delta T_{\mathrm{sky}}(\vec{\hat{\Omega}})
   = \Bigl( 1 + M(\vec{\hat{\Omega}})\Bigr)
     ~\delta T_{\text{s-i}}(\vec{\hat{\Omega}}),
\label{ModulationAnsatz}
\end{equation}
where $\vec{\hat{\Omega}}$ denotes a position on the celestial sphere and
$\delta T_{\text{s-i}}(\vec{\hat{\Omega}})$ is the outcome of the 
underlying statistically isotropic process before
modulation. Roughly speaking, where the modulating field $M(\vec{\hat{\Omega}})$ is positive,
power on scales smaller than the scale of variation of $M(\vec{\hat{\Omega}})$ is
enhanced, whereas where $M(\vec{\hat{\Omega}})$ is negative, power is suppressed.
We refer to this as a ``power asymmetry.''  If $M(\vec{\hat{\Omega}})
= A\vec{\hat{d}}\cdot\vec{\hat{\Omega}}$, we have a model of dipolar
modulation with amplitude $A$ and direction $\vec{\hat{d}}$, but higher-order or mixed
modulation may also be considered, such as a quadrupole modulation or modulation by a 
scale-invariant field
$M(\vec{\hat{\Omega}})$, to name just a few special cases.
Alternatively, and more closely tied to physical models, we can consider 
modulations of the position- or $k$-space fluctuations.

In \citet{planck2013-p09} and \citet{planck2014-a18}, the details of constructing
efficient estimators for statistical anisotropy, in particular in the presence
of realistic data involving sky cuts and possibly incompletely removed foreground
contamination, are considered in depth.  In addition, the question of the
statistical significance of any detected ``anomalies'' from the expectations of
base $\Lambda$CDM is examined in detail.  Importantly, in the {\em absence} of a particular
inflationary model for such an observed anomaly, the significance should be corrected
for the ``multiplicity of tests'' that {\em could} have resulted in 
similarly-significant detections (i.e., for the ``look elsewhere effect''), although applying
such corrections can be ambiguous.  In this paper, however, we consider only forms
of statistical anisotropy that are predicted by specific inflationary models, and
hence such corrections will not be necessary.

   Several important questions can be posed regarding the link between statistical isotropy
and inflation.  In particular, we can ask the following questions.  (1) Does 
a statistically significant finding
of a violation of statistical isotropy falsify inflation? (2) If not, what sort of non-standard
inflation could produce the required departure from statistical isotropy? (3) What other perhaps
non-inflationary models could also account for the violation of statistical isotropy?
In this section, we begin to address these questions by assessing the viability of an
inflationary model for dipolar asymmetry, as well as by placing new limits on the
presence of quadrupolar power asymmetry.

   For the case of the observed dipolar asymmetry examined in detail in \citet{planck2014-a18}, there
are two aspects that make inflationary model building difficult.  First is the problem of
obtaining a significant amplitude of dipole modulation.  In \citet{planck2014-a18}
the asymmetry was found to have amplitude $A \approx 6$--7\,\% on scales
$2 \le \ell \le 64$.  This compares with the expected value of $A = 2.9\,\%$ on
these scales due to cosmic variance in statistically isotropic skies.  One basic
strategy for incorporating the violation of statistical isotropy into inflation is to
consider some form of multi-field inflation and use one of the directions orthogonal to
the direction of slow roll as the field responsible for the modulation. Obtaining
the required modulation is problematic because most extra fields in multi-field
inflation become disordered in a nearly scale-invariant
way, just like the fluctuations in the field parallel to the direction of slow roll. What is
needed resembles a pure gradient with no fluctuations of shorter wavelength. In
\citet{liddle2013cosmic} it was
proposed that such a field could be produced using the supercurvature mode of open inflation.
(See, however, the discussion in \citet{kanno2013viable}.)  Also,
in order to respect the $f_{\mathrm{NL}}$ constraints, one must avoid that the
modulating field leave a direct imprint on the temperature anisotropy.

   The second aspect which makes model building difficult for dipolar asymmetry is
that the measured amplitude is strongly scale dependent, and on scales
$\ell \gsim 100$ no significant detection of a dipolar modulation amplitude is
made \citep{planck2014-a18}, once our proper motion has been taken into account
\citep{planck2013-pipaberration}.  On the other hand, the simplest models are
scale-free and produce statistical anisotropy of the type described by the ansatz
in Eq.~(\ref{ModulationAnsatz}), for which the bulk of the statistical weight
should be detected at the resolution of the survey.  To resolve this difficulty,
\citet{ehk09} proposed modulating CDI fluctuations generated within the
framework of a curvaton scenario, because, unlike adiabatic
perturbations, CDI perturbations entering the Hubble radius before last scattering
contribute negligibly to the CMB fluctuations (recall Fig.~\ref{fig:JVmodecomparison}).

   The situation for the quadrupolar power asymmetry is different from the dipolar
case in that no detection is currently claimed.  Model building is easier than
the dipolar case since no pure gradient modes are required, but also more
difficult in that anisotropy during inflation is needed.
While the isotropy of the recent expansion of the Universe (i.e., since the CMB fluctuations
were first imprinted) is tightly constrained, bounds on a possible anisotropic expansion
at early times are much weaker.
\citet{Ackerman:2007} proposed
using constraints on the quadrupolar statistical anisotropy of the CMB to probe the isotropy
of the expansion during inflation---that is, during the epoch when the perturbations
now seen in the CMB first exited the Hubble radius.
Assuming an anisotropic expansion during inflation, \citet{Ackerman:2007} computed its
impact on the three-dimensional power spectrum on super-Hubble scales
by integrating the mode functions for the perturbations
during inflation and beyond. Several sources of such anisotropy have been proposed, such as vector fields
during inflation
\citep{Dimastrogiovanni:2010,Soda:2012,Maleknejad:2012,Schmidt:2012,Bartolo:2013,Naruko2014},
or an inflating solid or elastic medium \citep{Bartolo:2013}.

\subsection{Scale-dependent modulation and idealized estimators}

The ansatz in Eq.~(\ref{ModulationAnsatz}) expressed in angular space
may be rewritten in terms of the multipole expansion and generalized to
include scale-dependent modulation by means of Wigner $3j$
symbols:
\begin{equation}
\left< a_{\ell m}^T~a_{\ell'm'}^T \right> = \sum _{L=0}^\infty \sum _{M=-L}^{+L}
C_{\ell; \ell'; L, M}^{TT}
\left(
\begin{matrix}
\ell & \ell' & L\cr
m    & m'    & M
\end{matrix}
\right) .\label{eq:gencovar}
\end{equation}
Because of the symmetry of the left-hand side, the coefficients $C_{\ell; \ell'; L, M}^{TT}$
acquire a phase $(-1)^{\ell+\ell'+L}$ under interchange of $\ell$ and $\ell'.$
This is the most general form consistent with the hypothesis of Gaussianity.
The usual isotropic power spectrum, which is the generic prediction
of simple models of inflation, includes only the $L = 0$ term, where
$C_{\ell; \ell'; 0, 0}^{TT}=C_\ell^{TT}$ and the Wigner $3j$ symbol provides the
$\delta_{\ell,\ell'}\delta_{m,m'}$ factor. The coefficients $C_{\ell; \ell'; L, M}^{TT}$ 
with $L>0$ introduce statistical anisotropy.

If we assume that there is a common vector (corresponding to $L=1$
on the celestial sphere) that defines the direction of the anisotropy of the
power spectrum for all the terms of $L=1$, we may adopt a more restricted
ansatz for the bipolar modulation, so that
\begin{equation}
C_{\ell; \ell'; 1, M}^{TT} = C^1_{\ell,\ell'} X_M^{(1)},
\label{eq:dipolarcovar}
\end{equation}
where we assume that $X_M$ is normalized (i.e., $\sum _M{X_M}{X_M}^*=1)$. In such a model, supposing that
$C^1_{\ell,\ell'}$ is theoretically determined, but the orientation of the unit vector
$X_M$ is random and isotropically distributed on the celestial sphere, we may
construct the following quadratic estimator for the direction:
\begin{equation}
\begin{aligned}
X_M^{(L)} =&
\sum_{\ell,m}\sum_{\ell',m'}
\frac{w_{\ell,\ell';L}}{(2L+1)(C_\ell)^{1/2}~(C_{\ell'})^{1/2}}\\
\times&
\left(
\begin{matrix}
\ell & \ell' & L\cr
m    & m'    & M
\end{matrix}
\right) ~~
a_{\ell m}^T~a_{\ell 'm'}^T,
\end{aligned}
\end{equation}
where the weights for the unbiased minimum variance estimator are given by
\begin{equation}
w_{\ell , \ell';L}= C_{\ell,\ell'}^L \left(
\sum _{\ell,\ell'}
C_{\ell,\ell'}^L \right) ^{-1}.
\end{equation}
This construction, which for the $L=1$ case may be found in
\citet{mszb11} and \citet{planck2014-a18},
may be readily generalized to $L>1$ in the above way.

\subsection{Constraining inflationary models for dipolar asymmetry}

In this section, we confront with \Planck\ data the modulated curvaton
model of \citet{ehk09}, which attempts to explain the observed large-scale
power asymmetry via a gradient in the background curvaton field.  In this model,
the curvaton decays after CDM freeze out, which results in a
nearly-scale-invariant isocurvature component between CDM and radiation.
In the viable version of this scenario, the curvaton contributes
negligibly to the CDM density.  A long-wavelength fluctuation in the
curvaton field initial value $\sigma_*$ is assumed, with amplitude
$\Delta\sigma_*$ across our observable volume.  This modulates the curvaton
isocurvature fluctuations according to $S_{\sigma\gamma} \approx
2\delta\sigma_*/\sigma_*$.  The curvaton produces all of the final
CDI fluctuations, which are nearly scale invariant, as well as a component of the
final adiabatic fluctuations.  Hence both of these components will be modulated,
and the parameter space of the model will be constrained by observations of
the power asymmetry on large and small scales, as well as the full-sky
CDI fraction.  In practice, the very tight constraints on
small scale power asymmetry obtained in \citet{planck2014-a18} imply a small
curvaton adiabatic component, which implies that the CDI and adiabatic
fluctuations are only weakly correlated.  This model easily satisfies
constraints due to the CMB dipole, quadrupole, and non-Gaussianity
\citep{ehk09}.

There are two main parameters that we constrain for this model.
First, the fraction of adiabatic fluctuations due to the curvaton $\xi$
is defined as
\begin{equation}
\xi \equiv \frac{\Sigma_\sigma^2{\cal P}_\sigma}
    {{\cal P}_{{\cal R}_{\mathrm{inf}}} + \Sigma_\sigma^2{\cal P}_\sigma}.
\end{equation}
Here, ${\cal P}_{{\cal R}_{\mathrm{inf}}}$ and ${\cal P}_\sigma$ are the
inflaton and curvaton primordial power spectra, respectively, and $\Sigma_\sigma$
is the coupling from curvaton isocurvature to adiabatic fluctuations.
(Up to a sign, $\xi$ is equal to the correlation parameter.)  Next, the
coupling of curvaton to CDI,
$M_{{\mathrm{CDI}}\sigma}$, is determined by the constant $\kappa \equiv
M_{{\mathrm{CDI}}\sigma}/R \gsim -1$, where
\begin{equation}
R \equiv \frac{3\Omega_\sigma}{4\Omega_\gamma + 3\Omega_\sigma
             + 3\Omega_{\mathrm{CDM}}}
\end{equation}
and all density parameters are evaluated just prior to curvaton decay.
The isocurvature fraction can be written in terms of these two parameters by
\begin{equation}
\beta_{\mathrm{iso}} = \frac{9\kappa^2\xi}{1 + 9\kappa^2\xi}.
\label{alphadef}
\end{equation}
These parameters determine the modulation of the CMB temperature fluctuations
via $\Delta C_\ell/C_\ell = 2K_\ell\Delta\sigma_*/\sigma_*$, where
\citep{ehk09}
\begin{equation}
K_\ell \equiv \xi\frac{C_\ell^\mathrm{ad} + 9\kappa^2C_\ell^\mathrm{iso}
         + 3\kappa C_\ell^\mathrm{cor}}
               {C_\ell^\mathrm{ad} + \xi\left(9\kappa^2C_\ell^\mathrm{iso}
         + 3\kappa C_\ell^\mathrm{cor}\right)}.
\label{Kelldef}
\end{equation}
Here $C_\ell^\mathrm{ad}$, $C_\ell^\mathrm{iso}$, and $C_\ell^\mathrm{cor}$
are the adiabatic, CDI, and correlated power spectra calculated for unity
primordial spectra.

Note that this modulated curvaton model contains some simple special
cases. For $\kappa = 0$, we have a purely adiabatic (i.e., scale-invariant)
modulation. This is equivalent to a modulation of the scalar amplitude, 
$A_\mathrm{s}$. On the other hand, if we take the limit $\kappa \to \infty$,
with fixed $\kappa^2\xi$ (i.e., with fixed isocurvature fraction
$\beta_{\mathrm{iso}}$), we obtain a pure CDI modulation.  For $\kappa =
\xi = 0$ we have no modulation, i.e., we recover base $\Lambda$CDM.  Therefore this
model is particularly useful for examining a range of possible modulations
within the context of a concrete framework.

   In order to constrain this model, we use a formalism which was developed
to determine the signatures of potential gradients in physical parameters in
the CMB \citep{mszb11}, and which is used to examine dipolar modulation and
described in detail in \citet{planck2014-a18}.  This approach is well-suited
to testing the modulated curvaton model since it can accommodate
scale-dependent modulations.  Briefly, we write the temperature anisotropy
covariance given a gradient $\Delta X_M$ in a parameter $X$ as
\begin{eqnarray}
\quad C_{\ell m\ell'm'} &=& C_{\ell}\delta_{\ell\ell'}\delta_{mm'}
   + (-1)^m\frac{\delta C_{\ell\ell'}}{2}
     \left[(2\ell + 1)(2\ell' + 1)\right]^{1/2}\nonumber\\
&\times& \left(\begin{matrix}
\ell & \ell' & 1\cr
0    & 0     & 0
\end{matrix}\right)
\sum_M \Delta X_M
\left(\begin{matrix}
\ell  & \ell' & 1\cr
-m    & m'    & M
\end{matrix}\right),
\end{eqnarray}
where $\delta C_{\ell\ell'} \equiv dC_{\ell}/dX + dC_{\ell'}/dX$.  Note that
this covariance takes the form of Eqs.~(\ref{eq:gencovar}) and
(\ref{eq:dipolarcovar}), with
\begin{equation}
C^1_{\ell;\ell'} = \frac{\delta C_{\ell\ell'}}{2}
     \left[(2\ell + 1)(2\ell' + 1)\right]^{1/2}
\left(\begin{matrix}
\ell & \ell' & 1\cr
0    & 0     & 0
\end{matrix}\right).
\end{equation}

We then construct a maximum likelihood estimator for the gradient
components.  We use $C^{-1}$ filtered data \citep{planck2014-a17} and perform
a mean-field subtraction, giving
\begin{eqnarray}
\quad \Delta\hat X_M
   &=& \frac{3}{f_{1M}}\sum_{\ell m\ell'm'}(-1)^m C^1_{\ell;\ell'}
\left(\begin{matrix}
\ell  & \ell' & 1\cr
-m    & m'    & M
\end{matrix}\right)\nonumber\\
&\times&
\left(T_{\ell m}T^*_{\ell'm'} - \left<T_{\ell m}T^*_{\ell'm'}\right>\right)
\left[ \sum_{\ell\ell'}\left(C^1_{\ell;\ell'}\right)^2 F_\ell F_{\ell'} \right]^{-1}.
\label{Eq:fulldipoleest}
\end{eqnarray}
Here $f_{1M}$ is a normalization correction due to the applied mask,
$M(\Omega)$, and is given by
\begin{equation}
f_{1M} \equiv \int d\Omega\,Y^*_{1M}(\Omega)M(\Omega).
\end{equation}
The $T_{\ell m}$ are the filtered data and $F_\ell \equiv \left< T_{\ell m}
T^*_{\ell m}\right>$. Note that the lack of aberration in the
\Planck\ Full Focal Plane simulations \citep{planck2014-a14} is expected to have negligible
effect on this analysis and on that of the quadrupolar modulation in the
next subsection, since the CDI modulation is heavily suppressed for
$\ell \gsim 500$, whereas the effect of aberration has a very different
$\ell$ dependence and will bias the modulation signal
for $\ell \lsim 1000$ at an insignificant level.

In practice, exploring the parameter space of the
model is sped up dramatically by binning the estimator defined in 
Eq.~(\ref{Eq:fulldipoleest}) into bins of width $\Delta\ell = 1$, which means
that the estimators only need to be calculated once \citep{planck2014-a18}.
Finally, for the modulated curvaton model we identify
\begin{equation}
\frac{dC_\ell}{dX} = 2K_\ell C_\ell.
\end{equation}
Note that for our constraints we fix the curvaton gradient to its maximum
value, $\Delta\sigma_*/\sigma_* = 1$.  Therefore, our constraints are
conservative, since smaller $\Delta\sigma_*/\sigma_*$ would only reduce the
modulation that this model could produce.

\begin{figure}
\begin{center}
\includegraphics[width=0.45\textwidth]{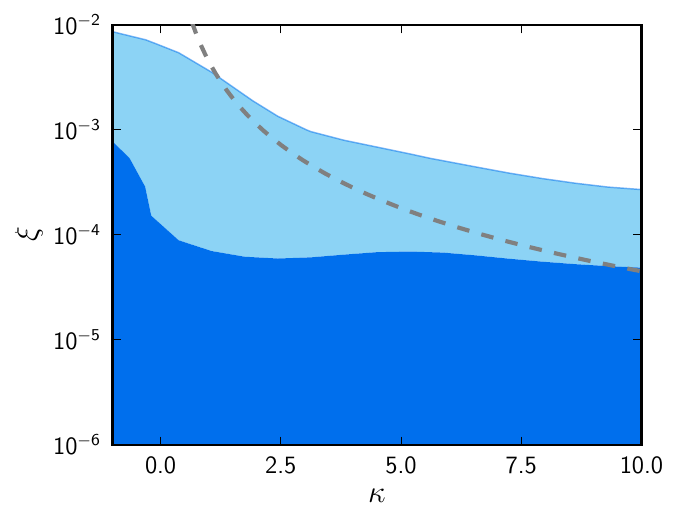}
\end{center}
\caption{68\,\% and 95\,\% CL regions in the modulated isocurvature model
parameter space using the \Planck\ temperature data up to $\ell_{\mathrm{max}}
= 1000$ (contours).  The region above the dashed curve is ruled out by
the \Planck\ constraint on an uncorrelated, scale-invariant isocurvature
component.}
\label{kappaxiplot}
\end{figure}

The temperature anisotropies measured by \Planck\ constrain the modulated
curvaton parameters $\kappa$ and $\xi$ via Eqs.~(\ref{Kelldef}) and
(\ref{Eq:fulldipoleest}).  Figure~\ref{kappaxiplot} shows the constraints in
this parameter space evaluating the estimator to $\ell_{\mathrm{max}} =
1000$.  The maximum likelihood region corresponds to a band at $\kappa
\gsim 3$.  For parameters in this region, the model produces a large-scale
asymmetry via a mainly-CDI modulation.  However, the {\em amplitude} of this
large-scale asymmetry is lower than the 6--$7\,\%$ actually
observed \citep{planck2014-a18}.  The reason is that, had a CDI modulation
produced all of the large-scale asymmetry, the consequent small-scale
asymmetry (due to the shape of the scale-invariant CDI spectrum) would be
larger than the \Planck\ observations allow.  The allowed CDI modulation is
further reduced by the \Planck\ 95\,\% upper limit on an uncorrelated,
scale-invariant (``axion''-type) isocurvature component,
$\beta_{\mathrm{iso}} < 0.033$, from Sect.~\ref{sec:iso}.  Imposing this
constraint reduces the available parameter space in the $\kappa$--$\xi$
plane via Eq.~(\ref{alphadef}), as illustrated in Fig.~\ref{kappaxiplot}.

The best fit in Fig.~\ref{kappaxiplot} corresponds to
$\Delta\chi^2 = -6.8$ relative to base $\Lambda$CDM, for two extra
parameters.  In order to assess how likely such an improvement would be in
statistically isotropic skies, we note that the best-fit CDI modulation
amplitude is very close to the mean amplitude expected due to cosmic
variance, as calculated directly from Eq.~(\ref{Eq:fulldipoleest}).  More
precisely, since the amplitude is $\chi^2$ distributed with three degrees
of freedom, i.e., Maxwell-Boltzmann distributed, we conclude that about
44\,\% of statistically isotropic skies will exhibit a measured [via
Eq.~(\ref{Eq:fulldipoleest})] isocurvature modulation larger than that of
the actual sky.

To summarize, the modulated curvaton model can only produce a small part
of the observed large-scale asymmetry, and what it can produce is entirely
consistent with cosmic variance in a statistically isotropic sky.  Hence
we must favour the base $\Lambda$CDM model 
over this model.  Finally, note that further generalizing the
model (e.g., to allow non-scale-invariant CDI spectra) may allow more
large-scale asymmetry to be produced and hence result in an improved
$\Delta\chi^2$, at the expense of more parameters.  On the
other hand, the neutrino isocurvature modes are not expected to fit the
observed asymmetry well due to their approximate scale invariance (see
Fig.~\ref{fig:JVmodecomparison}).

\subsection{Constraints on quadrupolar asymmetry generated during inflation}

In this section we assume a quadrupolar directional dependence of the
primordial scalar power spectrum about some axis $\pm \vec{\hat{d}}$ and having a 
scale-dependent amplitude $g(k).$ More specifically, we assume 
\begin{equation}
\mathcal{P}_{\mathcal{R}}({\vec k}) = \mathcal{P}_{\mathcal{R}}^0(k)\left\{1 + g(k)
   \left[\left(\vec{\hat k}\cdot\vec{\hat{d}}\right)^2 - \frac{1}{3}\right]\right\},
\label{Pkgk}
\end{equation}
which can be rewritten as 
\begin{equation}
\mathcal{P}_{\mathcal{R}}({\vec k}) = \mathcal{P}_{\mathcal{R}}^0(k)
   \left[1 + \sum_{M}g_{2M}(k)\,Y_{2M}(\vec{\hat k})\right],
\label{prim_power_glm}
\end{equation}
where
\begin{equation}
g_{2M}(k) \equiv \frac{8\pi}{15}g(k)\,Y_{2M}^*(\vec{\hat{d}}),
 \end{equation}
with $g_{2M}(k)$ satisfying $g_{2,-M}(k)=(-1)^M\,g^*_{2M}(k)$.  
In this analysis, we will treat the modulation scale dependence as a power
law, $g(k) = g_*(k/k_*)^{q}$, and consider five values of the spectral index,
namely $q = -2$, $-1$, $0$, $1$, and $2$.  Importantly, for $q \ne 0$ our
constraints on $g_*$ will depend on the pivot scale, chosen as
$k_* = 0.05\,\mathrm{Mpc}^{-1}$ as elsewhere in this paper.  
Models have been proposed predicting
both positive and negative $g_*$
\citep[see, e.g.][]{Tsujikawa:2014rta}, so we keep the sign of $g_*$ free.

Often in the literature the term $-g(k)/3$ is not included in the modulated
power spectrum, Eq.~(\ref{Pkgk}).  Our form sets the modulation monopole to
zero, so that there is no correction to the isotropic power spectrum dependent
on $g(k)$.  We do this because for large $|q|$ the correction would require
a joint analysis with the isotropic power spectrum likelihood.  Inflationary
models have been proposed which predict both forms. For example, the model in 
\cite{Ohashi:2013mka} includes the modulation monopole, while the model in \cite{Libanov:2010nk} 
does not.  For $q = 0$ our results apply 
to both forms due to the degeneracy of a scale-independent
correction to $\mathcal{P}_{\mathcal{R}}^0(k)$ with the scalar amplitude, 
$A_{\mathrm s}.$ However, for nonzero 
tilt a joint analysis would yield tighter constraints on $g_*$ when the
monopole correction is present, in which case our results will be conservative.

Given the anisotropic power spectrum of Eq.~(\ref{prim_power_glm}), the statistically
anisotropic part of the CMB temperature fluctuations has the following expectation
value \citep{Ma2011}:
\begin{equation}
\begin{aligned}
C_{\ell,\ell';2,M}
   &= i^{\ell-\ell'} D_{\ell \ell'} g_{2M}(k)\,
     \left[\frac{5(2\ell+1)(2\ell'+1)}{4\pi}\right]^{\frac{1}{2}}\\
  &\times \left(\begin{array}{ccc}2&\ell&\ell'\\0&0&0\end{array}\right),
\end{aligned}
\end{equation}
where $D_{\ell \ell'} \equiv 4\pi\int d\ln k\,
\Delta^{\mathrm s}_{\ell,T}(k)\,\Delta^{\mathrm s}_{\ell',T}(k)\,
\mathcal{P}_{\mathcal{R}}^0(k)(k/k_*)^{q}$ and $\Delta^{\mathrm s}_{\ell,T}(k)$ 
denotes the temperature radiation transfer function.

The analysis is carried out using the foreground-cleaned CMB temperature maps 
{\tt Commander}, {\tt NILC},
{\tt SEVEM}, and {\tt SMICA}, where we apply the extended common mask UT76.
The implementation details of the optimal estimator can be found in Sect.~5.3
of \citet{planck2014-a18}.  However, here we apply an inverse variance weighted filter
that assumes a simple white noise component, but optimally accounts for the
mask in the same manner as \citet{planck2013-p12} and Sects.~6.3 and 6.6 of 
\citet{planck2014-a18}.  
We estimate $g_{2M}$ from the data at multipoles $2\le \ell \le 1200$. The
range of multipoles is chosen such that the impact of foreground residuals on
the conclusions is insignificant.
Neglecting
very small scales, however, sacrifices little constraining power
because those scales are noise dominated.
This conclusion was based on 
realistic simulations containing residual foregrounds.
Moreover, we estimate the statistical uncertainty in $g_{2M}$ with
various $\ell_{\mathrm{max}}$ values using simulations.

Once we have obtained estimates for the five $g_{2M}$ coefficients, we must
determine values for the model parameters of interest, namely $g_*$ and 
$\vec{\hat{d}}$.  We assume the $g_{2M}$ 
coefficients to be Gaussian distributed due to cosmic variance. We have
explicitly checked this hypothesis with simulations.  Hence the likelihood
function is
\begin{eqnarray}
{\cal L} &\propto& |G|^{-1/2}\label{g2Mlike}\\
         &\times& \exp\left[-\frac{1}{2}
              \left(\hat{g}_{2M} - g_{2M}(g_*,\vec{\hat{d}})\right)^{\textsf T}
        G^{-1}\left(\hat{g}_{2M} - g_{2M}(g_*,\vec{\hat{d}})\right)\right],\nonumber
\end{eqnarray}
where $G$ is the $g_{2M}$ covariance matrix, which is estimated using isotropic
simulations.  One approach to determining the model parameters would be to 
use this likelihood to calculate marginalized posterior 
distributions for $g_*$, from which mean values and errors could be 
determined.  However, we find that $g_*$ is so poorly constrained that the 
means and widths thus calculated strongly depend on the prior 
for $g_*$. Two sensible priors are uniform in $g_*$ or proportional 
to $g_*^2$ [i.e., uniform in the Cartesian components of 
$(g_*,\vec{\hat{d}})$].  In addition, we find that the posterior means are 
much closer to zero then the widths, which is due to the approximate 
degeneracy between a modulation $(g_*,\vec{\hat{d}})$ and modulation 
$(-g_*,\vec{\hat{d}}')$, where $\vec{\hat{d}}'$ is orthogonal to 
$\vec{\hat{d}}$.  In such a situation the degree of consistency between 
the measured value of $g_*$ and the expectations of cosmic variance in 
statistically isotropic skies is unclear.

Instead we determine best-fit values for $g_*$ and $\vec{\hat{d}}$ 
by maximizing the likelihood over the three parameters using a grid approach.
To characterize how unexpected our best-fit values are in statistically
isotropic skies, we repeat the procedure replacing our estimates for $g_{2M}$
from the data with estimates from $1000$ isotropic simulations.  We finally
calculate $p$-values, which give the fraction of simulations with a larger 
value of $|g_*|$ than the actual data.  
Note that from the Bayesian perspective the maximum-likelihood values amount
to maximizing the posterior for $g_*$ given a uniform prior on $g_*$, so that
these values will change with a different prior.  However, we have checked
that the $p$-values depend only very weakly on the choice of prior.

\begin{table*}[htb!]
\begingroup
\newdimen\tblskip \tblskip=5pt
\caption{
Minimum-$\chi^2$ $g_*$ values for quadrupolar modulation,
determined from the \texttt{Commander}, \texttt{NILC}, \texttt{SEVEM}, and
\texttt{SMICA} foreground-cleaned maps.  Also given are $p$-values, defined as 
the fraction of simulations with larger $|g_*|$ than the data. These results 
demonstrate that the data are consistent with cosmic variance in statistically
isotropic skies.}
\label{tab:gstar}                            
\nointerlineskip
\vskip -3mm
\footnotesize
\setbox\tablebox=\vbox{
   \newdimen\digitwidth
   \setbox0=\hbox{\rm 0}
   \digitwidth=\wd0
   \catcode`!=\active
   \def!{\kern\digitwidth}
   \newdimen\signwidth
   \setbox0=\hbox{+}
   \signwidth=\wd0
   \catcode`+=\active
   \def+{\kern\signwidth}
\halign{\hbox to 1.5cm{#\leaderfil}\tabskip 1em&
\hfil#\hfil&\tabskip 0.8em&
\hfil#\hfil&
\hfil#\hfil&
\hfil#\hfil&
\hfil#\hfil\tabskip 0pt\cr
\noalign{\doubleline}
\omit&\multispan2 \hfil\texttt{Commander}\hfil & \multispan2 \hfil\texttt{NILC}\hfil
     &\multispan2 \hfil\texttt{SEVEM}\hfil & \multispan2 \hfil\texttt{SMICA}\hfil\cr
\noalign{\vskip -5pt}
\omit&\multispan8\hrulefill\cr
\omit\hfil$q$\hfil
   &\omit\hfil $g_*$ \hfil&\omit\hfil $p$-value [\%]\hfil
   &\omit\hfil $g_*$ \hfil&\omit\hfil $p$-value [\%]\hfil
   &\omit\hfil $g_*$ \hfil&\omit\hfil $p$-value [\%]\hfil
   &\omit\hfil $g_*$ \hfil&\omit\hfil $p$-value [\%]\hfil\cr
\noalign{\vskip 4pt\hrule\vskip 6pt}
\quad$-2$ & $-7.39\times10^{-5}$ & 79.2 & $-7.66\times10^{-5}$ & 79.8 &
$-7.43\times10^{-5}$ & 80.6 & $-7.52\times10^{-5}$ & 80.2 \cr
\noalign{\vskip 3pt}
\quad$-1$ & +$5.99\times10^{-3}$ & 97.3 & +$6.65\times10^{-3}$ & 95.8 &
+$6.27\times10^{-3}$ & 97.2  & +$6.22\times10^{-3}$ & 96.9 \cr
\noalign{\vskip 3pt}
\quad$+0$ & $-2.79\times10^{-2}$ & 12.5 & $-2.38\times10^{-2}$ & 26.9 &
$-2.56\times10^{-2}$ & 20.7 & $-2.56\times10^{-2}$ & 20.0 \cr
\noalign{\vskip 3pt}
\quad$+1$ & $-2.15\times10^{-2}$ & !8.2 & $-1.79\times10^{-2}$ & 23.7 &
$-1.93\times10^{-2}$ & 17.8 & $-1.93\times10^{-2}$ & 16.7 \cr
\noalign{\vskip 3pt}
\quad$+2$ & $-1.28\times10^{-2}$ & !9.7 & $-1.07\times10^{-2}$ & 23.7 &
$-1.13\times10^{-2}$ & 20.4 & $-1.15\times10^{-2}$ & 18.1 \cr
\noalign{\vskip 3pt\hrule\vskip 4pt}}}
\endPlancktable                    
\endgroup
\end{table*}

Table~\ref{tab:gstar} shows the $g_*$ values obtained by minimizing 
$\chi^2$ as well as
the $p$-values for the data compared to statistically isotropic simulations.  
Note that the constraints on $g_*$ are strongest for the most negative 
values of the exponent $q$.  This is because for fixed $g_*$ 
the largest asymmetry over the
range of observable scales occurs for $q = -2$, at the largest scales, due
to the location of the pivot scale, $k_*$.
Our limits provide a stringent test of rotational symmetry during inflation.
We find 
no sign of deviation from statistical isotropy.\footnote{The constraints
from the \Planck\ 2013 data by \citet{Kim2013} should be multiplied by a factor
of $\sqrt{2}$ in our normalization.}


%% file: section_thirteen.tex
In this section we discuss the implications of the recent constraints on the primordial 
$B$-mode polarization from the cross-correlation
of the BICEP2 and Keck Array data at 150\,GHz with the \Planck\ maps at higher
frequencies to characterize and remove the contribution 
from polarized thermal dust emission from our Galaxy 
\citepalias{pb2015}.  
On its own, the BKP likelihood leads to a 95\,\%~CL upper
limit of $r<0.12$, compatible with and independent of the constraints obtained using the 
2015 \Planck\ temperature and large angular scale polarization in Sect.~\ref{sec:modelcomp}. (Note, however, that 
the BKP likelihood uses the Hamimeche-Lewis approximation \citep{Hamimeche:2008ai}, which requires 
the assumption of a fiducial model.)  
The BKP results are also compatible with the \Planck\ 2013 Results \citep{planck2013-p11,planck2013-p17}. The
posterior probability distribution for $r$ obtained by BKP
peaks away from zero at $r \approx 0.05$, but the region of large posterior probability includes $r=0$.

\begin{figure*}[!t]
\begin{center}
\includegraphics[width=18cm]{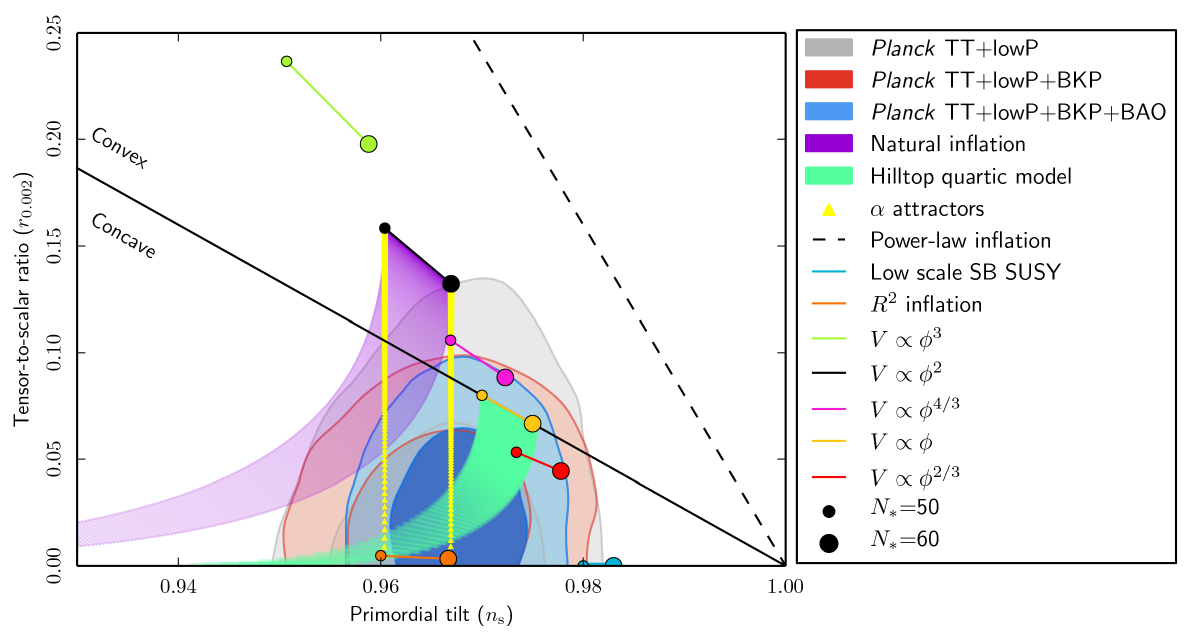}
\end{center}
\caption{Marginalized joint 68\,\% and 95\,\%~CL regions for $n_\mathrm{s}$ and $r$ at $k = 0.002$\,Mpc$^{-1}$
from \Planck\ alone and in combination with its cross-correlation with BICEP2/Keck Array and/or BAO data
compared with the theoretical predictions of selected inflationary models. Note that the marginalized 
joint 68\,\% and 95\,\%~CL regions have been obtained by assuming $d n_\mathrm{s}/d \ln k = 0$.}
\label{fig:nsvsr_BKP}
\end{figure*}

Here we combine the baseline two-parameter BKP likelihood 
using the lowest five $B$-mode bandpowers with the \Planck\ 2015 likelihoods.
The two BKP nuisance parameters are the 
$B$-mode amplitude and frequency spectral index of the polarized thermal dust emission. 
The combined analysis yields the following constraint on the tensor-to-scalar ratio:
\begin{equation}
\label{erre_t_bkp}
r_{0.002} <  0.08 \quad \text{(95\,\% CL, \Planck\ TT+lowP+BKP)} \,, 
\end{equation}
further improving on the upper limits obtained from the different data combinations presented
in Sect.~\ref{sec:modelcomp}.

By directly constraining the tensor mode, the BKP likelihood  
removes degeneracies between the tensor-to-scalar ratio and other parameters.
Adding tensors and running, we obtain
\begin{equation}
\label{erre_run_t_bkp}
r_{0.002} < 0.10 \quad \text{(95\,\% CL, \Planck\ TT+lowP+BKP)} \,,
\end{equation}
which constitutes almost a 50\,\% improvement over the \Planck\ TT+lowP constraint quoted in 
Eq.~(\ref{erre_run_t}). These limits on tensor modes are more robust than the limits using the 
shape of the $C_\ell^{TT}$ spectrum alone because scalar perturbations cannot generate $B$ modes irrespective
of the shape of the scalar spectrum.

\subsection{Implications of BKP on selected inflationary models}

Using the BKP likelihood further strengthens the constraints
on the inflationary parameters and models discussed in Sect.~\ref{sec:numerical}, as seen in 
Fig.~\ref{fig:nsvsr_BKP}.  If we set $\epsilon_3=0$, 
the first slow-roll parameter is constrained to $\epsilon_1 < 0.0055$ at 95\,\% CL by \Planck\ TT+lowP+BKP.
With the same data combination, concave potentials are preferred over convex potentials 
with $\ln B = 3.8$, which improves on the $\ln B = 2$ result obtained from the \Planck\ data alone.

\input TABLE4_BKPupdate_restricted.tex

Combining with the BKP likelihood strengthens the constraints on the 
selected inflationary models studied in Sect.~\ref{sec:numerical}. 
Using the same methodology as in Sect.~\ref{sec:numerical} and 
adding the BKP likelihood gives a Bayes factor preferring $R^2$ over chaotic inflation 
with monomial quadratic potential and natural inflation by odds of 403:1 and 270:1, respectively, 
under the assumption of a dust equation of state during the entropy generation stage.
The combination with the BKP likelihood further penalizes the double-well model compared to $R^2$ inflation.
However, adding BKP reduces
the Bayes factor of the hilltop models compared to $R^2$, because these models can predict a value of the 
tensor-to-scalar ratio that better fits the statistically insignificant peak at $r \approx 0.05$.
See Table~\ref{table:model_compar_BKP} for the $\Delta \chi^2$ and the Bayes factors of inflationary 
models with the same two cases of post-inflationary evolution studied in Sect.~\ref{sec:numerical}. 
Note, however, that the $\Delta \chi^2$ are computed with respect to the best fit of
baseline + tensors, unlike in Table~\ref{table:model_compar}. 

\subsection{Implications of BKP on scalar power spectrum}

\begin{figure}
\begin{center}
\includegraphics[width=0.45\textwidth]{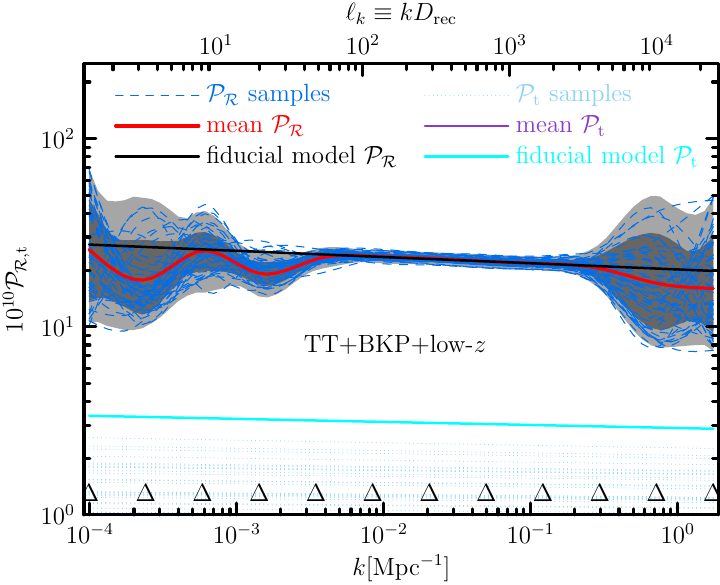}
\includegraphics[width=0.45\textwidth]{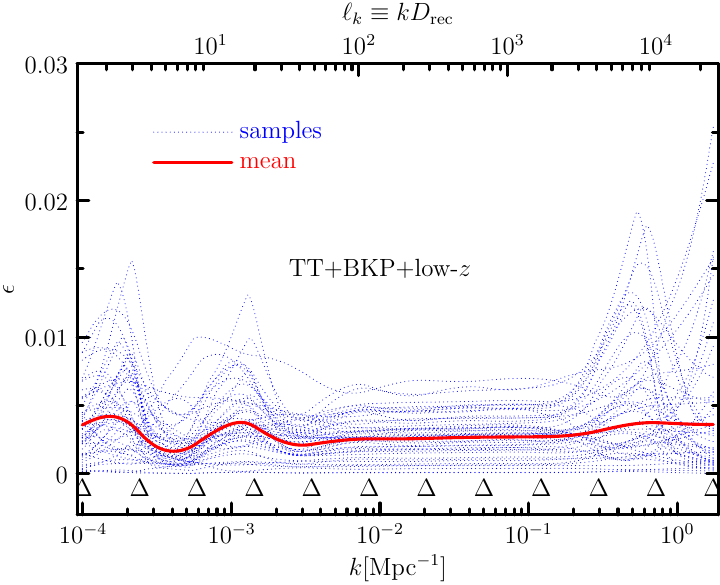}
\includegraphics[width=0.45\textwidth]{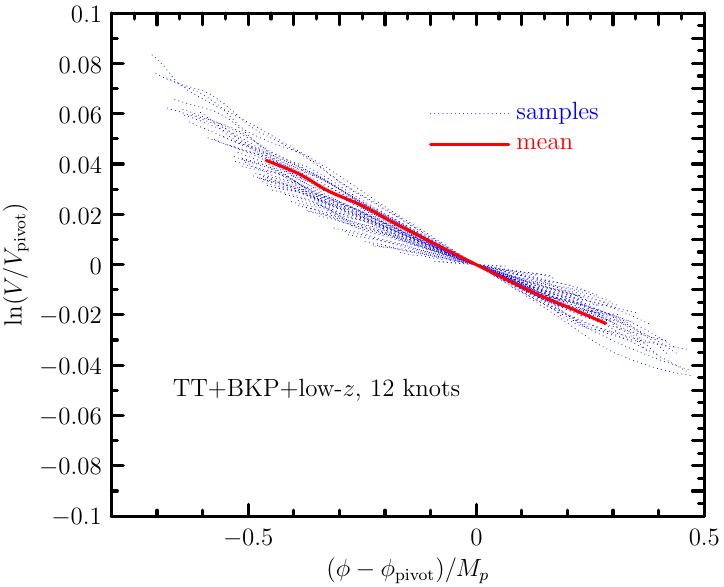}
\end{center}
\caption{Impact of BKP likelihood on scalar primordial power spectrum reconstruction.
We show how including the BKP likelihood affects
the reconstruction in Sect.~\ref{sec:csrecon}. The top panel is to be
compared with the reconstructions in Fig.~\ref{fig:traj_power_spline}, and we observe that
including BKP has a minimal impact given the uncertainty in the
reconstruction. The middle panel is to be compared with Fig.~\ref{SlowRollBond}, and
here we notice that including BKP excludes the trajectories with large values of $\epsilon .$
The bottom panel shows how the inflationary potential reconstructions are modified by BKP
(to be compared with Fig.~\ref{fig:traj_pot_spline}).\label{BondZhiqiWithBKP}}
\end{figure}

The presence of tensors would, at least to some degree, require an enhanced
suppression of the scalar power spectrum on large scales to account for the low-$\ell$
deficit in the $C_\ell^{TT}$ spectrum.
We therefore repeat the analysis of an exponential cutoff studied in Sect.~\ref{LargeScaleSuppression} with
tensor perturbations included and the standard tensor tilt (i.e., $n_\mathrm{t} = -r/8$).
Allowing tensors does not significantly degrade the
$\Delta \chi^2$ improvement found in Sect.~\ref{LargeScaleSuppression} for \Planck\ TT+lowP with a best fit at $r \approx 0$.
When the BKP likelihood is combined, we obtain $\Delta \chi^2 = -4$ with respect to the base
$\Lambda$CDM model with a best fit at $r \approx 0.04$. However,
since this model contains 3 additional parameters, it is not preferred over base $\Lambda$CDM.

In Fig.~\ref{BondZhiqiWithBKP} we show how the scalar primordial power spectrum reconstruction
discussed in Sect.~\ref{sec:csrecon} is modified when the BKP likelihood is also included.
While the power spectrum reconstruction hardly varies given the uncertainties in the method, the trajectories
of the slow-roll parameters are significantly closer to slow roll.
When the 12-knot reconstruction is carried out, the upper bound on the tensor-to-scalar ratio is
$r < 0.11$ at 95\,\% CL. 
The $\chi ^2$ per degree of freedom
for the 5 low-$k$ and 6 high-$k$ knots are
1.14 and 0.22, respectively, corresponding to $p$-values of $0.33$ and $0.97.$

\subsection{Relaxing the standard single-field consistency condition}

We now relax the consistency condition (i.e., $n_\mathrm{t} = - r/8$)
and allow the tensor tilt to be independent of the tensor-to-scalar ratio.  
This fully phenomenological analysis with the BKP likelihood is complementary 
to the study of inflationary models with generalized Lagrangians in Sect.~\ref{sec:putfnl}, which 
also predict modifications to the consistency condition $n_\mathrm{t} = - r/8$ for a nearly scale-invariant 
spectrum of tensor modes. In this subsection we adopt a phenomenological approach,
thereby including radical departures from $n_\mathrm{t} \lsim 0$, including values which are 
predicted in alternative models to inflation \citep{Gasperini:1992em,Boyle:2003km,Brandenberger:2006xi}.
In Sect.~\ref{sec:putfnl} we folded in the \Planck\ $f_\mathrm{NL}$ 
constraints \citep{planck2014-a19}, whereas here we consider \Planck\ and BKP likelihoods only.
Complementary probes such as pulsar timing, direct detection of gravitational waves,
and nucleosynthesis bounds could be used to constrain blue values for the tensor spectral index 
\citep{Stewart:2007fu}, but here we are primarily interested in what CMB data can tell us.

We caution the reader that in the absence of a clear detection of a tensor
component, joint constraints on $r$ and $n_\mathrm{t}$ depend strongly 
on priors, or equivalently on the choice of parameterization.  
Nevertheless, the BKP likelihood has some constraining power 
over a range of scales more than a decade wide around 
$k \approx 0.01$\,Mpc$^{-1}$, so the results are not entirely prior driven.

\begin{figure}[t!]
\begin{center}
\includegraphics[width=\columnwidth]{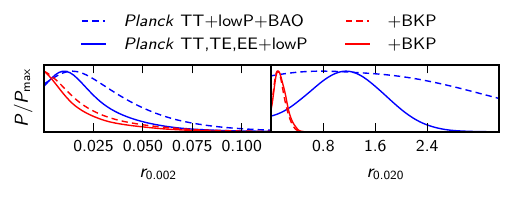}
\end{center}
\caption{Posterior probability density of the tensor-to-scalar ratio at two
different scales. The inflationary consistency relation is relaxed and $r_{0.002}$
and $r_{0.020}$ are used as sampling parameters,
assuming a power-law spectrum for primordial tensor perturbations. When the 
BKP likelihood is included in the analysis, the results
with \Planck\ TT+lowP+BAO and \Planck\ TT,TE,EE+lowP coincide 
(dashed and solid red curves, respectively).
\label{fig:freent_1d}
}
\end{figure}

\begin{figure}[b!]
\begin{center}
\includegraphics{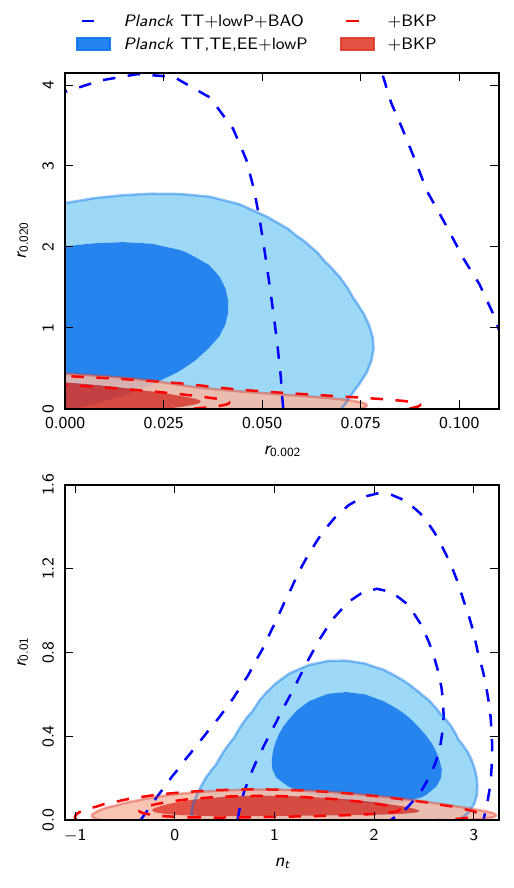}
\end{center}
\caption{68\,\% and 95\,\% CL constraints on tensors when the
inflationary consistency relation is relaxed, with \Planck\ TT+lowP+BAO (blue dashed
contours) and TT,TE,EE+lowP (blue shaded regions). The red colours are for the same
data plus the BKP joint likelihood. The upper panel shows our
independent primary parameters $r_{0.002}$ and $r_{0.020}$.
The lower panel shows the derived parameters $n_t$
and $r_{0.01}$. The scale $k=0.01\,$Mpc$^{-1}$ is near the
decorrelation scale of ($n_t$, $ r$) for the \Planck+BKP data.
\label{fig:freent_2d}
}
\end{figure}

\begin{figure}
\includegraphics[width=0.5\textwidth]{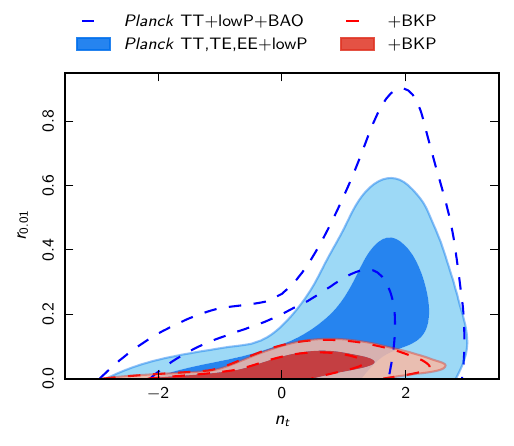}
\caption{The same as Fig.~\ref{fig:freent_2d} lower panel, but using $n_t$ and $r_{0.002}$ as
primary parameters. 
\label{fig:freent_2d_bis}
}
\end{figure}

The commonly used ($r$, $n_\mathrm{t}$) parameterization suffers
from pathological behaviour around $r=0,$ which could be problematic for statistical sampling.  
We therefore use a parameterization 
specifying $r$ at two different scales, ($r_{k_1},r_{k_2}$) 
(analogous to the treatment of primordial isocurvature in Sect.~\ref{sec:iso}) 
as well as the more familiar
($r$, $n_\mathrm{t}$) parameterization.
We present results based on 
$k_1 = 0.002$\,Mpc$^{-1}$ and $k_2 = 0.02$\,Mpc$^{-1},$ 
also quoting the amplitude at $k=0.01$\,Mpc$^{-1}$ for both parameterizations.
This scale is close to the decorrelation scale for ($r$, $n_\mathrm{t}$) for the 
\Planck+BKP joint constraints.
We obtain $r_{0.002} < 0.07$ ($0.06$) and $r_{0.02} < 0.29$ ($0.31$) at 95\,\% CL from the two-scale parameterization 
with \Planck\ TT+lowP+BAO+BKP (TT,TE,EE+lowP+BKP).  Figure~\ref{fig:freent_1d} illustrates the 
impact of the BKP likelihood on the one-dimensional posterior probabilities for these two parameters.
The derived constraint at $k=0.01$\,Mpc$^{-1}$ is $r_{0.01} < 0.12$ ($0.12$) at 95\,\% CL with 
\Planck\ TT+lowP+BAO+BKP (TT,TE,EE+lowP+BKP).
The upper panel of Fig.~\ref{fig:freent_2d} shows the relevant 2D contours 
for the tensor-to-scalar ratios at the two scales and the improvement due 
to the combination with the BKP likelihood.
The lower panel shows the 2D contours in ($r_{0.01},n_\mathrm{t}$) obtained by sampling with the two-scale 
parameterization. 
Figure~\ref{fig:freent_2d_bis} shows the 2D contours in ($r_{0.01},n_\mathrm{t}$)
obtained by the  ($r_{0.002},n_\mathrm{t}$)  parameterization.

We conclude that positive values of the tensor tilt, $n_\mathrm{t}$, are not statistically significantly 
preferred by the BKP joint measurement of $B$-mode polarization in combination with \Planck\ data, 
a conclusion at variance with results reported using the BICEP2 data \citep{Gerbino:2014eqa}. 
However, the now firmly established contamination by polarized dust emission easily explains the discrepancy.
Values of tensor tilt consistent with the standard single-field inflationary consistency relation 
are compatible with the \Planck+BKP constraints.

%% file: TABLE4_BKPupdate_restricted.tex
\begin{table}
\begingroup
\newdimen\tblskip \tblskip=5pt
\caption{Results of inflationary model comparison using the cross-correlation
between BICEP2/Keck Array and \Planck.  This table is similar
to Table~\ref{table:model_compar}, which did not use the BKP likelihood. 
Note, however, that the $\Delta \chi^2$ are computed with respect to the best fit of 
baseline + tensors, unlike in Table~\ref{table:model_compar}.}
\label{table:model_compar_BKP}
\nointerlineskip
\vskip -3mm
\footnotesize 
\setbox\tablebox=\vbox{
\newdimen\digitwidth
\setbox0=\hbox{\rm 0}
\digitwidth=\wd0
\catcode`*=\active
\def*{\kern\digitwidth}
\newdimen\signwidth
\setbox0=\hbox{+}
\signwidth=\wd0
\catcode`!=\active
\def!{\kern\signwidth}
\halign{#\hfil&\hfil#\hfil&\hfil#\hfil&\hfil#\hfil&\hfil#\hfil\cr
\noalign{\doubleline}
Inflationary model & \multispan2\hfil$\ln B_{0X}$\hfil\cr
\noalign{\vskip 2pt}
 & \ $w_\mathrm{int}=0$\ &
\ $w_\mathrm{int}\ne0$\cr
\noalign{\vskip 5pt\hrule\vskip 5pt}
$R + R^2/(6 M^2)$ & *$\dots$ & *$+0.3$ \cr
$n=2/3$ & *$-1.9$ & *$-1.2$ \cr
$n=1$ & *$-1.6$ & *$-1.8$ \cr
$n=4/3$ & *$-2.1$ & *$-2.5$ \cr
$n=2$ & *$-6.0$ & *$-5.6$ \cr
$n=3$ & $-16.0$ & $-15.6$ \cr
$n=4$ & $-30.1$ & $-29.9$ \cr
Natural & *$-5.6$ & *$-5.0$  \cr
Hilltop ($p=2$) & *$-0.7$ & *$-0.4$ \cr
Hilltop ($p=4$) & *$-0.6$ & *$-0.9$ \cr
Double well & *$-4.3$ & *$-4.2$ \cr
Brane inflation ($p=2$) & *$+0.2$ & *$\phantom{-}0.0$ \cr
Brane inflation ($p=4$) & *$+0.1$ & *$-0.1$ \cr
Exponential tails & *$-0.1$ & *$\phantom{-}0.0$ \cr
SB SUSY & *$-1.8$ & *$-1.5$ \cr
Supersymmetric $\alpha$-model & *$-1.1$ & *$+0.1$ \cr
Superconformal ($m=1$) & *$-1.9$ & *$-1.4$ \cr
Superconformal ($m\ne1$) & *$-2.5$ & *$-2.2$ \cr  
\noalign{\vskip 3pt\hrule}}}
\endPlancktable
\endgroup
\end{table}

%% file: conclusions.tex
\def\gtorder{\mathrel{\raise.3ex\hbox{$>$}\mkern-14mu
             \lower0.6ex\hbox{$\sim$}}}
\def\ltorder{\mathrel{\raise.3ex\hbox{$<$}\mkern-14mu
             \lower0.6ex\hbox{$\sim$}}}

The \Planck\ full mission temperature and polarization data are consistent 
with the spatially flat base $\Lambda$CDM model, whose perturbations are Gaussian and adiabatic 
with a spectrum described by a simple power law, as predicted by the simplest inflationary models.
For this release, the basic \Planck\ results do not rely on external data. The first \Planck\ 
polarization release at large angular scales from the LFI 70\,GHz channel determines an optical depth of $\tau = 
0.067 \pm 0.022$ (68\,\% CL, \Planck\ low multipole likelihood), a value smaller than the previous \Planck\ 2013 result
based on the WMAP9 polarization likelihood as delivered by the WMAP team. This \Planck\ value of $\tau$
is consistent with an analysis of WMAP9 polarization data cleaned for polarized dust emission using the
\Planck\ 353\,GHz data \citep{planck2013-p08,planck2014-a13}. 
The estimates of cosmological parameters from the full mission 
temperature data and polarization on large angular scales are consistent with those of the 
\Planck\ 2013 release. The $TE$ and $EE$ spectra at $\ell \ge 30$ together with the lensing power spectra lead to 
cosmological constraints in agreement with those obtained from temperature.

The \Planck\ full mission temperature and large angular scale polarization data rule out an exactly 
scale-invariant spectrum of curvature perturbations at 5.6\,$\sigma .$ For the base $\Lambda$CDM 
model, the spectral index is measured to be $n_\mathrm{s} = 0.965 \pm 0.006$ (68\,\% CL, \Planck\ TT+lowP). 
No evidence for a running of the spectral index is found, with $d n_\mathrm{s}/{d \ln k} = -0.008 \pm 0.008$ 
(68\,\% CL, \Planck\ TT+lowP).

The \Planck\ full mission data improve the upper bound on the tensor-to-scalar ratio to $r_{0.002} < 0.10$ 
(95\,\% CL, \Planck\ TT+lowP), a bound that changes only slightly when including the \Planck\ lensing likelihood, the high-$\ell$ 
polarization likelihood, or the likelihood from the WMAP large angular scale polarization map 
(dust-cleaned with the \Planck\ 353\,GHz map). We 
showed how the low-$\ell$ deficit in temperature contributes to the \Planck\ upper bound on $r_{0.002}$, but 
this deficit is not a statistically significant anomaly within the base $\Lambda$CDM cosmology. 
Using the full mission \Planck\ data, we find the upper bound on 
$r_{0.002}$ stable, even when extended cosmological models or models with CDM isocurvature are considered.
The \Planck\ bound on $r_{0.002}$ is consistent with the recent result $r_{0.002} < 0.12$ at 95\,\% CL 
obtained by the BKP cross-correlation analysis
which accounts for contamination 
by polarized dust emission \citep{planck2013-pip56}. 
By combining \Planck\ TT+lowP with the BKP cross-correlation likelihood, we obtain $r_{0.002} < 0.08$ at 95\,\% CL.

The increased precision of the \Planck\ full mission data reduces the area enclosed by the 95\,\% confidence 
contour in the $(n_\mathrm{s}, r)$-plane by 29\,\%. 
We performed a Bayesian model comparison with the same methodology as in 
\citetalias{planck2013-p17}, taking into account reheating uncertainties by marginalizing over two extra 
parameters: the energy scale at thermalization, $\rho_\mathrm{th}$, and the parameter 
$w_\mathrm{int}$ characterizing the average equation of state between the end of inflation and thermalization. 
Among the models considered using this approach, the $R^2$ inflationary 
model proposed by \cite{Starobinsky:1980te} is the most preferred.
Due to its high tensor-to-scalar ratio, the quadratic model is now strongly disfavoured with 
respect to $R^2$ inflation for \Planck\ TT+lowP in combination with BAO data. By combining wth the BKP likelihood,
this trend is confirmed, and natural inflation is also disfavoured.

We reconstructed the inflaton potential and the Hubble parameter evolution during the observable part of
inflation using a Taylor expansion of the inflaton potential or $H(\phi ).$ This analysis did not rely on the slow-roll 
approximation, nor on any assumption about the end of inflation. 
When higher-order terms were allowed, both reconstructions led to a change in the slope of the potential at the beginning of the 
observable range, thus better fitting the low-$\ell$ temperature deficit by turning on a non-zero running of 
running and accommodating $r_{0.002} \approx 0.2$. These models, however, are not significantly favoured compared
to lower order parameterizations that lead to slow-roll evolution at all times.

Three distinct methods were used to reconstruct the primordial power 
spectrum. All three methods strongly constrain deviations from a featureless power spectrum over the range of 
scales $0.008\,\mathrm{Mpc}^{-1} \ltorder k \ltorder 0.1\,\mathrm{Mpc}^{-1}$. More interestingly, they also independently 
find common patterns in the primordial power spectrum of curvature perturbations 
$\mathcal{P}_{\mathcal{R}}(k)$ at $k \ltorder  0.008\,\mathrm{Mpc}^{-1}$. These patterns are related to the
dip at $\ell \approx 20$--$40$ in the temperature power spectrum.
This deviation from a simple power-law spectrum has weak statistical significance due to the large cosmic variance 
at low $\ell .$

This direct reconstruction of the power spectrum is complemented by a search for parameterized 
features in physically motivated models. The models considered range from the minimal case of a 
kinetic energy dominated phase preceding a short inflationary stage (with just one extra parameter), to a 
model with a step-like feature in the potential and in the sound speed (with five extra parameters). 
As with the \Planck\ 2013 nominal mission data, these templates lead to an improved fit, up to 
$\Delta \chi^2 \approx 12$. However, neither the Bayesian evidence nor a frequentist simulation-based analysis 
shows any statistically significant preference over a simple power law.

We have updated the analysis that combines power spectrum constraints with those derived from the 
$f_\mathrm{NL}$ parameters \citep{planck2014-a19}. New limits on the sound speed inferred from the full 
mission temperature and polarization data further constrain the slow-roll parameters for generalized models, 
including DBI inflation. For the first time, we derived combined constraints on Galileon inflation, including 
the region of parameter space in which the predicted spectrum of gravitational waves has a blue spectral 
index.

Several models motivated by the axion monodromy mechanism in string theory predict oscillatory modulations 
and corresponding non-Gaussianities, potentially detectable by \Planck. A $TT$-only analysis picks up four 
possible modulation frequencies, which remain present when the high-$\ell$ polarization likelihood is included.
An inspection of frequency residuals in the high-$\ell$ $TT$ likelihood does not reveal evidence of 
foreground-related systematics at similar frequencies. 
However, a Bayesian evidence comparison prefers the 
smooth base $\Lambda$CDM model over modulated models, suggesting that the latter could simply be fitting the 
noise in the data. The monodromy model predicts resonant non-Gaussian features correlated to power spectrum 
features. A partial analysis beyond the power spectrum was presented.
We also constrained a 
possible pseudo-coupling of the axion to gauge fields by requiring that non-Gaussianities induced by inverse 
decay satisfy the \Planck\ bounds on $f_\mathrm{NL}$.

Section~\ref{sec:iso} reports on a search for possible deviations from purely adiabatic initial 
conditions by studying a range of models including isocurvature modes as well as possible
correlations with the adiabatic mode.
The \Planck\ full mission temperature data are consistent with adiabaticity. The \Planck\ 
$TT$ data place tight constraints on three-parameter extensions to the flat adiabatic base $\Lambda$CDM 
model, allowing arbitrarily correlated mixtures of the adiabatic mode with one isocurvature mode (of either 
the CDM, baryon, neutrino density, or neutrino velocity type). Adding the high-$\ell$ $TE$ and $EE$ 
polarization data further squeezes the constraints, since polarization spectra contain additional shape and 
phase information on acoustic oscillations. The likelihood with polarization included is in 
agreement with adiabatic initial conditions. However, the tightening of the constraints after including polarization must be 
interpreted with caution because of possible systematic effects. For this reason we emphasize 
the more conservative \Planck\ TT+lowP bounds in Table~\ref{tab:how_much_ic}. 
The constraints on the six base-$\Lambda$CDM cosmological model parameters remain stable when correlated 
isocurvature modes are allowed. The largest shifts occur for the neutrino density mode, but these shifts
are not significant (i.e., $<1\,\sigma$). The constraints on the 
tensor-to-scalar ratio also remain stable when isocurvature modes are allowed.

Finally we examined the connection between inflation and statistical isotropy, a key 
prediction of the simplest inflationary models. We tested separately the two lowest moments of an 
anisotropic modulation of the primordial curvature power spectrum. We found that a modulated curvaton 
model proposed to explain the observed large-scale dipolar power asymmetry cannot account for 
all of the asymmetry, and hence is not preferred over statistically isotropic base $\Lambda$CDM. 
The full mission temperature data place the tightest constraints to date on a quadrupolar modulation of curvature 
perturbations.
